\documentstyle[amssymb,epsf,psfig]{thesis}

\begin{document}

\pagenumbering{roman}

%
%

\titlepage

\begin{center}

\vspace{2\baselineskip}

{\large 
	Cauchy-Characteristic Matching In General Relativity }
\end{center}

\vspace{5\baselineskip}

\begin{center}
				by \\
			B. Szil\'agyi \\
                     M.S., University of Timi\c{s}oara, Romania, 1994 
\end{center}

\vspace{6\baselineskip}

\begin{center}
		Submitted to the Graduate Faculty 			\\
	   of Arts and Sciences in partial fulfillment 			\\
	      of the requirements for the degree of 			\\
		      Doctor of Philosophy 
\end{center}

\vspace{6\baselineskip}

\begin{center}
		    University of Pittsburgh 				\\
			      2000			
\end{center}

\endtitlepage

\setcounter{page}{2}

%
%

\begin{center}
                {\bf Cauchy-Characteristic Matching In General Relativity}
\end{center}
\begin{center}
                              B. Szil\' agyi, Ph.D.             \\
                            University of Pittsburgh, 2000
\end{center}

The problem of self-gravitating, isolated systems is undoubtedly
an important and intriguing area of research in General Relativity.
However, due to the involved nature of Einstein's equations physicists
found themselves unable to fully explore such systems.

From Einstein's theory we know that
besides the  electromagnetic spectrum,
objects like quasars, active galactic nuclei,
pulsars and black holes
also generate a physical signal of purely gravitational nature. 
Now scientists involved in the Laser Interferometric Gravitational
Observatory (LIGO) project are feverishly trying to build an instrument
that will detect it.

While the theory of gravitational radiation has been developed
using sophisticated mathematical techniques, the actual form
of the signal from a given source is impossible to determine
analytically. The  need  to investigate this
and related problems has led to the creation of the field  of
numerical relativity.

Immediately two major approaches emerged.
The first one formulates the gravitational radiation problem
as a standard Cauchy initial value problem. 
The advantage of this approach is that it is able to handle regions
of space-time where strong fields are present and caustics in the
wavefronts are likely to form. But it 
suffers from some inherent
disadvantages when it comes to  the
prediction of  gravitational radiation waveforms.

Another approach is the Characteristic Initial value problem.
This method is unable
to treat regions  of space-time where caustics form but it is uniquely
suited to study radiation problems because it describes
space-time in terms of radiation wavefronts.

The fact that the advantages and disadvantages of these two systems
are complementary suggests that one may want to use the two of them
together. In a full nonlinear problem it would be  advantageous to evolve
the inner (strong field) region using Cauchy evolution and the outer
(radiation) region with the Characteristic approach.
Cauchy Characteristic Matching enables one to evolve the whole
space-time matching the boundaries of Cauchy and Characteristic
evolution.
The methodology of Cauchy Characteristic Matching
 has been successful in numerical evolution
of the spherically symmetric Klein-Gordon-Einstein field equations as well as
for 3-D non-linear wave equations. In this thesis
the same methodology is studied in the context of the Einstein equations.

\begin{center}
\bf
		ACKNOWLEDGMENTS
\end{center}

First of all, I want to thank my advisor, Jeffrey Winicour.
His wisdom and patience made this work possible. Without
his continuous input I would not have been able to accomplish the work
being done  the last four years.  In him I found
not only an immense source of knowledge but also a source 
of guidance and encouragement so crucial in scientific research.

I would also like to thank Roberto G\'omez for the many things
I have learned from him in physics, 
computers, and many other things
while working together.  

During these years the Relativity Group of the University of Pittsburgh
made the Department of Physics a warm environment of work. 

I would like to thank Sascha Husa for important discussions on a variety of subjects.
Also, I would like to thank Yosef Zlochower
for his suggestions to improve the manuscript.

Finally, I want to express how much the unconditional 
support of my wife, Anita,
meant to me during these years. She and our children, \'Abrah\'am and Abig\'el,
are a refreshing environment. My parents, Imre and
M\'arta, have always thought of me with warm hearts, waiting for
the return of their son who wandered to such a remote school.
Without the input of these people I would have been a very depressed
and ineffective person.

This work has been supported in part by the Andrew Mellon Predoctoral
Fellowship, by the NSF PHY 9510895 and NSF PHY 9800731
to the University of Pittsburgh.
Computer time for this project has been primarily 
provided by the Pittsburgh Supercomputing Center under grant PHY860023P.  
Additional runs were performed at the National Center for Supercomputing
Applications. I thank the Albert-Einstein-Institut
for hospitality.

\vspace{2cm}

\noindent{\em To Anita, \'Abrah\'am and Abig\'el}

\tableofcontents
\listoffigures

\newpage
\pagenumbering{arabic}


\chapter{INTRODUCTION}

\section{The investigation of general relativistic astrophysical systems}

Self gravitating isolated systems represent an interesting and
challenging area of physics. The Einstein equations describe
how the presence of massive compact objects like quasars or black-holes
change the geometry of space-time. On the other hand, 
the exterior of these objects is free of matter and is 
affected to a much lesser extent by the interior
dynamics of these systems. 

These objects are too remote to allow for direct observation of
their structure. Instead, physicists and astronomers can build 
various   antennae designed to detect the electromagnetic
and the gravitational signals emitted by these isolated systems.
One of these antennae is LIGO, the Laser Interferometer 
Gravitational-Wave Observatory, built by scientists at Caltech and MIT, 
with funding from the National Science Foundation. 
LIGO consists of a pair of high-precision laser interferometers. 
The laser beams
are contained in an  L-shaped vacuum installation, with 4 km long arms.
Each arm contains mirrors that cause the laser beam to 
bounce back and forth between
them hundreds of times before it interferes with the beam from the other arm.  
As gravitational radiation passes, the structure of space-time between the mirrors
changes. This change amounts to a change in the distance between
the mirrors.
Thus the propagation of the laser-beam between these mirrors is perturbed,
causing a change in the interference pattern.  The change in distance
between these mirrors, however, could be as little as one part in 1000 of the size
of a proton. This illustrates the difficulty physicists face when 
trying to detect gravitational radiation by direct methods. 

The existence of gravity-waves was hypothetical until J. H. Taylor and R. A. Hulse
detected them indirectly in 1974. Using the radio telescope at Arecibo, Puerto Rico, 
they made a series of precise measurements of radio pulses emitted by two neutron
stars orbiting around one another.  They found that the orbital of  spin these 
objects was increasing. They also showed that the rate of this speed-up confirmed
the predictions of general relativity for massive stars shedding orbital energy in 
the form of gravity waves. For their discovery, 
the two scientists were awarded the 1993 Nobel Prize in physics.

A correct interpretation of the gravitational signals requires
understanding  how they  propagate and interact with the antenna.
One also needs to know how these signals are generated.
The strongly nonlinear character of the Einstein equations
prohibits us from solving them 
analytically except in the weak field limit and for some
highly symmetrical configurations. Thus approximation techniques
are used.

One of the most promising ways to tackle the field equations
of General Relativity is the simulation of curved space-time 
via a finite-difference approximation.
This proceeds as follows:

\begin{itemize}

\item Choose a coordinate system suitable to describe the
      physical system. 

\item Define  an initial surface with respect to the coordinate-choice.
The initial surface  represents a 3-dimensional subset of the
4-dimensional space-time to be simulated.
The fourth dimension
defines the direction in which
the numerical simulation proceeds to evolve space-time.

\item Represent the three-dimensional initial surface 
 by a discrete grid-structure.
Evolution in the direction of the fourth coordinate is done via
small, discrete ``time-steps''.  

\item Represent the field equations by a finite-difference evolution 
algorithm. When doing
so, the  partial differential operator
$\partial$ is approximated  by finite a  difference operator $D_\Delta$. The finite
difference approximation is consistent if $\lim_{\Delta\rightarrow 0} D_\Delta f = \partial f$.

\end{itemize}

Although the methodology might sound straight-forward, the real life
situation is much more involved. In general relativity 
there are no preferred coordinate
systems. The coordinates must be adapted to the
system. In the case of  the propagation of
gravitational radiation a natural coordinate system is defined by
the characteristics of space-time -- the curves along which 
disturbances propagate. 
 The characteristic approach
has been successful in numerical simulations of highly dynamical 
space-times. 
Furthermore, with use of radial compactification techniques,
the evolved space-time region can be extended to infinity where the antennae are 
effectively located.
However,  Einstein's equations imply that in
certain circumstances the light rays (characteristics)
refocus. Thus a coordinate system that is based on characteristics 
may become singular.

Another approach is based upon the Cauchy problem for general relativity. 
It provides initial data on an arbitrary space-like surface and
then evolves space-time as a function of an arbitrarily defined
time-coordinate.  The  Cauchy approach avoids the focusing problem  
but it has its own drawbacks. One is the
choice of the right coordinates, where ``right'' means ``one that works,'' i.e. does not encounter singularities.
Another  is that the Cauchy problem suffers from structural
disadvantages when evolving gravitational radiation.
Since the observer is practically at  infinity with respect
to the gravitational source, the Cauchy grid must 
be extended to infinity as well. This is not possible
in any finite  computational tool, because compactification 
cannot resolve the waveform of
the radiation. Since radiation propagates along light-like characteristics,
grid-points distributed in a spatial direction
are not efficient.

Another major problem in numerical relativity is the choice of 
evolution equations. The Einstein system, although written 
in a beautifully compact form, cannot be used in their original 
form for numerical simulation. Instead, one has to use 
combinations of various components of the Einstein tensor 
as numerical evolution equations.  Ideally one wants to 
recast the original Einstein system in a first-order 
symmetric-hyperbolic form since those systems have well 
understood mathematical properties 
which can be 
well-handled in a numerical context.  However
these first order forms become quite complicated, with a 
large number of auxiliary variables and  equations. Even 
today's supercomputers are slow when simulating space-time via 
those systems.  Thus 
it is preferable to  trade a first-order 
symmetric-hyperbolic system  for one with less
variables, but with mathematical properties which are harder to understand.
The evolution equations for the characteristic initial value 
problem, as first  provided by Bondi \cite{Bondi62}, form a set 
of equations well suited for numerical simulation. 
From the many available Cauchy systems we have adopted the 
Arnowitt-Deser-Misner (ADM) system \cite{Arnowitt62,York79}
which is based upon the  Hamiltonian formulation of general 
relativity.

\section{Cauchy-characteristic matching in general relativity}

Given the complementary strengths and weaknesses of the
characteristic and Cauchy formulations, the strategy
we pursue in this thesis is that of Cauchy-Characteristic Matching,
where the strong-field region is described by Cauchy evolution, 
the weak-field region is described by Characteristic evolution
and the interface between the two domains is handled by 
Cauchy-Characteristic Matching (CCM).

CCM has been successfully applied to the 
problem of nonlinear scalar waves propagating
in a 3-D Euclidean space \cite{Bishop97b},
 CCM has been used to evolve
 the spherical collapse of a self-gravitating scalar field onto
a black hole \cite{Gomez97b}.

In order to be able to build a routine that
interfaces the boundary of two evolution systems,
one needs to make sure that the the finite-difference
representation of the field-equations is numerically stable
in both domains. Furthermore, one needs to assure that 
these numeric field-equations are able to handle the matching boundary
conditions properly.  
These conditions are not satisfied if physically
spurious exponential modes are generated.
The characteristic
code is a robustly stable
evolution algorithm, which is able to handle 
boundaries. However, the ADM system needs
an appropriate treatment of the boundary in order to avoid 
spurious exponentially growing modes. At the boundary
one must prescribe 
the radiation degrees  of freedom  and then use an appropriate set
of boundary equations that determine the remaining components of the
metric tensor.

Furthermore, since the coordinates of the Cauchy system are
arbitrary while the coordinates of the characteristic system are 
based on the  light-cone structure of space-time, 
one needs to perform a non-trivial coordinate transformation
when matching the characteristic and Cauchy evolution equations.

At the end of this thesis
we lay out a problem for future research. Specifically, 
although we have identified a stable way of applying boundary 
conditions with the linearized ADM equations in Cartesian coordinates
for the case of a boundary aligned to the spatial grid-structure, 
application of a spherical boundary (a prerequisite of the CCM)
for a Cartesian ADM code is non-trivial.

\section{Preview}

The contents of this thesis is outlined as follows:

Chapter \ref{chap:intro} gives a brief description of the
Cauchy and Characteristic formulations of the equations of General Relativity.
At the end of the chapter
the concept of Cauchy Characteristic Matching (CCM) is briefly presented.
Next, in Chapter \ref{chap:nullcode}, the Pitt Null Code is described. 
First the underlying physics is presented, i.e. the 
characteristic slicing, the spin-weighted metric functions and the
equations describing the evolution of space-time. The end of the chapter
shows how one can use characteristic evolution to numerically evolve
black-hole space-times.

Chapter \ref{chap:ccm} defines the concept of Cauchy Characteristic Matching
first for a spherically symmetric scalar wave, next for a 3-D scalar field. Then 
the same concept is outlined for the case of general relativity. To make the understanding
of the details easier, first a geometrical description is given. 
Chapters \ref{chap:extraction} and \ref{chap:injection} describe the two modules
of CCM (extraction and injection) 
in detail. Along with the presentation of the matching modules, calibration tests are provided 
to show proper second-order accuracy for a number of test-beds.

Next, in  Chapter \ref{chap:ccmtest}, we study the stability  properties of CCM. 
As it is shown, the numerical noise of the individual modules do not excite any 
short-time instabilities. However, in order to analyze the
 long-term stability properties of matching, one needs to assure that both
the Cauchy and the characteristic evolution codes are able to deal with the 
discretization error that is inherent to numerical boundary algorithms. 
This issue is addressed in Chapter \ref{chap:ladm-cart}. 
As it is  shown, the characteristic code
is able to deal with constraint violating boundary modes of high frequency
without signs of numerical instabilities. However, as  shown in the same
chapter, the Cauchy code using the ADM equations is numerically unstable unless
the boundary conditions are treated properly. 
A major contribution of this thesis is that,
in the context of linearized
gravitational theory, Cartesian coordinates, 
 we have elucidated the appropriate boundary conditions on the faces of a cube
for the coupled set of partial-differential equations that form the principal part 
of the ADM equations. In particular, one should not specify boundary values
for six metric components but provide boundary data for 
the two radiation degrees of freedom and use a set of boundary
constraints to determine the remaining four components of
the spatial metric tensor.

The   stability of the  injection
module requires a spherical  boundary condition
for the Cartesian Cauchy code. As discussed in Chapter \ref{chap:ladm-sph}, 
for the ADM system
this implies use of boundary constraints
in spherical coordinates. The question of spherical
boundary constraints applied to a Cartesian grid
is a complicated  problem.

\section{New developments in this thesis}

The work presented in this thesis is the result of the work of a number of 
people. It is not easy to separate the individual contributions.
Jeffrey Winicour had
a most significant role in this work. 
Besides him Joe Welling, Roberto G\' omez, 
Philip Papadopoulos, Luis Lehner, and Nigel Bishop had important contributions as well.
The formalism and the code used in Cauchy-Characteristic Matching (CCM)
 is based on the work of these people.

The years that I was present in the Pittsburgh numerical relativity group
were marked by a close collaboration with Roberto G\'omez. 
During our collaboration, Roberto has taught me
much about the physical background of our work, 
scientific programming as well as 
code-optimizing on various supercomputer platforms.

When I started to work in the group 
most of the code developing of the extraction module was already done.
My contribution to the extraction was the calibration of the module. 
This work included the development of 
Maple algebra scripts that provided the analytic results
needed for convergence tests. 

The injection module in its current form was 
was developed by Roberto and myself. 
To assure the proper convergence rate for the 
various interpolation routines involved
as well as for the whole module,
I used a number of test-beds.

The stability runs of CCM had been the subject of our work for years.
Although  Roberto  had a significant input to 
the question of stability of CCM, 
the experiments described in Chapter \ref{chap:ccmtest} were
designed and performed by myself.

Another area of collaboration was the numerical 
study described in
Chapter \ref{chap:ladm-cart}. Here Roberto's part was 
the development of the linearized Cauchy evolution codes
for the various evolution algorithms. My part was the
analysis, design and implementation of the boundary routines as
well as the test-runs that concluded 
in a stable boundary-algorithm
for the linearized ADM system.
These are the first robustly stable Cauchy
evolutions of a bounded gravitational system.
This is my major contribution to the work described in this thesis.

I also made some contributions to the development of the formalism used
in the study of axisymmetric event horizons and of the characteristic code
itself.


\chapter{An Overview Of Numerical Relativity}
\label{chap:intro}

\section{Introduction}
\label{sec:intro.intro}

Since Einstein first wrote down the equations of general relativity (GR)
in 1916, only a limited class of analytical solutions have been found.  
The first 
solution, the space-time metric of a ``point charge'' was written
down not by Einstein but by Schwarzschild about one year after the field
equations were defined. 

As always, when exact analytic approaches become too cumbersome,
approximations are used to further investigate the space 
of solutions. Approximations on an analytic level give perturbative
solutions around known exact solutions. On the numeric level one can
approximate the space-time continuum by a discrete set of points, 
write down the field equations as a set of finite-difference
equations and use computer power to evolve regions of space-time.

Perhaps the first crisp definition of numerical relativity was given by
Charles Misner in 1957 \cite{Misner57}:

\begin{quote}
First we assume that have a computing machine better than anything we 
have now, and many programmers and a lot of money,
and you want to look at a nice pretty solution of the Einstein equations.
The computer wants to know from you what are the values of $g_{\mu\nu}$
and $\partial_t g_{\mu\nu}$ at some initial surface. Mme. Foures [Y. Choquet-Bruhat]
has told us that to get these initial condition you must specify something
else and hand over that problem, the problem of the initial values,
to a smaller computer first, before you start on what Lichnerowicz called 
the evolutionary problem. The small computer would prepare the initial
conditions for the big one.  Then the theory, while not guaranteeing 
solutions for the whole future, says that it will be some
finite time before anything blows up.
\end{quote}
Still, more than forty years after this statement, 
when one wants to build a general relativistic code, several problems arise:  

\begin{itemize}

\item First of all one has to determine
whether the field equations are well 
\label{page:wellposedness} posed 
or not, i.e. Does 
it make sense to treat the system as a set of coupled evolution equations
with some suitable initial data,
with boundary-values to be 
provided at the edge of the evolution zone?
Or does the system have attributes similar to 
 an elliptic equation where 
``evolution'' as a function of ``time'' makes no mathematical sense?

\item Next  one has to rewrite
the original field equations in a form that is suitable for numerical 
evolution. This question is related to the previous one in the
sense that if one recasts the equations as a first-order 
symmetric-hyperbolic system of equations, then 
the numerical implementation becomes
a lengthy but straightforward problem.

\item Another related problem is that, despite their
exponential growth, computer resources are still a limiting factor and so one
might want to trade a system with a large number of variables
and clear mathematical properties to one with six or maybe twelve variables,
but with mathematical properties yet to be understood.

\item A fourth issue that appears when doing 
finite-differencing is the question
of numerical instabilities. This problem, if not studied carefully for
the adopted discretization scheme, might give rise 
to non-physical exponential
modes 
which eventually make the physical content 
of the numerical simulation worthless. As it will be seen,
the stability properties of a code depend not only on the evolution scheme
being used but also on the treatment of the boundary.

\end{itemize}

Despite the difficulties that arise there have been great successes
attached to the domain of numerical relativity
which motivate further search for development.
One  discovery, which had not 
been anticipated purely by analytic approaches, is the critical phenomena
 found when simulating spherically symmetric
gravitational collapse. 
On the analytic side, Christodoulou made a penetrating study
of the existence and
uniqueness of solutions to the characteristic initial value problem
\cite{Christodoulou86a,Christodoulou86b,Christodoulou87a,Christodoulou87b}.
He showed that weak initial data
evolve to Minkowski space asymptotically in time, 
but that sufficiently strong data form
a horizon.\footnote{For a definition of the ``horizon'' of a black hole see
Section~\ref{sec:null.1BH}.} In the latter 
case, he showed that the geometry is
asymptotically Schwarzschild in the approach 
to ${\cal I}^+$ (future time infinity) from outside
the horizon.  What this
analytic tour-de-force did not reveal was the 
remarkable critical behavior in the
transition between these two regimes, which was 
discovered by Choptuik \cite{Choptuik92b, Choptuik93} using
computational simulation. 

Another major result of numerical relativity was 
the numerical simulations of axisymmetric space-times
that enabled evolution of dynamical black holes 
\cite{Shapiro92a,Anninos93b,Hughes94a,Abrahams94a,Anninos95a,Libson94a}.
Grid-sucking (gridpoints falling into black holes), the importance of 
the right choice of time slicing, 
numerical instabilities, and the question of 
boundaries are just some of the
issues that demanded attention.

Besides  these (and many other) 
numeric results of mainly theoretic significance there is another reason why
numerical relativity can have  a significant role in our understanding of
the fundamental laws of our universe. 
There is a good hope that through the building and use of 
laser interferometers (LIGO, VIRGO, etc.) gravitational radiation will
be detected.
This data needs to be confirmed and interpreted, at least in part,
 by numerical simulations. The events most likely to be detected
are  rotating black holes inspiraling into
each other  --  a problem that requires full 3-dimensional treatment.  
There are post-Newtonian 
approaches \cite{Damour99,Damour00} on the analytic side
that provide  approximate information for some 
phases of the inspiral, but the full picture is unlikely to be completed
without the use of accurate 3-D numerical simulation.

The question of evolution equations also demands attention.
If one counts the number of equations or 
field components, at first sight it seems that there are ten components
of the symmetric, $4\times 4$ Einstein tensor to be solved, a number that fits
perfectly the ten ``independent'' components of the metric tensor. The obvious
solution, then, would be that one should do numerical simulations solving for
all components of the Einstein equations. This approach, however, does not
take into account the different quantities involved in the Einstein
system. As it will be seen in subsequent sections, certain degrees of freedom
in the space-time metric correspond to gauge choices (choice of coordinates)
which have neither physical meaning  nor associated evolution 
equations. 
The equations governing these gauge degrees of freedom are constraints, which
if  satisfied in some  3-D subspace of the space-time manifold, will
be satisfied throughout the whole space-time, provided the rest of the Einstein
equations are satisfied. 

Thus the issue of coordinates and  the related question of separating the Einstein
equations into evolution equations and constraints are the first questions that will 
be addressed in the following sections.

\section{Coordinates}
\label{sec:intro.coord}

In general relativity the arc-length between two points 
in space-time  can be computed using the 4-metric. In the case
of two infinitesimally close points $A$ and $B$ the four-distance $ds^2$ can
be computed using \begin{equation}
ds^2 = g_{\mu \nu} dx^\mu dx^\nu
\label{eq:intro.ds2}
\end{equation}
where $g_{\mu \nu}$ is the $4\times 4$ symmetric metric tensor and
$dx^\mu = x^\mu(A)-x^\mu(B)$. (Here and throughout this thesis Greek indices
run from $0$ to $3$, while  lowercase Latin indices correspond to 
spatial dimensions and they run from $1$ to $3$.)
The distance between these two points is called {\em spacelike, null,} 
or {\em  timelike} if $ds^2$ is positive, zero, or negative. This definition can
be extended to coordinates as well: keeping three of the four 
coordinates fixed and
varying one  leads to 
a positive, zero or negative value of $ds^2$, defining
spacelike, null, and timelike coordinates. 
Vectors fall in similar categories
with respect to the sign of their length-square
\begin{equation}
v^\nu v_\nu = g_{\mu\nu} v^\nu v^\mu.
\end{equation}

The lack of physical meaning 
associated to the coordinates implies that there are  no 
restrictions in their choice  --  a freedom
that can be helpful or, at times, confusing. 
The simplest form of the metric tensor 
$g_{\mu \nu}$ at a given point corresponds to 
a coordinate system where $g_{\mu \nu}$ is diagonal,  
but this cannot, in general, be done locally in the neighborhood
of a point. In other words, at each point in 
space-time there is a Minkowski frame (in which space-time is described
by the Minkowski metric) but, except for the trivial cases, there is no
analogue of inertial coordinates in the neighborhood of the point.
 However, the number of positive and negative 
elements of the diagonalized metric is an invariant global property of the
space-time manifold. This  defines the {\em signature}
of the metric which in our convention is $[-,+,+,+]$. 

\section{The Cauchy formulation}

\subsection{The ``3+1'' slicing of space-time}

In the numerical simulation of a time-dependent system, one evolves some set
of physical quantities as a function of time, i.e. 
\begin{itemize}
\item provide some initial (Cauchy) data at  $t=0$
\item use the evolution equations to determine the fields at any 
point of the evolution domain
at a later time $t$.
\end{itemize}

One of the problems of setting up the equivalent 
problem for the Einstein equations
is that the notion of {\it time}  is coordinate-dependent
and has no unique
physical meaning. One chooses  it in some suitable 
manner and then sets up the Cauchy problem.
For this reason
we do the following \cite{Wald84}:

\begin{itemize}
\item pick a function $t$ and a vector field $t^\mu$ such that the surfaces, 
$\Sigma_t$, of constant $t$ are spacelike Cauchy surfaces satisfying 
$t^\mu \nabla_\mu t = 1$;
\item choose evolution equations based upon the choice of $(t, \Sigma_t)$;
\item provide initial data on $\Sigma_0$;
\item compute the metric fields at later times $t > 0$
via the evolution equations.
\end{itemize} 

\label{page:intro.timedef}

\begin{figure}
\centerline{\epsfxsize=4in\epsfbox{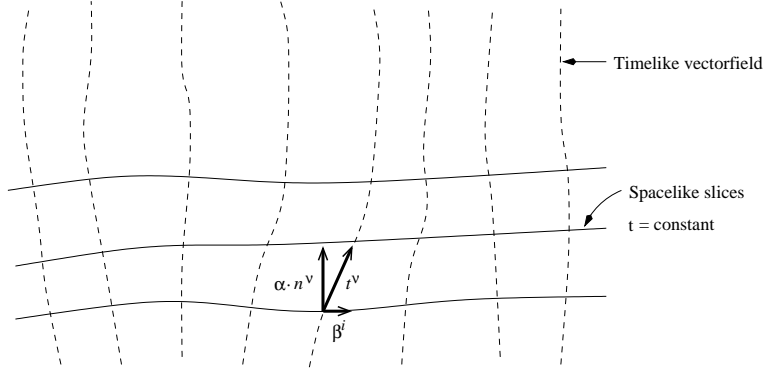}}
\caption{Illustration of the ``3+1'' slicing.}
\label{fig:intro.cauchy-slicing}
\end{figure}

Figure~\ref{fig:intro.cauchy-slicing} illustrates the resulting 
``3+1'' slicing approach. 

In the coordinate system $\{ x^\mu\}$, with $x^0 = t$,
the Einstein and Ricci tensors $G_{\mu \nu}$ and $R_{\mu \nu}$ are given by
\begin{eqnarray}
G_{\mu \nu} &=& R_{\mu \nu} - \frac 1 2 g_{\mu \nu} R 
\label{eq:intro.Einstein}\\
R_{\mu \nu} &=& - \frac 1 2 g^{\alpha \beta} \left\{ 
-2 \partial_\beta \partial_{(\nu} g_{\mu ) \alpha} 
+ \partial_\alpha \partial_\beta g_{\mu \nu} 
+ \partial_\mu \partial_\nu g_{\alpha \beta} 
\right\} 
\nonumber \\ &&
+ F_{\mu \nu}(g, \partial g),
\label{eq:intro.Ricci}
\end{eqnarray}
where $F_{\mu \nu}(g, \partial g)$ is a non-linear term in the metric and
its first derivatives.
As it can be seen from Eqs.~(\ref{eq:intro.Einstein}) - (\ref{eq:intro.Ricci}), 
the field
equations involve  up to second  time derivatives of the metric. 
For this reason initial data consists
of  specifying the metric $g_{\mu \nu}$ and its time-derivative 
$\partial_t g_{\mu \nu}$ on $\Sigma_0$.

Let $n^\nu$ be the unit normal to the $\Sigma_t$ surfaces.
The equations
\begin{equation}
G_{\mu \nu} n^\nu = 0
\end{equation}
contain no second time derivatives of any of the metric components, i.e. they 
are fully specified by the initial data. These four equations 
(or any combination
of them) can be viewed as initial value constraints. 
As a consequence of the Bianchi identity 
\begin{equation}
\nabla^\mu G_{\mu \nu} = 0,
\end{equation}
if the constraints are initially satisfied and the 
spatial components of the Einstein
tensor vanish everywhere, then the constraints are also globally satisfied.
\label{page:intro.CauchyConstraints}

Since four of the ten components of the metric tensor can be fixed
by the choice of gauge, it is a perfect match that the six remaining components
can be evolved by requiring that the six spatial components of the Einstein tensor
(or some combination of those with the constraints) be satisfied.

As already stated on page~\pageref{page:wellposedness}, 
when trying to ``evolve'' a system of equations one might
face problems of well-posedness. 
It is a classic result of Choquet-Bruhat \cite{Choquet62}
that the ``3+1'' splitting described above gives rise 
to a well-posed initial value problem 
in harmonic coordinates (i.e. in coordinates $x^\mu$ 
that satisfy the wave equation
$\nabla_\nu \nabla^\nu x^\mu = 0$).

\label{sec:lapse&shift}
Given a $t = $ constant surface $\Sigma_t$, 
its normal vector $n^a$, and the 4-metric 
$g_{\mu\nu}$, the 3-metric ${}^{(3)}g_{\mu\nu}$   
of the surface $\Sigma_t$ is given by
\begin{equation}
{}^{(3)}g_{\mu\nu} = g_{\mu\nu} + n_\mu n_\nu.
\end{equation}

The evolution proceeds in the $t$-direction,
tangent to the ``evolution'' vector $t^\alpha$ satisfying 
$t^\alpha \partial_\alpha = \partial_t$.

The lapse $\alpha$ and the shift vector $\beta^\mu$
 are defined by
\begin{eqnarray}
t^\mu = \alpha\, n^\mu + \beta^\mu,
\end{eqnarray}
i.e. the lapse is equal to the projection of 
 $t^\mu$ onto the unit normal $n^\mu$  to $\Sigma_t$, while
the shift is the projection of $t^\mu$ 
onto the Cauchy surface $\Sigma_t$.

Next we must choose the evolution equations. Given the ``3+1''
slicing,  one natural choice would be  the spatial
components of the Einstein tensor. However, 
as  seen from numerical experiments presented in Chapter
\ref{chap:ladm-cart}, these Einstein tensor components do not provide
a set of evolution equations suitable for numerical evolution.
A more suitable choice, referred
to as ``Standard ADM'' system,  comes from a Hamiltonian
formulation of the theory \cite{Arnowitt62}. 

\subsection{The ``Standard ADM'' system}
\label{sec:intro.ADM}

The ADM evolution equations are a  first-order in time, second-order
in space system, defined by
\begin{eqnarray}
\partial_t \; {}^{(3)}g_{ij} &-& {\cal L}_\beta \; {}^{(3)}g_{ij} 
= -2\alpha  \; {}^{(3)}K_{ij}, \label{eq:intro.adm.geq} \\
\partial_t \; {}^{(3)}K_{ij} &-& {\cal L}_\beta \; {}^{(3)}K_{ij} = - D_i D_j \alpha
\nonumber \\ 
&+& \alpha\left (\; {}^{(3)}R_{ij} + \; {}^{(3)}K \; {}^{(3)}K_{ij} 
- 2 \; {}^{(3)}K^l_i \; {}^{(3)}K_{lj} \right).\label{eq:intro.adm.keq} 
\end{eqnarray}

The symbol ${\cal L}_v$ stands for Lie-derivative along a vector field $v^\rho$. 
For a tensor field $T_{\mu\nu}$ it can
be computed using
\begin{eqnarray}
{\cal L}_v T_{\mu\nu} = v^\rho \partial_\rho T_{\mu\nu} 
+ T_{\rho\nu} \partial_\mu v^\rho +
T_{\mu\rho} \partial_\nu v^\rho. \label{eq:Lie-def} 
\end{eqnarray}
The covariant derivative $D$ is defined as the three-dimensional derivative operator
with respect to ${}^{(3)}g_{ij}$.

Equation~(\ref{eq:intro.adm.geq}) defines the extrinsic curvature of the Cauchy slice $\Sigma_t$
according to 
\begin{equation}
{}^{(3)}K_{ij} = \frac 12 {\cal L}_n\, {}^{(3)}g_{ij}.
\end{equation}
The Hamiltonian and the momentum constraints are defined by
\begin{eqnarray}
{\cal C} &=& \; {}^{(3)}R-\; {}^{(3)}K_{ij} \; {}^{(3)}K^{ij} + \; {}^{(3)}K^2, \\
{\cal C}^i &=& D_j \left( \; {}^{(3)}K^{ij} - \; {}^{(3)}g^{ij} \; {}^{(3)}K\right).
\end{eqnarray}

Although the evolution equations (\ref{eq:intro.adm.geq}) - (\ref{eq:intro.adm.keq})
appear to be compact, their explicit forms are  lengthy expressions. 
The number of variables (six metric components ${}^{(3)}g_{ij}$ 
and six extrinsic curvature components
${}^{(3)}K_{ij}$)
is appealing for numerical simulations but 
the system cannot be put into a first-order symmetric-hyperbolic form. Thus
its mathematical properties are difficult to understand.
Chapter~\ref{chap:ladm-cart} contains a study of the linearized
ADM system from a numerical point of view.

\section{The characteristic formulation}

The gauge freedom inherent in the ``3+1'' formalism
 can be extremely useful  or, in some cases, confusing. 
The spacelike surfaces can be chosen to facilitate the description
of general relativistic objects such as black holes, etc.

As an alternative, the {\em
Characteristic Formulation} uses a foliation of space-time
by null hypersurfaces which facilitates the description of waves.
This section gives a brief description
of the choice of the coordinates and the form of the metric. More
detailed discussion can be found in \cite{Bondi62, Sachs62a, Winicour98}
as well as in Chapter \ref{chap:nullcode}.

\subsection{The null foliation of space-time}
\label{sec:intro.nullcoord}

The construction proceeds as follows: 
\begin{itemize}

\item Choose a spacelike, closed, convex 2-surface.

\item Choose some angular coordinates, say $(\theta, \phi)$,
that uniquely describe the points of the 2-surface.

\item Transport the 2-surface along a timelike vector field
to obtain a 3-surface referred to as the  {\em timelike world-tube} $\Gamma$.

\item A ``2+1'' slicing of the world-tube (similar to the ``3+1'' method) 
provides a  time-coordinate $u$ on $\Gamma$.

So far the parameters $(u, \theta, \phi)$ 
describing the 3-surface $\Gamma$ have been established.

\item Unless caustics or other topological ``defects'' are present,
the outgoing null-rays emanating from the world-tube  
associates a unique point
to any space-time point in the exterior.
This provides a natural extension 
 of the coordinates $(u, \theta, \phi)$ to the exterior space-time.

\item Finally, choose  a radial coordinate by requiring
 that the metric of surfaces  
$u=$ constant, $r = $ constant have form $r^2 h_{AB} dx^A dx^B$,
with $\det (h_{AB}) = \det (q_{AB})$ where $q_{AB}$ is the unit-sphere 
metric.\footnote{Here and in subsequent expressions of this work 
capital Latin letters $A, B, C, \ldots$ are used to indicate angular
coordinates.}  The coordinate $r$ is sometimes referred to as the
surface-area coordinate.

\end{itemize}

Much of the gauge freedom of general relativity is
fixed in the characteristic formulation by geometric choices. 
In fact the only
gauge freedom present in this formulation is the choice and  
parameterization of the world-tube, 
i.e. the coordinates $(u, \theta, \phi)$. The only drawback of the characteristic
approach is that
the formalism is difficult to  
adapt to the description of space-time regions
with caustics.

\begin{figure}
\centerline{\epsfxsize=3.in\epsfbox{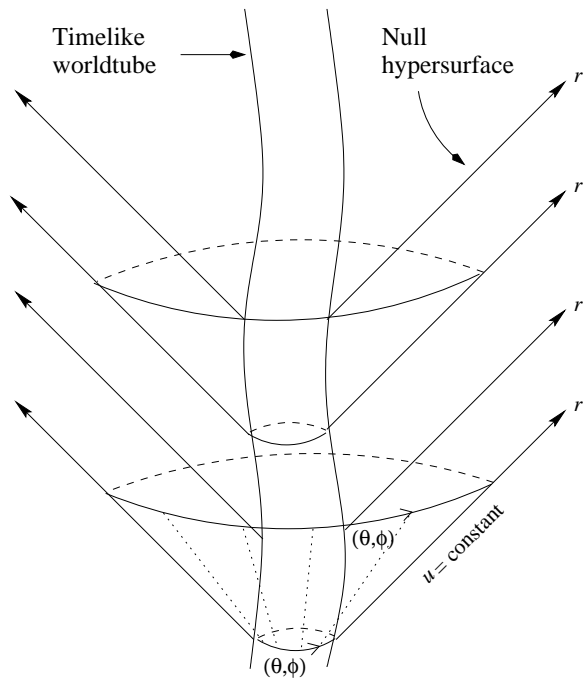}}
\caption{Null Foliation of space-time.}
\label{fig:intro.null-slicing}
\end{figure}

Figure~\ref{fig:intro.null-slicing} illustrates  the null foliation of space-time.

\pagebreak[3]
\subsection{The metric and its interpretation}

With coordinates described in Section~\ref{sec:intro.nullcoord}
the space-time metric takes the Bondi-Sachs form 
\cite{Bondi62, Sachs62a, Winicour98}
\pagebreak[3]
\begin{eqnarray}
ds^2 &=& - \left(e^{2 \beta} V/r - r^2 h_{AB} U^A U^B \right) du^2 
- 2 e^{2 \beta} du \, dr \nonumber \nopagebreak \\
&& - 2r^2 h_{AB} U^B du dx^A + r^2 h_{AB} dx^A dx^B.
\end{eqnarray}

Because the construction of the coordinates is based upon
 the lightcone structure of space-time, quantities built out of 
the metric components have geometrical meaning.
Six real field variables appear in the metric: $V, \beta, U^A$ and $h_{AB}$.
The symmetric 2-tensor  $h_{AB}$ represents the conformal geometry of the 
surfaces defined by $dr=du=0$. The requirement $\det (h_{AB}) = \det (q_{AB})$
reduces $h_{AB}$ to the two radiative degrees of freedom.
 The scalar field $\beta$  measures 
the {\em expansion} of the light rays as they propagate outward.  The function
$V$ is the analog of the Newtonian potential, and its asymptotic expansion contains
the mass aspect of the system. 

Further understanding of the  metric components 
(including the gauge functions $U^A$) stems from the  intrinsic 
metric of the $r = $ constant surfaces,
\begin{equation}
\gamma_{ij} dx^i dx^j = - e^{2 \beta} \frac V r du^2 + r^2 h_{AB} 
\left( dx^A - U^A du \right) \left( dx^B - U^B du\right).
\end{equation}

The ``2+1'' decomposition of the world-tube is
 used to define the coordinates $(u, \theta, \phi)$ and
identifies  $g_{AB} = r^2 h_{AB}$ as the metric of
the 2-surfaces of constant $u$ which foliate it.
The square of the lapse is  $e^{2 \beta} V/r$ while the 
2-dimensional shift vector 
on the world-tube is given by $(-U^A)$.

\subsection{Field equations}
\label{sec:intro.null.eqs}

The vacuum Einstein equations  $G_{\mu\nu} = 0$ divide into three
categories,
as pointed out by Bondi \cite{Bondi62}. 
The equations corresponding to the Ricci tensor components
 $R_{rr}, R_{rA}$ and $h^{AB} R_{AB}$
contain no derivatives with respect to $u$. These 
equations involve the metric  only 
within  $u=$ constant hypersurfaces, which  motivates calling them 
 {\em hypersurface equations}.
The  equations $R_{AB} - h_{AB} h^{CD} R_{CD}/2$
contain first derivatives with respect to $u$ as well as second radial and
angular derivatives. Their solution at a  space-time point involves use of the metric
components in a 4-dimensional neighborhood of that point.
They are  called  {\em  evolution equations} and
describe the propagation of
 gravitational radiation. 
They  correspond to the general-relativistic version of the scalar wave
equation in  retarded spherical coordinates 
$(u, r, \theta,\phi)$. 
The third category consists of  {\em conservation equations} 
which contain
no second radial-derivatives. These are the analogues, with respect to 
an $r$-foliation, of the traditional  constraints
in a Cauchy formalism (see page \pageref{page:intro.CauchyConstraints}).
Bondi used the Bianchi identities 
to show that one of them, the
conservation condition $R^r_r = 0$ is automatically fulfilled
by virtue of the other equations.  He further showed
that the remaining conservation equations
\begin{equation}
R^r_u = 0, \;\;\; R^r_A = 0
\end{equation}
are satisfied on a complete outgoing null cone if they hold 
on a single spherical cross-section.
Explicit expressions for the hypersurface equations the evolution
equation and the conservation laws can be found in \cite{Lehner98}.

Chapter \ref{chap:nullcode} is dedicated to the description and
 applications of the Pitt Null Code, a numerical implementation
of the characteristic formalism.

\section{Alternative  formulations}

In recent years various systems of hyperbolic 
equations deduced form Einstein's equations
have been proposed by 
Abrahams, Anderson, Choquet-Bruhat and York 
\cite{Abrahams95a, Abrahams97b},
Bona and Masso \cite{Bona92}, 
Choquet-Bruhat and Ruggeri 
\cite{Choquet83},
Friedrich \cite{Friedrich85}, 
Frittelli and Reula \cite{Frittelli96}, 
Iriondo, Lequizamon and Reula \cite{Iriondo98}. 
Mixed elliptic and hyperbolic systems have also been formulated by 
Christodoulou and Klainerman \cite{Christodoulou93},
Rendall \cite{Rendall95},
Choquet-Bruhat \cite{Choquet96}, Shibata and Nakamura \cite{Shibata95}.
For the Baumgarte-Shapiro system (derived from the ADM equations)
see \cite{Baumgarte99}.
Other hyperbolic systems related to \cite{Baumgarte99}
have been discussed by
Alcubierre, Br\" ugmann, Miller and Suen 
\cite{Alcubierre99c}, 
Frittelli and Reula \cite{Frittelli99},
and Friedrich and Rendall \cite{Friedrich00}.
H\"{u}bner has implemented numerical evolution
based upon a hyperbolic formalism for global Cauchy evolution of compactified
space-time \cite{Friedrich81a,Friedrich81b,Huebner96,Huebner98}. 
(Please note that
the list is intended to be
informative rather than exhaustive.)

While on a mathematical level the properties of first-order hyperbolic
systems are much better understood than those of higher-order 
non-hy\-per\-bo\-lic
ones, there are considerably more variables and equations to be evolved.  
This issue might eventually be solved as 
available computational resources increase.

\section{Cauchy-Characteristic Matching (CCM)}
\label{sec:intro.ccm}

As portrayed in the previous sections, the Cauchy and the characteristic
approaches have  complementary strengths and weaknesses.
Cauchy evolution has promising ability to   evolve strong field regions 
of space-time but it is limited to  a finite spatial region, which
introduces an artificial outer boundary. Characteristic evolution, if used
with a compactified radial coordinate,  
allows the incorporation of future null infinity within 
a finite computational grid. However, in turn, it suffers from complications due
to gravitational fields causing focusing of the light rays.  
Unification of both
 methods seems to be a promising way 
of taking advantage of each  formalisms'  strengths. The  methodology 
called Cauchy-Characteristic Matching (CCM) utilizes Cauchy evolution 
within some prescribed timelike world-tube, but replaces the need for an outer 
boundary condition by matching to a characteristic evolution in the exterior
of this world-tube. This provides a global 
solution and allows calculation of the  gravitational
wave-form reaching future null infinity \cite{Winicour98, Bishop93,
Bishop96, Bishop98a}.

The methodology of CCM has been successful in numerical evolution
of the spherically symmetric Klein-Gordon-Einstein field equations as well as 
for 3-D non-linear wave equations \cite{Gomez96,Bishop97b}. 

A detailed description of the full 3-D CCM for the case of Einstein equations
will be provided  in chapters~\ref{chap:ccm}-\ref{chap:ccmtest}.

\chapter{The Characteristic Code}
\label{chap:nullcode}

This chapter provides
a description of a 3-D characteristic code known as
the Pitt Null Code. The code was developed over a period of fifteen years
by the Pitt numerical relativity group. The final version was completed by 
Luis Lehner, whose thesis  \cite{Lehner98} is
used as primary reference throughout most of this chapter.

\section{The physical algorithm}

\subsection{Coordinates}
\label{sec:null.coordinates}

The choice of coordinates is based on
the characteristic foliation of space-time as 
described in Section~\ref{sec:intro.nullcoord}.
Although some of the coordinate freedom is fixed by 
geometric considerations, the choice of the angular coordinates
on the world-tube is arbitrary. Furthermore, radial compactification is
employed to describe space-time out to future null infinity.

\subsubsection{Angular coordinates}

Since both the topology of the world-tube $\Gamma$ and the topology of  
future null infinity ${\cal I}^+$ are $S^2$, it is natural to adopt a spherical
coordinate system $(\theta, \phi)$. However, this does not provide
a smooth covering of the $S^2$. Our solution is  to
use  two stereographic
coordinate patches to cover the sphere, an approach that has
been successful in several applications \cite{Bishop97b,Gomez97c}. 
This removes the singularity
at the poles but, since the two patches
overlap, there is a problem of consistency between
the two patches. 

Let $\theta$ and $\phi$ label the points on the sphere in the usual way.
The two angular degrees of freedom are combined into a single complex
stereographic coordinate $\zeta = q + I p$, defined on the north patch by
\label{page:stereo-coord}

\begin{equation}
\zeta_N = \tan \left( \frac \theta 2 \right) e^{I \phi},
\end{equation}
with $I = \sqrt{-1}$. The coordinate $\zeta_N$ provides a smooth
description of the sphere everywhere except at the point $\theta=\pi$.
Analogously $\zeta_S = 1/\zeta_N$ provides a smooth coordinatization
except for the north pole. The patches described by $\zeta_N, \zeta_S$
are sufficient to cover the sphere, with some overlap region.

Now let $\Psi$ be a smooth scalar field on the sphere. The
consistency condition  that  $\Psi(\zeta, \bar \zeta)$
(and  its derivatives) be equal on the two patches is 
\begin{eqnarray}
\Psi_N \left[ \zeta_N, \bar \zeta_N \right] = 
\Psi_S \left[ \zeta_S(\zeta_N, \bar \zeta_N), \bar \zeta_S(\zeta_N, \bar
\zeta_N) \right], \label{eq:null.crosspatch}
\end{eqnarray}
in the overlap region.

\subsubsection{Radial compactification}

Asymptotically flat space-times can be given rigorous interpretation in the 
limit $r \rightarrow \infty$ along a null hypersurface (holding $u$ constant) 
with the use of compactification techniques
\cite{Penrose}. The surface area
coordinate $r$ is mapped into a 
finite range by the transformation \cite{Gomez94a}
\begin{equation}
x := \frac r {r+1}.
\label{eq:null.xdef}
\end{equation}
Space-time fields are globally defined in the interval $x_1 \le x \le 1$,
where $x_1$ denotes the  location of the inner radial
grid-boundary. Future null infinity (${\cal I}^+$) 
is described by $x = 1$. 

\subsection{Equations}

\subsubsection{The {\em eth} formalism}
\label{sec:null.eth}

\noindent
{\em Spin-weighted fields}
\vspace{0.5em}
\label{sec:null.spin-fields}

\noindent
With the coordinates defined as described in Section~\ref{sec:null.coordinates},
one could write out the Einstein equations in terms
of partial derivatives of the metric functions. 
However, the  equations can be written  more conveniently  using the 
 ({\em eth\/}) $\eth$-formalism 
\cite{Newman66, Penrose84, Goldberg67, Gomez97c}.

In this formalism tensor fields on the sphere are replaced by  
spin-weighted fields. The result is a compact and efficient manner of treating
tensors on a sphere as well as their derivatives.

The distance squared between two infinitesimally close points on a unit sphere
can be written as
\begin{equation}
ds^2 = q_{AB} dx^A dx^B,
\end{equation}
where the 2-metric $q_{AB}$ is defined by the choice of angular coordinates:
\begin{equation}
q_{AB} = \mbox{diag}\left ( 4/(1+\zeta \bar \zeta)^2, 4/(1+\zeta \bar \zeta)^2 \right).
\label{eq:null.qABdef}
\end{equation}
One can rewrite  $q_{AB}$ in terms of a complex basis vector
$q_A$ (dyad) as
\begin{equation}
q_{AB} = (q_A \bar q_B + \bar q_A q_B) / 2,
\end{equation}
where $q^A$ satisfied $q^A q_A = 0$ and $q^A \bar q_A = 2$.\footnote{
Here we depart from other conventions 
\cite{Penrose84} to avoid factors of $\sqrt{2}$.}
A possible choice of $q^A$ (on the two patches) is given by
\begin{eqnarray}
q^A_S &=& (1 + \zeta_S \bar \zeta_S) \cdot (\delta^A_2 + I \delta^A_3), \\
q^A_N &=& (1 + \zeta_N \bar \zeta_N) \cdot (\delta^A_2 + I \delta^A_3). 
\end{eqnarray}
In the overlap between the patches the two dyads $q^A_N$ and $q^A_S$
are related by $q^A_N = e^{I \alpha} q^A_S$, where 
$e^{I \alpha} = - \bar \zeta_S / \zeta_S.$

Having introduced a complex basis one can represent any vector field
on the sphere in the form  $v = q^A v_A$. In the overlap region 
the vector transformation law between the two basis $q_S^A$ and $q_N^A$
translates into $v_N = e^{I \alpha} v_S$.
Furthermore, any tensor field $v_{A_1 \ldots A_n}$ can be represented
by $2^{n-1}$ different complex scalar fields of the form
\begin{equation}
v = t^{A_1} \ldots t^{A_n} \; v_{A_1} \ldots v_{A_n},
\label{eq:eth.tensorexpand}
\end{equation}
where $t^A$ stands either for $q^A$ or $\bar q^A$.
Assuming that in Eq.~(\ref{eq:eth.tensorexpand}) the dyad $q^A$ occurs
$p$ times (and its complex conjugate, $\bar q^A$ occurs $n-p$ times), 
the transformation law of the complex scalar $v$ between the two patches
is given by $v_N = e^{I s \alpha} v_S$, where $s=2p-n$. The spin-weight
of the field $v$ is given by the integer $s$.

\vspace{1em}
\noindent
{\em Spin-weighted derivatives}
\vspace{0.5em}

\noindent
The technique used to express tensor fields in terms of spin-weighted
fields extends to covariant derivatives, since a covariant
derivative of a tensor is  another tensor. Given a spin $s$ quantity $F$
the {\em eth} operators $\eth$ and $\bar \eth$ are defined by \cite{Bishop97c}:
\begin{eqnarray}
\eth F &=& q^A D_A F = q^A \partial_A F + \zeta \, s \, F, \\
\bar \eth F &=& \bar q^A D_A F = \bar q^A \partial_A F - \bar \zeta \, s \, F.
\end{eqnarray}
The spin of $\eth F$ is $s+1$ while  $\bar \eth F$ has spin $s-1$.

\subsubsection{Spin-weighted equations}
\label{sec:null.spin-eqs}

In order  to use the formalism described in the previous section
we introduce the complex spin-weighted quantities $J = h_{AB} q^A q^B / 2, 
\;\; K = h_{AB} q^A \bar q^B / 2$ and $U = U^A q_A$. In addition,
we exchange the real variable $V$ with the function $W = (V-r)/r^2$.
The determinant condition
$\det( h_{AB}) = \det( q_{AB})$ translates into $1 = K^2 - J \bar J$, 
which determines $K$ in terms of $J$. The spin-weight of $J$ is 2, and
the function $U$ has spin-weight 1, while $K$, $W$, and $\beta$  have zero
spin-weight. Complex conjugation of a spin-weighted quantity gives a
spin-weight of equal magnitude and opposite sign. Thus, for example, the
spin-weight of $\bar U$ is $-1$.

The vacuum Einstein equations  $G_{\mu\nu} = 0$ can be
decomposed into spin-weighted terms as 
well \cite{Winicour83,Winicour84,Bishop97c}.
As described in Section~\ref{sec:intro.null.eqs},
these divide into hypersurface equations, evolution equations 
and conservation equations.

\vspace{1em}
\noindent
{\em Hypersurface equations}
\vspace{0.5em}

\noindent
The Ricci tensor component $R_{rr}=0$ implies
\begin{eqnarray}
\beta_{,r} &=& N_\beta,\label{eq:null.betaeq}
\end{eqnarray}
where $N_\beta$ contains quadratically aspherical terms, i.e. 
terms that are quadratic in the deviation from spherical symmetry
\cite{Bishop96}.
For reference we give the full non-linear
 hypersurface equation for $\beta$ (see~\cite{Bishop97c}, Eq.~(A1)):
\begin{equation}
\beta_{,r} = \frac{r}{8} \left( J_{,r} \bar J_{,r} 
- \left( K_{,r} \right)^2 \right).
\label{eq:null.betaeq-full}
\end{equation}

The equations $R_{rA} = 0$ imply
\begin{eqnarray}
U_{,r} &=& r^{-2} e^{2 \beta} Q + N_U,\label{eq:null.Ueq}\\
(r^2 Q)_{,r} &=& -r^2 (\bar \eth J + \eth K)_{,r} 
+ 2 r^4 \eth (r^{-2} \beta)_{,r} + N_Q,\label{eq:null.Qeq}
\end{eqnarray}
where $Q$ is used as an intermediate variable to eliminate second radial
derivatives of $U$ and the quadratically 
aspherical terms are included in $N_U$ and $N_Q$.

The  equation for $W$ is given by $R_{AB} h^{AB}$ = 0, i.e.
\begin{eqnarray}
W_{,r} &=& \frac 1 2 e^{2 \beta} {\cal R} - 1 - e^\beta \eth \bar \eth e^\beta
+ \frac 1 4 r^{-2} (r^4 ( \eth \bar U + \bar \eth U))_{,r} + N_W,
\label{eq:null.Weq}
\end{eqnarray}
where the quadratically aspherical terms are included in $N_W$.
The quantity ${\cal R}$ is the curvature scalar of $u=$ constant $r=$ constant 
surfaces and it is computed using
\begin{equation}
{\cal R} = 2 K - \eth \bar \eth K + \frac 1 2 ( \bar \eth^2 J + \eth^2 J)
+ \frac {1}{4K} ( \bar \eth \bar J \eth J - \bar \eth J \eth \bar J).
\end{equation}

The hypersurface equations (\ref{eq:null.betaeq}) - (\ref{eq:null.Weq})
contain angular and radial derivatives but no time derivatives of the
metric functions. Given $J$ on a $u=$ constant hypersurface,
these  equations  propagate
$\beta, U, Q$ and $W$ along the radial direction, 
in terms of integration constants on an inner world-tube.

\vspace{1em}
\noindent
{\em Evolution equations}
\vspace{0.5em}

\noindent
The evolution equations $R_{AB}-h_{AB} h^{CD} R_{CD}/2$ together 
with the determinant condition $\det (h_{AB}) = \det (q_{AB})$ result
in the equation for $J$
\begin{eqnarray}
2 (rJ)_{,ur} - \left( r^{-1} V (rJ)_{,r}\right)_{,r} &=& 
- r^{-1} (r^2 \eth U)_{,r} + 2 r^{-1} e^\beta \eth^2 e^\beta  \nonumber \\
&&- (r^{-1}W)_{,r} J + N_J,
\label{eq:null.Jeq}
\end{eqnarray}
where the quadratically aspherical terms are included in $N_J$.
 
\vspace{1em}
\noindent
{\em  Conservation equations}
\vspace{0.5em}

\noindent
For a world-tube given by $r=$ constant 
the conservation equations
take the form \cite{Bishop97c}
\begin{eqnarray}
\beta_{,u} &=& {\cal K}_\beta, \\
Q_{,u} &=&  - 2 \eth \beta_{,u} - q^A {\cal K}_{A},
\end{eqnarray}
where  ${\cal K}_\beta, {\cal K}_{A}$
are purely null-hypersurface terms (composed out of
$\beta, U^A, V$ and $h_{AB}$ and their $r$ and $x^A$ derivatives). 
They determine the evolution of the integration constants on the inner
world-tube.

\section{The construction of the code}

The Pitt Null Code is a general relativistic code  solving
a discretized version of the Einstein equations in the context of
a null foliation of space-time, as described in
Section~\ref{sec:intro.nullcoord}.
In this section we outline the finite-difference implementation of the 
Pitt Null Code.  For further details see \cite{Lehner99b,Lehner98}.

\subsection{The structure of the numerical grid}
\label{sec:null.gridstructure}

The compactified radial coordinate $x$ is discretized as
$x_i = x_1 + (i-1) \Delta x$ where $i = 1 \ldots N_x$ and
$\Delta x = (1-x_1)/(N_x-1).$ The point $x_{N_x} = 1$ lies at 
null infinity. The point $x_1$ has to lie on or inside the 
world-tube,\footnote{Most of the time, when running
the characteristic code by itself, the innermost radial gridpoint will lay
on the world-tube. However, as it will be seen in Chapter \ref{chap:ccm},
in the framework of Cauchy-Characteristic Matching, the world-tube may have
a location that changes with respect to the characteristic grid structure 
during the code evolution,  but it is always outside the innermost radial
gridpoint. In this chapter we make the assumption that the innermost
radial gridpoint is on the world-tube $\Gamma$, i.e. $x_1 = x_{|\Gamma}$.} i.e. $x_1 \le
x_{|\Gamma}$.The stereographic coordinate $\zeta = q + I p$ is  discretized by
$q_j = -1 + (j-3) \Delta$ and $p_k = -1 + (k-3) \Delta$, where $j,k = 1 \ldots
N_\zeta$ and $\Delta = 2 / (N_\zeta-5)$. The evolution proceeds with time-step
$\Delta u$ subject to the  Courant-Friedrichs-Levy (CFL) condition which states
that the numerical  domain of dependence of the finite difference equations
must contain the analytic domain of dependence of the original equations.

The fields $J, \beta, Q,$ and $W$ are represented by their values
on this rectangular grid, e.g. $J^n_{ijk} = J(u_n,x_i,q_j,p_k)$.
As first shown in \cite{Gomez94a}, stability requirements suggest 
that the field $U$ is
represented by values at the midpoints 
$x_{i+\frac 1 2} = x_i + \Delta x /2 $
on a radially staggered grid (so that $U^n_{ijk} = U(u_n,x_{i+\frac 1 2},
q_j,p_k)$). In addition the variables must be chosen such that they 
are regular functions of $x$ 
throughout the whole numerical grid, including the outermost
gridpoint. This requires \cite{Isaacson83} 
that  $W(x)$ be replaced by 
$\tilde W(x) = r^{-2} W(x)$ in the numerical scheme.

\subsection{Boundary values}
\label{sec:null.boundary}

The boundary data to be supplied on the 
timelike world-tube are the metric functions $J,\beta,U,W$.
Furthermore, one needs to supply $J$ on
the first null hypersurface (initial data), 
which can be provided free
of constraints in the null--timelike boundary problem.
The world-tube data must satisfy
conservation laws.  Numerically one can either provide constrained boundary
data or, to test the robustness of the numerical algorithm, one can
put any data onto the world-tube and see whether the 
code handles the constraint-violating modes.

\subsection{The {\em eth} module}

Computing the action of $\eth$ and $\bar \eth$ on various fields involves
angular derivatives. For the sake of clarity we illustrate how the 
module works for the case of computing $\eth U$, with $U$ given on all
angular gridpoints on both patches. First one computes the necessary
partial derivatives using second-order centered finite
difference expressions, e.g. 
$U_{,q}$ = $\frac{U_{j+1}-U_{j-1}}{2 \Delta} + O(\Delta^2)$.
These are computed everywhere except at the edges of the angular grid.
Then one  constructs the quantity $\eth U$ up to $O(\Delta^2)$ accuracy
throughout
the angular grid except for the edges. The missing values
at the grid boundaries are obtained from the values
 computed on the other patch, using the appropriate 
spin-weighted transformation rule combined with a 2-D quadratic polynomial
interpolator. The width of the overlap region between the two angular
grid patches is designed specifically so that this interpolation can be done
properly.

\subsection{Hypersurface equations}

Having eliminated second radial derivatives with the use of the 
intermediate variable $Q$,
the radial discretization of the hypersurface equations 
becomes straightforward. In the following we shall give explicit
finite-difference expressions for the radial algorithm of the hypersurface
equations in the quasispherical approximation \cite{Bishop97c,Lehner98}. 
The additional terms that enter when solving for the full 3-D problem do not
introduce any qualitatively new terms.

\begin{enumerate}

\item {\it Equation for $\beta$:} 
In the quasispherical approximation $\beta$ remains independent of $r$.
Thus, its value is completely defined by its boundary value:
\begin{equation}
\beta_i = \beta_{|\Gamma}.
\end{equation}

\item {\it Equation for $Q$:}  Re-expressing radial derivatives 
in terms of derivatives
with respect to $x$, one obtains the rule for propagating $Q$
\begin{equation}
2 Q + x(1-x) Q_{,x} = - x(1-x) (\bar \eth J + \eth K)_{,x} - 4 \eth \beta,
\end{equation}
where the quadratically aspherical terms are omitted.
Replacing $x$-derivatives with their second-order
finite-difference approximation and the value $Q_{i-1/2}$ by the
average $(Q_i + Q_{i-1})/2$, one obtains an expression that involves
values of $Q, J, K$ and $\beta$ at the points $x_i$ and $x_{i-1}$:
\begin{eqnarray}
Q_i &+& Q_{i-1} + x_{i-\frac 12}\left(1-x_{i-\frac 12}\right) 
\frac{Q_i-Q_{i-1}}{2 \Delta x} = 
\nonumber \\&&
- x_{i- \frac 12} \left( 1 - x_{i- \frac 12}\right) 
\left( 
\bar \eth \frac{J_i - J_{i-1}}{2 \Delta x}
+ \eth \frac{K_i - K_{i-1}}{2 \Delta x}
\right) 
\nonumber \\&&
- 2 \eth \left(\beta_i + \beta_{i-1}\right) + O(\Delta^3),
\end{eqnarray}
which can easily be solved for $Q_i$. After a radial march that proceeds
from $\Gamma$ to ${\cal I}^+$, the global  truncation error in $Q$ is
$O(\Delta^2)$.

\item {\it Equation for $U$:} In the quasispherical approximation, in terms of the
variable $x$, the hypersurface equation for $U$ is
\begin{equation}
U_{,x} = \frac {e^{2 \beta} Q}{r_1 x^2},
\end{equation}
where $r_1$ is the location of the innermost radial gridpoint
 in terms of the coordinate $r$.
The discretized rule for updating $U$ is
\begin{equation}
U_i = U_{i-1} + \frac  {e^{2 \beta_i} Q_i}{r_1 x_i^2} \Delta x
+ O(\Delta^3)
\end{equation}
for all points except the point $x_{\frac 3 2}$, lying just
outside $\Gamma$. At that point $U$, is written as a Taylor expansion
around the world-tube: 
\begin{equation}
U_1 = U_{|\Gamma} + U_{,x|\Gamma} \left( x_{1+\frac 1 2} -
x_{|\Gamma} \right) + O(\Delta^2).
\end{equation}
The algorithm
provides values of $U$ with $O(\Delta^2)$ global error.

\item {\it Equation for $W$:} In terms of 
$\tilde W = W/r^2$ and its $x$-derivatives,
the equation for $W$ within the quasispherical approximation reduces to
\begin{eqnarray}
x^2 \tilde W_{,x} + 2 \frac x {1-x} \tilde W &=& 
\frac 1 2 e^{2 \beta} {\cal R}  - 1 - e^\beta \eth \bar \eth e^\beta
\nonumber \\ 
&+& \frac 1 4 x^2 \left( \eth \bar U + \bar \eth U \right)_{,x} 
\nonumber \\ 
&+& \frac x {1-x} \left( \eth \bar U + \bar \eth U \right).
\end{eqnarray}
The corresponding finite-difference  rule for propagating
$\tilde W$ is
\begin{eqnarray}
&&x^2_{i-\frac 12} \left( 1 - x_{i-\frac 12} \right)
\frac{\tilde W_i - \tilde W_{i-1}}{\Delta x} +
x_{i-\frac 12} \left( \tilde W_i + \tilde W_{i-1} \right) = 
\nonumber \\ && \;\;\;\;\;
\frac 12 \left(1 - x_{i-\frac 12} \right)
\Big(
\frac 12 e^{2 \beta_i} {\cal R}_i + \frac 12 e^{2 \beta_{i-1}}
{\cal R}_{i-1} 
\nonumber \\ && \;\;\;\;\;
- 2 - e^{\beta_i} \eth \bar \eth e^{\beta_i}
- e^{\beta_{i-1}} \eth \bar \eth e^{\beta_{i-1}} 
\Big)
\nonumber \\ && \;\;\;\;\;
+ \frac 14 x^2_{i-\frac 12} \left(1 - x_{i-\frac 12} \right)
\left(
\eth \frac{\bar U_i - \bar U_{i-2}}{2 \Delta x}
+ \bar \eth \frac{U_i - U_{i-2}}{2 \Delta x}
\right)
\nonumber \\ && \;\;\;\;\;
+ x_{i-\frac 12} \left( \eth \bar U_{i-1} + \bar \eth U_{i-1}\right)
+ O(\Delta^3).
\end{eqnarray}

In carrying out the radial march the error in $\tilde W$ is
$O(\Delta^2)$ except for the outermost radial gridpoint where
a numerical analysis implies that  the error is 
$O(\Delta^2 \log \Delta)$.

\end{enumerate}

\subsection{Evolution equation}
\label{sec:null.eveq}

The numeric implementation of the evolution equation for $J$ is
based on an algorithm that has proved successful in the axisymmetric 
case \cite{Gomez94a,Papadopoulos94}.

\begin{figure}
\centerline{\epsfxsize=3in\epsfbox{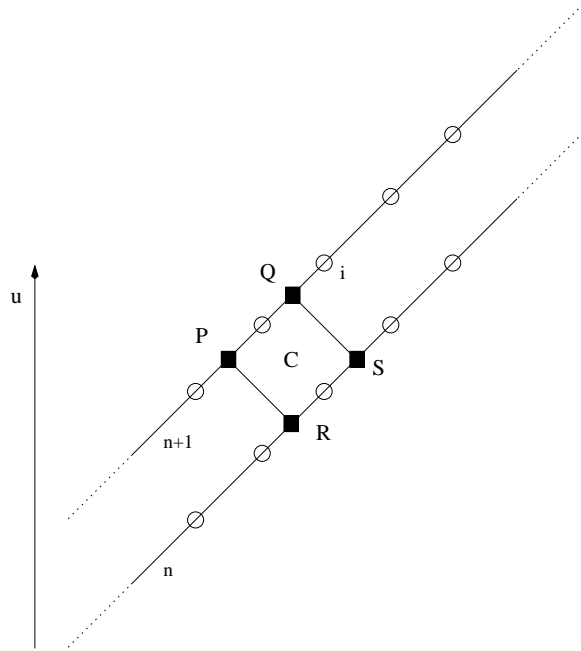}}
\caption{The null parallelogram.}
\label{fig:null.parall}
\end{figure}

Let $\Psi$ be a scalar field with a source on a flat background. 
In spherical, retarded-time
coordinates, the scalar wave equation $\Box \Psi = S$  can be written as
\begin{equation}
\Box^{(2)} \psi = - \frac {L^2 \psi}{r^2} + rS,
\label{eq:null.swe2d}
\end{equation}
where $\psi = r \Psi$, $L$ is the angular momentum operator and
$\Box^{(2)}$ is the (flat) 
2-dimensional wave operator intrinsic to the $(u,r)$ plane.
The operator $\Box^{(2)}$ can be written as
\begin{equation}
\Box^{(2)} \psi = 2 \psi_{,ur} - \psi_{,rr} = \psi_{,uv},
\end{equation}
where $v = u + 2r$.

Next 
a null parallelogram ${\cal A}$ is built out of a set of two outgoing 
null geodesic segments $(PQ, RS)$, lying on two null cones that are 
$\Delta u$ apart, and two incoming null geodesic segments $(RP, SQ)$ 
that are $\Delta r$ apart. Integration of Eq.~(\ref{eq:null.swe2d}) 
over ${\cal A}$ gives
\begin{equation}
\psi_Q = \psi_P + \psi_S - \psi_R + \frac 12 \int_{\cal A} du \; dr
\left[  - \frac{L^2 \psi}{r^2} + r S   \right]. 
\end{equation}

To extend the null-parallelogram algorithm to the case of the gravitational evolution
equation \cite{Lehner99b,Lehner98}
we first rewrite Eq.~(\ref{eq:null.Jeq}) in the form
\begin{equation}
e^{2 \beta} \Box^{(2)} (rJ) = {\cal H},
\label{eq:null.Box2Jeq}
\end{equation}
where $\Box^{(2)}$ is now the 2-D wave operator in the curved space-time
\begin{equation}
 \Box^{(2)} \psi = e^{-2 \beta} \left[ 2 \psi_{,ru} 
- \left( \frac V r \psi_{,r}\right)_{,r} \right],
\end{equation}
and 
\begin{equation}
{\cal H} = - r^{-1} \left( r^2 \eth U \right)_{,r} + 2 r^{-1} e^\beta \eth^2 e^\beta
- \left( r^{-1} W\right)_{,r} J + N_J.
\end{equation}

Integrating Eq.~(\ref{eq:null.Box2Jeq}) over the null parallelogram ${\cal A}$
one obtains
\begin{equation}
(rJ)_Q = (rJ)_P + (rJ)_S - (rJ)_R + \frac 12 \int_{\cal A} du \, dr {\cal H}.
\end{equation}
Since the segments $(RP, SQ)$ are ingoing null geodesics, their
orientation with respect to the fixed numerical grid  vary as the background
metric changes. This means that one cannot construct the parallelogram ${\cal A}$ with
corners lying on numerical gridpoints.  
Numerical analysis and experimentation  show \cite{Gomez92b} 
that a stable algorithm results by placing
this parallelogram so that the sides formed by the incoming rays
intersect adjacent $u$-hypersurfaces at equal but opposite $x$-displacement
from the neighboring gridpoints. The values of $rJ$ at the vertices of the parallelogram
are approximated to second-order accuracy by linear interpolation between nearest
neighbor gridpoints on the same outgoing characteristic.  The integrand is 
approximated by its value at the center $C$ of the parallelogram. The resulting
equation is
\begin{equation}
(rJ)_Q =(rJ)_P + (rJ)_S - (rJ)_R + \frac 12 \Delta u\left( 
r_Q - r_P + r_S - r_R
\right) {\cal H}_C.
\end{equation}
After the appropriate interpolations,
the value of $rJ$ at the $i$-th gridpoint is updated according to the
finite-difference expression \cite{Lehner98}
\begin{eqnarray}
(rJ)^{n+1}_i &=& {\cal F} \left[ 
(rJ)^{n+1}_{i-1}, 
(rJ)^{n+1}_{i-2}, 
(rJ)^{n}_{i+1}, 
(rJ)^{n}_{i}, 
(rJ)^{n}_{i-1}
\right] \nonumber \\
&+& \frac 12 \Delta u \left( 
r_Q - r_P + r_S - r_R
\right) {\cal H}_C,
\label{eq:null.Jmarch}
\end{eqnarray}
where the symbol ${\cal F}$ is a linear function in the $(rJ)$'s
and $J$ is  given at all gridpoints at level $u_n$ and at gridpoints $x_1 \ldots x_{i-1}$
at level $u_{n+1}$.
In the quasispherical
approximation the expression
Eq.~(\ref{eq:null.Jmarch}) provides an explicit algorithm for updating $(rJ)$
via an outgoing radial march. 
However, with the quadratically
aspherical terms included, the right-hand-side of Eq.~(\ref{eq:null.Jmarch})
contains $J_{,u}$ which requires the unknown value $J^{n+1}_i$ when computing 
${\cal F}$ in the center of the computational cell. 
Thus an iterative approach is used \cite{Lehner99b}:
\begin{enumerate}
\item First the value $J^{n}_i$ is copied into $J^{n+1}_i$. This provides
an initial guess value in the point $(n+1,i)$ 
with an error of $O(\Delta)$.
\item Next the function ${\cal F}$ is computed at the center of 
the computational cell. As already stated, this requires use of $J^{n+1}_i$. 
\item Using Eq.~(\ref{eq:null.Jmarch}) a corrected  value for  $J^{n+1}_i$
is computed.
\item Points (2) and (3) are repeated 
a sufficient number of times to ensure convergence.
\end{enumerate}
 
On level $u_n$ the
finite-difference stencil  contains the four points 
$x_{i-2}$ $\ldots$  $x_{i+1}$;
 on level $u_{n+1}$, the three points $x_{i-2} \ldots x_{i}$.
The use of four points at level $u_n$ introduces dissipation which
cures  non-linear instabilities that
otherwise would occur. In order to obtain an accurate
global discretization at infinity, the evolution variable $\Phi = xJ$ is used.
For further details please refer to 
\cite{Lehner99b,Lehner98}.

\subsection{Stability and accuracy tests}

The code was tested for stability and  accuracy. A
harsh stability test was performed by
providing random initial and  boundary data of
$O(\Delta^2 \cdot 10^{-6})$ and running it for 2000 ``crossing times,'' where one crossing
time is the time it takes light to go across the
world-tube in a flat background. 
The code showed no signs of instability \cite{Szilagyi99}.

To check the accuracy of the code 
runs were performed with larger and larger grid-sizes,
using the same physical parameters. The numeric results
were compared to the analytic solution.
The following test-beds were used 
to check convergence: 
\begin{itemize}
\item Quasispherical waves. A linearized solution of the quasispherical
equations (involving a spherical harmonic with angular momentum $\ell = 6$) 
was run with a very small amplitude ($|J| = 10^{-9}$). The inner boundary was
set to $r_1=1$ and the wave was evolved numerically between $u_i=0$ and
$u_f = 0.5$.

\item Boost and rotation symmetric solutions. A family of non-linear solutions
called SIMPLE, \label{page:null.SIMPLE}
with exact boost and rotation symmetry \cite{Gomez94a} was also used. The
field variables are
\begin{eqnarray}
J &=& \frac{(1+\Sigma)^4 - 16}{8 (1+\Sigma)^2},\\
\beta &=& \log \frac{(1+\Sigma)}{2 \sqrt{\Sigma}},\\
U &=& - \frac{a^2\, r\, \Lambda}{\Sigma},\\
\tilde W &=& \frac 1 r \left( -1 + \frac 1 \Sigma \right) 
+ a\, r^2\, \frac{(2\, \Xi - 1)}\Sigma, 
\end{eqnarray}
where
\begin{equation}
\Sigma = \sqrt{1 + a^2 \, r^2 \, \Xi}, \;\; \Lambda =  \frac{2 \, \zeta
\left(1 - \zeta \bar \zeta \right)}{\left(1+\zeta \bar \zeta \right)^2}
\;\; \mbox{and} \;\; \Xi = \frac{4 \zeta \bar \zeta}{\left( 1 + \zeta \bar
\zeta \right)^2}. 
\end{equation}  
Because of its cylindrical symmetry this solution is not asymptotically
flat, but it is used to construct an asymptotically flat solution by smoothly
pasting it to asymptotically flat null data outside some radius $R_0$.  The
resulting solution  provides a non-linear test inside  the domain of
dependence of the analytic (unmixed) solution.

\item Schwarzschild in rotating coordinates. In a null frame that rotates
with angular velocity $\omega$
according to $\phi \rightarrow \tilde \phi + \omega u$, the Schwarzschild
metric can be written as
\begin{eqnarray}
ds^2 &=& - \left( 1 - \frac{2m}r - \omega^2 r^2 \sin^2 \theta \right) du^2
- 2 \, du \, dr \nonumber \\ &&
+ 2 \, \omega \, r^2 \sin^2 \theta \, du \, d \phi 
+ r^2 q_{AB} \, dx^A \, dx^B.
\end{eqnarray}
The test was performed with the choices $r_1 = 3 m,\; \omega = 0.5, \;m=1$. The
metric fields were  evolved numerically between $u_i = 0$ and $u_f = 0.5$.

\end{itemize}
These tests show that the code is second-order accurate.
See \cite{Lehner98} for details.

\section{The news}
\label{sec:null.news}

\begin{figure}
\centerline{\epsfxsize=4.2in\epsfbox{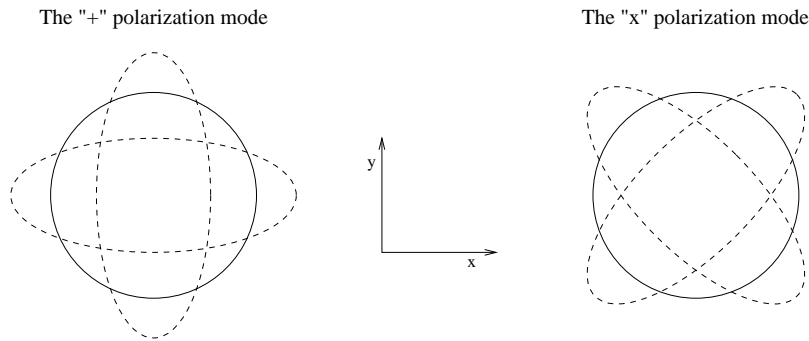}}
\caption{The ``$+$'' and the ``$\times$'' polarization modes
of a gravitational plane-wave propagating in a flat background along the $z$ axis.}
\label{fig:null.ttmodes}
\end{figure}

As mentioned in Section~\ref{sec:intro.intro},
a prime motivation for numerical relativity is the
prediction/interpretation of waveforms for laser interferometer detectors of
gravitational radiation such as LIGO, VIRGO, etc. 
So far we have discussed 
the construction of a code that 
evolves the gravitational radiation
to future null infinity ${\cal I}^+$
if appropriate boundary data is given on 
an inner timelike boundary $\Gamma$.
One needs to connect quantities evolved by this
code  to data that the detectors will actually measure.
In short, the effects of gravitational radiation on experimental
apparatus can be understood as follows:

Consider linearized gravitational waves propagating
through a flat, empty region of space-time. Let $(t,x,y,z)$
be Cartesian coordinates, with metric 
$g_{\mu\nu} = \eta_{\mu\nu}+\varepsilon h_{\mu\nu}$. As a further
restriction, require that $h_{\mu\nu}$  satisfy the
``TT'' (tracefree-transverse) gauge constraints,  
$h^{TT}_{0\mu} = 0, \;\;\delta^{jk} h^{TT}_{ij,k}=0$, 
and $\delta^{ij} h^{TT}_{ij} = 0$.
A gravitational plane-wave propagating in the $z$-direction 
is defined by the functions \cite{MisnerGRWaves}
\begin{eqnarray}
\label{eq:null.ttplus}
&&h^{TT}_{xx} - h^{TT}_{yy} = \Re\left(A_{+} e^{-I\omega (t-z)}\right),
\\
\label{eq:null.ttcross}
&&h^{TT}_{xy} = h^{TT}_{yx} =  \Re\left(A_{\times} e^{-I\omega (t-z)}\right).
\end{eqnarray}
Consider a ring that, in absence of gravitational fields, is a perfect 
circle.
Both polarization modes transform the ring into ellipses. 
The length of the axes of the ellipse
oscillate according to the oscillating metric components
(\ref{eq:null.ttplus}), (\ref{eq:null.ttcross}).
The main axes will be along the Cartesian $x$- and $y$-directions 
for the ``$+$'' mode and at an angle of
$\pi/4$ from the $x$- and $y$-axis for the ``$\times$'' mode.
 Figure~\ref{fig:null.ttmodes} illustrates  the effects of the
two polarization modes.

Even though in a flat background gravitational radiation
can be described analytically, one faces difficulties 
in a general relativistic context.
In classical mechanics energy is defined as the quantity
whose conservation is 
related to the homogeneity of time \cite{Landau91}. In
electromagnetism, radiation is defined as a phenomena that carries energy-momentum.
Since {\em time} in general relativity has no unique 
physical meaning, one faces difficulties when 
defining energy.
The problem was solved by the definition of asymptotically flat
space-times \cite{Penrose}. Space-times are referred to as asymptotically flat
if an appropriate boundary representing the ``points at infinity'' can be
``added'' to the space-time in a suitable way. At these ``points'' the induced
metric is in fact flat and quantities like the energy carried away by
gravitational radiation, and the notion of total energy, can be defined
unambiguously.

The formalism of calculating the gravitational 
radiation 
and a  description of the 
numerical module that does this is described in 
\cite{Tamburino66,Isaacson83,Winicour87,Gomez97c,Bishop97c,Lehner98}.  
An important module of the Pitt Null Code  
provides the waveforms detected by a gravitational radiation antenna.

In the case of gravitational radiation antenna such as LIGO or VIRGO 
the test object used to measure the effects of gravitational radiation is
not a sphere but a set of mirrors and laser beams. Specifically, in the LIGO detectors,
one uses a pair of 4 km long laser-beams that are at right angle.
These beams define the $x$ and $y$ axes of the reference frame. Thus if
a ``$+$'' mode is present, it causes a change in 
 the relative lengths of the arms of the detector. 
Although this change is typically orders of magnitude below the size of
the proton, a precise interferometer might be able to detect the effects of the radiation.
Furthermore, using data from the two 
LIGO sites and from the other similar detectors
from Europe, Japan and Australia, the background noise can be filtered out 
in a very effective way. Comparing the time-delay between the signals as detected at the various
sites, one can also determine the 
position of the source of the gravitational radiation.

\subsection{Sources of gravitational radiation}

\begin{figure}
\centerline{
\psfig{figure=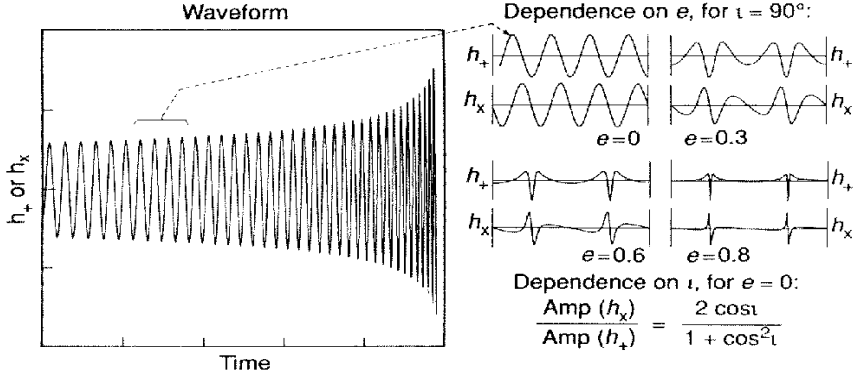,height=2.25in,width=4.5in}}
\caption{Chirp waveform from an inspiral event of a compact binary
system \cite{Abramovici92}. On the right the dependency of the waveform 
of the orbital eccentricity $e$ and the orbital inclination $\i$
is shown. }
\label{fig:null.chirp}
\end{figure}

Although the universe contains a countless number of
sources of gravitational radiation,
most of them cannot be detected because of
the weakness of the gravitational
signal that reaches the Earth.
However, dynamic astrophysical systems 
such as coalescing compact binaries
are good candidates for emitting gravitational 
waves which can be detected
by an observer on earth.  
These systems  consist of either
two neutron stars, two black holes or one of each. Due to their
small size ($\sim 20$ km in case of a neutron star), they can orbit
each other at close range and a high orbital frequency (up to $\sim
500$ Hz). In these coalescing binaries
gravitational waves are 
emitted with a high efficiency.
Thus, for instance, a double neutron star system which is
500 km apart radiates away most of its potential energy within minutes.
The rate at which gravitational radiation is emitted 
increases as the co-orbiting objects approach each other.
The typical waveform produced during this accelerated
inspiral is a chirp signal \cite{Shutz86} 
(see Figure~\ref{fig:null.chirp}). To first order the chirp
signal can be described by the change of its frequency
over time $\dot f$ and by its amplitude $A$:
\begin{eqnarray}
\dot f &\propto& M_c^{5/3}\, f^{11/3} + 
\left(
\begin{array}{c}
\mbox{relativistic corrections} \\
M_1, M_2, S_1, S_2
\end{array}
\right) \\
A &\propto& k_{\mbox{\scriptsize orbit}}\, M_c^{5/3}\, \frac{f^{2/3}}{r}
\end{eqnarray}
with $M_c$ the chirp mass
\begin{equation}
M_c =\frac{\left(M_1 \, M_2 \right)^{3/5}}{\left(M_1+M_2\right)^{1/5}},
\end{equation}
$f$ the orbital frequency, $M_1, M_2, S_1$ and $S_2$ the mass and spin of the 
two compact objects, respectively, $k_{\mbox{\scriptsize orbit}}$
a constant accounting for the inclination of the 
source orbital plane and $r$ the distance of the detector to the source.
The details of the exact waveform of the inspiral event provide 
further information about the system itself such as the  eccentricity of
the orbit, the spin and the mass of the objects 
(see, for example, \cite{Cutler94}).

While the coalescence of two neutron stars can be described
by  analytic approximations, the case of two black holes
poses a much more complex problem.
 The coalescence of the binary black-hole system
is roughly divided into three phases \cite{Flanagan98}:
\begin{itemize}
\item During the first phase, the {\em inspiral}, 
the two black holes are well separated 
and the waveform of the emitted gravitational waveform
can be determined  via post-Newtonian approaches
(see, e.g. \cite{Blanchet95a,Blanchet95b,Clifford96a,Blanchet96a}).
During this phase the gravitational radiation reaction time-scale is much longer
than the orbital period. The inspiral ends when the binary orbit
becomes relativistically dynamically  unstable at an orbital separation of
$r \sim 6 M$, where $M$ is the total mass of the binary.
\cite{Kidder93,Cook94}. The gravitational waves from the inspiral 
carry encoded within them the masses and spins of the two black holes,
some of the orbital elements of the binary, and the distance to the
binary \cite{Thorne87,Thorne95}.
\item Towards the end of the inspiral, the black holes
 make a gradual transition from 
a radiation-reaction driven inspiral to a freely-falling plunge
\cite{Kidder93,Ori00,Lai96}.  This phase is referred to as  the {\em merger}.
Gravitational waves emitted during this process  are expected 
to be rich in information about the dynamics of relativistic gravity in 
a highly nonlinear, highly dynamic regime.
This is the phase in 
which numerical relativity has an important role, since the
highly non-linear character of the merger prohibits use of
 perturbative approaches.
\item In the final state the system settles down to a stationary Kerr
state. Thus the nonlinear dynamics of the merger gradually become more and more
describable as oscillations of this final black hole's
quasi-normal modes \cite{Teukolsky74,Chandrasekhar75}. The corresponding
emitted gravitational waves consist of 
a superposition of exponentially damped sinusoids. This phase is often
referred to as the {\em ring-down} phase.
The waves from the ring-down carry information about the mass and spin
of the final black hole \cite{Echeverria88,Finn92}.
\end{itemize}

In order to obtain a rough estimate of the amplitude
 of the observed strain (relative length-change)
$h=\Delta L/L$ caused by a 
gravitational wave from a typical 
astrophysical source one can use a dimensional argument
together with the approximation 
that gravitational radiation couples to the quadrupole moment only.
Denoting the quadrupole of the mass distribution
of a source by $Q$, we write \cite{Sigg98}:
\begin{equation}
h \sim \frac{G \ddot Q}{c^4 r} \sim 
\frac{G (E^{\mbox{\scriptsize non-symm.}}%
_{\mbox{\scriptsize kin}}/c^2)}{c^2 r}
\end{equation}
with $G$ the gravitational constant and 
$E^{\mbox{\scriptsize non-symm.}}_{\mbox{\scriptsize kin}}$ 
the non symmetric part of the
kinetic energy. Setting the non-symmetric energy equal to one
solar mass
$$
E^{\mbox{\scriptsize non-symm.}}%
_{\mbox{\scriptsize kin}}/c^2 \sim M_{\odot},
$$
and assuming that the source is located at inter-galactic or cosmological
distances, one obtains a strain estimate of order
\begin{eqnarray}
h &\lesssim& 10^{-21} \quad \quad \mbox{Virgo cluster,} \\
h &\lesssim& 10^{-23} \quad \quad \mbox{Hubble distance.} 
\end{eqnarray}
For a detector with a baseline of $10^4\, {\rm m}$ the relativistic
length changes become of order
\begin{equation}
\Delta L = h L \lesssim 10^{-19}\, {\rm m} 
\quad \mbox{to} \quad 10^{-17}\, {\rm m}.
\end{equation}
This is a rather optimistic estimate. 
Most sources will radiate significantly
less energy in gravitational waves.

Similarly, one can estimate the upper bound for the 
frequencies of gravitational waves \cite{Sigg98}. A gravitational
wave source cannot be much smaller than its Schwarzschild 
radius $2 G M/c^2$, and cannot emit strongly at periods shorter
than the light travel time $4\pi G M/c^3$ around its circumference.
This yields a maximum frequency of 
\begin{equation}
f \lesssim \frac{c^3}{4 \pi G M} \simeq 10^4 \,{\rm Hz}\, \frac{M_\odot}{M}.
\end{equation}

The rate at which black-hole coalescences occur is discussed, 
for instance, in \cite{Flanagan98}.
For {\em solar mass} binaries with total masses in the range 
$10 M_{\odot} \lesssim M \lesssim 50 M_\odot$
the rate of coalescence in the Universe is expected to
be about $1/100, 000$ years in our Galaxy, or several
per year within a distance of $200 {\rm Mpc}$ 
\cite{narayan,phinney,heuvel,tutukov,yamaoka}. 
For {\em intermediate mass} black hole binaries,
with total masses in the range $50 M_\odot 
\lesssim M \lesssim (\mbox{a few}) \times 10^3 M_\odot$,
there is little observational evidence. Still, even if 
the coalescence rate of intermediate mass black-hole binaries
is $\sim 10^{-4}$ that of neutron-star binaries (which is
thought to be $\sim  10^{-5} \mbox{yr}^{-1}$ in our Galaxy),
the black-hole sources would still be seen more often by LIGO
than the neutron-star sources, and thus could be the first detected type
of source. As regard as {\em supermassive} black hole binaries,
there is a variety of strong circumstantial evidence that supermassive
black holes, in the mass range $10^6-10^9 M_\odot$ exist 
in quasars and active galactic nuclei. Also, 
$\sim 25 \% - 50 \%$ of
 nearby massive spiral and elliptical galaxies are expected
to harbor quiescent supermassive black holes.
A main scientific goal of space-based gravity wave detectors, such as LISA,
is to detect various phenomena related to such black holes. These
include capture of compact stars 
\cite{Schutz89,LISAreport,Thorne95,HilsBender,Ericnew},
the formation of supermassive black holes 
\cite{Schutz89,LISAreport}, as well as the collision of these
in the context of Galaxy mergers 
\cite{Blandford,Thorne87,Schutz89,LISAreport,Thorne95,Wahlquist,Haehnelt}.
Such events  would be detectable throughout the observable
universe with large signal to noise ratios
\cite{LISAreport,Thorne95}.   The
overall event rate for these phenomena is
uncertain, but could be one or more events per year.

\section{Black hole evolution}
\label{sec:null.1BH}

Consider a space-time containing a singularity.
According to the cosmic censorship principle the singularity
is hidden by an event horizon that causally disconnects
the singularity from future null infinity ${\cal I}^+$.
The region delimited by the event horizon is called a black hole.
Inside the black-hole horizon
 the lightcones are tilted inwards so that no light ray can escape.

\begin{figure}
\centerline{\epsfxsize=3in\epsfbox{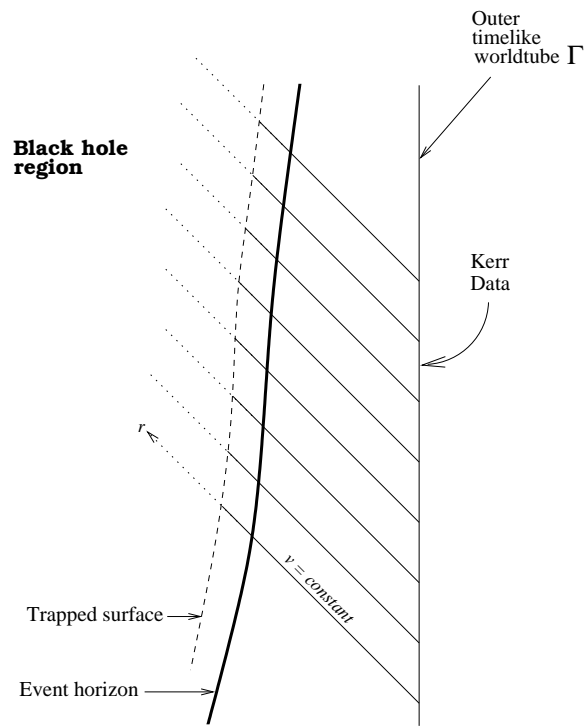}}
\caption{The space-time region surrounding the black hole as described by
an ingoing null foliation. The region close to the singularity is excised from
the computational domain by means of locating a trapped surface, 
as described in Section~\ref{sec:null.BH.excision}.}
\label{fig:null.ingoingBH}
\end{figure}

The issue of black hole evolution is one of the most demanding problems
in numerical relativity. 
As it was first suggested in \cite{Thornburg87}, in order to avoid the singularity,
one can excise its neighborhood. Recently 
\cite{Gomez97b,Gomez97a} 
it was shown that the black-hole region can be evolved 
by a characteristic evolution
based upon ingoing null cones. 
These null cones are truncated at an inner
boundary (see Figure~\ref{fig:null.ingoingBH}).

The construction of the ingoing null coordinate system is the analog of the
outgoing case (see Section~\ref{sec:intro.nullcoord})
except that after the construction and coordinatization of the timelike
world-tube one uses ingoing null rays (instead of outgoing ones)
to coordinatize the space-time region inside  $\Gamma$.
The space-time metric takes  the ingoing Bondi-Sachs form
\begin{eqnarray}
ds^2 &=& \left( e^{2 \beta} \frac V r + r^2 h_{AB} U^A U^B \right)
dv^2 + 2 e^{2 \beta} dv \, dr \nonumber \\ &&
- 2 r^2 h_{AB} U^B dv \, dx^A
+ r^2 h_{AB} dx^A \, dx^B.
\end{eqnarray}
The analytical and numerical toolkit used for the outgoing formulation
can be refurbished for the incoming case via the substitution \cite{Gomez97b}
\begin{equation}
\beta \rightarrow \beta + I \pi / 2.
\end{equation}

The black hole is  contained inside the timelike world-tube $\Gamma$.
In order to perform black-hole simulations, in addition to
initial and boundary data  one needs an excision algorithm
that locates an inner boundary such that the code avoids  the singularity.

\subsection{Outer boundary data}

As stated in Section~\ref{sec:null.boundary}, 
the characteristic code needs
the  values $J,\beta,U,W$ (as well as $U_{,r}$) at the outer boundary $\Gamma$. 
In a number of cases
the space-time metric is known in Cartesian coordinates but not in
Bondi coordinates. A global coordinate transformation from Cartesian to Bondi
coordinates  requires knowledge
of the light-cones emanating from $\Gamma$, for the given space-time. 
In most cases this is complicated. 
Here we use an alternative approach. With 
the Cartesian metric provided 
around the world-tube $\Gamma$, a
numeric module -- the extraction 
module (see Chapter \ref{chap:extraction}, also \cite{Bishop96,Bishop98a})
 -- is used to perform
the coordinate transformation locally, 
computing the characteristic boundary 
quantities $J,\beta,U$ and $W$ around $\Gamma$.

The space-time metric
of the Kerr (spinning) black hole in Cartesian Kerr Schild 
coordinates takes the form \cite{Kerr65}
\begin{equation}
ds^2 = - dt^2 + dx^2 + dy^2 + dz^2 + 2 H k_{\mu} k_{\nu} dx^\mu dx^\nu,
\label{eq:null.kerrBH}
\end{equation}
where $k_\mu$ is tangent to an ingoing congruence of twisting light rays 
and $H$ is a potential given by
\begin{equation}
H = \frac {M_k r^3_k} { r^4_k + a^2 z^2},
\end{equation}
with
\begin{equation}
\frac{x^2 + y^2}{r_k^2 + a^2} + \frac{z^2}{r_k^2} = 1.
\end{equation}
The parameter $a$ defines the angular momentum of the spinning black hole.
The Schwarzschild metric can be obtained by setting $a=0$.

Boundary data for a Kerr black-hole simulation
is constructed by using the metric (\ref{eq:null.kerrBH})
as input for the extraction module and using the output
as boundary data for the characteristic code.

Alternatively, for the case of the Schwarzschild metric,
a gauge transformation $(t,x,y,z) \rightarrow (t',x',y',z')$ 
is first performed where the primed frame 
$(t',x',y',z')$
wobbles, rotates, or is boosted
with respect to the unprimed frame $(t,x,y,z)$. 
Then the extraction module 
computes the coordinate transformation locally (around $\Gamma$)
from the primed Cartesian 
coordinates to 
Bondi coordinates. It should be noted that the physics described by the
primed Cartesian frame is the same as for the unprimed Cartesian frame. 
The gauge transformation
amounts to making the world-tube wobble, rotate, or  boost with respect to 
the singularity.

For the wobbling case the coordinates $x'^{\alpha}$ are defined by
\begin{eqnarray}
t' &=&  t, \;\; z' \;= \;  z, \\
x' &=& ( x + b) \cos \omega t -  y \sin \omega t, \\
y' &=& ( x + b) \sin \omega t +  y \cos \omega t.
\end{eqnarray}

A very similar coordinate transformation gives the space-time 
metric of a black hole in rotating coordinates:
\begin{eqnarray}
\label{eq:null.rotating.gauge.FIRST}
t' &=&  t, \;\; z' \; =\;   z, \\
x' &=&  x \cos \omega (t + r_c) -  y \sin \omega (t + r_c), \\
y' &=&  x \sin \omega (t + r_c) +  y \cos \omega (t + r_c), 
\label{eq:null.rotating.gauge.LAST}
\end{eqnarray}
with $r_c^2 = x^2 + y^2 + z^2$. In this case the null metric quantities
are known analytically:
\begin{eqnarray}
J &=& 0, \;\;\; \beta \; = \; 0, \\
U &=& I \omega \sin \theta \; e^{I \phi}, \\
V &=& -r + 2 M_s.
\end{eqnarray}

The boosted frame is defined by 
\begin{eqnarray}
\label{eq:null.boosted.gauge.FIRST}
 x' &=& x, \;\;  y' \; = \; y,
\\  z' &=& z \cosh \alpha - t \sinh \alpha, 
\\  t' &=& t \cosh \alpha - z \sinh \alpha.
\label{eq:null.boosted.gauge.LAST}
\end{eqnarray}
When using the gauge conditions 
(\ref{eq:null.boosted.gauge.FIRST}) - (\ref{eq:null.boosted.gauge.LAST}), 
the world-tube $\Gamma$ is boosted with respect to the singularity.
Thus one must stop the characteristic evolution before 
 $\Gamma$ reaches the horizon of the black-hole.

\subsection{Initial data}

In addition to boundary data on $\Gamma$, the characteristic evolution needs
initial data, i.e. the constraint-free function $J$ on the initial null-hypersurface.

\begin{figure}
\centerline{\epsfxsize=3in\epsfbox{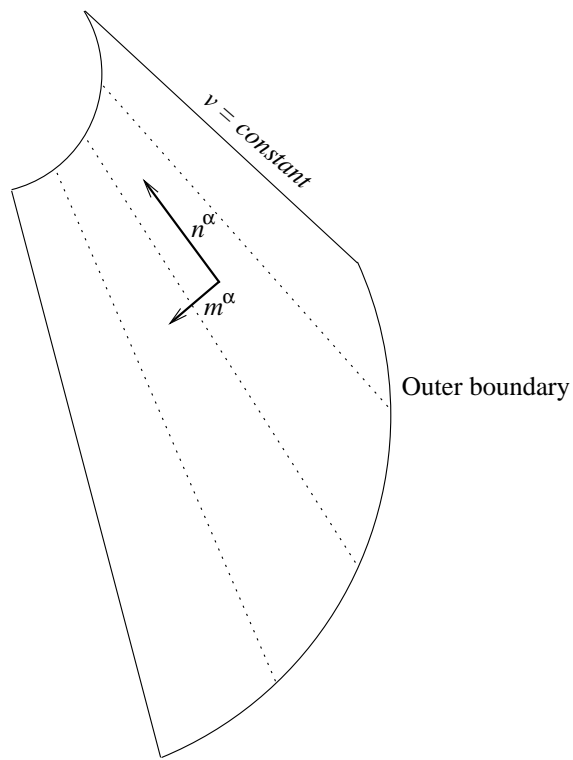}}
\caption{Section of an ingoing null hypersurface.}
\label{fig:null.incoming.vect}
\end{figure}

Since the black hole  is either rotating or moving with respect
to $\Gamma$  the initial null-hypersurface is not necessarily
spherically symmetric and initial data
cannot be given analytically. Let the incoming null vector
$n^a$ and the complex spacelike vector
$m^a$ span the tangent space to the null hypersurface 
(see Figure~\ref{fig:null.incoming.vect}).

In constructing the initial data for a moving
 black-hole we require that the Weyl tensor component 
$C_{abcd}\, n^a n^c m^b m^d$ (corresponding to $\Psi_4$ in Newman-Penrose
terminology \cite{Newman92})
 vanish on the initial (non-symmetric) null
hypersurface
\cite{Gomez97a}:
\begin{equation}
\left( r^2 J_{,r} \right)_{,r} - 2 \beta_{,r} \left(r^2 J\right)_{,r}
- \frac{r^2}2 J \left( J_{,r} \bar  J_{,r} - K^2_{,r} \right) = 0.
\end{equation}
In the linearized regime this corresponds to the condition that
the outgoing radiation is set to zero.
Combining this condition with the hypersurface equation 
(\ref{eq:null.betaeq-full})  for $\beta$
 one obtains
\begin{equation}
r^2 \left( r^2 J_{,r} \right)_{,r} 
- 2 \beta_{,r} \left( r^4 J \right)_{,r} = 0,
\end{equation}
an equation that can be solved numerically in terms of boundary values
for $J$ and $J_{,r}$ on the world-tube.

We set $J=0$ to prescribe
initial data for a rotating black-hole space-time 
(see Eq.~(\ref{eq:null.rotating.gauge.FIRST}) 
- (\ref{eq:null.rotating.gauge.LAST})) 
in which no gravitational radiation is present.

In the case of a Kerr (spinning) black hole, construction of characteristic 
data on the  initial null-hypersurface is more complicated.
One approach is to give Kerr boundary data on $\Gamma$
and let the initial null hypersurface describe a space-time
region containing a distorted Kerr black-hole with some radiation.
The characteristic evolution causes  some of this  radiation
to fall into the black hole and some to
propagate outward. The radiation that hits $\Gamma$ is reflected 
since the world-tube corresponds to a  stationary Kerr space-time.
Eventually  all (initial) radiation  falls into the black hole.

\subsection{Singularity excision}
\label{sec:null.BH.excision}

Given the boundary data and initial data, the ingoing characteristic code
still needs an inner boundary algorithm, when evolving black-hole space-times.
This algorithm is necessary so that the evolution code does not reach the
singularity.

Since singularities are causally disconnected from ${\cal I}^+$,
one can   cut out 
a space-time region inside the black-hole horizon
without
introducing unphysical effects on the phenomena 
 outside the  horizon. This is called
 singularity excision \cite{Thornburg87} -- a technique that has 
become standard in black-hole simulations.

Normally for a convex, topologically $S^2$ surface   
the light rays
 emitted in the outward normal direction  form a divergent beam. 
However gravitational lensing can refocus light rays so that in special 
circumstances (such as  the neighborhood of singularities) 
the outgoing
light cones from the surface  converge. 
Such a surface is called  trapped.  
In the limit in which the outgoing light cone
neither expands nor converges we obtain a marginally trapped surface (MTS). Under
reasonable physical circumstances these trapped surfaces 
always lie inside black-hole
event horizons. Thus (marginally) trapped surfaces  
are a guide for determining the
region of space-time to be excised.

For a given slice ${\cal S}$ of an ingoing null hypersurface ${\cal N}_v$, 
defined by $r = R(v, x^A)$, the divergence $\Theta_l$ of the outgoing null 
normals is given by \cite{Gomez97b}
\begin{eqnarray}
\frac {r^2 e^{2 \beta}} 2 \Theta_l &=& - V - \frac 1 {\sqrt{q}} \left[ \sqrt{q} 
\left( e^{2 \beta} h^{AB} R_{,B} - r^2 U^A \right)
\right]_{,A} \nonumber \\
&&- r \left( r^{-1} e^{2 \beta} h^{AB} \right)_{,r} R_{,A} R_{,B}
+ r^2 U^A_{,r} R_{,A}.
\label{eq:null.mtseq}
\end{eqnarray}
The slice is marginally trapped if $\Theta_l = 0$.

Solving for Eq.~(\ref{eq:null.mtseq}) is numerically difficult.
Alternatively \cite{Lehner98}
 one can identify the largest $r=$ constant slice of ${\cal N}_v$ that
satisfies the algebraic inequality $Q \leq 0$ where
\begin{equation}
Q = -V + \frac{r^2}{\sqrt{q}} \left(\sqrt{q} \, U^A\right)_{,A}.
\label{eq:null.qbdry}
\end{equation}
This surface is either trapped or marginally trapped, and 
is called the Q-boundary.
Solving  Eq.~(\ref{eq:null.qbdry}) 
is computationally more efficient  but the Q-boundary
makes less efficient use of gridpoints then a MTS. Both approaches 
were implemented for the characteristic code. For details  see \cite{Lehner98}.

\subsection{Results}

In all the above test cases the characteristic code proved able 
to stably and accurately 
evolve dynamic space-time regions containing a single black hole. 
The tests show
not only that the evolution code can handle highly non-linear space-time
dynamics, but they also indicate
 that the extraction  module  is able
to provide the necessary boundary data in a time-dependent,
non-linear regime. For further details see \cite{Gomez98a,Gomez97a}.

\section{The binary problem}

\begin{figure}
\centerline{\epsfxsize=3in\epsfbox{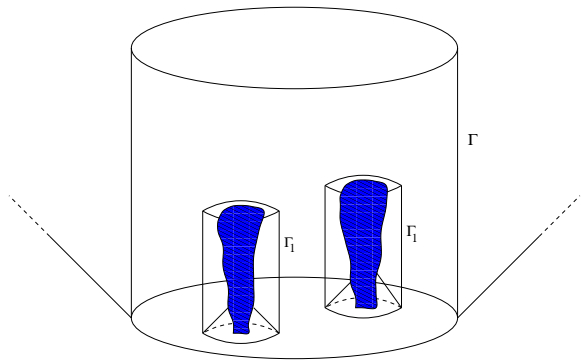}}
\caption{A possible scenario for early stage binary evolution.}
\label{fig:null.null3CCM}
\end{figure}

Section~\ref{sec:null.1BH} discusses simulation of single-black-hole space-times
via characteristic evolution. 
However, the binary problem is still unsolved.
 The lightcone structure of the space-time region between two spinning, co-orbiting
black holes is far from trivial. It is not possible 
to adapt the  characteristic code to handle
the caustics and 
crossovers
that are formed during the coalescence. 
Still,  one could  use characteristic evolution
matched to a Cauchy evolution  such that the Cauchy code does not have to deal
with excision. (See Figure~\ref{fig:null.null3CCM}.) 
The characteristic code can solve both the outer and the inner
boundary problems.  The missing element is 
a matching algorithm between the Cauchy and characteristic codes,
as described in chapters \ref{chap:ccm}-\ref{chap:ccmtest}.

There is another scenario in which  characteristic evolution can be used.
As first demonstrated  for the axisymmetric case  \cite{Lehner99a}
and then for more generic cases  \cite{Husa99},
horizon structures for  black-hole merger  
can be constructed as stand-alone items with almost no assumption
about the surrounding space-time. The time-reversed picture gives a white-hole horizon
that in the remote past is asymptotically in equilibrium as a Kerr white hole.
An  initial  perturbation of the white-hole 
is amplified and causes a fission of the horizon, the 
time-reversed scenario of a black-hole collision. 
The binary horizon structure obtained by these analytic means
is consistent with the numerical results
in \cite{Shapiro92a,Anninos93b,Hughes94a,Abrahams94a,Anninos95a,Libson94a,Matzner95a,Shapiro95a}.
The construction of such event horizons is outlined in the following section.

\subsection{Construction and structure of binary event horizons}

\begin{figure}
\centerline{\hbox{
\psfig{figure=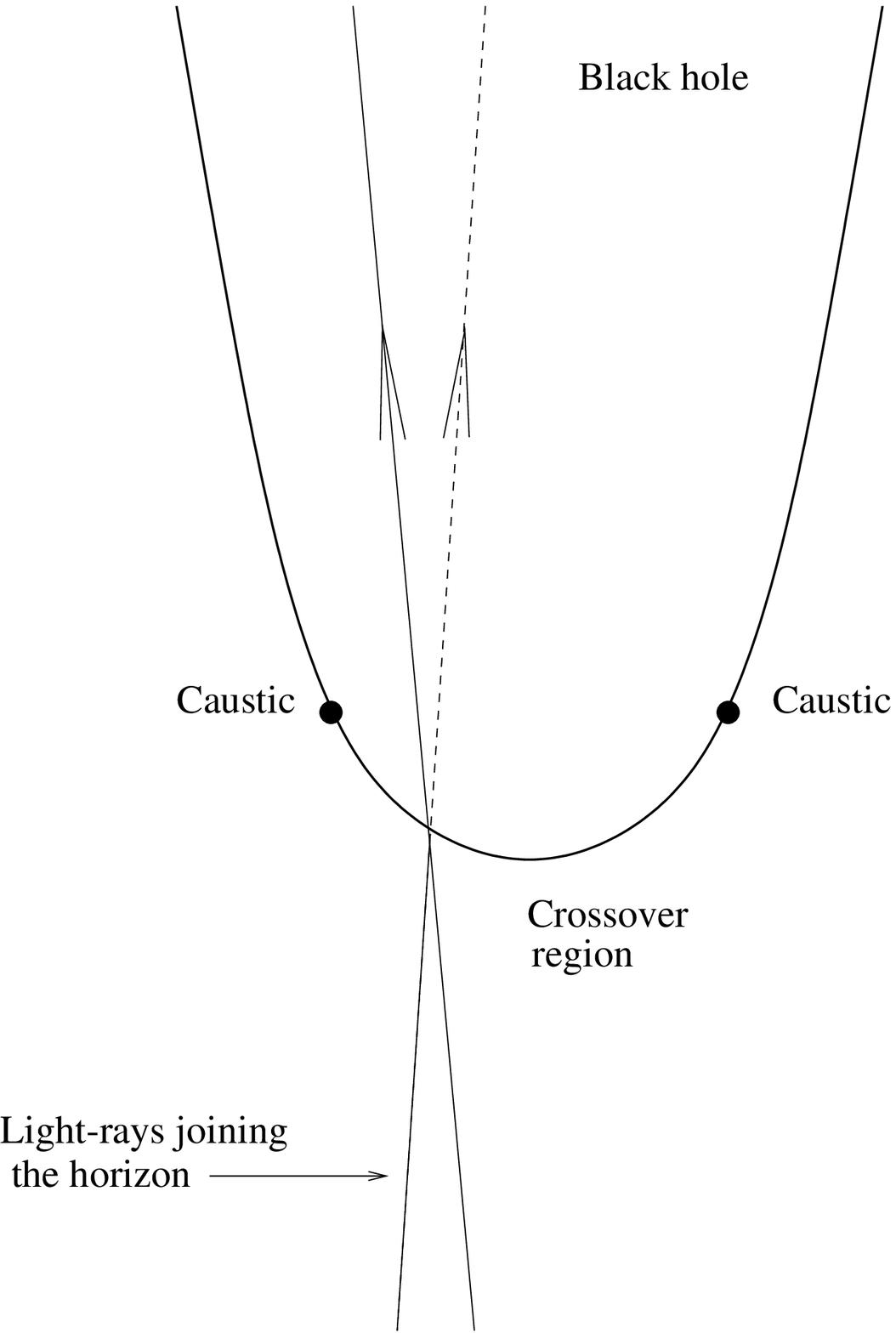,height=3.2in,width=1.8in}}
\hspace{0.2in}
\psfig{figure=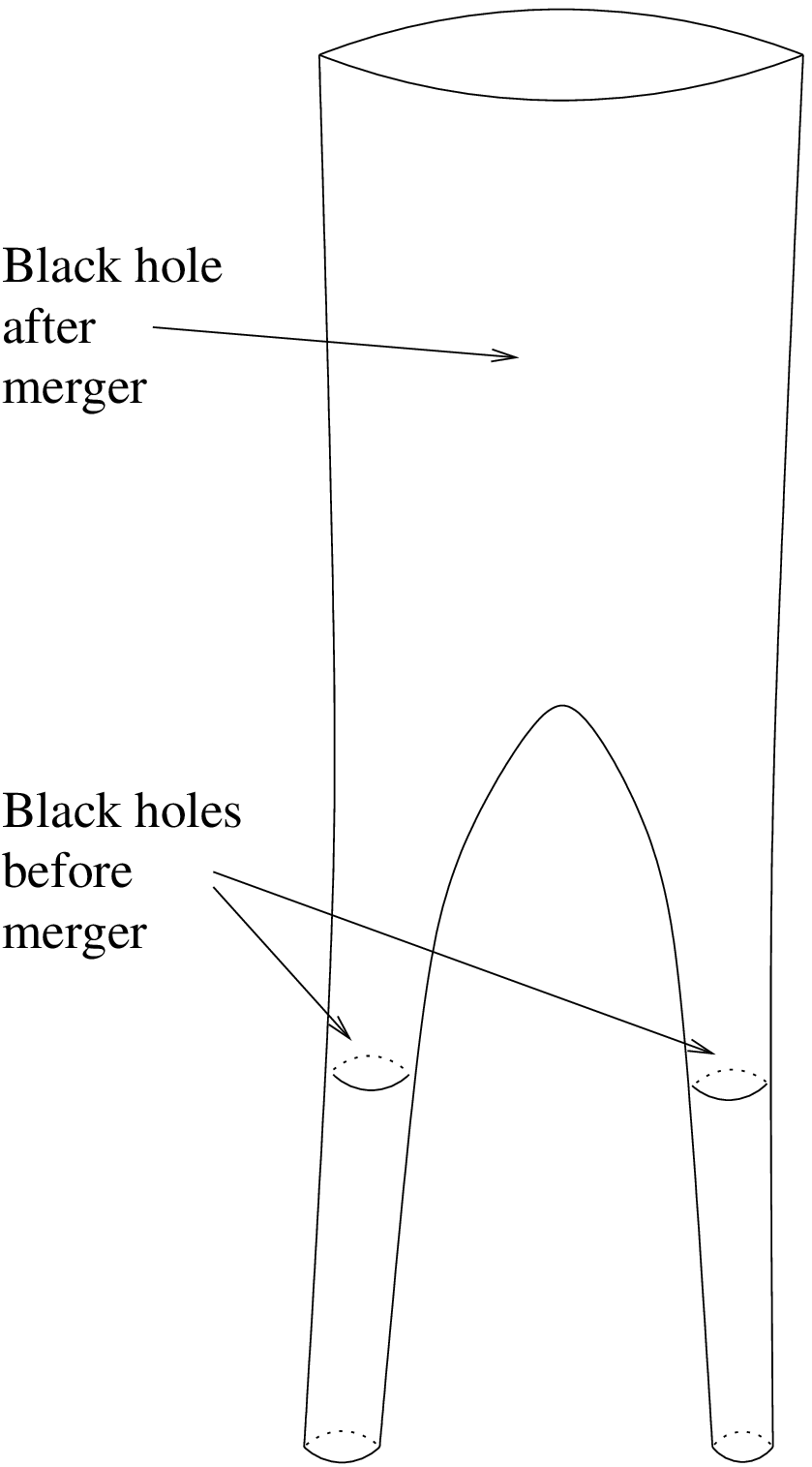,height=3.2in,width=1.8in}}
\caption{On the left: a single black-hole event horizon. 
On the right: head-on collision of black holes.}
\label{fig:null.horiz.null2BH}
\end{figure}

A black hole event horizon is a special null hypersurface with light rays 
emerging from an initial caustic-crossover region, where the horizon forms, and then 
expanding asymptotically to a constant surface area. 
The  caustics that play a role in generic horizon formation correspond to
 elementary  caustics studied in catastrophe optics.
Caustics are 2-dimensional surfaces where focusing of light rays results in infinite intensity.
In the case of event horizons the intensity is finite on the crossover set, where distinct 
light rays traced back on the horizon collide.  
Please refer to Figure~\ref{fig:null.horiz.null2BH} (left)
for an illustration of a black hole event horizon.

\begin{figure}
\centerline{\vbox{
\psfig{figure=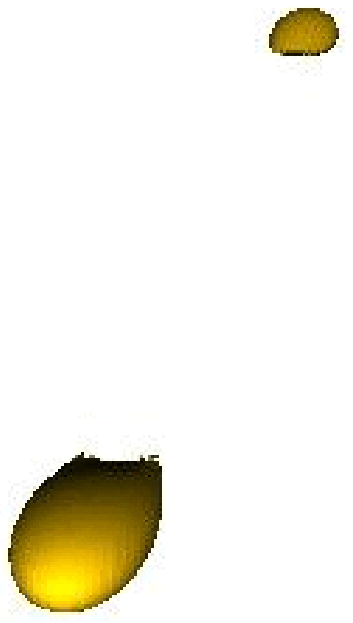,height=1.5in,width=1.2in}
\vspace{0.4in}
\psfig{figure=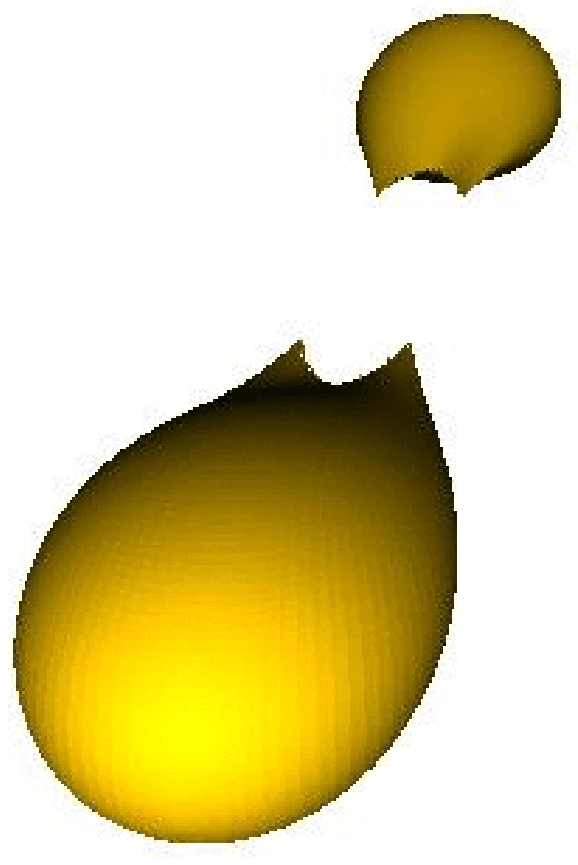,height=1.5in,width=1.2in}
\vspace{0.4in}
\psfig{figure=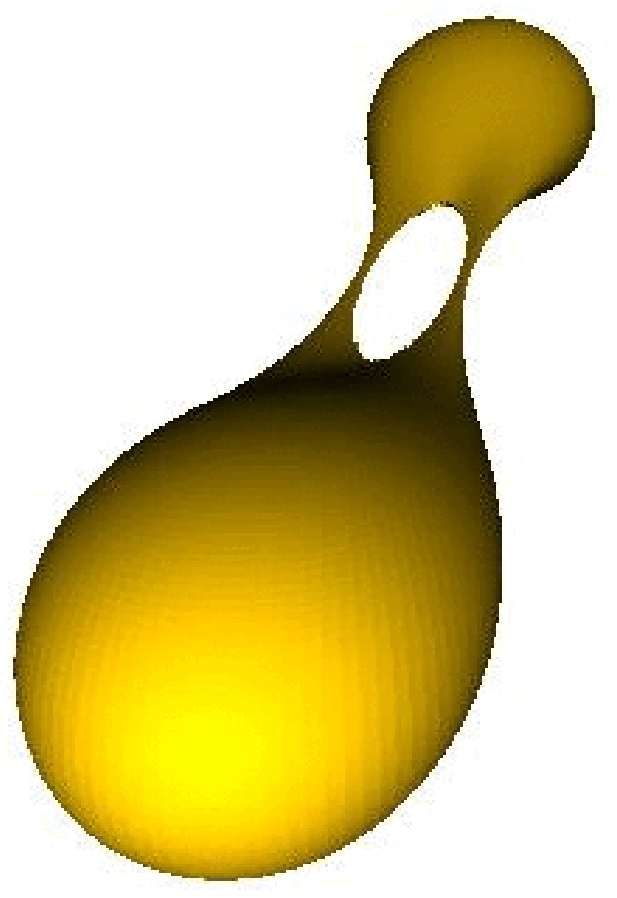,height=1.5in,width=1.2in}}}
\caption{Space-like, 2-D slices of a generic merger event horizon. 
The slice on the bottom shows a late, toroidal phase.}
\label{fig:null.horiz.gen3D}
\end{figure}

In the case of an axisymmetric head-on collision of two black holes (see Figure~\ref{fig:null.horiz.null2BH}, right),
the event horizon has the shape of a trouser where the legs represent the holes before collision.
In the vacuum case these extend forever into the past, although their cross-sectional area becomes vanishingly small
\cite{Lehner99a,Winicour99}. In a different scenario,  
gravitational collapse of a rotating cluster
leads to formation of a torus-shaped horizon which eventually 
closes up forming an asymptotically spherical
black-hole. The toroidal phase is also present in generic black hole mergers.
A sequence of  3-D slices of a generic merger horizon can be seen on
Figure~\ref{fig:null.horiz.gen3D}.

In \cite{Lehner99a} such axisymmetric horizons are constructed as stand-alone objects,
without prescribing the properties of the surrounding space-time. In a following paper \cite{Husa99}
the methodology is extended to the case of more generic event horizons. 
In these papers the direction of time is reversed, 
thus providing geometries of white-hole horizons. 
As stand-alone objects, white-hole horizons correspond to 
black-hole horizons, the difference being the orientation 
of the time-coordinate axis. 
Black hole horizons have an ever increasing 
area while the surface area of white-hole horizons
is continuously decreasing. 
The geometric optics construction  outlined below
is based on the caustics  and
the crossover sets that determine the formation of event horizons.

Let ${\cal S}_0$ be a smooth convex surface embedded in Minkowski space
at constant 
time $\hat t = 0$. 
Let $\hat {\cal H}^-$
be the null surface described by a  light beam that comes  from past null
infinity, crosses ${\cal S}_0$ perpendicularly, and  pinches off
to the future of ${\cal S}_0$. By convention $\hat {\cal H}^-$ ends 
at points 
where two light-rays cross each other. 
These endpoints consist of a set of caustic points
${\cal C}$, where neighboring rays focus, and a set of 
non-focal crossover points ${\cal X}$,
where distinct null rays meet. 

\begin{figure}
\centerline{\epsfxsize=3.5in\epsfbox{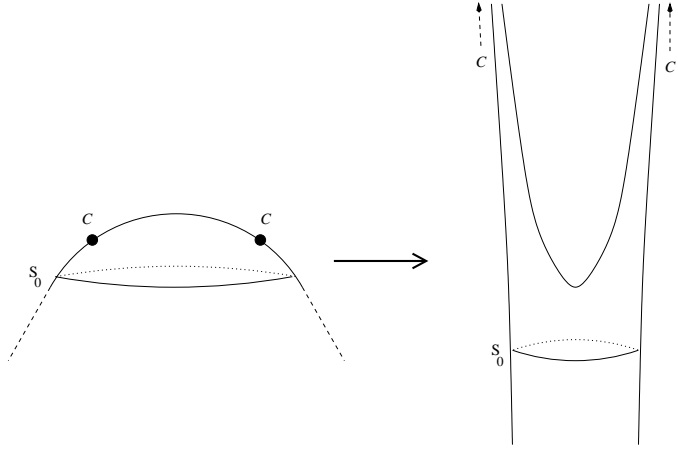}}
\caption{The effect of re-mapping $u = u(\hat u)$.}
\label{fig:null.nullWH}
\end{figure}

Let $\hat \gamma_{ab}$ be the degenerate metric of the 3-dimensional
null surface $\hat {\cal H}^-$ coordinatized by $(\hat u, \theta, \phi)$.
A conformal metric $\gamma_{ab} = \Omega^2 \hat \gamma_{ab}$ is constructed 
which describes the intrinsic geometry of a null hypersurface ${\cal H}^-$
with an affine parameter $u$ along its null rays.   
The dependence $u = u(\hat u)$ is defined by the requirement that 
the projection of the Einstein tensor onto ${\cal H}^-$ vanishes. (Here the
projected Einstein tensor  is computed viewing $\gamma_{ab}$ as
the metric of ${\cal H}^-$ in the frame $(u,\theta,\phi)$.
 In addition the conformal factor $\Omega^2$
is chosen such that the resulting conformal null hypersurface ${\cal H}^-$ is smooth and its 
radius $R$  has a finite limit $R_\infty$ as $u \rightarrow - \infty$.
Once $\Omega$ is chosen the remaining gauge 
freedom in $u$ can be fixed
by setting $u = \hat u = u_0$ on ${\cal S}_0$. 
The interesting feature of this
mechanism is that in this new parametrization of ${\cal H}$
 the caustic points ${\cal C}$ will not
be reached by any null rays before $u \rightarrow
\infty$ while crossover points belonging to ${\cal X}$ at any finite distance from
the caustic points will be reached by null rays at a finite $u$.
As it can be seen on Figure~\ref{fig:null.nullWH}, a trouser-shaped 
null hypersurface is generated that represents the event-horizon
of a fissioning white hole. In the reversed time-direction, it is
a horizon of two merging black holes.

\subsection{Evolution of exterior geometry}

A prime application of the conformal horizon
model is the calculation of the radiation emitted by colliding black holes
\cite{Winicour99}. 

\begin{figure}
\centerline{\psfig{figure=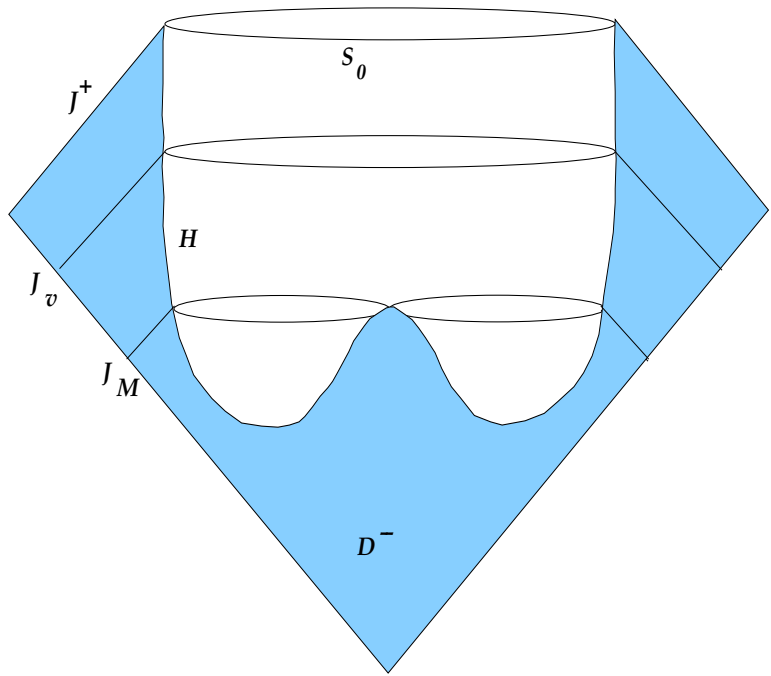,height=3.5in,width=4in}}
\caption{Space-time evolution from merger to ring-down via a time-reversed 
characteristic evolution.}
\label{fig:null.horizon-evol}
\end{figure}

The time-reversal of the bifurcating white-hole horizon ${\cal H}^-$ is 
a horizon of colliding black holes, ${\cal H}$. This 3-dimensional null
hypersurface serves as inner boundary for a characteristic evolution.
The evolution is carried out along a family of ingoing null hypersurfaces
$J_v$
which intersect the horizon in topological spheres.  The evolution 
is restricted to the period from merger to ring-down, or else the
inner boundary would be 
intersected by ingoing null hypersurfaces in disjoint pieces.
The evolution proceeds backward.  The first ingoing null hypersurface
$J^+$ intersects ${\cal H}^+$ in ${\cal S}_0$, 
as illustrated in Figure~\ref{fig:null.horizon-evol}.
Locating ${\cal S}_0$
at a late quasi-stationary time implies that the ingoing null
hypersurface $J^+$ approximates  ${\cal I}^+$.
Thus the conformal model of the horizon of a black hole 
collision can be used as characteristic initial data to construct
a vacuum space-time covering a very interesting 
nonlinear domain from merger to ring-down.

\chapter{Cauchy-Characteristic Matching}
\label{chap:ccm}

The characteristic description of space-time
 has proven to be a very appropriate
framework for evolving gravitational radiation from some timelike world-tube
$\Gamma$ to future null infinity ${\cal I}^+$. The drawback of the approach is
in its inability to treat space-time regions where refocusing occurs for the
 null geodesics that define its coordinate system.
By implication, the problem of
two coalescing black holes will not be solved by solely characteristic
evolution. In a  Cauchy algorithm, coordinates are chosen arbitrarily
so that caustics do not necessarily present a coordinate problem.
However, a Cauchy algorithm based upon a spacelike foliation
cannot be defined globally and requires an outer boundary where
an artificial boundary condition can introduce spurious back
radiation that contaminates the evolution.

The methodology of Cauchy-Characteristic Matching (CCM)  
adopted here is one that is very much proper to 
general relativity: that of describing space-time with multiple coordinate
patches. CCM is designed to handle the interface between a Cauchy
and a characteristic patch \cite{Bishop90,Bishop93}
in a scenario where the black-hole dynamics is described in the interior 
of a timelike world-tube $\Gamma$ by a
Cauchy code while the radiation is carried from the world-tube 
$\Gamma$   to future null infinity ${\cal I}^+$ 
by a characteristic code.

CCM has been successfully applied in a number of instances. The first of
these was the problem of non-linear scalar waves propagating 
in a 3-D Euclidean space \cite{Bishop97b}. Performance for 
the matching algorithm was compared with local and nonlocal radiation
boundary conditions proposed in the computational physics literature. For
linear problems CCM outperformed all local boundary conditions and
was about as accurate (for similar grid resolution)
 as the best nonlocal conditions. In terms of computational expense
CCM has far outperformed the nonlocal approaches. The same holds true with
regard  to accuracy in the non-linear regime, where CCM turned out
to be significantly more accurate than all other methods tested.

In another instance, a two-fold version of CCM has been used to evolve 
globally the spherical collapse of a self-gravitating scalar field onto
a black hole \cite{Gomez97b}. The black hole (with excised singularity)
 and its surrounding region
were described by an ingoing null foliation. This patch was matched to 
a Cauchy region, which in turn was matched to an 
exterior, outgoing characteristic
code carrying out the radiation to ${\cal I}^+$. 
Although it has spherical symmetry, the model has shown the fitness of 
the multi-patch approach (including CCM)
to deal with singular space-times.

The Southampton
group has worked out the formalism of CCM
for the case of Einstein equations with axial symmetry
\cite{Clarke95,Dubal95,dInverno96,dInverno97,Pollney00}. 
However, a numerically stable implementation 
of the axisymmetric general relativistic
CCM is yet to be borne out.

In the following sections 
the principles of CCM are described
-- first for a scalar field \cite{Bishop97b}, 
then for the case of general relativity \cite{Bishop98a}.
The two modules of the general relativistic
CCM -- extraction and injection -- are described 
in further detail in the following chapters.

\section{Cauchy-characteristic evolution in a flat background}

Let $\Psi$ be a smooth solution of the scalar wave equation. 
In Cartesian coordinates $(t,x,y,z)$ it satisfies
\begin{equation}
\partial_{tt} \Psi= \left(\partial_{xx}+\partial_{yy}+\partial_{zz}\right) 
\Psi.
\label{eq:matching.swe.3d.Cauchy}
\end{equation}
In stereographic spherical coordinates $(r, q, p)$
(see page \pageref{page:stereo-coord})  and a retarded
time coordinate $u=t-r$ the field equation 
(\ref{eq:matching.swe.3d.Cauchy}) takes the form
\begin{equation}
2 \partial_{ur} \psi = \partial_{rr}\psi +
\frac{P^2}{4 r^2} \left( \partial_{qq} + \partial_{pp} \right) 
 \psi,
\label{eq:matching.swe.3d.Null}
\end{equation}
where $\psi = r \Psi$ and $P = 1 + q^2 + p^2$.
In CCM, Eq.~(\ref{eq:matching.swe.3d.Cauchy}) is evolved up to some
radius $R_m$ using a Cauchy algorithm, while a characteristic algorithm
integrates Eq.~(\ref{eq:matching.swe.3d.Null}) for $r \geq R_m$.  
The matching procedures ensure that, in the continuum limit, $\Psi$ and its 
derivatives are continuous across the spherical boundary $r=R_m$ so that
spurious back reflection does not occur.

\subsection{The spherically symmetric case}

\subsubsection{Analytic discussion}

For the simple case of spherical symmetry 
the Cauchy version of the wave equation (\ref{eq:matching.swe.3d.Cauchy}) reduces to
\begin{equation}
\partial_{tt} \psi = \partial_{rr} \psi,
\label{eq:matching.swe.1d.Cauchy}
\end{equation}
while the characteristic version (\ref{eq:matching.swe.3d.Null})
becomes
\begin{equation}
2 \partial_{ur} \psi = \partial_{rr} \psi.
\label{eq:matching.swe.1d.Null}
\end{equation}

\begin{figure}
\centerline{\epsfxsize=3.5in\epsfbox{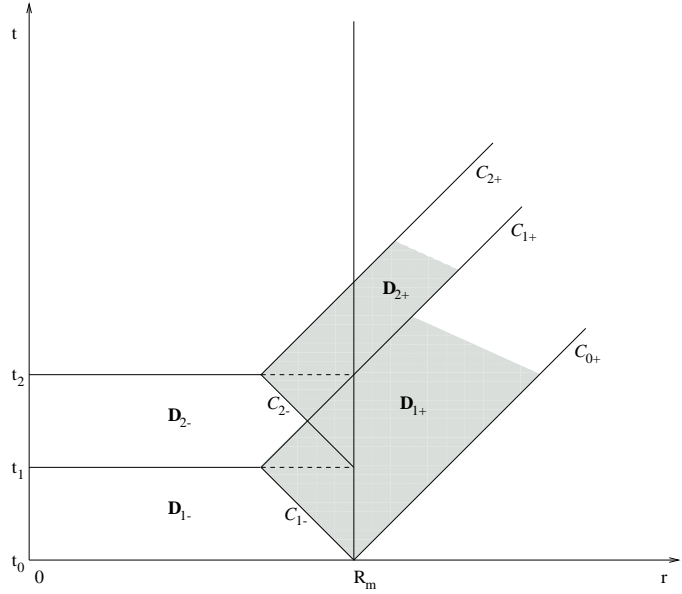}}
\caption{The matching scheme for the spherically symmetric case.
The symbols ${\bf D}_{1\pm}, {\bf D}_{2\pm}$ stand for domains of dependence
while $C_{0+}, C_{1\pm}, C_{2\pm}$ stand for outgoing 
(ingoing) characteristics.}
\label{fig:matching.swe.1d.anal}
\end{figure}

The evolution algorithm proceeds schematically as follows 
(see Figure~\ref{fig:matching.swe.1d.anal}):
\begin{itemize}
\item At the setup of the evolution one needs to provide initial
data in the domain $t=t_0,r\le R_m$ and on the outgoing characteristic
$C_{0+}$.
\item Using the Cauchy initial data the evolution  proceeds
throughout the domain of dependence ${\bf D}_{1-}\,$, determining
the data on the ingoing characteristic  $C_{1-}\,$.
\item Using the initial data $C_{0+}$ and the boundary data $C_{1-}$
the characteristic evolution determines data in the 
domain ${\bf D}_{1+}\,$. This  allows completion of 
the Cauchy data at $t=t_1$ all the way out to $r=R_m$. 
\item These steps can now be iterated: Cauchy data is now available to evolve
throughout the domain ${\bf D}_{2-}$ including  $C_{2-}$,
which determines the characteristic data to evolve
through  ${\bf D}_{2+}$, etc.
\end{itemize}

\subsubsection{Finite differencing}

In the discretized version of Cauchy-characteristic evolution the
crisscross pattern of characteristics inside the radius $R_m$ is at the
scale of grid spacing.
 The matching algorithm in this case is a 
 cross-grid interpolation scheme which in the continuum  limit
makes the sphere $R_m$ transparent to the wave propagation.

The Cauchy equation~(\ref{eq:matching.swe.1d.Cauchy})
is discretized using a standard second-order 
finite-difference scheme:
\begin{equation}
\frac{\psi^{n+1}_{i}-2\psi^{n}_{i}+\psi^{n-1}_{i}}{(\Delta t)^2} =
\frac{\psi^{n}_{i+1}-2\psi^{n}_{i}+\psi^{n}_{i-1}}{(\Delta r)^2}
+ O(\Delta^2).
\label{eq:matching.swe.1d.discrete}
\end{equation}
The characteristic evolution uses the null parallelogram algorithm
(see Section~\ref{sec:null.eveq}).

\begin{figure}
\centerline{\epsfxsize=4in\epsfbox{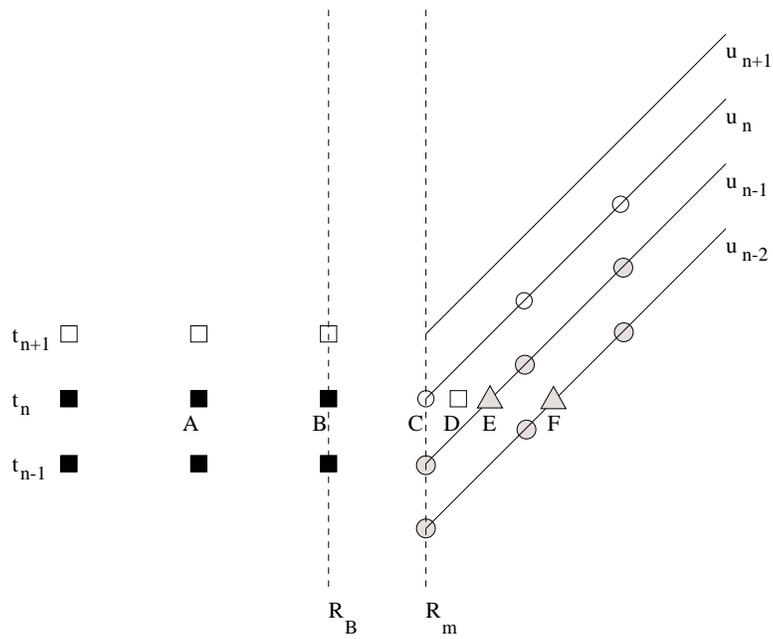}}
\caption{The matching interpolation scheme for the spherically symmetric case.
The points $A,B,D$ are Cauchy gridpoints with $D$ at the boundary.
The point $C$ is  the boundary gridpoint of the outgoing characteristic 
$u_{n}$. Points $E$ and $F$ are at the intersection of 
the Cauchy time level $t_n$ and the outgoing characteristics 
$u_{n-1}, u_{n-2}$.
}
\label{fig:matching.swe.1d.interp}
\end{figure}

Figure~\ref{fig:matching.swe.1d.interp} is 
a diagram illustrating the interpolation scheme.
The radius $R_m$ is defined by $R_m - R_B = \kappa \Delta r$,
where $\kappa$ is an arbitrary parameter.
Although in the spherically symmetric case the introduction of $\kappa$ is 
not crucial, it becomes important in higher dimensions.
In 3-D the Cauchy domain is represented by a
Cartesian grid and the characteristic domain 
is described by a spherical grid-structure.
By implication the boundary gridpoints of the two evolution domains do not
coincide. Thus it is important that the 1-D matching algorithm works
for a range of $\kappa$ that leaves room for applying a generalization
 of the same algorithm for 3-D non-aligned  grid-boundaries.
Given  $\Psi$ 
at all Cauchy points on level $t_n$ except
for the boundary point $D$ and at all characteristic gridpoints
up to level $u_{n-1}$, the field values $\Psi_E$ and $\Psi_F$
are obtained 
by radial interpolation along the null characteristics $u_{n-1}$ 
and $u_{n-2}$. Next the characteristic and Cauchy
boundary values $\Psi_C$ and $\Psi_D$ are computed by
interpolation along  the Cauchy surface $t_n$. 
With new boundary values provided, the evolution schemes 
can proceed to
update points at the next levels $t_{n+1}$ and $u_n$.
Interpolators
are quadratic such that the resulting error is $O(\Delta^4)$.
This will assure second-order accuracy for the overall evolution
scheme. 
The algorithm is numerically stable for a wide range of gap
sizes $0 \le \kappa \le 2$.

\subsection{The 3-D case}
\label{sec:matching.SWE.3d}

In a 3-dimensional application, 
the Cauchy field is represented on a 
Cartesian grid that consists
of points equally spaced in all directions:
\begin{equation}
x_i = -a + (i-1) h, \;\; y_j = -a + (j-1) h, \;\; 
z_k = -a + (k-1) h,
\end{equation}
($1 \le i, j, k \le 2 M$) where $h = 2 a / (2 M -1)$.
The angular and radial dimensions in 
the characteristic grid are discretized in the same fashion
as for the 
GR characteristic code (see 
Section~\ref{sec:null.coordinates}).
The inner boundary
of the characteristic grid is a spherical shell of radius $R_m = a - h/2$, 
centered at the origin of Cartesian coordinates. Cartesian gridpoints
are classified according to their position with respect to the 
boundary sphere of radius $R_m$. Points  inside the sphere are called
{\em interior} or {\em evolution points\/}; nearest neighbors
of evolution points which are on or outside the sphere $R_m$ are called
{\em boundary points}. The remaining gridpoints are not used.

The Cauchy evolution algorithm is the 3-D extension of
Eq.~(\ref{eq:matching.swe.1d.discrete}). The characteristic 
algorithm is a 3-D null parallelogram algorithm. 

As in the spherically symmetric case, the 3-D 
matching algorithm reduces
to a cross-grid interpolation scheme. Generalizing 
Figure~\ref{fig:matching.swe.1d.interp} into three dimensions,
the points $C,E,$  and $F$ are spheres at level $t_n$, with
radii $R_C = R_m, \;\; R_E = R_m + K \Delta t, \;\; R_F = R_m + 2K \Delta t$,
while $A,B$ and $D$ stand for Cartesian grid-points.
Field values on the outer
two spheres $E$ and $F$ are obtained  from interpolation along  characteristic
hypersurfaces. Next follows a 3-D 
 routine that updates the Cartesian boundary points $E$
and the sphere $C$ using 
a quadratic interpolation scheme between 
the outer spheres $E,F$ and the Cartesian points $A,B$. 
(This 3-D
interpolation scheme is described 
in detail in Section~\ref{sec:ladm-sph.SWE}.)

The parameter $K$ must be chosen to  assure that the field
values at point $D$ are provided by interpolation instead of
extrapolation. In \cite{Bishop97b} the choice 
\begin{equation}
K = [h/2 \Delta t] + 1
\end{equation}
is used, where the brackets denote the integer part.

It should be noted that in \cite{Bishop97b}
an external source-term and a non-linear self-coupling term was added to 
the right-hand-side of
the scalar wave equation (\ref{eq:matching.swe.3d.Cauchy}) and 
(\ref{eq:matching.swe.3d.Null}) without degrading
the performance of CCM.

\section{Cauchy-characteristic evolution in general relativity}

Having described the architecture of CCM for the case of a
scalar wave in
a flat background, the following section examines the 
application to general relativity.
The algebraic details
of CCM are  discussed in  Chapters \ref{chap:extraction}
and~\ref{chap:injection}. Here it is presented 
from a more geometric point of view.
Once CCM is understood on this geometric level,
the details become more transparent.

\begin{figure}
\centerline{\epsfxsize=4.2in\epsfbox{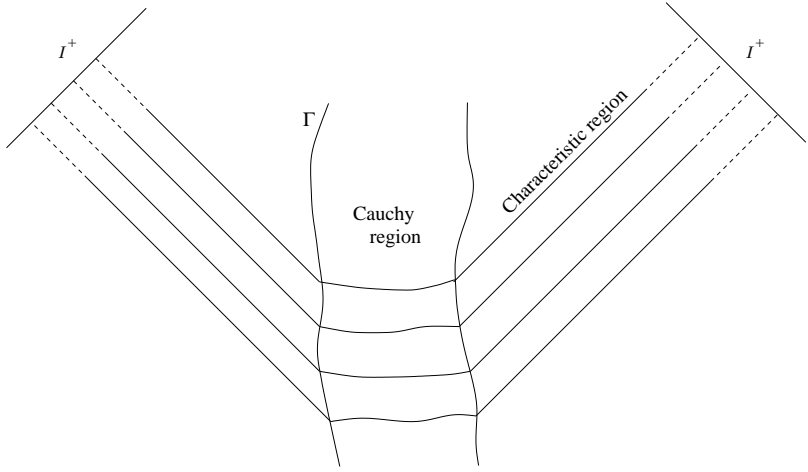}}
\caption{A sketch of the matching scheme for the case of
general relativity. The inside of the timelike world-tube
$\Gamma$ is evolved by a Cauchy code while its exterior
is evolved by a characteristic code.}
\label{fig:matching.GR.sketch}
\end{figure}

The main concept of CCM for  general relativity is 
the same as for other field equations: a timelike world-tube is
used to match  between a Cauchy and a characteristic region.
The two evolution codes provide boundary data for each other, similar
to the flat-background scalar wave case.

There are two major new features that appear only 
when evolving the Einstein equations 
 (as opposed to evolving some scalar field).
One is  a nontrivial, dynamic coordinate transformation
that has to be done numerically 
time-step after time-step.
The other is the  
consistency between  boundary conditions and
the coupled tensorial field equations. 
The first issue is discussed in the remainder of this chapter
with further details in Chapters \ref{chap:extraction}
and~\ref{chap:injection}, while the second issue
is addressed in
Chapters \ref{chap:ladm-cart} and \ref{chap:ladm-sph}.

CCM proceeds as follows:
Let $t^\alpha$ be a timelike vector field.
Using ``3+1'' slicing one can define the Cartesian coordinates $(t,x,y,z)$ 
in a region of the space-time. 
Let $\Gamma$ be a timelike world-tube defined by
$ (x^2+y^2+z^2)_{|\Gamma} = R_\Gamma^2$, with $R_\Gamma^2$ constant.
The coordinates $(x,y,z)$ induce a 
natural 
stereographic coordinate system $(q,p)$ on $\Gamma$ by the transformation
\begin{eqnarray}
q(x^i) = \frac{x}{R_\Gamma \pm z}, \;\;\; p(x^i) = \pm \frac{y}{R_\Gamma\pm z}
\label{eq:matching.qp=fxi}
\end{eqnarray}
on the north $(+)$ and south $(-)$ patches. These coordinates
are related to the
complex stereographic coordinate $\zeta$ used in the characteristic code
(see Chapter \ref{chap:nullcode})
by $\zeta = q + I \cdot p$. 
This completes the ``2+1'' description of the timelike world-tube $\Gamma$
and  the necessary ingredients 
 to construct an outgoing 
characteristic coordinate system $y^\beta = (u,r,y^A)$.\footnote{Here and in
further instances the notation $y^A$ denotes
angular coordinates: $y^2 = q, y^3 = p$.} The retarded time-coordinate
$u$ is fixed by $u_{|\Gamma} = t_{|\Gamma}$, while $r$ stands
for the Bondi surface-area coordinate (see Section~\ref{sec:null.coordinates}).
The coordinates $y^\beta$  provide a description of the
exterior of $\Gamma$, provided  light rays leaving
$\Gamma$ in the outgoing normal direction do not refocus.

The coordinatization of the Cauchy region inside $\Gamma$ is largely
arbitrary while the description of the outside
characteristic region is uniquely determined by the choice of lapse and shift
in the parametrization of $\Gamma$.
This fundamental difference between the two coordinate systems explains why
the issue of coordinate transformation at $\Gamma$ is
a nontrivial problem.
If one recalls that the characteristic
coordinates are radially aligned to null-geodesics and 
that the light-ray paths are defined by the (dynamic)
space-time metric, one can easily see that the 
Jacobian determining the coordinate transformation
involved in CCM is dynamic as well.

The CCM algorithm for general relativity 
consists of the following two parts:
\begin{itemize}
\item {\em Extraction.} Cauchy data is passed onto the characteristic
evolution. This involves a coordinate transformation from Cartesian to
Bondi coordinates.
\item {\em Injection.} Characteristic data is passed onto the
Cauchy evolution. This involves a coordinate transformation
  from Bondi to Cartesian coordinates.
\end{itemize}

\subsection{Extraction}
\label{sec:matching.extraction}

In extraction, the coordinate transformation 
must be carried out
 in a
neighborhood of the world-tube, not just  on it.
This is because  the surfaces $\Gamma_t$
do not, in general, correspond to 
surfaces of constant Bondi $r$.
Thus $\Gamma$ 
typically intersects characteristic radial gridlines between gridpoints. 
Since the overall desired computational accuracy 
is $O(\Delta^2)$ and since the distance between $\Gamma_t$ and 
nearest radial gridpoints is $O(\Delta)$, one needs
to compute  not only the null metric on $\Gamma$ but its
 radial derivatives as well.
The  transformation between  Cartesian and 
Bondi coordinates is done first by transforming from Cartesian
to affine null coordinates, then from affine null to Bondi coordinates.

Among  the building elements of the Jacobian between the Cauchy and
 affine null coordinates  are the generators $\ell^\alpha$ of the
outgoing null geodesics that are normal to $\Gamma_t$.
The vector $\ell^\alpha$ 
can easily be computed at the world-tube 
using the Cartesian metric, lapse, and shift:
\begin{equation}
\ell^\alpha_{|\Gamma} = 
\left( \frac{n^\alpha + s^\alpha}{\alpha - {}^{(3)}g_{ij} 
\beta^i s^j} \right)_{|\Gamma},
\label{eq:matching.elldef}
\end{equation}
where $s^\alpha$ is a spatial unit vector that is
 normal  to $\Gamma_t$ in the outward direction,
while $n^\alpha$ is a timelike vector field normal
to the $t=$ constant Cauchy slice $\Sigma_t$. 
The denominator in Eq.~(\ref{eq:matching.elldef})
is determined by the normalization condition
 $\ell^\alpha t_\alpha = -1$.
In the flat space-time case, using Minkowski coordinates, the components 
of the null vector at the world-tube
 are $\ell^\alpha_{|\Gamma} = (1, \frac{x^i}{R_\Gamma})$.

The  null geodesics leaving the world-tube along $\ell^\alpha$ are
described in the neighborhood of $\Gamma$, in Cartesian coordinates, 
by
\begin{equation}
x^\alpha = x^{(0)\alpha }+ \ell^{(0)\alpha} \lambda + 
 \ell^{(0)\alpha}_{,\lambda} \lambda^2 + 
O(\lambda^3),
\end{equation}
where the notation 
$x^{(0)\alpha} \equiv x^\alpha_{|\Gamma}$, etc. is employed.
The $\lambda$-derivative of the null vector $\ell^\alpha$ is computed
using the geodesic equation
\begin{equation}
\ell^{\alpha}_{,\lambda}
+ \Gamma^{\alpha}_{\mu \nu}
\ell^{\mu}\ell^{\nu} = 0,
\label{eq:matching.geodesic-equation}
\end{equation}
where
\begin{equation}
\Gamma^{\alpha}_{\mu \nu} =
\frac{1}{2} \;g^{\alpha \beta} \left [
\;g_{\mu \beta, \nu} + \;g_{\nu \beta, \mu} 
- \;g_{\nu \mu, \beta}
\right ].
\label{eq:matching.gamma-def}
\end{equation}

The metric in affine null coordinates 
$\tilde y^{\tilde \alpha} = (u, \lambda, y^A)$ 
is computed using
\begin{equation}
\tilde \eta _{\tilde \alpha \tilde \beta} =
\frac{\partial x^\mu}{\partial \tilde y^{\tilde \alpha}}
\frac{\partial x^\nu}{\partial \tilde y^{\tilde \beta}}\; 
g_{\mu \nu}.
\end{equation}
Given the coordinate transformation 
$x^{\mu} = x^\mu(\tilde y ^{\tilde \alpha})$ 
up to $O(\lambda^3)$, the Jacobian (which includes 
$\lambda$-derivatives of $x^\alpha$) is known around the
world-tube up to $O(\lambda^2)$.

The second  transformation
from affine coordinates $(u,\lambda,q,p)$ to 
Bondi coordinates $y^\alpha = (u, r, q, p)$ can be understood as follows:

The area $dA$ of a 2-D surface-element $dq \cdot dp$ 
is determined by the the square root of the
determinant of the angular metric 
$\tilde \eta_{\tilde A \tilde B}$. This quantity 
clearly depends on the radial coordinate. (E.g. in the case of 
flat space-time and the usual $(r, \theta, \phi)$ coordinates
the angular metric is $\mbox{diag}(r^2, r^2 \sin^2 \theta))$.
Let $dA_1$ stand for the area  $dq \cdot dp$ on the unit sphere.
The Bondi surface area coordinate $r$ is defined as the square root
of the ratio $dA/dA_1$, i.e.:
\begin{equation}
r = \left( \frac{\det \left(\tilde \eta_{\tilde A \tilde B} \right)} 
{\det \left(q_{A B} \right)}\right)^{\frac 14}
= \frac{P}{2} \left(\det \left(\tilde \eta_{\tilde A \tilde B} 
\right)\right)^{\frac 14}.
\label{eq:matching.rBondi}
\end{equation}
where we have used Eq.~(\ref{eq:null.qABdef}).
The derivative of the surface-area coordinate with respect to the
affine parameter along a null geodesic labeled by some $(q,p)$ 
is related to the expansion of the light-rays, i.e. 
\begin{equation}
r_{,\lambda} = e^{-2 \beta}.
\end{equation}

Equation~(\ref{eq:matching.rBondi})
provides  $r$ around the world-tube up to $O(\lambda^2)$ -- the
same order to which the metric 
elements $\tilde \eta_{\tilde \alpha \tilde \beta}$ are computed.
Computing the Bondi metric $\eta_{\alpha \beta}$
 involves terms with $r_{,\lambda}$. For this
reason, without additional information, 
the radial derivative of the Bondi metric cannot be computed on $\Gamma$.
The missing information $r_{,\lambda \lambda}$ 
is obtained using the $\beta$-hypersurface equation (\ref{eq:null.betaeq-full})
and the relation
\begin{equation}
\beta_{,\lambda} = - \frac{\eta^{ru}_{,\lambda}}{2 \eta^{ru}} = 
- \frac{r_{,\lambda \lambda}}{2 r_{,\lambda}}.
\label{eq:matching.betaleq}
\end{equation}
For further details see Chapter \ref{chap:extraction}.

As a last step, given  the Bondi metric $\eta_{\alpha \beta}$
in the neighborhood of $\Gamma$ up to $O(\lambda^2)$
one needs to read off the Bondi functions $J, \beta, U, W$
at the world-tube, along with their radial derivatives. This can be done
immediately given  the definition of these functions 
($U \equiv q_A U^A, J \equiv q^A q^B h_{AB}, W \equiv  (V-r)/r^2$), and
the form of the contravariant Bondi metric:
\begin{equation}
\eta^{\alpha\,\beta} =
\left [
\begin {array}{cccc}
0 &
-e^{-2\,\beta} &
0 &
0 
\\
\noalign{\medskip}
-e^{-2\,\beta} &
 e^{-2\,\beta} \displaystyle{\frac{V}{r}} &
-e^{-2\,\beta} U^{2} &
-e^{-2\,\beta} U^{3}
\\
\noalign{\medskip}
0 &
-e^{-2\,\beta} U^{2} &
 r^{-2}\, h^{22} &
 r^{-2}\, h^{23}
\\
\noalign{\medskip}
0 &
-e^{-2\,\beta} U^{3} &
 r^{-2}\, h^{32} &
 r^{-2}\, h^{33}
\\
\noalign{\medskip}
\end {array}
\right] .
\label{eq:matching.bondimetric}
\end{equation}

\subsection{Injection}

With  extraction providing the coordinate transformation
in a neighborhood of $\Gamma$, injection
does not involve much additional algebra. 
When reversing the coordinate transformation
(i.e. from Bondi coordinates $y^\beta = (u,r,q,p)$  to Cartesian 
coordinates $x^\alpha = (t,x,y,z)$) one can invert the 
Jacobian and obtain the necessary information.
 An even simpler approach is to compute the 
contravariant Cartesian metric $g^{ij}$,
which involves the same Jacobian matrix as needed for computing the
covariant Bondi metric (see \cite{Wald84}).
The reverse coordinate transformation is performed on 
a set ${\cal S}_{Null}$ of 
null gridpoints in the neighborhood of the world-tube $\Gamma$.
The position of these points is fixed in Bondi coordinates.
However, due to the dynamic nature of the coordinate transformation,
the same set of points has a time-dependent location in the
Cartesian frame. Thus,
in addition to computing the Cartesian metric components 
on  ${\cal S}_{Null}$,
these gridpoints are labeled by values $x^\alpha = f(y^\beta)$ computed
as a Taylor expansion around $\Gamma$.

The next problem is purely numeric:
we need to transfer  data from ${\cal S}_{Null}$
to the set of  
boundary points of the Cartesian grid, ${\cal S}_{Cauchy}$. 
Similar to the
case of  scalar matching, this is done  
in two steps: first interpolate metric information onto a spherical
grid contained in $\Sigma_t$ 
that surrounds the set of Cartesian points ${\cal S}_{Cauchy}$, then use
a 3-D interpolation algorithm to transfer the boundary information
from this spherical grid onto the Cartesian boundary points.

Let $\Lambda$ be a Cartesian spherical world-tube
 with (constant) radius $R_\Lambda = R_\Gamma + O(\Delta x)$, concentric with
$\Gamma$.
Let $\Lambda_t$ be the intersection of $\Lambda$ with 
the  Cauchy slice $\Sigma_t$. 
The injection world-tube $\Lambda$ must be chosen such that 
the sphere $\Lambda_t$ surrounds ${\cal S}_{Cauchy}$ at all $t$.

Information needs to be transferred from ${\cal S}_{null}$ to 
$\Lambda_t$, then from $\Lambda_t$ to ${\cal S}_{Cauchy}$.
The elements of  ${\cal S}_{null}$ are labeled
by  their Bondi $(u,r, y^A)$  and Cartesian $(t,x,y,z)$ coordinates. 
They form a 
subset of a regular Bondi grid. In terms of their Cartesian coordinates, 
however, they form an irregular 4-D
grid which changes from time-step to time-step.

\begin{figure}
\centerline{\epsfxsize=4.2in\epsfbox{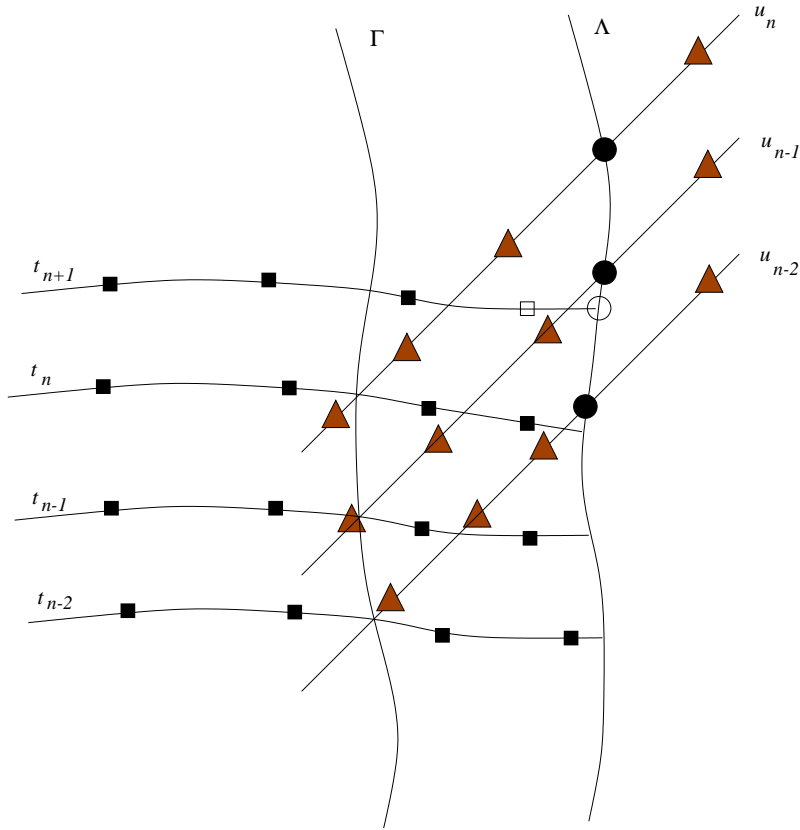}}
\caption{A sketch of the interpolation scheme for the the injection.
Cauchy evolution and boundary points are represented by 
filled and empty squares.  
Null gridpoints are represented by triangles. The filled
circles stand for points on $\Lambda$ where 
the Cartesian metric has been computed from null 
data at neighboring gridpoints.
The empty circle indicates the sphere $\Lambda_t$ where one needs to obtain
Cauchy boundary data by time interpolation.}
\label{fig:matching.GR.injection}
\end{figure}

In order to get data onto the sphere $\Lambda_t$, 
first data is interpolated onto the world-tube
$\Lambda$ from neighboring null gridpoints.
Then a time-interpolation is performed to determine 
Cauchy boundary data on $\Lambda_t$. Finally, a 3-D interpolation 
is performed within the Cauchy slice $\Sigma_t$ in order
to obtain the Cauchy metric at the boundary-points ${\cal S}_{Cauchy}$
of the Cartesian grid.
The interpolation algorithm and the algebra
employed in the injection are described in
detail in Chapter~\ref{chap:injection}.
An illustration of the interpolation schemes involved in the injection
can be seen in Figure~\ref{fig:matching.GR.injection}.

\chapter{The Extraction Module}
\label{chap:extraction}

This chapter describes  the extraction module, a
 numerical algorithm designed to extract characteristic boundary data
from a Cauchy evolution domain.

The first step in the extraction module is getting data from the Cauchy grid
onto the world-tube $\Gamma$ by means of an  interpolation algorithm. Then
a lengthy algebraic calculation is performed to determine the Bondi
variables $J, \beta, U, W$ and their radial derivatives on $\Gamma$. 
During extraction, the world-tube 
 $\Gamma$ is located in Bondi coordinates 
and then the set of characteristic boundary gridpoints is defined.
Finally, the functions 
$J,\beta,U$ and $W$ are  placed onto these boundary points.
(The characteristic evolution proceeds from these points to
future null infinity ${\cal I}^+$.)
A detailed description of these steps is given in the following sections.
The latter part of the chapter is devoted 
to calibration tests and their results.

\section{Interpolation schemes}

There are several ways to transfer data from
a Cartesian grid to a spherical grid. Issues that need to 
be considered include desired accuracy, numerical 
stability and computational efficiency.
In addition, filtering techniques might be 
useful for producing smooth output.

\subsection{Cubic, 3-D interpolator}

\begin{figure}
\centerline{\epsfxsize=4in\epsfbox{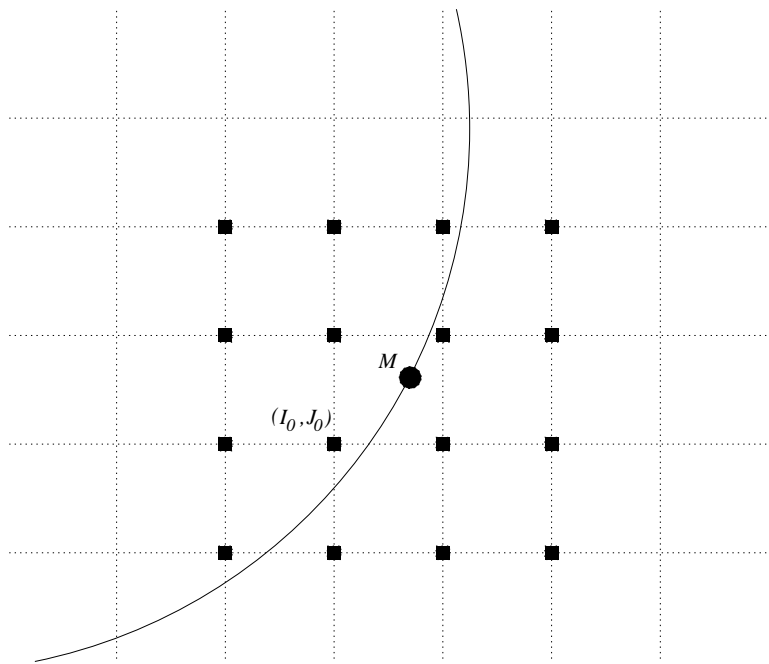}}
\caption{A 2-D version of the
 cubic interpolation scheme. Data from 
the  Cartesian gridpoints marked with filled squares is used
to obtain interpolated data at a point $M$ on the sphere.}
\label{fig:extract.interp.cubic}
\end{figure}

A straightforward approach to the problem is 
a 3-D cubic interpolator. Given a smooth
function $F(x,y,z)$ that is known on a 3-D grid
 $F_{[I,J,K]} = F(x_{[I]},y_{[J]},z_{[K]})$, the algorithm 
for computing $F$ and its Cartesian derivatives ${\partial F}/{\partial x^i} $
on a stereographic gridpoint $M(q,p)$
is the following:
\begin{itemize}
\item First locate the Cartesian cell containing
$M$, i.e. find $(I_0,J_0,K_0)$ such that
\begin{equation}
x_{[I_0]} \le x_M < x_{[I_0+1]}, \;\;
y_{[J_0]} \le y_M < y_{[J_0+1]}, \;\;
z_{[K_0]} \le z_M < z_{[K_0+1]}.
\end{equation}
The numerical stencil used for the interpolation is a set of
64 Cartesian gridpoints $(x^i)_{[I,J,K]}$,
$I_0-1 \le I \le I_0+2,\;\;$ $J_0-1 \le J \le J_0+2,\;\;$ $K_0-1 \le K \le K_0+2$,
that form the cell $(I_0,J_0,K_0)$ and its $26$ neighboring cells.

\item Next, using the grid values $F_{[I,J,K]}$ at the gridpoints
of the interpolation stencil,
construct a polynomial
\begin{equation}
{\cal P}_F(x,y,z) = \sum_{i,j,k = 0}^3 c_{i,j,k} (x-x_{[I_0]})^i 
(y-y_{[J_0]})^j (z-z_{[K_0]})^k
\end{equation}
with the coefficients $c_{i,j,k}$ determined by
 ${\cal P}_F(x_{[I]},y_{[J]},z_{[K]}) = F_{[I,J,K]}$
at the 64 gridpoints.

\item Last, the function $F$ and its derivatives
are constructed
at the stereographic gridpoint $M$ from 
the interpolation polynomial according to
\begin{eqnarray}
F(x_M, y_M, z_M) &=& {\cal P}_F(x_M,y_M,z_M) + O(\Delta^4), \\
\frac{\partial}{\partial x^i} F(x_M, y_M, z_M)
 &=& \frac {\partial}{\partial x^i} {\cal P}_F(x_M,y_M,z_M) + O(\Delta^3).
\end{eqnarray}
\end{itemize}
A 2-D version of the cubic interpolation scheme can be seen
in Figure~\ref{fig:extract.interp.cubic}.

The approach is straightforward to implement, easy to test, and has
a well defined accuracy. However, its
extended interpolation stencil could potentially lead to instability of
the matching algorithm.
 The interpolation stencil gridpoints 
 must be inside the injection world-tube
$\Lambda$. 
These gridpoints are at most $2 \sqrt{3} \Delta x$ away from $\Gamma$.
Thus, in order to be able to use the 
cubic polynomial interpolator described above, the
injection world-tube must have a radius 
$R_\Lambda \approx R_\Gamma + 2 \sqrt{3} \Delta x$ or larger. 
Since the injection strategy is based on a 
Taylor expansion around the extraction
world-tube $\Gamma$, this can effect the accuracy
 of  matching.

\subsection{Best fit algorithm}
\label{sec:extract.fit}

As an alternative to polynomial interpolation  another
algorithm was tested to  transfer data between 
the Cartesian and spherical grid-structure.
This algorithm is based on least-square fit.

\subsubsection{General linear least-squares}

Let $\varphi$ be a function known at a discrete set of points
$(x_l,y_l,z_l)_{l=1\ldots N_p}$,
and let $\{F_a(x,y,z)\}_{a=1..M_f}$ be 
a set of  $M_f$ linearly independent functions. The set 
of coefficients $\{C_a\}_{a=1..M_f}$ that minimize the
 merit function
\begin{equation}
\chi^2 = \sum_{l=1}^{N_p} \left [ \varphi(x_l,y_l,z_l) - 
\sum_{a=1}^{M_f} C_a F_a(x_l,y_l,z_l)\right]^2
\end{equation}
 can be computed
solving a linear algebraic equation \cite{Press86}:
\begin{equation}
\sum_{a=1}^{M_f} \alpha_{ab}C_b = \beta_{b},\;\;\;\;\; b = 1\ldots M_f,
\label{eq:extract.fit.linear-eq}
\end{equation}
where
\begin{eqnarray}
\label{eq:extract.fit.alpha-def}
\alpha_{ab} &=& \sum_{l=1}^{N_p} F_a(x_l,y_l,z_l) \cdot F_b(x_l,y_l,z_l), \\
\label{eq:extract.fit.beta-def}
\beta_{a}   &=& \sum_{l=1}^{N_p} F_a(x_l,y_l,z_l) \cdot \varphi(x_l,y_l,z_l).
\end{eqnarray}

Note that the matrix $\alpha_{ab}$ does not depend on the actual data $\varphi(x_l,y_l,z_l)$, 
it depends only on the coordinates of the 
data-points $(x_l,y_l,z_l)_{l=1\ldots N_p}$.
The matrix equation (\ref{eq:extract.fit.linear-eq})
is solved using LU decomposition.\footnote{The term LU 
decomposition stands for an algorithm in which one decomposes a matrix into a
lower triangular and an upper triangular matrix. (See \cite{Press86}, pp.31ff.)}

\subsubsection{Basis-functions}

The  set of basis-functions is chosen to be:
\begin{equation}
F_{iklp}(r_0;x,y,z) = 
\left(\frac{x}{r}\right)^i 
\left(\frac{y}{r}\right)^{(l-i-k)} 
\left(\frac{z}{r}\right)^k 
\left(\frac{r_0}{r} - 1 \right)^p,
\end{equation}
where 
\begin{eqnarray}
k &=& 0 \ldots 1, \\
i &=& 0 \ldots l-k,\\
l &=& 0 \ldots l_{max}, \\ 
p &=& 0 \ldots p_{max}, \\
r &=& \sqrt{x^2+y^2+z^2}.
\end{eqnarray}

For any given $r_0$, the set of basis-functions are determined by the
set of three parameters: $(l_{max}, p_{min}, p_{max})$.

The fitting zone is always a spherical shell of thickness of a few grid-zones.
Defining $r_0$ as the average of the outer and inner radii of the fitting zone
the renormalized radial coordinate $r_0/r-1$ takes values around 
zero which is
advantageous if we think of the 
radial fitting as a kind of expansion in series.

Making a least-square fit of the numerical data to a linear combination
of this set of functions gives an analytic function that smoothly
approximates our data.

\subsubsection{Fitting in the extraction}

In order to transfer data from the Cauchy grid to the world-tube  the following
needs to be done:

\begin{itemize}

\item choose a set of basis-functions 
(by specifying $(l_{max}, p_{min}, p_{max})$);

\item define a fitting zone (a spherical shell surrounding the sphere $\Gamma_t$);

\item use Cauchy data points within this shell to construct a
smooth representation of the ``3+1'' data:

\begin{eqnarray}
	g_{ij} &\rightarrow& \sum_{a=1}^{M_f} C_a^{g_{ij}} F_a(x,y,z), \\
	\alpha &\rightarrow& \sum_{a=1}^{M_f} C_a^{\alpha} F_a(x,y,z), \\
	\beta^i &\rightarrow& \sum_{a=1}^{M_f} C_a^{\beta^i} F_a(x,y,z);
\end{eqnarray}

\item evaluate the above constructed smooth functions on the sphere $\Gamma_t$.

\end{itemize}

Since the location of $\Gamma_t$ in coordinates $(x,y,z)$ 
is independent of $t$,  
 the set of points used in the extraction-fitting is the same.
Thus  the matrix $\alpha_{ab}$ given by Eq.~(\ref{eq:extract.fit.alpha-def})
does not need to be computed at each iteration. 
This also means that  the LU 
decomposition of $\alpha_{ab}$ needs to be done only once, at the 
first time-step.

The time derivative of the Cauchy data
on the world-tube is smoothly represented by
\begin{eqnarray}
	\dot{g}_{ij} &\rightarrow& \sum_{a=1}^{M_f} \dot{C}_a^{g_{ij}} F_a(x,y,z), \\
	\dot{\alpha} &\rightarrow& \sum_{a=1}^{M_f} \dot{C}_a^{\alpha} F_a(x,y,z), \\
	\dot{\beta}^i &\rightarrow& \sum_{a=1}^{M_f} \dot{C}_a^{\beta^i} F_a(x,y,z),
\end{eqnarray}
where $\dot{C}_a^{g_{ij}}, \dot{C}_a^{\alpha}, \dot{C}_a^{\beta^i}$ 
 are obtained by finite differencing in time.

Space-derivatives are computed either by first 
computing the space-derivatives of
the Cauchy functions $\alpha, \beta^i, g_{ij}$ at the Cartesian gridpoints and then
fitting them onto $\Gamma_t$, or by computing the quantities
\begin{eqnarray}
	g_{ij,k} &\rightarrow& \sum_{a=1}^{M_f} C_a^{g_{ij}}
					\frac{\partial F_a}{\partial x^k}, \\
	\alpha_{,k} &\rightarrow& \sum_{a=1}^{M_f} C_a^{\alpha}  
					\frac{\partial F_a}{\partial x^k}, \\
	\beta^i_{,k} &\rightarrow& \sum_{a=1}^{M_f} C_a^{\beta^i}   
					\frac{\partial F_a}{\partial x^k}. 
\end{eqnarray}

A weakness is that the algorithm does not have a
well defined  accuracy. An argument for using it is that it
can work with a Cartesian ``stencil'' much more compact than the
3-D cubic polynomial interpolator.

A similar algorithm was adopted to replace some of the
polynomial interpolation from the injection routine.  However,
at least for the chosen settings, this fitting technique
induced short-time instabilities in the injection module.
This led to abandoning the use of fitting.

\section{Algebra}

Next we describe in detail the algebraic 
calculations performed
in the extraction module. The presentation is based on \cite{Bishop98a}.

\subsection{Parameterization of the world-tube}

The $t=$ constant surfaces of the world-tube $\Gamma$
 are topologically spherical,
and they can be para\-metrized 
by labels $\tilde y^{\tilde A}, \tilde A = [2,3]$.
Future oriented null rays are parametrized by their  labels 
$\tilde y^{\tilde A}$
on $\Gamma_t$
and an affine parameter $\lambda$ along the
radial direction, with $\lambda=0$ on the world-tube. 
The angular coordinates
are $\tilde y^2 = q = \Re(\zeta)$ and 
$\tilde y^3 = p = \Im(\zeta)$ with
$\zeta$ being the complex stereographic coordinate as described in
Section~\ref{sec:null.coordinates}. The two-surface $\Sigma_t$ is
represented by a  
discrete numeric grid-structure
identical to the angular grid-structure 
of the
characteristic code (see Section \ref{sec:null.gridstructure}).

The Cartesian 
coordinates $(x,y,z)$ of a particular point on $\Gamma_t$
can be written as functions of the angular coordinates
$\zeta=q+Ip$ and the radius $R_\Gamma$
of the extraction world-tube using
\begin{eqnarray}
\label{eq:extract.xi=f(yi).FIRST}
x(\tilde y^A) &=& 2 R_\Gamma \cdot \left( \frac{\Re(\zeta)}
{1 + \zeta \bar \zeta}\right),\\
y(\tilde y^A) &=& \pm 2 R_\Gamma \cdot \left( \frac{\Im(\zeta)}
{1 + \zeta \bar \zeta}\right),\\
z(\tilde y^A) &=& \pm R_\Gamma \cdot \left( \frac{1 - \zeta \bar \zeta}
{1 + \zeta \bar \zeta}\right).
\label{eq:extract.xi=f(yi).LAST}
\end{eqnarray}

The reverse expressions $\tilde y^{\tilde A} = f(x^i)$ are given by
Eq.~(\ref{eq:matching.qp=fxi}).

\subsection{4-D geometry around the world-tube}

The geometry around the world-tube is fully specified by the 4-D
metric. 
These determine the unit normal $n^\alpha$ to the Cauchy 
slices $\Sigma_t$
as well as the outward pointing normal $s^\alpha$ to the world-tube 
$\Gamma$. The vectors $n^\alpha$ and $s^\alpha$  are used 
to determine the directions
of  the outgoing null radial geodesics -- an element necessary to 
compute the coordinate transformation between the Cartesian and the 
(affine) null coordinates.

From a numerical point of view the first step is getting the Cauchy metric
information onto $\Gamma_t$ by means of interpolation, fitting, etc.
as described in the previous section. The
ADM code operates with the the 3-D metric
${}^{(3)}g_{ij}$, lapse function $\alpha$, and the shift vector $\beta^i$.
In order  
to reconstruct the full 4-D metric $g_{\mu \nu}$
and its derivatives $g_{\mu \nu,\rho}$ we use
\begin{eqnarray}
g_{ij} &=& {}^{(3)}g_{ij}, \\
g_{it} &=& {}^{(3)}g_{ij} \beta^j, \\
g_{tt} &=& - \alpha^2 + g_{it} \beta^i,\\
g_{it,\rho} &=& {}^{(3)}g_{ij,\rho} \beta^j + {}^{(3)}g_{ij} \beta^j_{,\rho}, \\
g_{tt,\rho} &=& - 2 \alpha \alpha_{,\rho} +
{}^{(3)}g_{ij,\rho} \beta^i \beta^j + 2 \,{}^{(3)}g_{ij} \beta^i \beta^j_{,\rho}.
\end{eqnarray}

It follows from the definition of lapse and shift (see 
Section~\ref{sec:lapse&shift}) that the unit normal $n^\mu$
to the $t=$ constant hypersurfaces can be written 
as
\begin{equation}
n^\mu = \frac 1 \alpha \left( 1, - \beta^i \right).
\end{equation}

Let $s^\alpha = (0, s^i)$ be the outward pointing unit normal
to the sphere $\Gamma_t$. By construction $s^i$ lies in 
$\Sigma_t$, and it can be expressed in terms of
\begin{equation}
q^i = \frac{\partial x^i}{\partial \tilde y^2},\;\;
p^i = \frac{\partial x^i}{\partial \tilde y^3},
\label{eq:extract.qpidef}
\end{equation}
which can be computed analytically from 
Eqs.~(\ref{eq:extract.xi=f(yi).FIRST}) - (\ref{eq:extract.xi=f(yi).LAST}).
Using the unit antisymmetries $\epsilon_{ijk}$,
we obtain the spatial
components of the normal 1-form $\sigma_i$ 
\begin{equation}
\sigma_i = \epsilon_{ijk} q^j p^k,
\end{equation}
with norm $\sigma = \sqrt{\sigma_i \sigma_j \,{}^{(3)}g^{ij}}$. Then the vector $s^i$ 
is equal to the contravariant, renormalized vector $\sigma^i$:
\begin{equation}
s^i =  {}^{(3)}g^{ij} \frac {\sigma_j}{\sigma}.
\label{eq:extract.sidef} \end{equation}
The generators $\ell^\mu$ of the outgoing null cone through
$\Gamma_t$ are given on the world-tube by the vectors
$n^\mu$ and $s^\mu$, normalized such that 
$\ell^\mu t_\mu = -1$, that is,
\begin{equation}
\ell^\mu = \frac{n^\mu + s^\mu}{\alpha - {}^{(3)}g_{ij} \beta^i s^j}.
\end{equation}

This supplies
all the necessary elements 
on the world-tube $\Gamma_t$ 
 to perform the first
coordinate transformation from Cartesian coordinates
$x^\mu = (t,x,y,z)$ into affine null coordinates
$\tilde y^{\tilde \nu} = (u, \lambda, q, p)$.
Once we have the metric in affine null coordinates it takes 
only a few more algebraic steps to obtain the Bondi metric,
from which we can extract the Bondi metric variables $J,\beta,U$ and $W$
in terms of a Taylor expansion around the world-tube $\Gamma$.

\subsection{Coordinate transformation}
\label{sec:extract.coordtrans}

The coordinate transformation
$x^\alpha \rightarrow \tilde y^{\tilde \beta}$ needs to be
performed  
in a neighborhood of the world-tube $\Gamma$.

The standard approach in carrying through coordinate transformations
is to compute the Jacobian 
${\partial x^{\mu}}/{\partial \tilde y^{\tilde \alpha}}$, then
use the tensorial transformation rule 
\begin{equation}
\tilde \eta _{\tilde \alpha \tilde \beta} =
\frac{\partial x^\mu}{\partial \tilde y^{\tilde \alpha}}
\frac{\partial x^\nu}{\partial \tilde y^{\tilde \beta}}\;
{}^{(4)}g_{\mu \nu}.
\end{equation}
However, the geometric properties of the affine null coordinate
system make this task easier  because  four out of the ten
components of the $4\times 4$ symmetric tensor are given by
\begin{equation}
\tilde \eta_{\lambda \lambda}=0,\;\;\tilde \eta_{\lambda \tilde A}=0,
\;\; \tilde \eta_{\lambda u} = -1.
\label{eq:extract.knowneta}
\end{equation}
The first  of these conditions indicates that the radial coordinate
$\lambda$ is null, the second one 
is implied by the fact that the null vector $\ell^\mu$ is
normal to $\Gamma$, 
while the last condition is a consequence of  $\lambda$ being
an affine parameter. 
The additional degree of freedom in the choice of $\lambda$ 
is eliminated by 
\begin{equation}
\lambda_{|\Gamma} = 0.
\end{equation}

At this point   six independent metric
functions remain. The part of the coordinate transformation that is still
missing is
\begin{equation}
J^{\mu}_{\tilde \alpha} \equiv
\frac{\partial x^\mu}{\partial \tilde y^{\tilde \alpha}} = 
x^{(0) \mu}_{,\tilde \alpha} +
x^{(1) \mu}_{,\tilde \alpha} \lambda + O(\lambda^2), \;\mbox{for}\; 
\tilde y^{\tilde \alpha} = (u,q,p). 
\label{eq:extract.x_expansion}
\end{equation}
By construction
the coordinates $t$ and $u$ are  related
by $u_{|\Gamma} = t_{|\Gamma}$ and 
$\left(\partial t / \partial u \right)_{|\Gamma} = 1$.
Furthermore, since  the location of $\Gamma_t$
 in Cartesian coordinates is time-independent,  only the
angular derivatives of $x^{(0)} (\equiv x_{|\Gamma})$  survive in
Eq.~(\ref{eq:extract.x_expansion}).
These can easily be computed using
Eqs.~(\ref{eq:extract.xi=f(yi).FIRST}) - (\ref{eq:extract.xi=f(yi).LAST}).

The harder task in the coordinate transformation is the evaluation
of the $O(\lambda)$ part. This can be done starting from 
\begin{equation}
x^{\mu}_{,\lambda \tilde A} = \ell^{\mu}_{,\tilde A}\, ,\;\;
x^{\mu}_{,\lambda u} = \ell^{\mu}_{,u}\,.
\label{eq:extract.Olambda}
\end{equation}

Since the null vector $\ell^\mu$ depends on the Cauchy metric components,
the terms listed in Eq.~(\ref{eq:extract.Olambda})
involve derivatives of the Cauchy metric as well as zeroth order
Jacobian terms.
In order to compute the derivatives of $\ell^\mu$ first 
derivatives of $n^\mu$ are computed. Then derivatives 
of the spacelike vector $s^i$ follow.

Since
the vector $n^\mu$ is the unit normal to the Cauchy surfaces $\Sigma_t$,
it is entirely determined by the choice of lapse and shift.
The Cartesian derivatives
\begin{eqnarray}
   n^{i}_{,\nu}  =  {1 \over \alpha^{2}} 
   \left( \alpha_{,\nu} \beta^{i} - \alpha \beta^{i}_{,\nu} \right),
   \;\;\;
   n^{t}_{,\nu}  =  - {1 \over \alpha^{2}} \alpha_{,\nu} 
   \label{eq:extract.dna}
\end{eqnarray}
can be used to compute derivatives with respect to $(u,q,p)$ via
\begin{eqnarray}
   n^{\mu}_{,\tilde{A}} &=& n^{\mu}_{,j} x^{j}_{,\tilde{A}},
   \label{eq:extract.naA}\\
n^{\mu}_{,u|\Gamma} &=& n^{\mu}_{,t|\Gamma}.
\end{eqnarray}

The vector $s^i$ is constructed from the time-independent vector $\sigma^i$
and the metric tensor components. Its time derivatives are
\begin{eqnarray}
   s^{i}_{,t} &=& g^{ik}_{,t} {\sigma_{k} \over \sigma}
              - g^{ik} {{\sigma_{k} \sigma_{,t}} \over \sigma^{2}} 
              = -g^{im} g^{kn} g_{mn,t} {\sigma_{k} \over \sigma}
              - s^{i} {\sigma_{,t} \over \sigma}  \nonumber \\
              &=& \mbox{} -g^{im} g_{mn,t} s^{n} 
              - s^{i} {\sigma_{,t} \over \sigma}.
\label{eq:extract.dsdt}
\end{eqnarray}
Here the quantity $\sigma$ (the norm of $\sigma^i$)
is time-dependent because the metric tensor used to build
the norm of a vector
varies with time. Thus
\begin{equation}
   2 \sigma \sigma_{,t} = \left( \sigma^2 \right)_{,t} =
   g^{kl}_{,t} \sigma_{k} \sigma_{l} =
   - g^{km} g^{ln} g_{mn,t} \sigma_{k} \sigma_{l} =
   - s^{m} s^{n} g_{mn,t} \sigma^{2}.
   \label{eq:extract.dsigmadt}
\end{equation}
Substituting Eq.~(\ref{eq:extract.dsigmadt}) into 
Eq.~(\ref{eq:extract.dsdt}) one obtains
\begin{equation}
   s^{i}_{,t} = \left( -g^{im} + s^{i} \frac{1}{2} s^{m} \right) 
   g_{mn,t} s^{n} .
   \label{eq:extract.sit}
\end{equation}
Similarly, it follows from Eq.~(\ref{eq:extract.sidef}) that 
\begin{eqnarray}
   s^{i}_{,\tilde{A}} &=& g^{ik}_{,j} x^{j}_{,\tilde{A}} {\sigma_{k} \over \sigma}
   + g^{ik} {\sigma_{k,\tilde{A}} \over \sigma}
   - g^{ik} {\sigma_{k} \sigma_{,\tilde{A}} \over \sigma^{2}} \nonumber \\
   &=& \mbox{} -g^{in} g^{km} g_{mn,j} \,x^{j}_{,\tilde{A}} {\sigma_{k} \over \sigma}
   + g^{ik} {\sigma_{k,\tilde{A}} \over \sigma} 
   - s^{i} {\sigma_{,\tilde{A}} \over \sigma},
   \label{eq:extract.siA1}
\end{eqnarray}
where the $\sigma_{k,\tilde{A}}$ are obtained from the analytic
expressions (\ref{eq:extract.qpidef}), and $\sigma_{,\tilde{A}}$ from
\begin{eqnarray}
   2 \sigma \sigma_{,\tilde{A}} & = & \left( \sigma^2 \right)_{,\tilde{A}}
   = \left( g^{kl} \sigma_{k} \sigma_{l} \right)_{,\tilde{A}}
   = g^{kl}_{,j} x^{j}_{,\tilde{A}} \sigma_{k} \sigma_{l}
   + 2 \, g^{kl} \sigma_{l} \sigma_{k,\tilde{A}} \nonumber \\
   &=& \mbox{} - g^{km} g^{ln} g_{mn,j} x^{j}_{,} \sigma_{k} \sigma_{l}
   + 2 \, g^{kl} \sigma_{l} \sigma_{k,\tilde{A}}
   \nonumber \\
   & = & \mbox{} - s^{m} s^{n} g_{mn,j} x^{j}_{,\tilde{A}} \sigma^{2}
   + 2 s^{k} \sigma \sigma_{k,\tilde{A}}.
   \label{eq:extract.sigmaA}
\end{eqnarray}
Collecting Eqs.~(\ref{eq:extract.siA1}) and (\ref{eq:extract.sigmaA}), the
angular derivatives of $s^i$ are:
\begin{eqnarray}
   s^{i}_{,\tilde{A}} & = & -g^{in} s^{m} g_{mn,j} \,x^{j}_{,\tilde{A}} 
   + g^{ik} {\sigma_{k,\tilde{A}} \over \sigma}
   + s^{i} \left( \frac{1}{2} s^{m} s^{n} g_{mn,j} x^{j}_{,\tilde{A}} 
                  - s^{k} {\sigma_{k,\tilde{A}} \over \sigma} \right)
   \nonumber \\
   & = & \left( g^{in} - s^{i} s^{n} \right) {\sigma_{n,\tilde{A}} \over \sigma}
   + \left( - g^{in} + \frac{1}{2} s^{i} s^{n} \right) 
     s^{m} g_{mn,j} x^{j}_{,}
   \label{eq:extract.siA}
\end{eqnarray}

\subsection{The affine null metric}
\label{sec:extract.nullmetric}

One of the useful quantities in computing 
the null metric $\tilde \eta_{\tilde \alpha \tilde \beta}$ is
the $\lambda$-derivative of the Cartesian tensor $g_{\alpha \beta}$:
\begin{equation}
   g_{\alpha\beta,\lambda}{}_{|\Gamma} = 
\left(
\frac{ \partial g_{\alpha \beta} } {\partial x^\mu} \right)_{|\Gamma}
\left(
\frac{ \partial x^\mu }{\partial \lambda}\right)_{|\Gamma} = 
g^{(0)}_{\alpha\beta,\mu} 
  \ell^{(0)}{}^{\mu}.
   \label{eq:extract.g1}
\end{equation}

The null metric can be expanded around $\Gamma$ as follows
\begin{equation}
   \tilde{\eta}_{\tilde{\alpha}\tilde{\beta}} 
= \tilde{\eta}^{(0)}_{\tilde{\alpha}\tilde{\beta}}
   + \tilde{\eta}^{(0)}_{\tilde{\alpha}\tilde{\beta},\lambda} \lambda + O(\lambda^{2}),
\end{equation}
where the zeroth order coefficients are given by
\begin{eqnarray}
   \tilde{\eta}^{(0)}_{{u}\tilde{u}} & = & g_{tt}{}_{|\Gamma},
   \nonumber \\
   \tilde{\eta}^{(0)}_{{u}\tilde{A}} & = & x^{i}_{,\tilde{A}} g_{it}{}_{|\Gamma},
   \nonumber \\
   \tilde{\eta}^{(0)}_{\tilde{A}\tilde{B}} & = & x^{i}_{,\tilde{A}} x^{j}_{,\tilde{B}} g_{ij}{}_{|\Gamma},
   \label{eq:extract.eta0}
\end{eqnarray}
and the first-order terms are
\begin{eqnarray}
   \tilde{\eta}^{(0)}_{{u}{u},\lambda} & = & \left[g_{tt,\lambda} 
   + 2\, \ell^{\mu}_{,{u}} g_{\mu t}\right]_{|\Gamma} + O(\lambda),
   \nonumber \\
   \tilde{\eta}^{(0)}_{{u} \tilde{A},\lambda} & = & \left[ x^{k}_{,\tilde{A}} \left(
      \ell^{\mu}_{,{u}} g_{k\mu} + g_{kt,\lambda} \right)
      + \ell^{k}_{,\tilde{A}} g_{kt} + \ell^{t}_{,\tilde{A}} g_{tt} \right]_{|\Gamma} 
      + O(\lambda) , 
   \nonumber \\
   \tilde{\eta}^{(0)}_{\tilde{A}\tilde{B},\lambda} & = & 
   \left[ x^{k}_{,\tilde{A}} x^{l}_{,\tilde{B}} g_{kl,\lambda}
   + \left(  \ell^{\mu}_{,\tilde{A}} x^{l}_{,\tilde{B}} + \ell^{\mu}_{,\tilde{B}} x^{l}_{,\tilde{A}} \right) 
     g_{\mu l} \right]_{|\Gamma}
   + O(\lambda) .
   \label{eq:extract.eta1}
\end{eqnarray}
Recall that the components $\tilde \eta_{\lambda \tilde \beta}$  are
given by Eq.~(\ref{eq:extract.knowneta}).

Given the covariant null metric 
$\tilde \eta_{\tilde \alpha \tilde \beta}$ as an expansion 
up to $O(\lambda^2)$, its contravariant form
$\tilde \eta^{\tilde \alpha \tilde \beta}$ is known to the
same order 
\begin{equation}
   \tilde{\eta}^{\tilde{\mu}\tilde{\nu}} = \tilde{\eta}^{(0)}{}^{\tilde{\mu}\tilde{\nu}}
   + \tilde{\eta}^{(0)\tilde{\mu}\tilde{\nu}}_{,\lambda} \lambda
   + O(\lambda^{2}) ,
\end{equation}
using
\begin{equation}
   \tilde{\eta}^{\tilde{\mu}\tilde{\alpha}} 
   \tilde{\eta}_{\tilde{\alpha}\tilde{\nu}} = \delta^{\tilde{\mu}}_{\tilde{\nu}} , \quad
   \tilde{\eta}^{\tilde{\mu}\tilde{\nu}}_{,\lambda}  = - \tilde{\eta}^{\tilde{\mu}\tilde{\alpha}} \,
   \tilde{\eta}^{\tilde{\beta}\tilde{\nu}} \, \tilde{\eta}_{\tilde{\alpha}\tilde{\beta},\lambda} .
\end{equation}

Similar to the case of the covariant metric, the components
 $\tilde{\eta}^{\lambda \tilde \alpha}$ are 
determined by the null coordinate conditions
\begin{equation}
   \tilde{\eta}^{\lambda {u}} = -1, \quad
   \tilde{\eta}^{{u}\tilde{A}} = \tilde{\eta}^{{u}{u}} = 0, 
   \label{eq:extract.tildeup}
\end{equation}
a consequence of Eq.~(\ref{eq:extract.knowneta}).
As a result, the contravariant null metric can be computed by
\begin{eqnarray}
   \tilde{\eta}^{\tilde{A}\tilde{B}} \tilde{\eta}_{\tilde{B}\tilde{C}} & = & \delta^{\tilde{A}}_{\, \,\tilde{C}}, 
   \nonumber \\
   \tilde{\eta}^{\lambda \tilde{A}} & = & \tilde{\eta}^{\tilde{A}\tilde{B}} \tilde{\eta}_{\tilde{B}{u}},
   \nonumber \\
   \tilde{\eta}^{\lambda\lambda} & = & - \tilde{\eta}_{{u}{u}}
   + \tilde{\eta}^{\lambda \tilde{A}} \tilde{\eta}_{\tilde{A}{u}},
   \label{eq:extract.etaup0}
\end{eqnarray}
with the first-order terms given by
\begin{eqnarray}
   \tilde{\eta}^{\tilde{A}\tilde{B}}_{,\lambda} & = & - \tilde{\eta}^{\tilde{A}\tilde{C}} \tilde{\eta}^{\tilde{B}\tilde{D}}
   \tilde{\eta}_{\tilde{C}\tilde{D},\lambda}\,,
   \nonumber \\
   \tilde{\eta}^{\lambda \tilde{A}}_{,\lambda} & = & \tilde{\eta}^{\tilde{A}\tilde{B}}
   \left( \tilde{\eta}_{{u}\tilde{B},\lambda} - \tilde{\eta}^{\lambda \tilde{C}} 
   \tilde{\eta}_{\tilde{C}\tilde{B},\lambda} \right),
   \nonumber \\
   \tilde{\eta}^{\lambda\lambda}_{,\lambda} & = & - \tilde{\eta}_{{u}{u},\lambda} 
   + 2\, \tilde{\eta}^{\lambda \tilde{A}} \tilde{\eta}_{{u}\tilde{A},\lambda} 
   - \tilde{\eta}^{\lambda \tilde{A}} \tilde{\eta}^{\lambda \tilde{B}} 
   \tilde{\eta}_{\tilde{A}\tilde{B},\lambda}\,.
\end{eqnarray}

\subsection{Metric in Bondi coordinates}\label{sec:bondimetric}

By convention the angular coordinates used in the Bondi frame
are defined to be the same as the angular coordinates used in
the affine null frame, i.e. $ y^{A} \equiv \tilde y^{\tilde A} = (q,p)$.
As a consequence,
switching from affine null coordinates $\tilde y^{\tilde \mu} = 
(u,\lambda,\tilde y^{\tilde A})$
to Bondi coordinates $y^\mu =(u,r,y^A)$ amounts to trading
the affine parameter $\lambda$ for the surface area coordinate
$r = r(u,\lambda,{y}^{A})$ defined by
\begin{equation}
   \label{eq:extract.rofeta}
   r  =  \left( \frac{\det(\tilde{\eta}_{\tilde{A}\tilde{B}})} 
{\det(q_{AB})} \right)
   ^{\frac{1}{4}} = \frac 2 P \det(\tilde{\eta}_{\tilde{A}\tilde{B}})%
^{\frac{1}{4}},
   \label{eq:extract.r}
\end{equation}
where we have used
$\det(q_{AB})=16/(1+q^{2}+p^{2})^{4} = 16/P^4$.

The only nontrivial elements of the Jacobian 
${\partial y^{\alpha}}/{\partial \tilde y^{\tilde \beta}}$ are
  $r_{,\lambda}$, $r_{,\tilde{A}}$ and
$r_{,{u}}$, which can be computed using
\begin{eqnarray}
   r_{,\lambda} &=& \frac{r}{4} 
\tilde{\eta}^{\tilde{A}\tilde{B}} \tilde{\eta}_{\tilde{A}\tilde{B},\lambda},
   \label{eq:extract.rl} \\
   r_{,\tilde{C}} &=& \frac{r}{4} \left(\tilde{\eta}^{\tilde{A}\tilde{B}} \tilde{\eta}_{\tilde{A}\tilde{B},\tilde{C}} 
   - \frac{\det(q_{\tilde{A}\tilde{B}})_{,\tilde{C}}}{\det(q_{\tilde{A}\tilde{B}})} \right), \\
   r_{,{u}} &=&  \frac{r}{4}\, \tilde{\eta}^{\tilde{A}\tilde{B}} \tilde{\eta}_{\tilde{A}\tilde{B},{u}}.
\label{eq:extract.ru}
\end{eqnarray}

The terms involved in computing $r_{,\tilde{C}}$ can be obtained via
\begin{eqnarray}
   \frac{\det(q_{\tilde{A}\tilde{B}})_{,\tilde{C}}}{\det(q_{\tilde{A}\tilde{B}})} & = & 
   - {\frac{8\;\tilde{y}^{\tilde{C}}}{P}}\,,
   \nonumber \\
   \tilde{\eta}_{\tilde{A}\tilde{B},\tilde{C}} & = & 
   \left( x^{i}_{,\tilde{A}\tilde{C}}\, x^{j}_{,\tilde{B}}
        + x^{i}_{,\tilde{A}}\,  x^{j}_{,\tilde{B}\tilde{C}} \right) g_{ij}
\nonumber \\ &&
   + x^{i}_{,\tilde{A}}\, x^{j}_{,\tilde{B}}\, x^{k}_{,\tilde{C}}\, g_{ij,k}
\end{eqnarray}
where the $x^{i}_{,\tilde{A}\tilde{C}}$ are given functions of $(q,p)$. 
Furthermore, when computing $r_{,u}$, we need
the $u$-derivative of the affine null metric, which,
according to  Eq.~(\ref{eq:extract.eta0}), is given by
\begin{equation}
   \tilde{\eta}_{\tilde{A}\tilde{B},{u}} 
= \left[ x^{i}_{,\tilde{A}}\, x^{j}_{,\tilde{B}}\, g_{ij,t} 
\right]_{|\Gamma} + O(\lambda).
\end{equation}

The Bondi metric $\eta^{\alpha\beta}$ 
is then obtained from
\begin{equation}
   \eta^{\alpha\beta} = 
   {{\partial y^{\alpha}} \over {\partial \tilde{y}^{\mu}}}
   {{\partial y^{\beta}} \over {\partial \tilde{y}^{\nu}}}
   \tilde{\eta}^{\tilde{\mu}\tilde{\nu}}.
   \label{eq:extract.etaup}
\end{equation}

Since the only difference between the affine and the Bondi null 
frames is the  radial coordinate, we only need to compute the metric elements  
$\eta^{rr}$, $\eta^{rA}$ and $\eta^{ru}$, the rest of the tensor
$\eta^{\alpha \beta}$ being identical with 
$\tilde \eta^{\tilde \alpha \tilde \beta}$.  Using
Eq.~(\ref{eq:extract.tildeup}),

\begin{eqnarray}
   \eta^{rr} & = & r_{,\tilde{\alpha}}\, r_{,\tilde{\beta}}\, \tilde{\eta}^{\tilde{\alpha}\tilde{\beta}} 
   = \left(r_{,\lambda}\right)^2 \tilde{\eta}^{\lambda \lambda} 
   + 2\, r_{,\lambda}\, \left(r_{,\tilde{A}}\,\tilde{\eta}^{\lambda \tilde{A}} - r_{,{u}} \right)
   + r_{,\tilde{A}}\,r_{,\tilde{B}}\, \tilde{\eta}^{\tilde{A}\tilde{B}}
   \nonumber \\
   \eta^{rA} & = & r_{,\tilde{\alpha}}\, \tilde{\eta}^{\tilde{\alpha} \tilde{A}} 
   = r_{,\lambda}\, \tilde{\eta}^{\lambda \tilde{A}} + r_{,\tilde{B}}\, \tilde{\eta}^{\tilde{A}\tilde{B}}
   \nonumber \\
   \eta^{ru} & = & r_{,\tilde{\alpha}}\, \tilde{\eta}^{\tilde{\alpha} {u}} = - r_{,\lambda}
   \label{eq:extract.bondim}
\end{eqnarray}

At this point we are  one step away from obtaining the Bondi
metric variables $J,U,\beta,W$  and passing them to the
characteristic code as boundary data.

\subsection{Boundary data for the characteristic code}
\label{sec:nullbdry}

Recall from Section~\ref{sec:matching.extraction}
that the null gridpoints do not necessarily lie {\em on} the world-tube $\Gamma$. 
Moreover,
the position 
of individual null gridpoints with respect to $\Gamma$
can differ from time-step to time-step.
In particular, gridpoints that are just inside of $\Gamma$ 
at $t=t_{[N]}$ could end up outside of it at $t=t_{[N+1]}$.
Thus, in order to provide boundary data for the characteristic code,
the following steps must be made:
\begin{itemize}
\item compute the Bondi metric variables around the world-tube 
up to $O(\lambda^2)$
\item localize the world-tube-slice $\Gamma_t$ in Bondi coordinates
\item insert the Bondi functions onto the
characteristic gridpoints neighboring $\Gamma_t$.
\end{itemize}

Then, the characteristic code has the necessary data to
 evolve the metric functions
from the world-tube to the outer boundary.\footnote{In the case of an
ingoing null simulation, the evolution proceeds from $\Gamma$
to the inner (possibly excised) boundary. See Section~\ref{sec:null.1BH} for an example.}

In order to compute the Bondi metric variables, 
the contravariant Bondi metric is written in the form
given by Eq.~(\ref{eq:matching.bondimetric}).
The functions $J,\beta,U,$ and $W$ can now be expressed
in terms of known quantities on the world-tube, as described in the next sections.
Although these functions have already been defined in Chapter~\ref{chap:nullcode},
their definition is repeated here for the ease of reading.

\subsubsection{The metric of the sphere $J$}

Given $r$ and $r_{,\lambda}$  and noting that
\begin{equation}
\eta_{AB} = \tilde{\eta}_{AB} \equiv r^2 h_{AB},
\end{equation}
 $h_{AB}$ and its derivative can be obtained from
\begin{eqnarray}
   h_{AB} &=& \frac{1}{r^{2}} \eta_{AB},
   \nonumber \\
   h_{AB,\lambda} &=& \frac{1}{r^{2}}
   \left( \eta_{AB,\lambda} 
   - \frac{2\,r_{,\lambda}}{r}\,\eta_{AB}\right).
\end{eqnarray}
In terms of $q^{A}$, $\bar{q}^{A},$ and 
$h_{AB}$, the metric functions $J$ and $K$  are then defined as follows:

\begin{equation}
   J \equiv \frac{1}{2} q^{A} q^{B} h_{AB}, \quad
   K \equiv \frac{1}{2} q^{A} \bar q^{B} h_{AB}.
   \label{eq:extract.jdef}
\end{equation}
As stated in Section~\ref{sec:null.spin-eqs}, 
the determinant condition
$\det( h_{AB}) = \det( q_{AB})$ implies  $1 = K^2 - J \bar J$.
Thus, 
it suffices to evaluate $J$, which can be computed as an expansion
around $\Gamma$ 
\begin{equation}
	J(y^\alpha) = J^{(0)} + J^{(0)}_{,\lambda} \lambda + O(\lambda^{2}),
\end{equation}
with $J^{(0)}$ and $J^{(0)}_{,\lambda}$ given by
\begin{eqnarray}
   J^{(0)} & = & \left(\frac{1}{2\,r^{2}} q^{A} q^{B} \eta_{AB} \right)_{|\Gamma},
   \nonumber \\ 
   J^{(0)}_{,\lambda} & = & \left(\frac{1}{2\,r^{2}} q^{A} q^{B} \eta_{AB,\lambda}
   - 2\,\frac{r_{,\lambda}}{r} J\right)_{|\Gamma}.
   \label{eq:extract.jl}
\end{eqnarray}

\subsubsection{The ``expansion factor'' $\beta$}

From inspection of 
Eq.~(\ref{eq:matching.bondimetric}) one can see that 
\begin{equation}
\eta^{ru} = -e^{-2 \beta}.
\label{eq:extract.etaru}
\end{equation}
Substituting the last of  Eqs.~(\ref{eq:extract.bondim}) 
into Eq.~(\ref{eq:extract.etaru}), one obtains
\begin{equation}
   \beta = -\frac{1}{2}\log(r_{,\lambda}). 
\label{eq:extract.beta=logrl}
\end{equation}

The first-order
derivative term $\beta^{(0)}_{,\lambda}$ 
involves $r^{(0)}_{,\lambda \lambda}$ which in turn involves second
derivatives of the Cauchy metric (or curvature terms).
 In order to avoid the associated algebraic
(and numeric) complications, we use the first differential order
characteristic equation (\ref{eq:null.betaeq-full}).

At a constant angle $(q,p)$  the relation $\partial_{\lambda}
= r_{,\lambda} \partial_{r}$ holds. This allows an
exchange of the $r$-derivatives in Eq.~(\ref{eq:null.betaeq-full})
with $\lambda$-derivatives:
\begin{equation}
   \beta_{,\lambda} = \frac{r}{8\, r_{,\lambda}} 
   \left( J_{,\lambda} \bar{J}_{,\lambda} - \left(K_{,\lambda}\right)^{2} 
   \right).
   \label{eq:extract.betal}
\end{equation}

Equation (\ref{eq:extract.betal}),
 shows how the derivative with 
respect to the affine parameter of the expansion of light rays 
is a quadratic term, i.e. it vanishes in linearized theory.

Substituting 
Eq.~(\ref{eq:extract.jdef}) into 
Eq.~(\ref{eq:extract.betal}), we obtain 
\begin{equation}
   \beta_{,\lambda} = \frac{r}{8\, r_{,\lambda}} 
   \left( J_{,\lambda} \bar{J}_{,\lambda} - 
   \frac{1}{1 + J \bar{J}} 
   \left[\Re \left(\bar{J} J_{,\lambda} \right)\right]^2  \right).
   \label{eq:extract.nbetal}
\end{equation}
Equations (\ref{eq:extract.beta=logrl}) and (\ref{eq:extract.nbetal})
hold globally  and thus
provide a way to compute 
$\beta^{(0)} = \beta_{|\Gamma}$ and 
$\beta^{(0)}_{,\lambda} = \beta_{,\lambda|\Gamma}$ 
Then, the expansion for $\beta$ around $\Gamma$ can  
be written 
\begin{equation}
	\beta(y^\alpha) = \beta^{(0)} + \beta^{(0)}_{,\lambda} \lambda + O(\lambda^{2}).
\end{equation}

In addition, using Eq.~(\ref{eq:extract.beta=logrl})
$r_{,\lambda \lambda}$ 
can be computed 
in terms of the known quantities 
$\beta_{,\lambda}$ and $r_{,\lambda}$:
\begin{equation}
\beta_{,\lambda} = \partial_\lambda \left( - \frac 1 2
\log \left( r_{,\lambda} \right) \right) =
- \frac {r_{,\lambda \lambda}}{2 r_{,\lambda}}.
\label{eq:extract.rlls}
\end{equation}

\subsubsection{The ``shift'' $U$}

The metric function $U$  is expressed, using Eq.~(\ref{eq:extract.bondim}), as
\begin{equation}
   U \equiv U^{A} q_{A} = \frac{\eta^{rA}}{{\eta^{ru}}} q_{A}
     = - \left(\tilde{\eta}^{\lambda \tilde{A}} 
       + \frac{r_{,\tilde{B}}}{r_{,\lambda}} \tilde{\eta}^{\tilde{A}\tilde{B}} \right) q_{\tilde{A}} ,
\label{eq:extract.Ueq}
\end{equation}
The  $\lambda$ derivative  of $U$ will be given by

\begin{eqnarray}
   U_{,\lambda} & = & - \left[ \tilde{\eta}^{\lambda \tilde{A}}_{,\lambda}
       + \left( \frac{r_{,\lambda \tilde{B}}}{r_{,\lambda}} 
              - \frac{r_{,\tilde{B}}\,r_{,\lambda\lambda}}{r^2_{,\lambda}}
         \right) \tilde{\eta}^{\tilde{A}\tilde{B}}
       + \frac{r_{,\tilde{B}}}{r_{,\lambda}} \tilde{\eta}^{\tilde{A}\tilde{B}}_{,\lambda} 
                    \right] q_{\tilde{A}} ,
   \nonumber \\
   & = & - \left( \tilde{\eta}^{\lambda \tilde{A}}_{,\lambda}
                 + \frac{r_{,\lambda \tilde{B}}}{r_{,\lambda}} \tilde{\eta}^{\tilde{A}\tilde{B}}
                 + \frac{r_{,\tilde{B}}}{r_{,\lambda}} \tilde{\eta}^{\tilde{A}\tilde{B}}_{,\lambda} 
                    \right) q_{\tilde{A}}
   \nonumber \\
&&
         + 2 \, \beta_{,\lambda} \left( U 
                     + \tilde{\eta}^{\lambda \tilde{A}} q_{\tilde{A}} \right) ,
\label{eq:extract.Uleq}
\end{eqnarray}
where Eq.~(\ref{eq:extract.rlls}) was used to eliminate
$r_{,\lambda\lambda}$ in the last line.

Then  applying 
Eqs.~(\ref{eq:extract.Ueq}) - (\ref{eq:extract.Uleq})
around the world-tube,
$U$ is found to second-order accuracy by:
\begin{equation}
	U(y^\alpha) = U^{(0)} + U^{(0)}_{,\lambda} \lambda + O(\lambda^{2}).
\end{equation}

\subsubsection{The ``potential'' $W$}

Recall from Eq.~(\ref{eq:matching.bondimetric}) that 
\begin{equation}
\eta^{rr} = e^{-2 \beta} \frac{V}{r}, \;\;\; \eta^{ur} = -e^{-2 \beta}.
\end{equation}
Thus the function $V$ is given by
\begin{equation}
V \equiv - r \eta^{rr}/\eta^{ru}.
\end{equation}
For Minkowski space, 
$\eta^{rr} = - \eta^{ru} = 1$ so that
the asymptotic value
of $V$ is given by $V=r$.
Thus, in the characteristic code, $V$ is replaced
 with the function  $W \equiv (V-r)/r^2$. In
terms of the affine contravariant null metric 
$\tilde \eta^{\tilde \alpha \tilde \beta}$, $W$ can be expressed as

\begin{equation}
   W = \frac{1}{r} \left(\frac{\eta^{rr}}{r_{,\lambda}} - 1 \right)
     = \frac{1}{r} \left( r_{,\lambda} \tilde{\eta}^{\lambda\lambda}
   + 2 \, \left( r_{,\tilde{A}}\,\tilde{\eta}^{\lambda \tilde{A}} - r_{,u} \right)
   + \frac{r_{,\tilde{A}}\,r_{,\tilde{B}}}{r_{,\lambda}}\, 
\tilde{\eta}^{\tilde{A}\tilde{B}} - 1 \right),
\label{eq:extract.W0}
\end{equation}
with the radial derivative term $W_{,\lambda}$ given by
\begin{eqnarray}
   W_{,\lambda} & = & - \frac{r_{,\lambda}}{r} W
   + \frac{1}{r} \left( r_{,\lambda} \tilde{\eta}^{\lambda\lambda}
   + 2 \, \left( r_{,\tilde{A}}\,\tilde{\eta}^{\lambda \tilde{A}} - r_{,u} \right)
   + \frac{r_{,\tilde{A}}\,r_{,\tilde{B}}}{r_{,\lambda}}\, \tilde{\eta}^{\tilde{A}\tilde{B}} - 1 
   \right)_{,\lambda} \nonumber
   \\
   & = &  - \frac{r_{,\lambda}}{r} 
   \left(
         \left( \frac{r_{,\lambda}}{r} + 2\, \beta_{,\lambda} \right)
         \tilde{\eta}^{\lambda\lambda}
         - \tilde{\eta}^{\lambda\lambda}_{,\lambda} 
         - \frac{1}{r}
   \right)
   + \frac{2}{r} 
     \left( \frac{r_{,\lambda}r_{,u}}{r} - r_{,\lambda u} \right)
   \nonumber \\
   &  & + \frac{2}{r} 
         \left( 
               r_{,\lambda \tilde{A}}
               - \frac{r_{,\lambda}r_{,\tilde{A}}}{r} 
         \right)
         \tilde{\eta}^{\lambda \tilde{A}}
   + 2\frac{r_{,\tilde{A}}}{r}\,\tilde{\eta}^{\lambda \tilde{A}}_{,\lambda} 
   \nonumber \\
   &  & +  \frac{r_{,\tilde{B}}}{r\,r_{,\lambda}} 
     \left(
           2 \, r_{,\lambda \tilde{A}} \tilde{\eta}^{\tilde{A}\tilde{B}}
         + 2 \, \beta_{,\lambda} r_{,\tilde{A} }
         + r_{,\tilde{A}} \tilde{\eta}^{\tilde{A} \tilde{B}}_{,\lambda} 
     \right)
   - \frac{r_{,\tilde{A}} \, r_{,\tilde{B}}}{r^2}\, \tilde{\eta}^{\tilde{A}\tilde{B}}.
\label{eq:extract.W1}
\end{eqnarray}

With Eqs.~(\ref{eq:extract.W0}) - (\ref{eq:extract.W1})
evaluated on the world-tube $\Gamma$,  
 $W$ is found to second-order accuracy by 
\begin{equation}
	W(y^\alpha) = W^{(0)} + W^{(0)}_{,\lambda} \lambda + O(\lambda^{2}).
\end{equation}

At this point the four
 Bondi metric functions $J,\beta,U,W$ have been obtained as Taylor-expansions
around $\Gamma$, up to $O(\lambda^2)$.
The location of the world-tube $\Gamma$
in the Bondi frame is now known, since
at any point of $\Gamma$ the four Bondi
coordinate values $(u,r,q,p)$ are known from the extraction.
Let $M$ be a gridpoint on $\Gamma$, labeled by the
angular coordinates $(q_{[I]}, p_{[J]})$.
For a given time-step $t_{[N]}$
let $r_{[K]}$ be the nearest neighbor 
of $M$ in the radial direction.
Then the extraction provides boundary 
data for the characteristic code
by setting\footnote{In fact, at this point
 the extraction module
uses the compactified coordinate $x$ (as defined in 
Eq.~(\ref{eq:null.xdef})) instead of the Bondi radial 
coordinate $r$, i.e. it first locates $\Gamma$ in terms
of the $x$-grid, then expands the world-tube values
of $J,\beta,U,W$ in terms of $(x-x^{(0)})$. However, the
difference between the two approaches is $O(\Delta^2)$, 
which is insignificant.}

\begin{eqnarray}
J_{[I,J,K]} &=& J^{(0)}_{[I,J]} +
\frac{r_{[K]} - r^{(0)}_{[I,J]}}{r^{(0)}_{,\lambda[I,J]}} \cdot
J_{,\lambda [I,J]}^{(0)} + O(\Delta^2),\\
\beta_{[I,J,K]} &=& \beta^{(0)}_{[I,J]} +
\frac{r_{[K]} - r^{(0)}_{[I,J]}}{r^{(0)}_{,\lambda[I,J]}} \cdot
\beta_{,\lambda [I,J]}^{(0)} + O(\Delta^2), \\
W_{[I,J,K]} &=& W^{(0)}_{[I,J]} +
\frac{r_{[K]} - r^{(0)}_{[I,J]}}{r^{(0)}_{,\lambda[I,J]}} \cdot
W_{,\lambda [I,J]}^{(0)} + O(\Delta^2),
\end{eqnarray}
and
\begin{eqnarray}
U_{[I,J,K]} &=& U^{(0)}_{[I,J]} +
\frac{r_{[K+\frac 12]} - r^{(0)}_{[I,J]}}{r^{(0)}_{,\lambda[I,J]}} \cdot
U_{,\lambda [I,J]}^{(0)} + O(\Delta^2),
\end{eqnarray}
where we have taken in account the fact that the field $U$ is represented
on a radially staggered grid.
Furthermore, since the characteristic equations
involve second derivatives of $U$, its boundary data must
be provided at an additional
radial gridpoint neighboring the world-tube $\Gamma$.

\section{Calibration of the extraction module}

Given the amount of algebraic calculations involved in extraction,
one needs a number of test-beds to assure that they are implemented correctly.
Furthermore, finite differencing and interpolation
 involved in the module also needs to be checked for
proper convergence. 

In calibration we have first worked out analytically 
all quantities involved in
the extraction. Then, to check proper second-order convergence of the
extraction module, we have compared the numerical output of the
module with the analytic functions for finer and finer resolution,
keeping all physical parameters fixed.

It is worth noting that the Jacobian of the Cartesian-Bondi 
transformation is used not only in the extraction but also in the injection.
Thus we have checked convergence not only for the numeric Bondi metric but 
also for the  Cartesian-Bondi Jacobian as computed by the extraction module.

Two classes of analytic solutions were used. 
The first class consisted of
solutions known in Cauchy coordinates. Given the Cauchy metric we
used the algebraic steps prescribed in the extraction to obtain
the Bondi-Cartesian coordinate transformation as well as the Bondi 
metric as an expansion around the world-tube $\Gamma$. The second 
class of analytic solutions was provided in Bondi coordinates. 
In this case the Cartesian coordinates were defined by hand, i.e.
we defined the Cartesian frame by prescribing the functions 
$x^\alpha = f(y^\beta)$.
The approach is 
straightforward and algorithmic for both classes, 
but the actual calculations are
quite lengthy. 
Thus convergence tests not only validate the numerical
implementation, but also provide 
a consistency check between 
the numeric and algebraic calculations.

\subsection{Minkowski space}
\label{sec:extract.calib.mink}

For the case of Minkowski space-time the lightcone structure is known
 analytically. Thus the coordinate transformation
can be given globally (instead of as an expansion
around $\Gamma$). The Cauchy metric in Cartesian coordinates $(t,x,y,z)$ is

\begin{equation}
g_{\mu \nu} = \mbox{diag}[-1,1,1,1].
\end{equation}
The first coordinate transformation $(t,x,y,z) \rightarrow (u,\lambda,q,p)$
is defined by
\begin{eqnarray}
t & = & u + \lambda, \\
x &=& 2 (R_\Gamma + \lambda) \frac{q}{P}, \\
y &=& \pm 2 (R_\Gamma + \lambda) \frac{p}{P}, \\
z &=& \pm   (R_\Gamma + \lambda) \frac{2-P}{P},
\end{eqnarray}
where $P=1+q^2+p^2$ and the $+ (-)$ sign
corresponds to the north (south) patch.
The null vector in the two coordinate patches is
given by
\begin{equation}
\ell^\mu = \left( 1,\;\; \frac{2q}{P},\;\;  \pm \frac{2p}{P}, \;\; 
\pm \frac{2-P}{P} \right).
\end{equation}

The second coordinate transformation $(u,\lambda,q,p) \rightarrow (u,r,q,p)$
is defined by $r = R_\Gamma + \lambda$. The Bondi metric variables are
known globally:  $J = \beta = U  = W = 0$.
The analytic results were reproduced by the numerical calculations 
to machine precision.

\subsection{Teukolsky waves}
\label{sec:extract.teuk}

\subsubsection{The Cauchy metric}
\label{sec:extract.teuk.cauchy}

Next results are presented for a linearized solution
that describes propagation of a time-symmetric 
quadrupole wave (Teukolsky wave) for unit lapse, zero shift. 
 The Cauchy 3-metric
in coordinates $(z,\rho,\phi)$ is given by
(\cite{Eppley79}, p. 282f):
\begin{eqnarray}
g_{zz} &=& 1 + 3 \, D \, \cos^4 \theta
+ 6 \,(B-C)\, \cos^3 \theta + 3\, C - A, \\
g_{\rho \rho} &=& 1 + 3 \,D\, \sin^2 \theta\, \cos^2 \theta  - A, \\
g_{z \rho} &=& 3 \,\sin \theta\, \cos \theta\, ( D \,\cos^2 \theta + B - C ),\\
g_{\phi \phi} &=& \rho^2\, (1 + 3 \,( A - C)\, \sin^2 \theta - A),
\end{eqnarray}
where $\theta = \arccos(z/r)$. The functions $A, B, C$ and $D$ are given by
\begin{eqnarray}
A &=& 3 \left[
\frac{F^{(2)}}{r^3} + \frac{3F^{(1)}}{r^4} + \frac{3F}{r^5}
 \right], \\
B &=& - \left[ 
\frac{F^{(3)}}{r^2} +
\frac{3 \,F^{(2)}}{r^3} +
\frac{6 \,F^{(1)}}{r^4} +
\frac{6 \,F}{r^5}
\right], \\
C &=& \frac 14 \left[ 
\frac{F^{(4)}}{r} +
\frac{2 \,F^{(3)}}{r^2} +
\frac{9 \,F^{(2)}}{r^3} +
\frac{21 \,F^{(1)}}{r^4} +
\frac{21 \,F}{r^5}
\right], \\
D &=& A + C - 2 \, B, \\
F &=& f(t+r) - f(t-r) , \\
F^{(n)} &\equiv& \left[ \frac {d^n f(x)} {d x^n} \right]_{x=t+r} 
- \left[ \frac {d^n f(x)} {d x^n} \right]_{x=t-r},
\end{eqnarray}
with $r = \sqrt{x^2+y^2+z^2}$ and
\begin{equation}
f(x) = \varepsilon x \, e^{-(x/\varpi)^2},
\end{equation}
so that at $t=0$ we obtain a Gaussian pulse of width $\varpi$
around the origin
of the coordinate system.

The coordinate transformation $(z,\rho,\phi) \rightarrow
(x,y,z)$ is defined in the usual way $\rho^2 = x\cdot x + y \cdot y, \;
\phi = \arctan (y/x)$, providing the 3-metric in Cartesian coordinates.

\subsubsection{The extraction quantities}
\label{sec:extract.teuk.extract}

The algebraic expressions
involved in extraction are worked out up to $O(\varepsilon^2)$.

For the quadrupole solution described above 
the null vector $\ell^{(0)\mu}$ has the form

\begin{eqnarray}
\label{eq:extract.teuk.ell.FIRST}
\ell^{(0)t} &=& 1, \\
\ell^{(0)x} &=&  2 \frac{q}{P} +
\varepsilon  \frac{q}{P}
\left( \gamma_1 +  \frac{\gamma_2}{P} -  \frac{\gamma_2}{P^2} 
\right) + O(\varepsilon^2), \\
\ell^{(0)y} &=& \pm 2\frac{p}{P} \pm \varepsilon \frac{p}{P}
\left( 
 \gamma_1 +  \frac{\gamma_2}{P} -  \frac{\gamma_2}{P^2} 
\right) + O(\varepsilon^2),\\
\ell^{(0)z} &=&
\pm \frac{2-P}{P} + \varepsilon 
\left(
\gamma_3 + \frac{\gamma_4}{P} \pm \frac{3 \gamma_2}{2 P^2}
\mp \frac{\gamma_2}{ P^3}
\right) + O(\varepsilon^2).
\label{eq:extract.teuk.ell.LAST}
\end{eqnarray}
The $\lambda$-derivative of the null vector at the world-tube $\Gamma$ 
is given by
\begin{eqnarray}
\label{eq:extract.teuk.ell_l.FIRST}
\ell^{(0)t}_{,\lambda} &=& \varepsilon\;  \frac{6 - 6 P + P^2}{P^2} \; \gamma_9, \\
\ell^{(0)x}_{,\lambda} &=& \varepsilon \frac{q}{P}
\left( \gamma_5 + \frac{\gamma_6}{P} - \frac{\gamma_6}{P^2} \right), 
\\
\ell^{(0)y}_{,\lambda} &=& \pm \varepsilon \frac{p}{P}
\left( \gamma_5 + \frac{\gamma_6}{P} - \frac{\gamma_6}{P^2} \right), 
\\
\ell^{(0)z}_{,\lambda} &=& \varepsilon \left(
\gamma_7 + \frac{\gamma_8}{P} \pm \frac{3 \gamma_6}{2 P^2}
\mp  \frac{\gamma_6}{P^3}
\right).
\label{eq:extract.teuk.ell_l.LAST}
\end{eqnarray}

Equations~(\ref{eq:extract.teuk.ell.FIRST}) - (\ref{eq:extract.teuk.ell.LAST})
and (\ref{eq:extract.teuk.ell_l.FIRST}) - (\ref{eq:extract.teuk.ell_l.LAST})
 give some of the components of the Jacobian 
$\partial x^\alpha / \partial \tilde y^{\tilde \beta}$. The ones that 
still need to be specified are  angular  and time derivatives
of $\ell^\mu$ and  $\ell^\mu_{,\lambda}$.

The expansion of the Bondi coordinate
 $r(u,\lambda,q,p)$ around $\Gamma$  is given by
\begin{eqnarray}
r(u,q,p,\lambda) = R_\Gamma &+&
\varepsilon \, \frac {6 - 6 P + P^2}{P^2} \, \gamma_{10}\nonumber \\  
&+&
\left(1 - \varepsilon \, \frac {6 - 6 P + P^2}{P^2} \, \gamma_{11} \right) \lambda
\nonumber \\ 
&+& O(\varepsilon^2) \, O(\lambda^2).
\end{eqnarray}
The  characteristic functions $J, U,W$ can be expanded around the world-tube as follows:
\begin{eqnarray}
J &=& \varepsilon  \frac{q^2 - p^2 + 2 I q p}{P^2} \left( 
\gamma_{12} + \lambda \gamma_{13} + O(\lambda^2)
\right) + O(\varepsilon^2),
\\
U &=& \varepsilon  \frac{(P-2)(q+I\,p)}{P^2} \left( 
\gamma_{14} + \lambda \gamma_{15} + O(\lambda^2)
\right) + O(\varepsilon^2),
\\
W &=& \varepsilon  \frac{6 - 6 P + P^2}{P^2} \left( 
\gamma_{16} + \lambda \gamma_{17} + O(\lambda^2)
\right) + O(\varepsilon^2).
\end{eqnarray}
The function $\beta^{(0)}$ is known 
once $r_{,\lambda}$ is given (see Eq.~(\ref{eq:extract.beta=logrl})).
In linearized theory, one can write 
$\beta = \beta^{(0)} + O(\varepsilon^2)$
since, as it can be seen in Eq.~(\ref{eq:extract.betal}), 
the expression $\beta_{,\lambda}$ is fully non-linear.

The symbols $\gamma_{1\ldots 17}$ stand for functions defined on $\Gamma$ that 
depend on the retarded-time coordinate $u$ but not on the  angular coordinates $(q,p)$.
Explicit expressions for $\gamma_{1\ldots 17}$ 
can be found in Appendix~\ref{app:teuk-ext}.

\subsubsection{Numeric results}

\begin{figure}
\centerline{\epsfxsize=4in\epsfbox{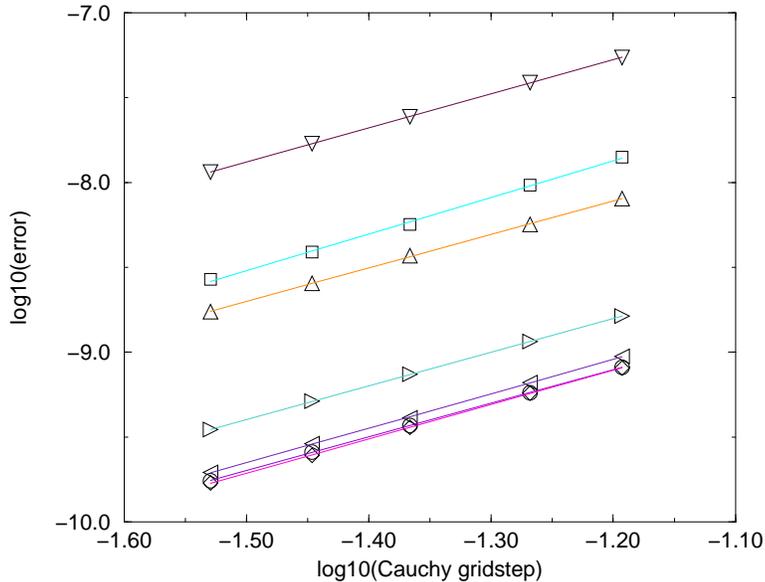}}
\caption{Convergence test of the Bondi metric quantities 
$J,$ $J_{,\lambda},$ $ \beta,$ $U,$ 
$U_{,\lambda},$ $W,$ $W_{,\lambda}$ for the quadrupole
linear waves. The function
$\beta_{,\lambda}$ has been omitted 
because it is $O(\varepsilon^2)$.
The error is measured 
with the $\ell_\infty$ norm of the difference 
between numeric and analytic values, which was found to be $O(\Delta^{1.97})$.
Parameters are $t=2, R_\Gamma = 1, \varepsilon = 10^{-8}, \varpi = 0.5$. }
\label{fig:extract.calib.teuk}
\end{figure}

The numeric extraction quantities were compared against the analytic results
for grid sizes $32,48,64,80$ and $96$. All extraction 
quantities have been found to be
second-order accurate, as seen from the slope of the 
convergence plot of the Bondi metric variables on the world-tube
 in Figure~\ref{fig:extract.calib.teuk}.

\subsection{SIMPLE}
\label{page:extract:SIMPLE}

Another test-bed was constructed starting from the non-linear, static,
axisymmetric solution SIMPLE
as described in \cite{Gomez94a}. The solution has already been mentioned 
on page \pageref{page:null.SIMPLE}  as a test-bed for the
characteristic code. Here we  use it in a different gauge so that the
axis of symmetry lies in the $y$-direction. This results in a more thorough
test of the extraction algebra since the stereographic coordinate
patches  are symmetrized around the $z$-axis. 

The space-time metric in Bondi null
coordinates $(u,r,\theta,\phi)$ takes the form
\begin{eqnarray}
ds^2 \ ((\hat V/r) e^{2 \hat \beta} 
- \hat U^2 r^2 e^{2 \hat \gamma} ) du^2 + 2 e^{2 \hat \beta} du \,
dr + 2\hat  U r^2 e^{2 \hat \gamma} du \, d\theta \nonumber \\
- r^2 (e^{2 \hat \gamma} d \theta^2 + e^{-2 \hat \gamma} sin^2 \theta 
d \phi^2),
\end{eqnarray}
where
\begin{eqnarray}
2 e^{\hat \gamma} = 1 + \Sigma, \;\;\;
e^{2 \hat\beta} = \frac{(1 + \Sigma)^2}{4 \Sigma}, \\
\hat U = - \frac{a^2 \rho \sqrt{r^2 - \rho^2}}{r \Sigma}, \\
\hat V = \frac{r (2 a^2 \rho^2 - a^2 r^2 + 1)}{\Sigma},
\end{eqnarray}
 $\rho = r \sin \theta, \Sigma = \sqrt{1 + a^2 \rho^2}$,
and $a$ is a free scale parameter.

A coordinate transformation 
$(u,r,\theta,\phi) \rightarrow (u, \lambda,\theta, \phi)$ 
is first performed by solving the differential equation
\begin{equation}
d \lambda  = e^{2 \hat \beta} dr.
\label{eq:extract.calib.simple.dldr}
\end{equation}
Even though $\hat \beta$ is known in terms of $r$, 
Eq.~(\ref{eq:extract.calib.simple.dldr}) cannot be integrated in
closed form. Therefore 
$\hat \beta$ (and all subsequent expressions) are expanded in terms of $a$. Then, one obtains
\begin{equation}
\lambda + R_\Gamma = r + \frac{1}{80} r^5 sin(\theta)^4 a^4 + \cdots,
\end{equation}
which determines $r$ as a function of $\lambda$:
\begin{equation}
r = (\lambda + R_\Gamma) - \frac{1}{80}(\lambda 
+ R_\Gamma)^5 sin(\theta)^4 a^4 + \cdots .
\end{equation}
Here the integration constant of Eq.~(\ref{eq:extract.calib.simple.dldr}) 
has been fixed by the gauge condition
\begin{equation}
\lambda_{| a = 0} = r - R_\Gamma.
\end{equation}
This coordinate transformation provides an affine null metric. 
To change the axis of symmetry of the space-time first
a Cartesian frame  $(t, \hat x, \hat y, \hat z)$
is defined by
\begin{eqnarray}
\label{eq:extract.simple.gauge1.first}
u &=& t - \lambda, \;\;\;
 (\lambda + R_\Gamma)^2 \,=\, \hat x^2 + \hat y^2 + \hat z^2, \nonumber \\ 
\theta &=& \arccos \left( \frac{\hat z}{\lambda + R_\Gamma} \right),\;\;\;
\phi \,=\, \arccos \left( \frac{\hat x}{\sqrt{\hat x^2 + \hat y^2}} \right),
\label{eq:extract.simple.gauge1.last}
\end{eqnarray}
then the $y$ and $z$ axes are rotated by $90^0$ according to
\begin{equation}
\hat x = x,\;\;\; \hat y=  z,\;\;\; \hat z = -y. 
\label{eq:extract.simple.gauge2}
\end{equation}

Thus a Cartesian solution is constructed (in coordinates $(t,x,y,z)$),
 symmetric 
with respect to the $y$-axis. Using this Cauchy metric
the algebraic steps prescribed by the extraction routine are followed to
 compute in the neighborhood of $\Gamma$ the 
lightcone generators $\ell^\alpha$, the affine and the Bondi null metric, 
and   the metric variables $J, \beta, U, W$.

In order to be able to give values for $a$ that exploit the non-linear
feature of the test-bed, we have worked out the expansion terms up to
$O(a^{12})$. Expressions for $r, J, \beta, U, W$ and their 
$\lambda$-derivatives can be found in Appendix~\ref{app:simple}.

The extraction code 
showed proper, second order convergent behavior to the analytic results.

\subsection{Schwarzschild in differentially 
rotating null coordinates}
\label{sec:extract.calib.schrot}

The Schwarzschild metric in outgoing Kerr-Schild coordinates 
${\hat x}^{\alpha}$ is given by
\begin{equation}
     ds^2 = -d{\hat t}^2 +d{\hat x}^2+d{\hat y}^2+d{\hat z}^2
            +{2m\over {\hat r}} (- d{\hat t} +
            {{\hat x}d{\hat x}+{\hat y}d{\hat y}+{\hat z}d{\hat z}
             \over {\hat r}})^2,
\label{eq:extract.sch.kerrshild}
\end{equation}
where ${\hat r}^2 ={\hat x}^2 +{\hat y}^2+{\hat z}^2$. 
(Equation~(\ref{eq:extract.sch.kerrshild}) is obtained from 
the metric of a Kerr black hole Eq.~(\ref{eq:null.kerrBH}), with angular
momentum set to zero.)
This metric and its derivatives on the world-tube ${\hat x}^2 +{\hat
y}^2+{\hat z}^2= R_\Gamma^2$ can be used to provide the boundary values for a
null coordinate system based upon the family of outgoing null cones
emanating from the ${\hat t}=$ constant foliation. In a simple choice of
null coordinates ${\hat y}^{\alpha}=({\hat u},{\hat r},{\hat \theta},{\hat \phi})$
 associated with ${\hat x}^{\alpha}$ 
we have ${\hat u}= {\hat t}-{\hat r}+R_\Gamma, \, 
\hat r = \sqrt{\hat x^2 + \hat y^2 + \hat z^2}, \,
\cos(\hat \theta) = \hat z /\sqrt{\hat x^2 + \hat y^2 + \hat z^2} , \, \tan ( \hat \phi) = \hat y / \hat x$. This
leads to the Eddington-Finkelstein version of the Schwarzschild metric,
\begin{equation}
     ds^2=-(1-2m/{\hat r})d{\hat u}^2 -2d{\hat u}d{\hat r}-{\hat r}^2
            q_{AB}d{\hat y}^A d{\hat y}^B.
\end{equation}

The world-tube and the outgoing null cones have intrinsically
spherically symmetric geometries. We obtain a
non-spherically symmetric null metric, suitable for code testing, by
introducing gauge freedom.  We set ${\hat u}=u+\psi (u)$, 
${\hat r}=r$, ${\hat \theta}=\theta$ and
${\hat \phi}=\phi + \xi (u,\theta)$, which leads to time-dependent 
$\beta, J$ and $U$.

Unfortunately, no $r$-dependence in $J$ can be introduced this way.
The spherical symmetry of the null cone requires that its shear vanish,
i.e. that $m^{\alpha} m^{\beta} k_{{\alpha};{\beta}} =0$, where
$k^{\alpha}$ and the complex spacelike vector $m^\alpha$ span the space
tangent to the null hypersurface. In a null coordinate system
$(u,r,y^A)$ with angular metric $h_{AB}$, the shear free condition can
be formulated as
\begin{equation}
   q^A q^B (k_{A;B} -{1\over 2} h_{AB}h^{CD}k_{C;D})=0,
\end{equation}
which is equivalent to the requirement that $J_{,r}=0$. So gauge
transformations that preserve the world-tube and its foliation can
introduce a non-vanishing $J$ but $J_{,r}$ and consequently
$\beta_{,r}$ vanish.

The Cartesian coordinates ${\hat x}^{\alpha}$ are also related to a
differentially rotating Cartesian frame by $x^{\alpha}$ by ${\hat
t}=t+\psi$, ${\hat x}=x\cos \xi -y\sin \xi$, ${\hat y}=x\sin \xi+y\cos
\xi$ and ${\hat z}=z$. The $x^{\alpha}$ and $y^{\alpha}$ frames are
related by the standard construction $u=t-r +R_\Gamma$, $r^2=x^2+y^2+z^2$,
$\theta(x,y,z) = {\rm arccos} (z/r)$,  and $\phi ={\rm arccot} (x/y)$.

The code tests are based upon extraction from the $x^{\alpha}$ frame to
obtain the null metric in the  $y^{\alpha}$ frame and on the reverse
injection process.  The explicit form of the Cartesian and
null metrics are listed below.

\subsubsection{Analytic results in Cartesian and Bondi coordinates}

The Cartesian 3-metric, the lapse, and the shift are 

\begin{eqnarray}
g_{xx} = 1&+&{\frac {2\,{x}^{2}m}{{{\it {r}}}^{3}}}
-{\frac {2\,xyz{\it {{\xi}_{\theta}}}}{{{\it {r}}}^{2}{\rho}}}
-{\frac {2\,{x}^{2}z \rho \, {\it {{\xi}_{\theta}}}\,{\it {{\xi}_{u}}}}{{{\it {r}}}^{3}}}
+{\frac {{x}^{2}{\rho}^{2}{{\it {{\xi}_{u}}}}^{2}}{{{\it {r}}}^{2}}}
+{\frac {2\,xy{\it {{\xi}_{u}}}}{{\it {r}}}}
\nonumber \\
&+&{\frac {{x}^{2}{z}^{2}{{\it {{\xi}_{\theta}}}}^{2}}{{{\it {r}}}^{4}}}
-{\frac {{x}^{2}\left ({\it {r}}-2\,m\right ){{\it \psi'(u)}}^{2}}{{{\it {r}}}^{3}}}
+{\frac {4\,{x}^{2} m {\it \psi'(u)}}{{{\it {r}}}^{3}}},
\\
g_{yy} = 1 &+& {\frac {2\,{y}^{2}m}{{{\it {r}}}^{3}}}
+{\frac {2\,xyz{\it {{\xi}_{\theta}}}}{{{\it {r}}}^{2}{\rho}}}
-{\frac {2\,y^2 z \rho {\it {{\xi}_{\theta}}}\,{\it {{\xi}_{u}}}}
 {r^{3}}}
+{\frac {{y}^{2}{\rho}^{2}{{\it {{\xi}_{u}}}}^{2}}{{{\it {r}}}^{2}}}
- {\frac {2\,xy{\it {{\xi}_{u}}}}{{\it {r}}}}
\nonumber \\
&+&{\frac {y^2 {z}^{2}{{\it {{\xi}_{\theta}}}}^{2}}{{{\it {r}}}^{4}}}
-{\frac {{y}^{2}\left ({\it {r}}-2\,m\right ){{\it \psi'(u)}}^{2}}{{{\it {r}}}^{3}}}
+{\frac {4\,{y}^{2} m {\it \psi'(u)}}{{{\it {r}}}^{3}}},
\\
g_{zz} = 1 &+& {\frac {2\,{z}^{2}m}{{{\it {r}}}^{3}}}
+{\frac {\rho^{4}{{\it {{\xi}_{\theta}}}}^{2}}{{{\it {r}}}^{4}}}
+{\frac {2\,z \rho^{3}
 {\it {{\xi}_{\theta}}}{\it {{\xi}_{u}}}}{{{\it {r}}}^{3}}}
+{\frac {{z}^{2}\rho^2{{
\it {{\xi}_{u}}}}^{2}}{{r}^{2}}}
\nonumber \\
&+&{\frac {{z}^{2}\left (2\,m-{\it {r}}\right ){{\it \psi'(u)}}^{2}}{{{\it {r}}}^{3}}}
+{\frac {4\,{z}^{2}m{\it \psi'(u)}}{{{\it {r}}}^{3}}},
\\
g_{xy} = & \; &{\frac {2\,xym}{{{\it {r}}}^{3}}}
+{\frac {x y {z}^{2}{{\it {{\xi}_{\theta}}}}^{2}}{r^{4}}}
+{\frac {\left ({x}^{2}-{y}^{2}\right ) \, z \, {\it {{\xi}_{\theta}}}}
{{{\it {r}}}^{2}\rho }}
\nonumber \\
&-&{\frac {2\,xyz\rho {\it {{\xi}_{\theta}}}\,{\it {{\xi}_{u}}}}
{r^3}}
+{\frac {xy {\rho}^{2}{{\it {{\xi}_{u}}}}^{2}}{{{\it {r}}}^{2}}}
-{\frac {\left ({x}^{2}-{y}^{2}\right ){\it {{\xi}_{u}}}}{{\it {r}}}}
\nonumber \\
&-&{\frac {xy\left ({\it {r}}-2\,m\right ){{\it \psi'(u)}}^{2}}{{{\it {r}}}^{3}}}
+{\frac {4\,x y m{\it \psi'(u)}}{{{\it {r}}}^{3}}},
\\
g_{xz} = & \; &{\frac {2\,xzm}{{{\it {r}}}^{3}}}
-{\frac {x z \rho^2{{\it {{\xi}_{\theta}}}}^{2}}{{{\it {r}}}^{4}}}
+{\frac {y\rho{\it {{\xi}_{\theta}}}}{{r}^{2}}}
\nonumber \\
&+&{\frac {x\rho\left 
 ({\rho}^{2}-{z}^{2}\right ){\it {{\xi}_{\theta}}}\,{\it {{\xi}_{u}}}}{{{\it {r}}}^{3}}}
+{\frac {xz\rho^2{{\it {{\xi}_{u}}}}^{2}}{{{\it {r}}}^{2}}}
\nonumber \\
&+&{\frac {x z \left (2\,m-{\it {r}}\right ){{\it \psi'(u)}}^{2}}{{{\it {r}}}^{3}}}
+{\frac {4\,x z m{\it \psi'(u)}}{{{\it {r}}}^{3}}}
+{\frac {y z\,{\it {{\xi}_{u}}}}{{\it {r}}}},
\\
g_{yz} = & \; & {\frac {2\, y z m}{{{\it {r}}}^{3}}}
-{\frac {y z \rho^2 {{\it {{\xi}_{\theta}}}}^{2}}{{{\it {r}}}^{4}}}
-{\frac {x \rho{\it {{\xi}_{\theta}}}}{{r}^{2}}}
\nonumber \\
&+&{\frac {y \rho \left ({\rho}^{2}-{z}^{2}\right )
 {\it {{\xi}_{u}}}\,{\it {{\xi}_{\theta}}}}{{{\it {r}}}^{3}}}
+{\frac {y z {\rho}^{2}{{\it {{\xi}_{u}}}}^{2}}{{{\it {r}}}^{2}}}
\nonumber \\
&+&{\frac {y z \left (2\,m-{\it {r}}\right ){{\it \psi'(u)}}^{2}}{{{\it {r}}}^{3}}}
+{\frac {4\,y z m{\it \psi'(u)}}{{{\it {r}}}^{3}}}
-{\frac {x z {\it {{\xi}_{u}}}}{{\it {r}}}},
\\
\alpha = & \; & \sqrt{\frac{1 + \psi'(u)}{1 + \frac{2 m}{r} - 
\psi'(u) \left( 1 - \frac{2 m}{r} \right) }},
\\
\beta^x = & \; & {\frac {-2\,xm+x\left (r-2\,m\right ){\it \psi'(u)}-y r^2 
{\it {{\xi}_{u}}}}{{\it {r}}\,\left [-\left ({\it {r}}-2\,m\right )\, 
{\it \psi'(u)}  +2\,m+{\it {r}}\right ]}},
\\
\beta^y = & \; & {\frac {-2\,ym+y\left (r-2\,m\right ){\it \psi'(u)}+x{r}^{2}
{\it {{\xi}_{u}}}}{{\it {r}}\,\left [-\left ({\it {r}}-2\,m\right )\, {\it \psi'(u)}  +2\,m+{\it {r}}\right ]}},
\\
\beta^z = & \; & {\frac {-2\,zm+z\left (r-2\,m\right ){\it \psi'(u)}}{{\it {r}}\,\left [-
\left ({\it {r}}-2\,m\right )\, {\it \psi'(u)} +2\,m+{\it {r}}\right ]}},
\end{eqnarray}
where 
\begin{eqnarray}
\rho  &=& \sqrt{x^2+y^2}, \nonumber \\
r &=& \sqrt{x^2+y^2+z^2}, \nonumber \\
{\xi}_{\theta} &=& \frac{\partial}{\partial \theta} \xi(u,\theta),
\nonumber \\
{\xi}_{u} &=& \frac{\partial}{\partial u} \xi(u,\theta),
\nonumber
\end{eqnarray}
with $\theta(x,y,z) = {\rm arccos} (z/r)$, and  $u = t - r + R_\Gamma$. 

Here $(u, r, \theta, \phi =\arccos (x/y))$ are the differentially
rotating null coordinates $y^{\alpha}$.  The Bondi metric
variables are 

\begin{eqnarray}
J &=& \frac {\left [2\,\left ( q^2-p^2 \right) + 4 i qp \right ] }
{P} 
\left [ \frac{\xi_\theta^2}{P} + \frac{i \xi_\theta}{\sqrt{p^2+q^2}} \right ],  \\
J_{,\lambda} &=& 0,
\\
\beta &=& {\frac {\ln (1+{\it {\psi'(u)}})}{2}},
\\
\beta_{,\lambda} &=& 0,
\\
U &=& \pm {\frac {2\,{\it {{\xi}_{u}}}\,\left ( p - I q \right )}{P}},
\\
U_{,\lambda} &=& 0,
\\
W &=& {\frac {\left (r-2\,m\right )\left (1+{\it {\psi'(u)}}\right )-r}{{r}^{2}}},
\\
W_{,\lambda} &=& {\frac {\left (-r+4\,m\right ){\it {\psi'(u)}}+4\,m}{{r}^{3}\left (1+{\it {\psi'(u)}}\right )}},
\end{eqnarray}
with $P = 1+p^2+q^2$, using the positive (negative) 
sign for the north (south) patch. Also it is known that 
\begin{eqnarray}
\theta_N (q,p) &=&  2 \, {\rm arctan} ( \sqrt{p^2+q^2}), \nonumber \\
\theta_S (q,p) &=&  2 \, {\rm arccot} ( \sqrt{p^2+q^2}). \nonumber
\end{eqnarray}

The coordinate transformation 
$( u, \lambda, q, p ) \rightarrow ( t,x,y,z )$
 is given by
\begin{eqnarray}
x &=& 2 \frac{\left( R_\Gamma + \frac{\lambda}{1 + \psi'(u)}\right) q }{1 + q^2 + p^2}, \nonumber \\
y &=& \pm 2 \frac{\left( R_\Gamma + \frac{\lambda}{1 + \psi'(u)}\right) p }{1 + q^2 + p^2}, \nonumber \\
z &=& \pm  \frac{\left( R_\Gamma + \frac{\lambda}{1 + \psi'(u)}\right), 
\left(1 - p^2 - q^2\right) }{1 + q^2 + p^2} \\
t &=& u + \frac{\lambda}{1 + \psi'(u)}.
\end{eqnarray}
All components of the null vector $\ell^\alpha$ corresponding to this 
coordinate transformation are independent of $\lambda$:
\begin{equation}
 \ell^\alpha_{,\lambda} = 0.
\end{equation}

\subsubsection{Test Results}

With all extraction quantities worked out analytically, the solution
provides a useful test-bed in calibrating the extraction module.
Thus convergence rates were measured not only for the Bondi functions 
but also for the various terms involved in the coordinate transformations.
The  numeric results of the extraction module exhibited second order convergence
to the analytic solution.

\subsection{Dynamic black-hole space-times}

In addition to the tests described in 
Sections~\ref{sec:extract.calib.mink} - \ref{sec:extract.calib.schrot},
there are a number of test-beds used for calibrating the
 characteristic evolution whose boundary
conditions
 were provided by the extraction module 
(see  Section~\ref{sec:null.1BH}).
The performance of those tests is not only a successful test
 of the characteristic code,
but it also proves that the extraction module is able to provide 
boundary data for the characteristic code.

\chapter{The Injection Module}
\label{chap:injection}

The injection module consists of a set of numerical routines
that are designed to construct Cauchy boundary data. The module
 starts from the characteristic
metric functions, performs
 the coordinate transformation $$(u, r, q, p) \rightarrow (t, x, y, z),$$
and uses a number of interpolation routines to obtain the Cauchy
metric at the Cartesian boundary gridpoints.

This chapter describes the injection module and its calibration in detail.

\section{The physical algorithm}

In injection, one first needs to define a 4-D region of space-time
(with respect to the 
Bondi frame $(y^\beta)$)
that surrounds the boundary of the Cauchy evolution domain.
This is done using the location of the extraction world-tube
$\Gamma$ as computed in the extraction and  the fact that injection will occur in 
an $O(\Delta)$ neighborhood of $\Gamma$.

In this region the Cartesian coordinates 
$x^\alpha = (t,x,y,z)$ need to be expressed 
as functions of the Bondi coordinates, i.e. we need the dependence
$x^\alpha = f(y^\beta)$. We also need the Jacobian terms 
$\partial x^i / \partial y^\beta$.
First the parameter $\lambda$ is computed as an integral
starting from $\Gamma_{u=t}$:
\begin{equation}
\lambda(y^\beta) =
\int_{\Gamma_{u}}^{r}\frac{dr}{r_{,\lambda}}
= \int_{\Gamma_{u}}^{r} e^{2 \beta} dr
\label{eq:inject.anal.lambda_int}
\end{equation}
where $\beta$ is known from the characteristic evolution.
During the integration of Eq.~(\ref{eq:inject.anal.lambda_int})
the coordinates $(u,q,p)$ are held fixed.

Next the values  ($x^\alpha$) are evaluated as an expansion around
$\Gamma$:
\begin{equation}
x^\alpha = x^{(0)\alpha} + \lambda \,\ell^{(0)\alpha} 
+ O(\lambda^2).
\label{eq:inject.anal.x}
\end{equation}

The Jacobian terms $\partial x^i / \partial y^A$ and $\partial x^i / \partial u$
can be computed up to $O(\lambda^2)$, by taking angular and time derivatives 
of Eq.~(\ref{eq:inject.anal.x}).
The remaining Jacobian terms  $\partial x^i / \partial r$ are computed 
in the same manner as $x^{\alpha}$, i.e. as an expansion around $\Gamma$:
\begin{equation}
x^i_{,r} = 
\frac{\partial x^i}{\partial \lambda}\, 
\frac{ \partial \lambda}{\partial r} 
= e^{2 \beta}\left( \ell^{(0)i} + \lambda \ell^{(0)i}_{,\lambda}
+ O(\lambda^2) \right).
\label{eq:inject.anal.x_l}
\end{equation}
In Eqs.~(\ref{eq:inject.anal.x}) - (\ref{eq:inject.anal.x_l})
the quantities $x^{(0)\alpha}, \ell^{(0)\alpha}$ and $\ell^{(0)\alpha}_{,\lambda}$
are provided by the extraction module.

Next the Bondi metric tensor $\eta^{\alpha\beta}$ is constructed using
the values $J, \beta, U$ and $W$:
\begin{eqnarray}
\label{eq:inject.anal.etaup.FIRST}
\eta^{rr} &=& e^{-2\beta} \left(r \cdot W + 1 \right), \\
\eta^{rq} &=& - \frac P2 \, e^{-2\beta} \cdot \Re(U),  \\
\eta^{rp} &=& - \frac P2 \, e^{-2\beta} \cdot \Im(U),  \\
\eta^{qq}  &=& \left( \frac{P}{2 r} \right)^2 \left( K - \Re(J) \right), \\
\eta^{qp}  &=& \left( \frac{P}{2 r} \right)^2 \left(   - \Im(J) \right), \\
\eta^{pp}  &=& \left( \frac{P}{2 r} \right)^2 \left( K + \Re(J) \right), \\
\eta^{ru} &=& - e^{-2\beta}, \;\; \eta^{qu} = \eta^{pu} = \eta^{uu} = 0.
\label{eq:inject.anal.etaup.LAST}
\end{eqnarray}
The spatial components of the Cauchy metric are then given by
\begin{equation}
g^{ij} = \frac{\partial x^i}{\partial y^{\alpha}}
            \frac{\partial x^j}{\partial y^{\beta}} \eta^{\alpha \beta}.
\label{eq:inject.anal.g4ij}
\end{equation}

Recall that Eqs.~(\ref{eq:inject.anal.lambda_int}) - (\ref{eq:inject.anal.g4ij})
are applied in an infinitesimal,   4-D neighborhood of 
the spacelike Cauchy boundary region. Thus the 4-metric components $g^{ij}$
can be evaluated at 
the Cauchy injection time  $t_{inj}$. Furthermore, at this time-level 
the gauge functions $\alpha, \beta^i$ (lapse and shift) 
are known. 
Also, recall from Section \ref{sec:lapse&shift} that for a Cauchy slice $\Sigma_{t_{inj}}$
with time-like unit normal $n^{\alpha} 
= \left(\frac{1}{\alpha}, \frac {-\beta^i}{\alpha}\right)$
the space-time metric components
$g^{ij}$ and  the intrinsic metric ${}^{(3)}g^{ij}$ of the  3-D 
Cauchy surface are related by
\begin{equation}
{}^{(3)}g^{ij}_{|t_{inj}} = g^{ij}_{|t_{inj}} + n^i n^j.
\label{eq:inject.anal.g3ij}
\end{equation}
Thus, we have obtained the contravariant 3-metric ${}^{(3)}g^{ij}$
at the injection time $t_{inj}$.
Given that ${}^{(3)}g^{ij}\, {}^{(3)}g_{jk} = \delta^i_k$, 
the covariant 3-metric ${}^{(3)}g_{ij|t_{inj}}$ 
is obtained via an elementary matrix inversion.

\section{Implementation}

Let $\left\{ \Lambda^{[I]}\right\}_{I  = 1 \cdots M}$ be a
 set of spherical world-tubes concentric with $\Gamma$ with radii $R^{[I]}$, 
defined by
$\sqrt{x^2+y^2+z^2} = R^{[I]}$. Let $\Lambda^{[I]}_t$ denote
the intersection of $\Lambda^{[I]}$ 
with the Cauchy slice  $\Sigma_t$.
Let the set of Cartesian radii $ R^{[I]}$
be  equally spaced, with $R^{[I+1]}-R^{[I]} = O(\Delta)$,
 such that all Cauchy boundary points
are contained between the two spheres $\Lambda^{[1]}_t$ and 
$\Lambda^{[M]}_t$ for all $t$. Note that in previous chapters, for simplicity, 
 a single injection world-tube  $\Lambda$ was introduced  
which in the current
context is identified with $\Lambda^{[M]}$, i.e. $R_\Lambda \equiv R^{[M]}$.

Defining a 4-D space-time region
that surrounds the Cauchy boundary region amounts to selecting a 
subset ${\cal S}_{Null}$ of the characteristic grid that
 surrounds the set of world-tubes $\left\{ \Lambda^{[I]} \right\}_{I=1\cdots M}$
in a 4-D sense.

This subset 
  ${\cal S}_{null}$ extends from
the extraction world-tube $\Gamma$ to a few grid-zones outside 
the injection world-tube $\Lambda^{[M]}$. Since the location of 
$\Lambda^{[M]}$
is defined with respect to Cartesian coordinates, in the Bondi
frame it changes shape and location during the numerical evolution.
Thus the choice of ${\cal S}_{null}$ can be done in two ways:
\begin{itemize}
\item Choose ${\cal S}_{null}$ once (at the first iteration)
such that it surrounds not only 
$\Lambda^{[M]}$ but a sphere of a larger radius $(1+f) \cdot R^{[M]}$
with $f>0$ a safety factor. Then the correctness of the choice of $f$ needs to be 
monitored time-step after time-step.
\item Another alternative is to choose ${\cal S}_{null}$ dynamically, 
i.e. at each time-step localize $\Lambda^{[M]}_t$ in Bondi coordinates and
use null gridpoints that adequately cover the Cauchy injection points. 
\end{itemize}

The current implementation  of the injection module uses 
the first way.

Let $u_n$ label the last time level of the
characteristic evolution.
Then 
computation of the Cartesian coordinate values $(x^\alpha)$ and 
of the Jacobian terms $\partial x^i / \partial y^\beta$ 
is done
at the retarded time $u_{n-1/2} \equiv u_n - \frac 12 \Delta u$, 
where $\Delta u = \Delta t$.
Furthermore, the $O(\lambda)$ terms of  $\partial x^i / \partial y^A$
are computed via second-order finite-difference formulae
for the angular derivatives, with the $O(1)$ terms known analytically.
In a similar way, with a vanishing $O(1)$ term in $\partial x^i / \partial u$,
the $O(\lambda)$ term 
is computed using
\begin{equation}
\left(\frac{\partial x^i}{\partial u}\right)_{|u_{n-1/2}} =
\frac{ x^i_{u_n} - x^i_{u_{n-1}}}{\Delta u} + O(\Delta^2).
\label{eq:inject.numer.jac_qp}
\end{equation}

The contravariant Bondi and Cauchy metric components
are computed using 
 Eqs.~(\ref{eq:inject.anal.etaup.FIRST}) - (\ref{eq:inject.anal.etaup.LAST})
and (\ref{eq:inject.anal.g4ij}) applied at $u=u_{n-1/2}$.

At this point the Cauchy metric components $g^{ij}$ are 
known on a set of gridpoints $(r_k,q_i,p_j)$,
at a number of retarded time levels $u_{n-1/2}, u_{n-3/2},$ etc.
(Results from previous time-steps are assumed to be known.)
Furthermore, all characteristic gridpoints are labeled not only by their
Bondi coordinates $(u,r,q,p)$, but also by their Cartesian coordinates
$x^\alpha=(t,x,y,z)$.

In order to compute the quantity ${}^{(3)}g_{ij}$ at a Cartesian boundary grid
point labeled by
$(t_{inj}, x^*, y^*, z^*)$, the injection module decomposes the
problem of 4-D interpolation into a number of simpler problems:

First a radial interpolation is performed in terms of 
$\hat r = \sqrt{x^2+y^2+z^2}$.
The 1-D interpolation is done along 
characteristics 
labeled by $(u,q,p)$, computing the quantities $g^{ij}$ and
$(x^\alpha)$ at the intersection points between these characteristics 
and the world-tubes $\Lambda^{[I]}$. 
The  results of this first interpolation  
are the functions $g^{ij}$ and
$(x^\alpha)$  at the 2-D surfaces $\Lambda^{[I]}_{u_{n-1/2}}$.

Next the injection module uses the outcome of the same interpolation from
previous time-steps, given on $\Lambda^{[I]}_{u_{n-3/2}}, 
\Lambda^{[I]}_{u_{n-5/2}}$, etc. to perform a 1-D 
time-interpolation (keeping $(q,p)$ and $\hat r$ constant)
and obtain the functions  $g^{ij}$ and $(x^i)$ 
at the spheres $\Lambda^{[I]}_{t_{inj}}$.

Given the lapse function $\alpha$ and the shift vector $\beta^i$, 
the contravariant 3-D Cauchy metric ${}^{(3)}g^{ij}$ is constructed using
Eq.~(\ref{eq:inject.anal.g3ij}). Then
the covariant 3-D Cauchy metric ${}^{(3)}g_{ij}$ is obtained.

It remains to transfer data from the spheres $\Lambda^{[I]}_{t_{inj}}$
to the actual Cartesian injection points  $(t_{inj},x^*,y^*,z^*)$.
Recall that the gridpoints of $\Lambda^{[I]}_{t_{inj}}$
are labeled by  Cartesian coordinate values $(x,y,z)$. These values come
from a radial and a time interpolation that is performed
 starting from a Bondi grid. Thus the spherical
grid of $\Lambda^{[I]}_{t_{inj}}$ is irregular in the Cartesian frame. 
The angular dimensions
are resolved via spherical decomposition  that represents
the Cauchy metric ${}^{(3)}g_{ij}$ with a set of coefficients $c_{\ell m}^{[I]ij}$
for each sphere $\Lambda^{[I]}_{t_{inj}}$. The angular coordinates used
in the decomposition are defined on $\Lambda^{[I]}$ by 
\begin{equation}
\hat q(x^i) = \frac{x}{R^{[I]} \pm z}, \;\;\;
\hat p(x^i) = \pm \frac{x}{R^{[I]} \pm z}
\label{eq:inject.numer.qp}
\end{equation}
on the north $(+)$ and south $(-)$ patches.

Finally, the Cauchy metric ${}^{(3)}g_{ij}$ 
is evaluated
at the injection points $(t_{inj},x^*,y^*,z^*)$. When doing so 
first the angular coordinates $(q^*, p^*)$
of the injection point are computed using the analog of 
Eq.~(\ref{eq:inject.numer.qp}). Then the metric functions ${}^{(3)}g_{ij}$ are 
reconstructed from the coefficients $c_{\ell m}^{[I]ij}$, at the angular
coordinates given by  $(q^*, p^*)$, at a number of spheres 
$\Lambda^{[I]}_{t_{inj}}$surrounding 
the Cartesian injection point. Lastly a radial interpolation is performed.

The injection routine has the option to perform all three 1-D
interpolations using  two, three, or four gridpoints.

\section{Calibration}

All test-beds described for the extraction module have been used
to test the CCM algorithm as a whole. When doing so 
we have proceeded as follows:
At  $ t_{[0]} = u_{[0]} = 0$ give initial data analytically 
on the Cauchy and characteristic slices.
Then give analytic Cauchy data 
at the next time-level $t_{[1]} = \Delta t$. 
Use the numeric extraction module, 
characteristic evolution code, and injection module to compute 
the Cauchy boundary values, as provided by CCM, in the injection
domain surrounding $\Gamma$. Then iterate the procedure, i.e.
give Cauchy data at $t_{[2]} = 2 \Delta t$, use extraction, null 
evolution and injection, and so on until some final time $t=t_f$.
At this final time-step the error in the injection is compared
to the analytic Cauchy metric. Repeating the procedure for higher 
and higher grid-sizes provides information about the convergence rate
of CCM as $\Delta \rightarrow 0$.  
An overall second-order convergence was confirmed.

In addition, to obtain a stricter test of the injection module, 
the following approach was used.
Take a solution that is known in Bondi and in Cartesian coordinates,
as well as the coordinate transformation as computed in the extraction
module. Using these, build analytically the results that
the injection module will produce with an error of $O(\Delta^2)$.
Recall that injection is designed to be an $O(\lambda^2)$ algorithm.
Since $\lambda > \Delta x$,  for certain grid sizes 
an error larger than $O(\Delta^2)$ might be masked by
the $O(\lambda^2)$ error. By performing the operations prescribed for the 
injection in some computer algebra utility, one can construct
the functions $x^\alpha = f(y^\beta) + O(\lambda^2)$ in the same
 fashion as in the injection. Then 
the Cauchy metric components  can be constructed.
The computer algebra result must be identical with the numerical
result except for the $O(\Delta^2)$ discretization error  
 in the injection module.

The test was performed for the linear quadrupole  waves. Obtaining
the solution in Bondi coordinates is a nontrivial problem by itself 
that is addressed in Appendix~\ref{app:teuk-null}. Here only
the outcome of the numerical convergence test is quoted.

\begin{figure}
\centerline{\psfig{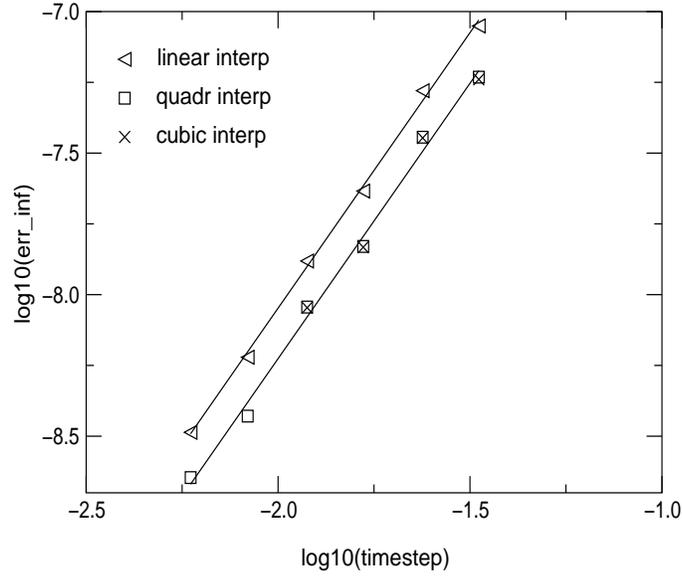}}
\caption{$g_{xy}$ as provided by the injection 
module at Cauchy boundary points. The test-bed is based on the quadrupole
linear waves with parameters 
$t = 1, R_\Gamma = 1, \varpi = 0.5$, and $\varepsilon = 10^{-8}$.
The spherical decomposition was done up to $\ell=4$,
with a discretization error of $O(\Delta^2)$. 
The error in $g_{xy}$ was measured with the
$\ell_\infty$ norm to be
less than $O(\Delta^{1.94})$.} 
\label{fig:inject.numer.conv}
\end{figure}

All interpolation routines and the spherical decomposition-reconstruction
routines were checked for convergence and showed
the correct convergence rates. It should be noted that  although
interpolation can be done up to $O(\Delta^4)$, there are a number
of operations in injection that are performed to $O(\Delta^2)$.
See Eq.~(\ref{eq:inject.numer.jac_qp}) for an example.
A convergence plot of the whole injection module is given
in Figure~\ref{fig:inject.numer.conv}, 
using the various interpolation routines. The graph shows
second order convergence. 
Besides testing the outcome of the whole injection module, 
tests were performed to ensure that the Cartesian coordinates of
the characteristic gridpoints $(x^\alpha)$, 
the Jacobian terms $\partial x^i / \partial y^\beta$, the
Bondi metric $\eta^{\alpha \beta}$, and the Cauchy metric components
$g^{ij}$, ${}^{(3)}g^{ij}$ and ${}^{(3)}g_{ij}$ are computed to $O(\Delta^2)$
with respect to the analytic  $\lambda$-expansion.

\chapter{Stability Of CCM}
\label{chap:ccmtest}

In this chapter a variety of numerical experiments are presented
to test the CCM modules in a
numeric environment.   While both the extraction and the injection modules
 show second-order accuracy with respect to a number of test-beds, 
this does not guarantee
the numerical stability of CCM as an interface between the
general relativistic Cauchy and
characteristic evolution codes.
The Cauchy system used in this chapter is the ADM formulation
(see section \ref{sec:intro.ADM}, also \cite{Arnowitt62,York79}),
developed by the Binary 
Black Hole Grand Challenge Alliance 
\cite{Cook97a}.

\section{Blending}
\label{sec:numeric:blend}

After the implementation and calibration of the CCM modules,
it was found that unstable (exponentially growing) numerical
modes arise  at the interface between
the two evolution codes.  Since the wavelength of those modes was typically
$O(\Delta)$,
various smoothing techniques were
tested. Least-square fitting (see Section~\ref{sec:extract.fit})
did not bring the expected improvement.
A technique that improved the fore-stated exponential growths
in certain circumstances is the following ``blending'' 
technique \cite{Gomez97d}.

The 3-D metric ${}^{(3)}g^{(Cauchy)}_{ij}$ is determined by
the Cauchy evolution at all Cartesian evolution gridpoints.
In order to provide boundary data
${}^{(3)}g^{(Bdry)}_{ij}$, one can apply the condition 
\begin{equation}
 {}^{(3)}g_{ij} =  f(x^i) \cdot {}^{(3)}g^{(Cauchy)}_{ij}+ (1-f(x^i)) \cdot {}^{(3)}g^{(Bdry)}_{ij}
\label{eq:stability.blenddef}
\end{equation}
where $f(x^i)$ is a smooth function that vanishes at the boundary grid-points
and has the value $1$ for all evolution gridpoints that are at least a distance $w$ 
away from the boundary point. The zone over which $1<f<0$ is the blending zone, with a 
blending width $w$.
A blending zone of width zero amounts to Dirichlet boundary conditions.

Applying analytic boundary conditions with no blending 
results in an unstable ADM  evolution for the linearized quadrupole waves. 
Analytic boundary conditions
blended over an $O(\Delta)$ region give 
an improved behavior for the same analytic solution.
Also, the full non-linear ADM system
with smooth initial data of linear order
and blended flat boundary conditions (i.e.,
in Eq.~(\ref{eq:stability.blenddef}) 
set ${}^{(3)}g_{ij}^{(Bdry)} = \delta_{ij}$)
 run for hundreds of crossing times without any signs of instability
\cite{Gomez97d}. 
Blending has also brought improvement 
in matching Cauchy evolution to  perturbative 
spherical evolution  \cite{Abrahams97a}.

Even though the use of blending 
postpones the appearance of exponential modes in most settings, 
it does not remove the numerical instability in all cases. For instance, 
constraint violating random initial data with blended flat boundary
conditions resulted in a numerically unstable run.

\section{``No feedback'' experiments}

\begin{figure}
\centerline{\epsfxsize=4.2in\epsfbox{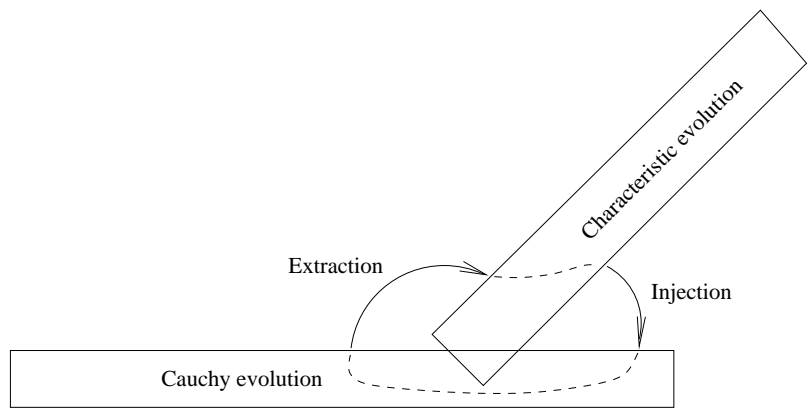}}
\caption{The loop created by CCM. Extraction data is
taken  from the Cauchy code and used as boundary data for the
characteristic code. This  is evolved along the characteristics
and  feeds into the injection module
which constructs the Cauchy boundary data. Lastly, the Cauchy
evolution propagates data from its boundaries back to the 
domain of dependence
of the extraction module.}
\label{fig:numeric.loop}
\end{figure}

As shown in Figure~\ref{fig:numeric.loop},
matching consists of a closed loop.
If the numerical amplification matrix of the
finite-difference algorithm  has any
eigenvalue with  absolute value greater than one,
an exponential mode is generated that poisons the evolution of physical
data. A finite-difference stability analysis 
of the coupled code is too complicated
to be carried out. Thus the numerical stability  of the system is 
 investigated via numerical experiments.

To investigate the potential problems arising from  feedback,
a series of numerical experiments was designed where the loop is
broken at various points with analytic data.

The experiments were based on the 
 quadrupole linearized solution. The Cauchy metric 
is defined in Section~\ref{sec:extract.teuk}. The characteristic
metric is provided as an expansion in terms of the affine parameter
$\lambda$ and is given in Appendix~\ref{app:teuk-null}.

The radius of the extraction world-tube is set
to $R_\Gamma = 1$. The amplitude and the width of the time-symmetric wave packet
are set to $\varepsilon = 10^{-8},\; \varpi = 0.5$.  The CFL ratio is given by
$\Delta t / \Delta x = 0.25$. 

In all of these experiments the injected Cauchy metric
is worked out up to $O(\lambda^3)$. Blending is used over a region of three Cauchy gridpoints.
Since runs with a closed loop show 
rapidly growing instabilities, most 
runs were performed up to $t = 20 R_\Gamma$.
Each set of experiments was run with a number of different grid sizes.

\subsection{Analytic injection data}

\begin{figure}
\centerline{\hbox{
\epsfxsize=2.1in\epsfbox{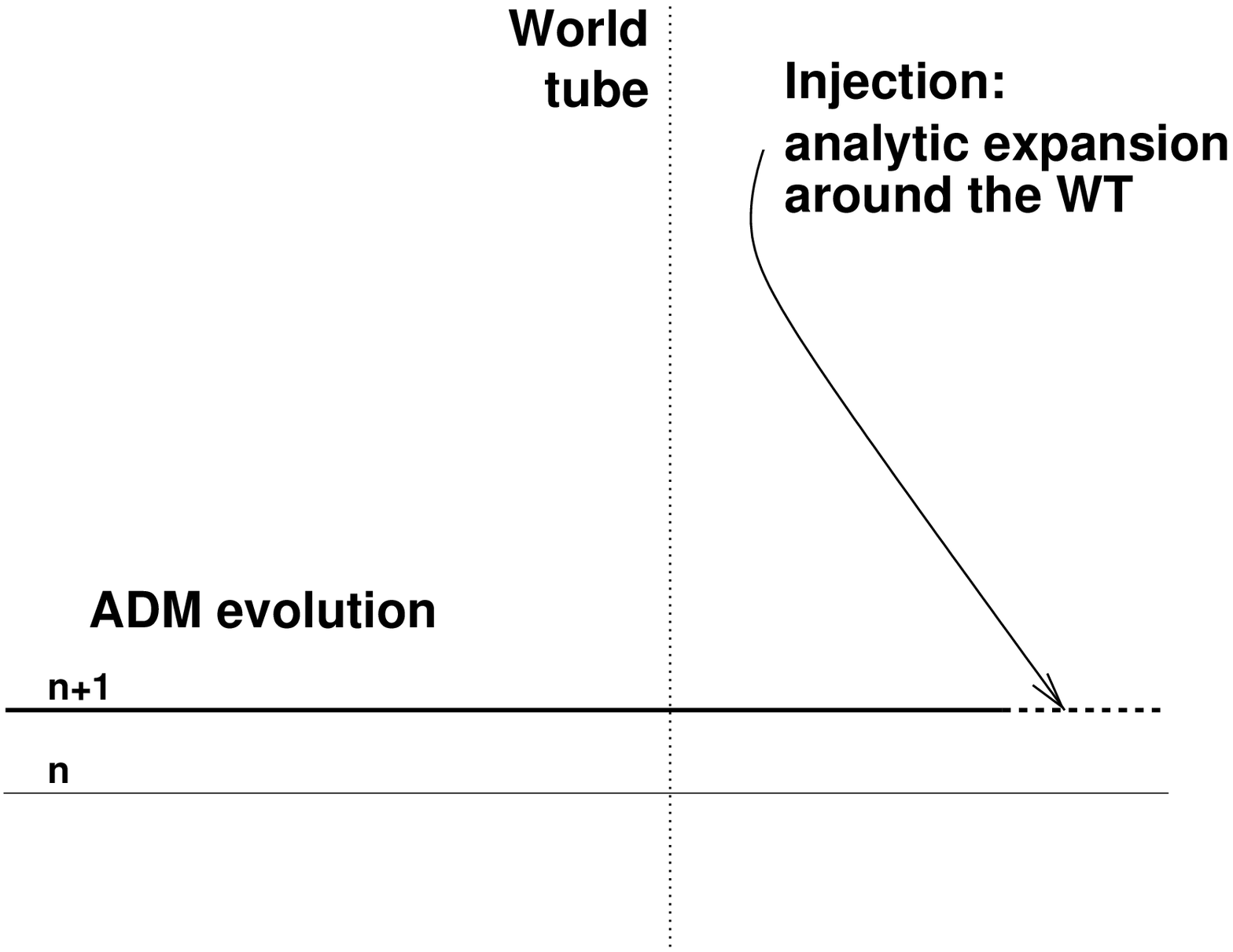}
\hspace{0.1in}
\epsfxsize=2.1in\epsfbox{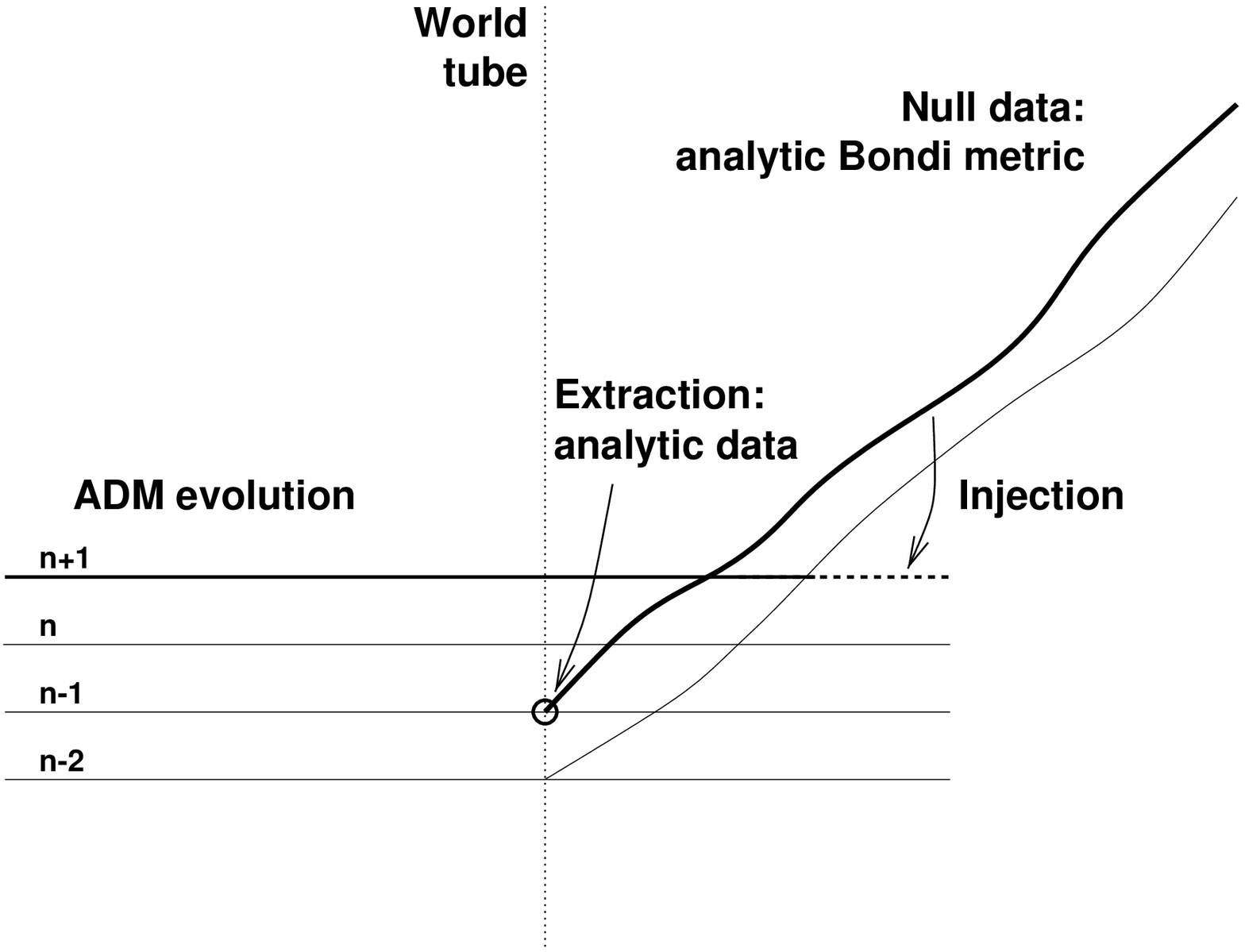}}}
\caption{Left: the scheme of the runs performed with analytic $O(\lambda^3)$ injection data.
Right:  the scheme of the runs performed with analytic extraction and characteristic data,
and numeric injection data.}
\label{fig:stability.ladm.setup}
\label{fig:stability.iadm.setup}
\end{figure}

In the first set of experiments the Cauchy boundary data is an 
analytic expansion up to $O(\lambda^3)$ of the exact solution
\begin{equation}
g^{injected}_{ij} = g^{exact}_{ij} + O(\lambda^3).
\end{equation}
A schematic diagram of the setup of the run can be seen on the 
left side of
Figure~\ref{fig:stability.ladm.setup}.

\begin{figure}
\centerline{\vbox{\hbox{
\psfig{figure=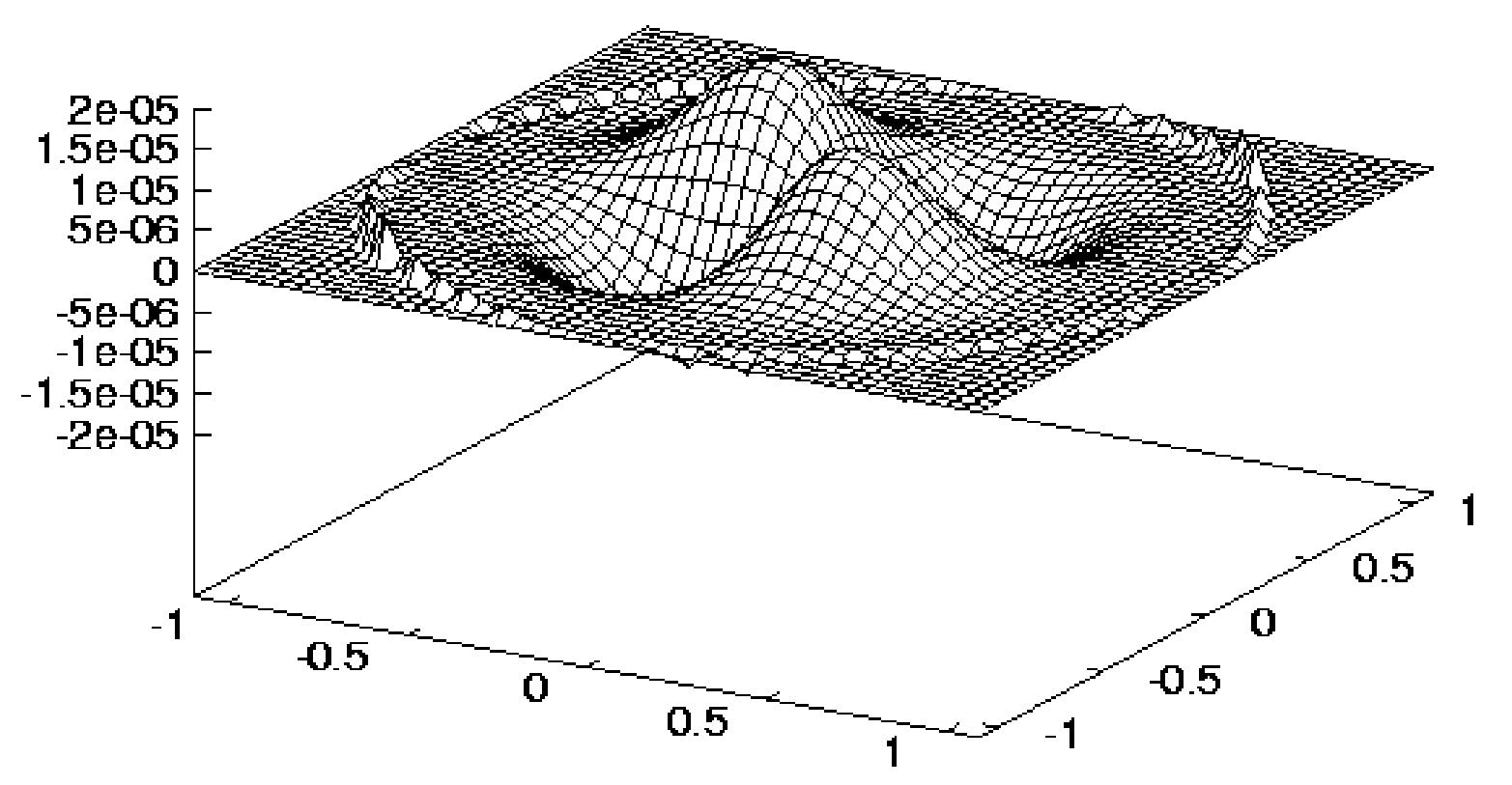,height=1.3in,width=2.2in}
\hspace{0.1in}
\psfig{figure=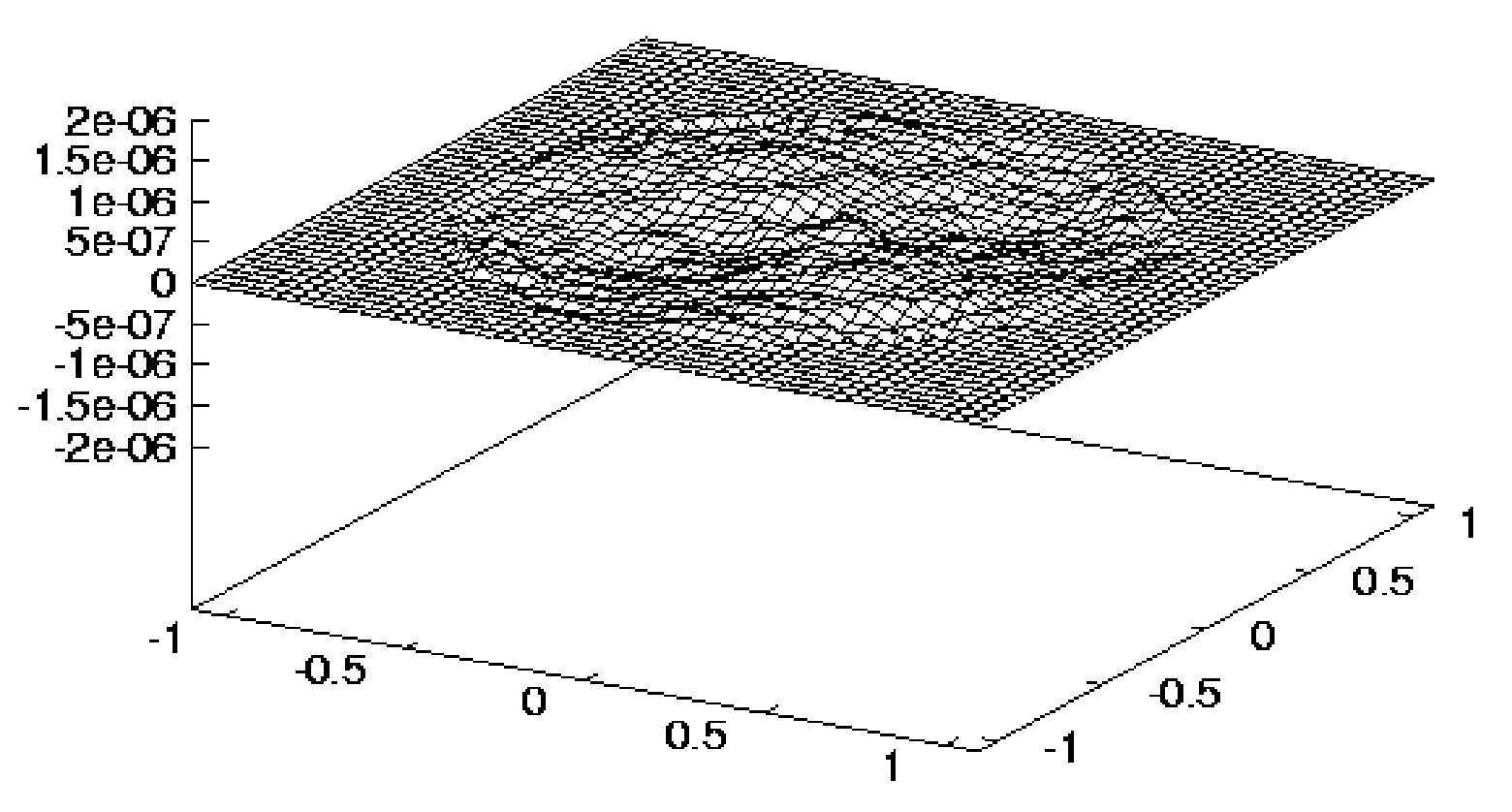,height=1.3in,width=2.2in}}
\vspace{0.5in}
\hbox{
\psfig{figure=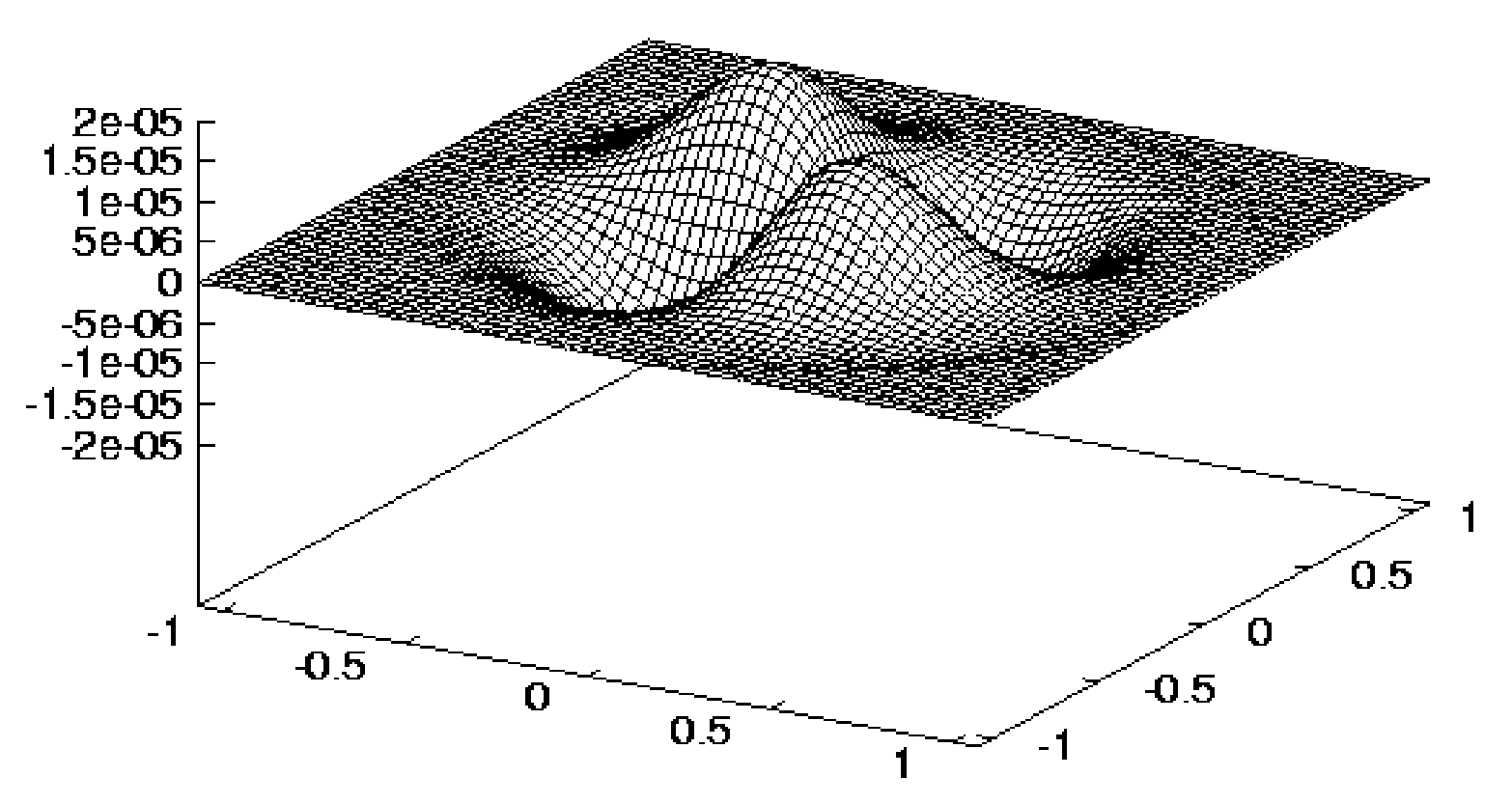,height=1.3in,width=2.2in}
\hspace{0.1in}
\psfig{figure=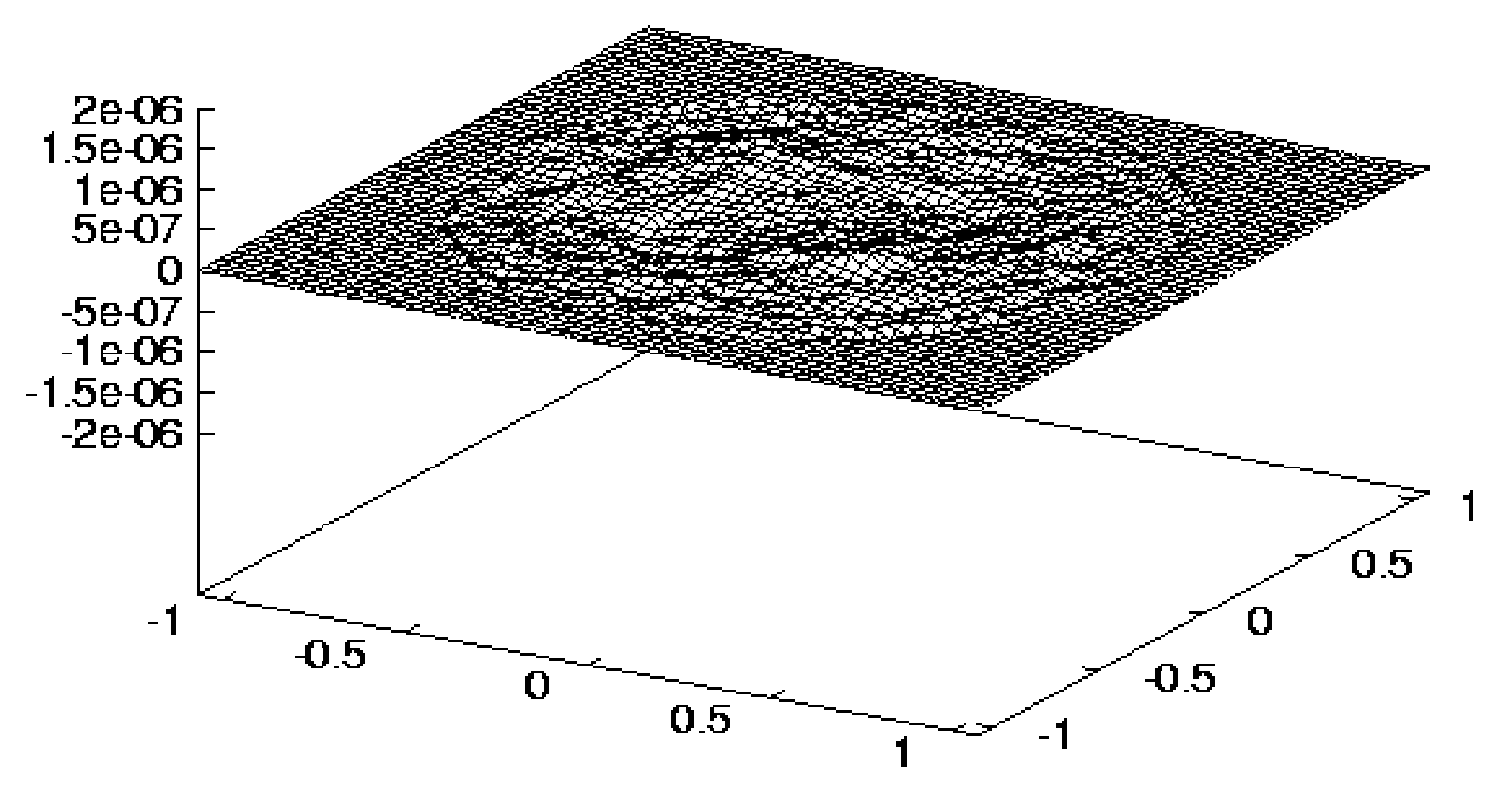,height=1.3in,width=2.2in}}}}
\caption{Cauchy run with analytic boundary data as an expansion around the extraction world-tube.
All plots represent $z=0$ slices of the metric component ${}^{(3)}g_{xy}$. Top left: the Cauchy data is
shown after one time-step for a grid size of $48^3$. The same run, at time $t=20$ can be seen 
on the top right plot. The bottom left and right plots are the first and the last time-step of a
$64^3$ grid run up to $t=20$. Comparing the initial time-step for the two different grids, one can see
that as the injection zone converges to the extraction world-tube, the effects of the $\lambda$-expansion
converge to zero.}
\label{fig:stability.ladm.plots}
\end{figure}

The experiment tests whether the $O(\lambda^3)$ error induces any
short-range instabilities. Runs were performed with grid sizes $32^3, 48^3, 64^3, 80^3$, 
up to $t=20$. As it can be seen in Figure~\ref{fig:stability.ladm.plots}, 
short-term instabilities did not develop.

\subsection{Analytic extraction and characteristic data}

\begin{figure}
\centerline{\vbox{\hbox{
\psfig{figure=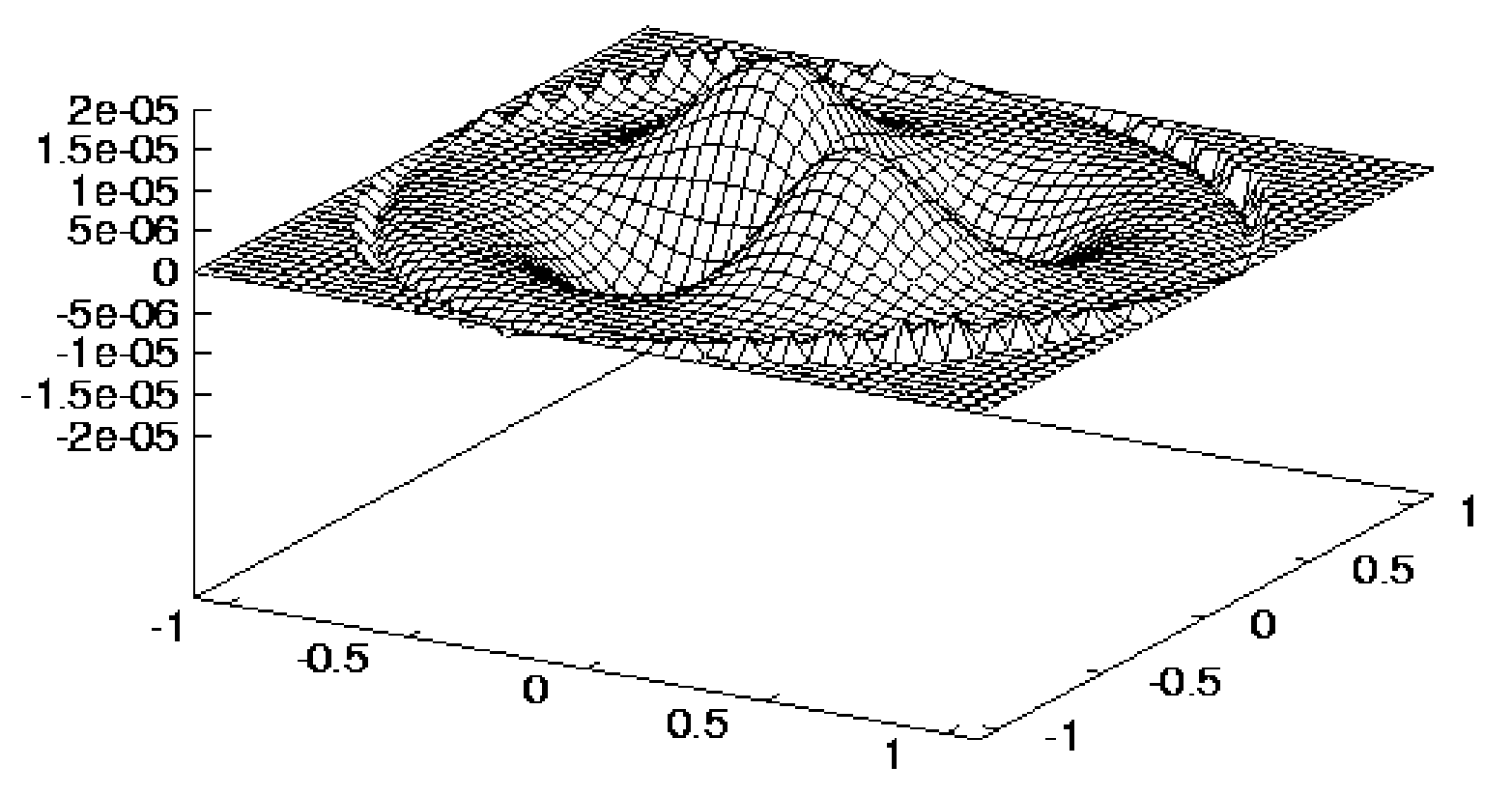,height=1.3in,width=2.2in}
\hspace{0.1in}
\psfig{figure=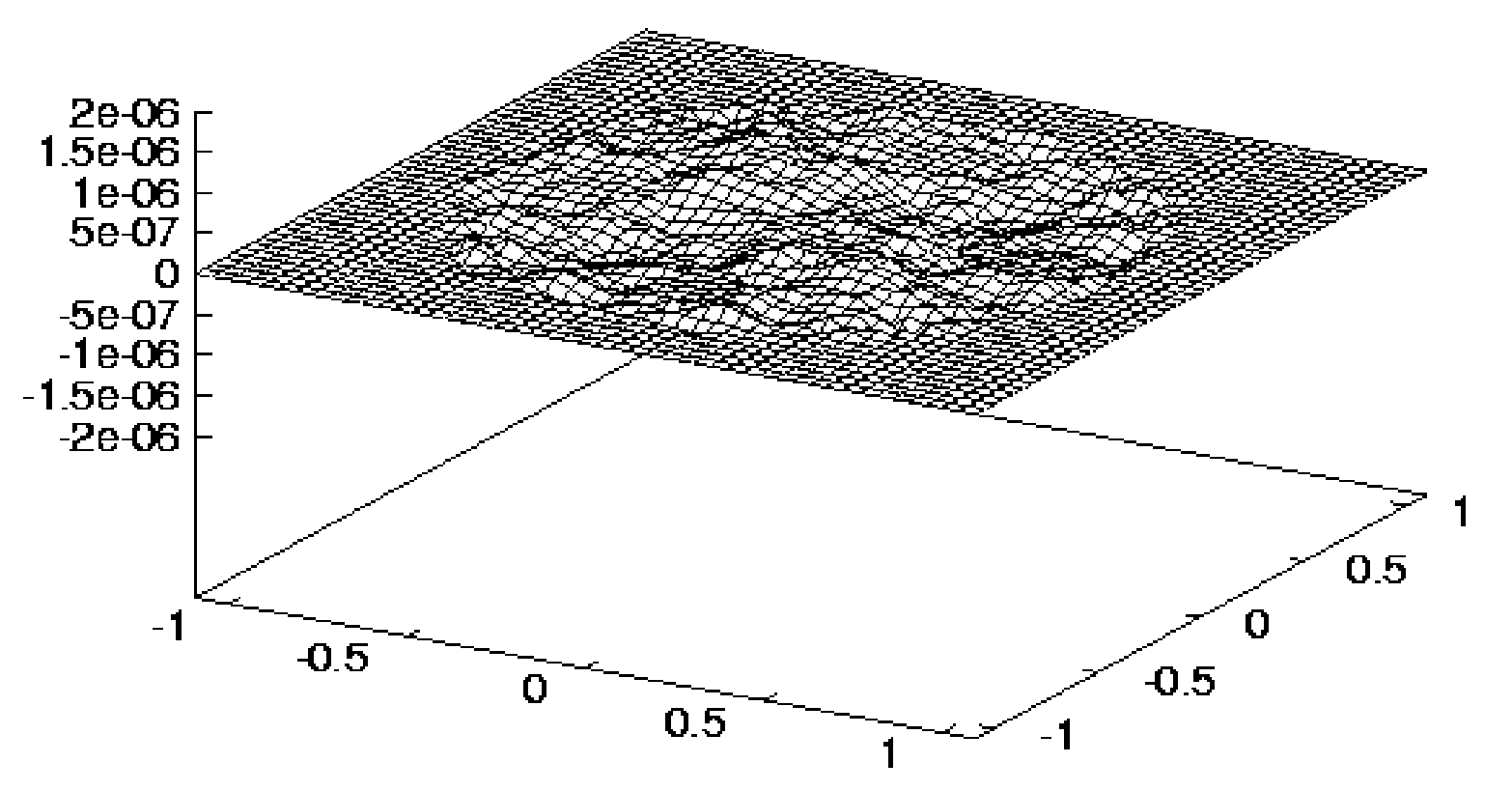,height=1.3in,width=2.2in}}
\vspace{0.5in}
\hbox{
\psfig{figure=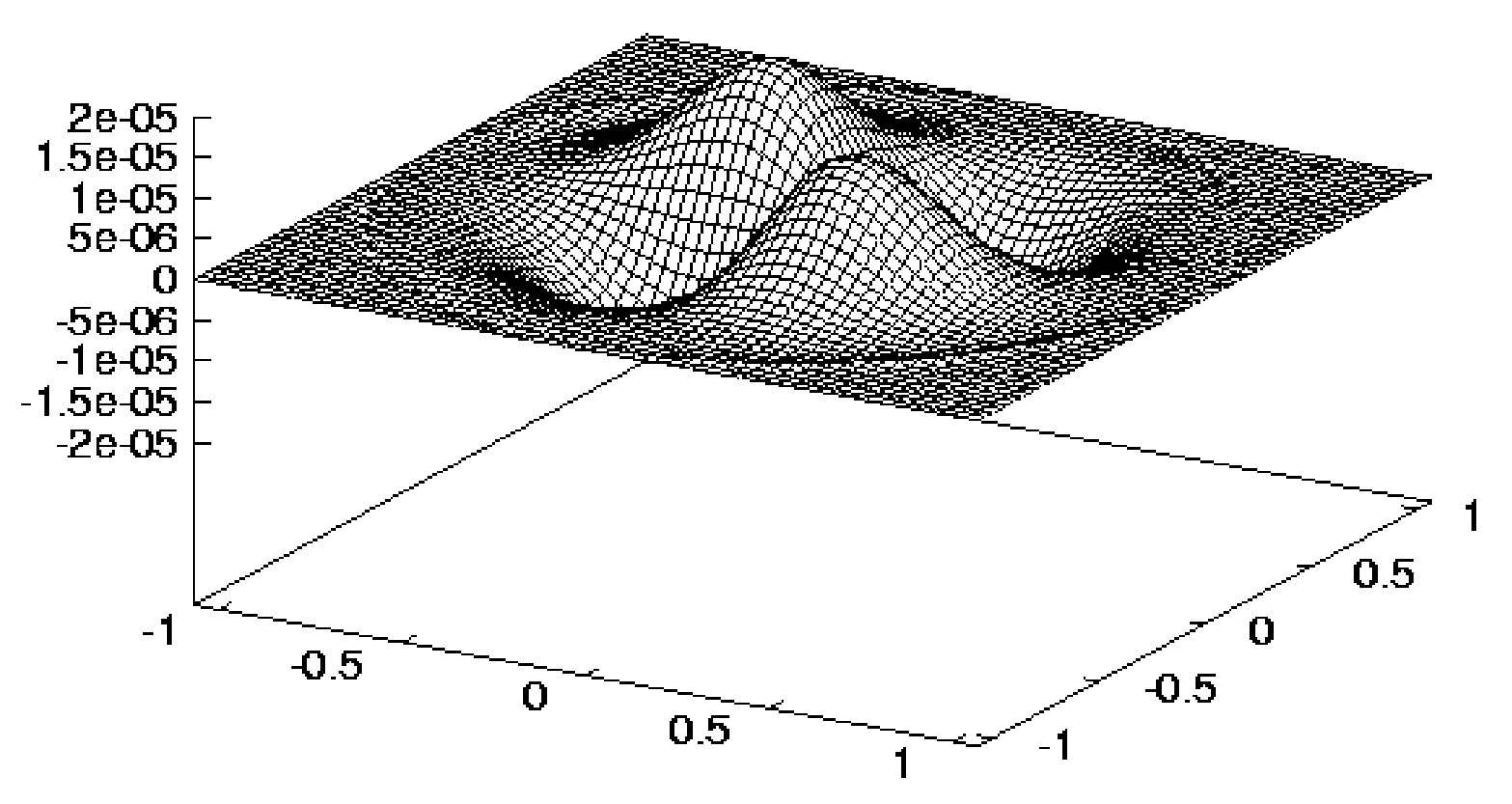,height=1.3in,width=2.2in}
\hspace{0.1in}
\psfig{figure=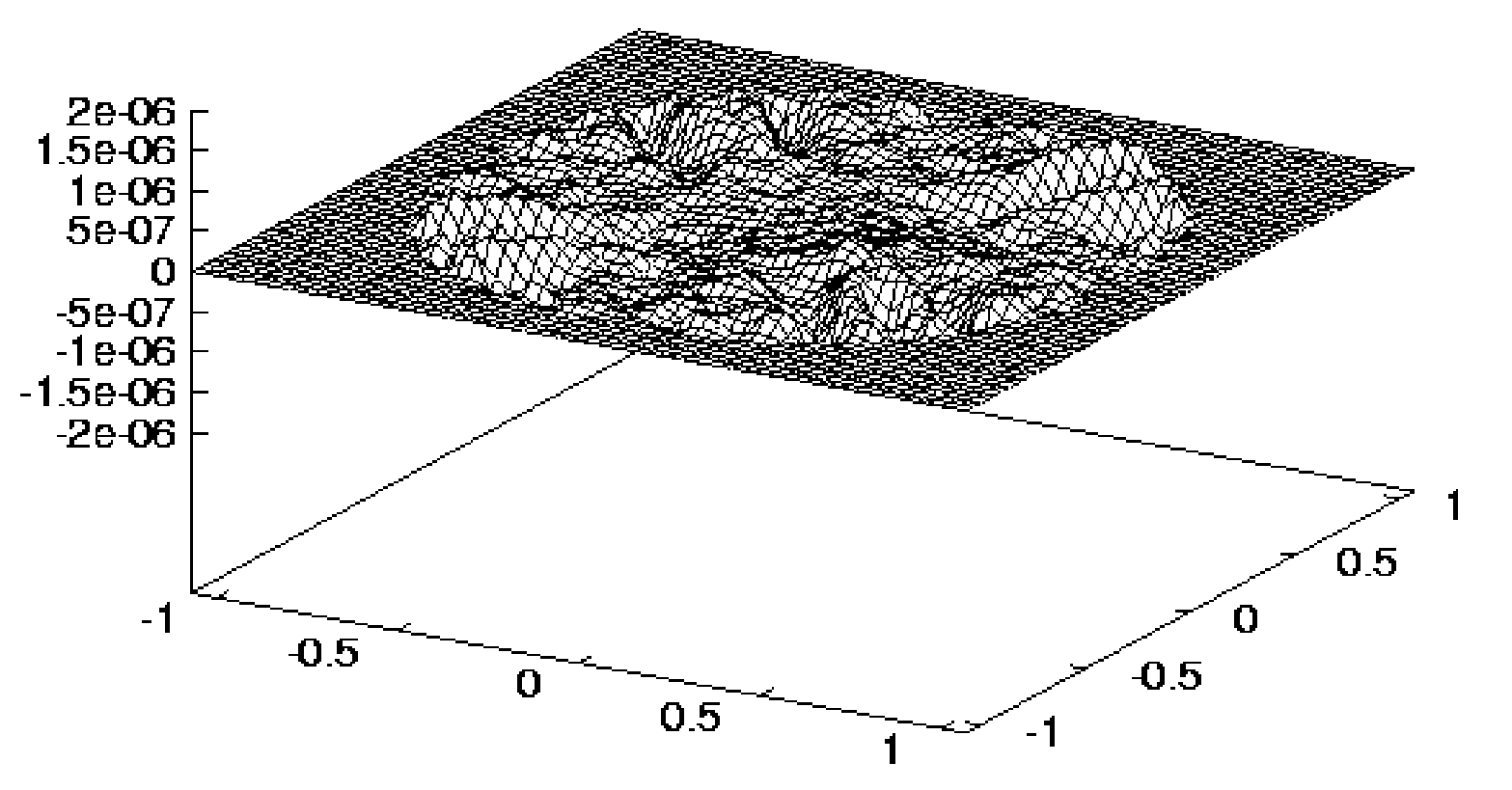,height=1.3in,width=2.2in}}}}
\caption{Cauchy run with analytic extraction and characteristic data, with numeric injection.
All plots represent $z=0$ slices of the metric component ${}^{(3)}g_{xy}$. Top left: the Cauchy data is
given after one time-step, for a grid size of $48^3$. The same run, at time $t=20$ can be seen 
on the top right plot. The bottom left and right plots are the first and the last time-step of a
$64^3$ grid run up to $t=20$. }
\label{fig:stability.iadm.plots}
\end{figure}

In the next set of runs the injection is numeric, but it is fed by analytic extraction and characteristic
data. The Bondi metric variables $(J,\beta,U,W)$ are given as an $O(\lambda^7)$ expansion around the extraction
world-tube $\Gamma$. Similar to the case of analytic injection data, no short-time instabilities were seen.
Runs were performed for grid sizes $48^3$ and $64^3$.
See the right-hand plot of Figure~\ref{fig:stability.iadm.setup} for a schematic diagram of the numerical setup.
Surface plots of ${}^{(3)}g_{xy}$ after the first 
time-step and at $t=20$ are provided in Figure~\ref{fig:stability.iadm.plots}. At the time-scale of the run no boundary
instabilities were seen.

\subsection{Analytic extraction data}

\begin{figure}
\centerline{\hbox{\vbox{
\epsfxsize=2.1in\epsfbox{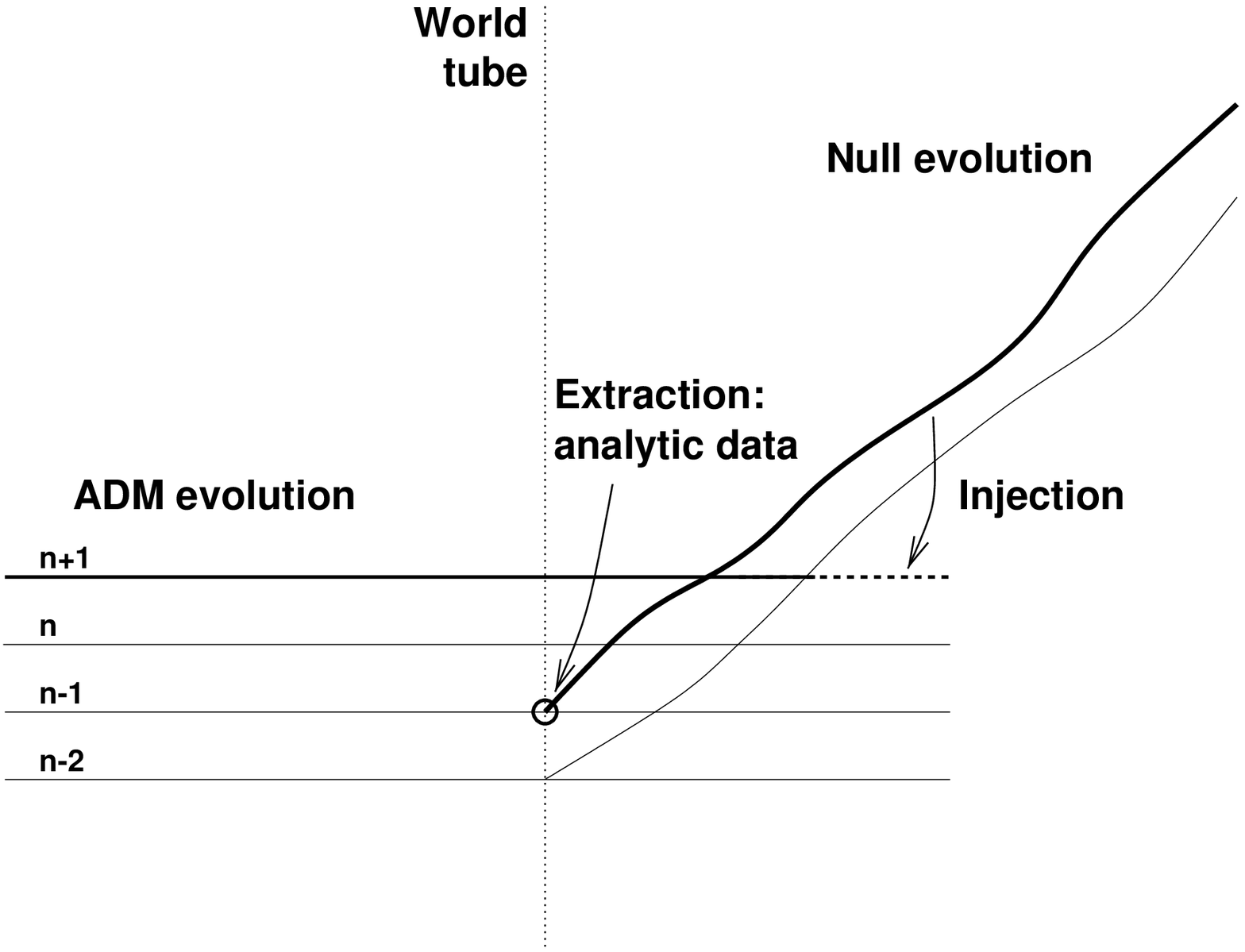}\vspace{0.8in}}
\hspace{0.1in}
\epsfxsize=2.1in\epsfbox{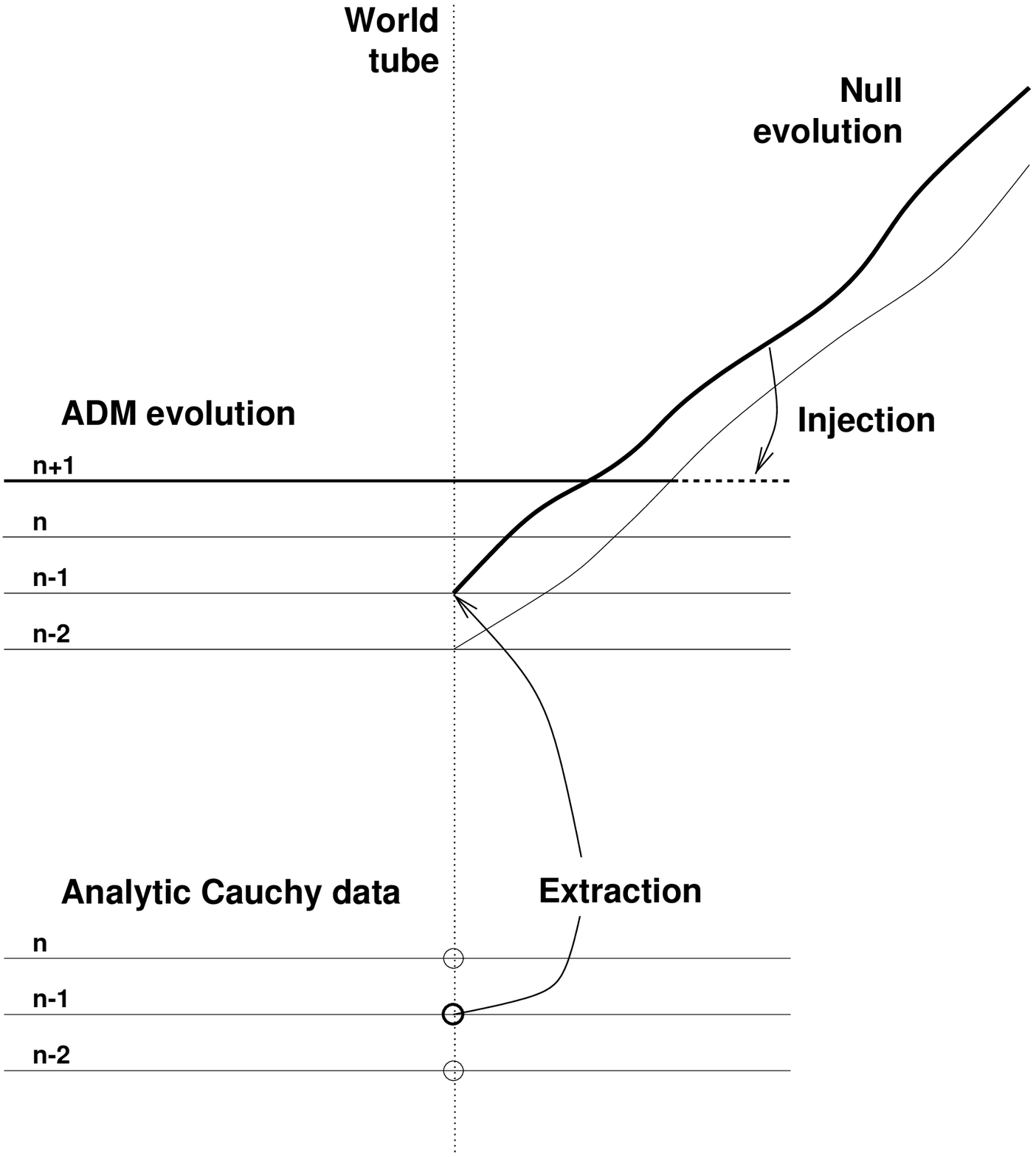}}}
\caption{Left: the scheme of the runs performed with analytic extraction data,
numeric characteristic evolution and injection.
Right:  the scheme of the runs performed with numeric extraction fed by analytic Cauchy
data.}
\label{fig:stability.nadm.setup}
\label{fig:stability.eadm.setup}
\end{figure}

\begin{figure}
\centerline{\vbox{\hbox{
\psfig{figure=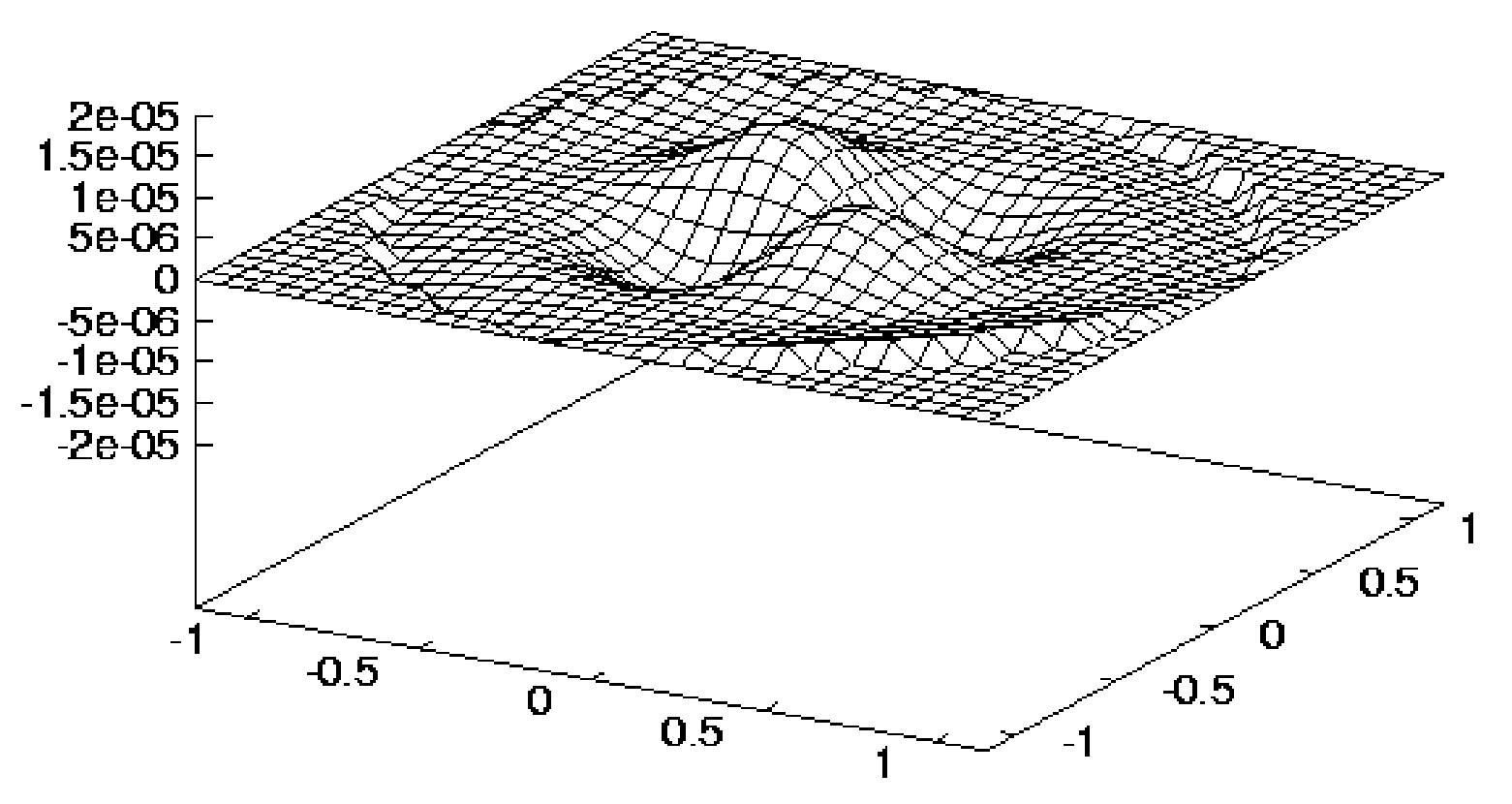,height=1.3in,width=2.2in}
\hspace{0.1in}
\psfig{figure=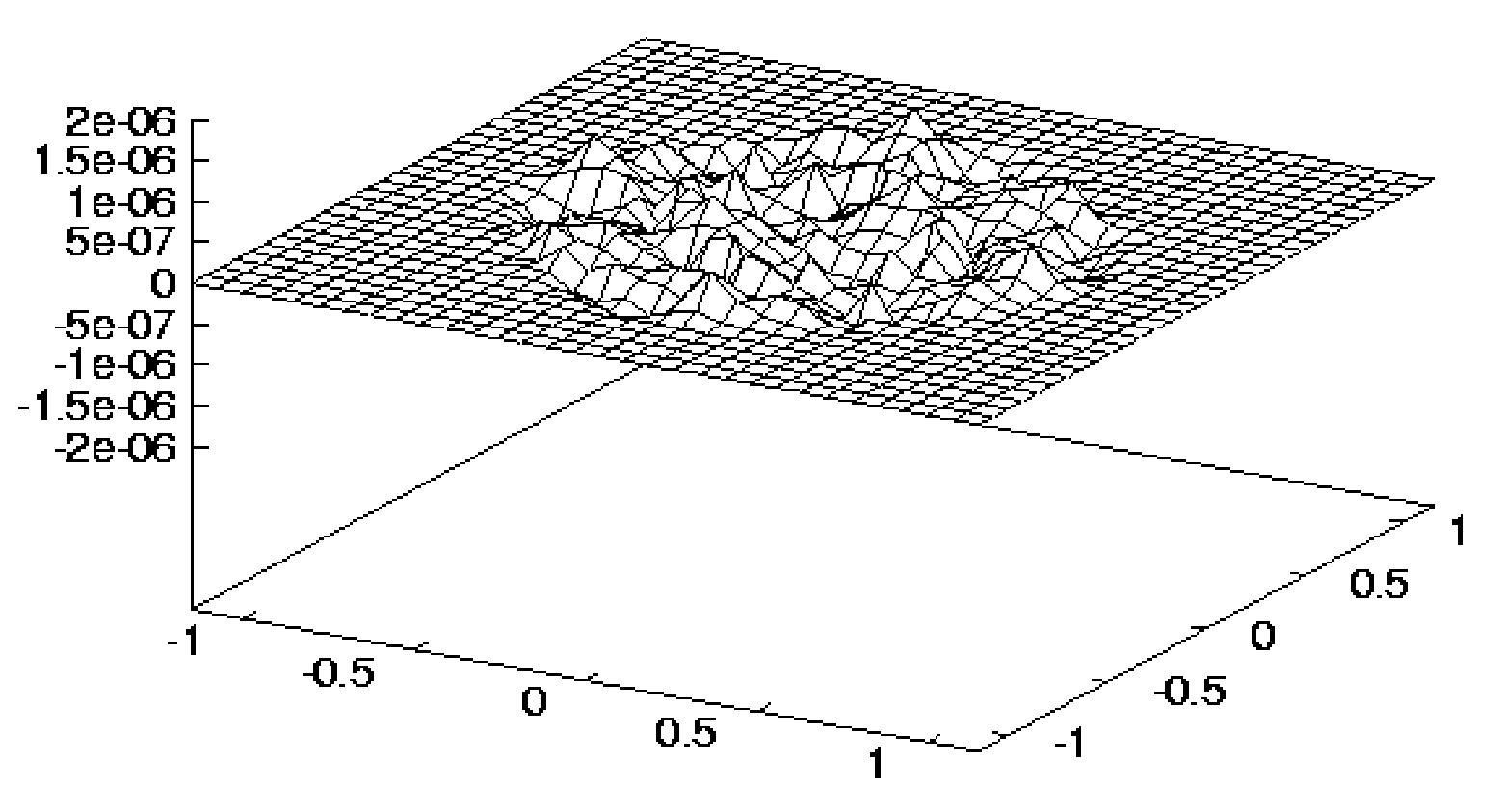,height=1.3in,width=2.2in}}
\vspace{0.5in}
\hbox{
\psfig{figure=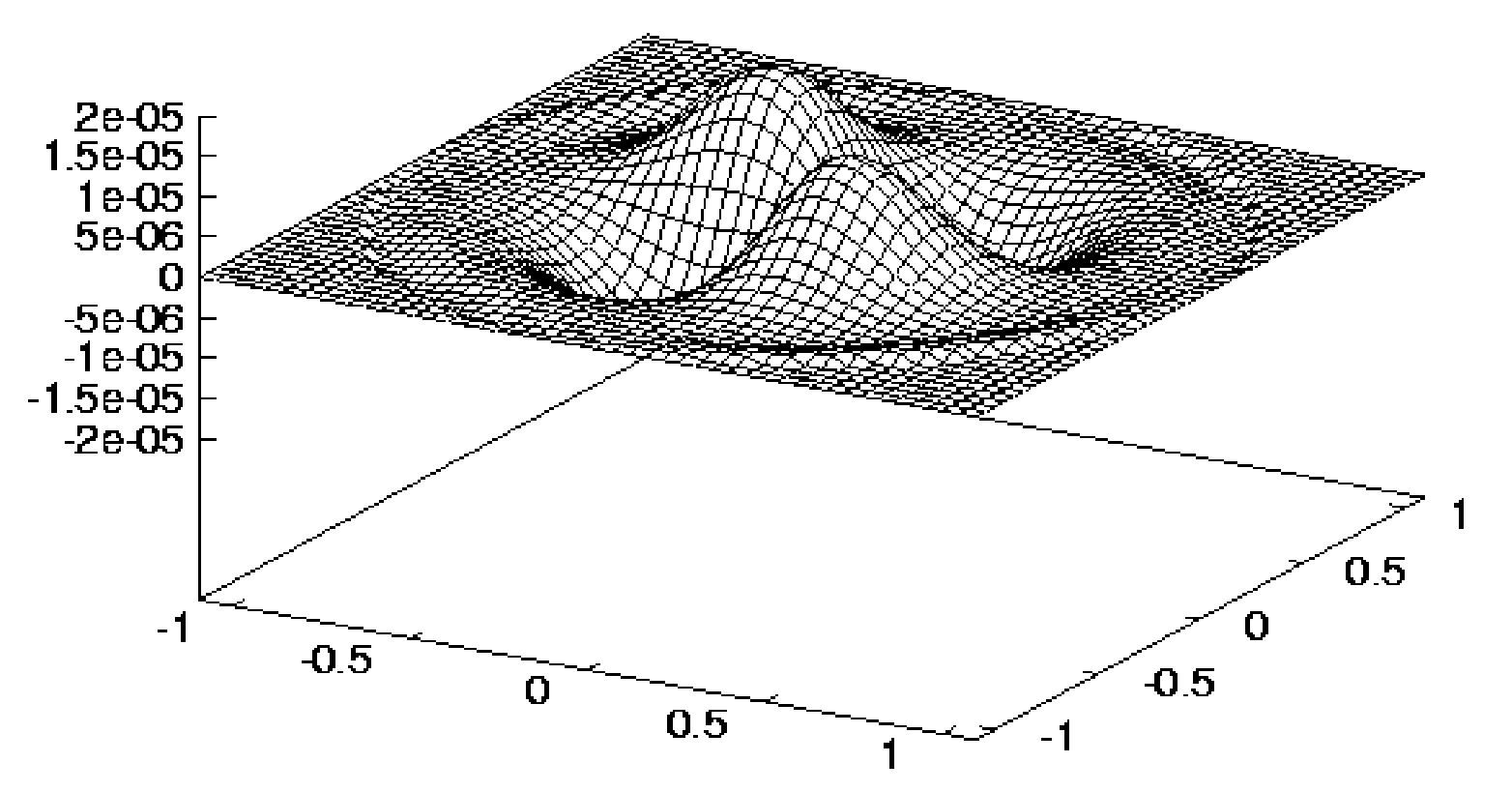,height=1.3in,width=2.2in}
\hspace{0.1in}
\psfig{figure=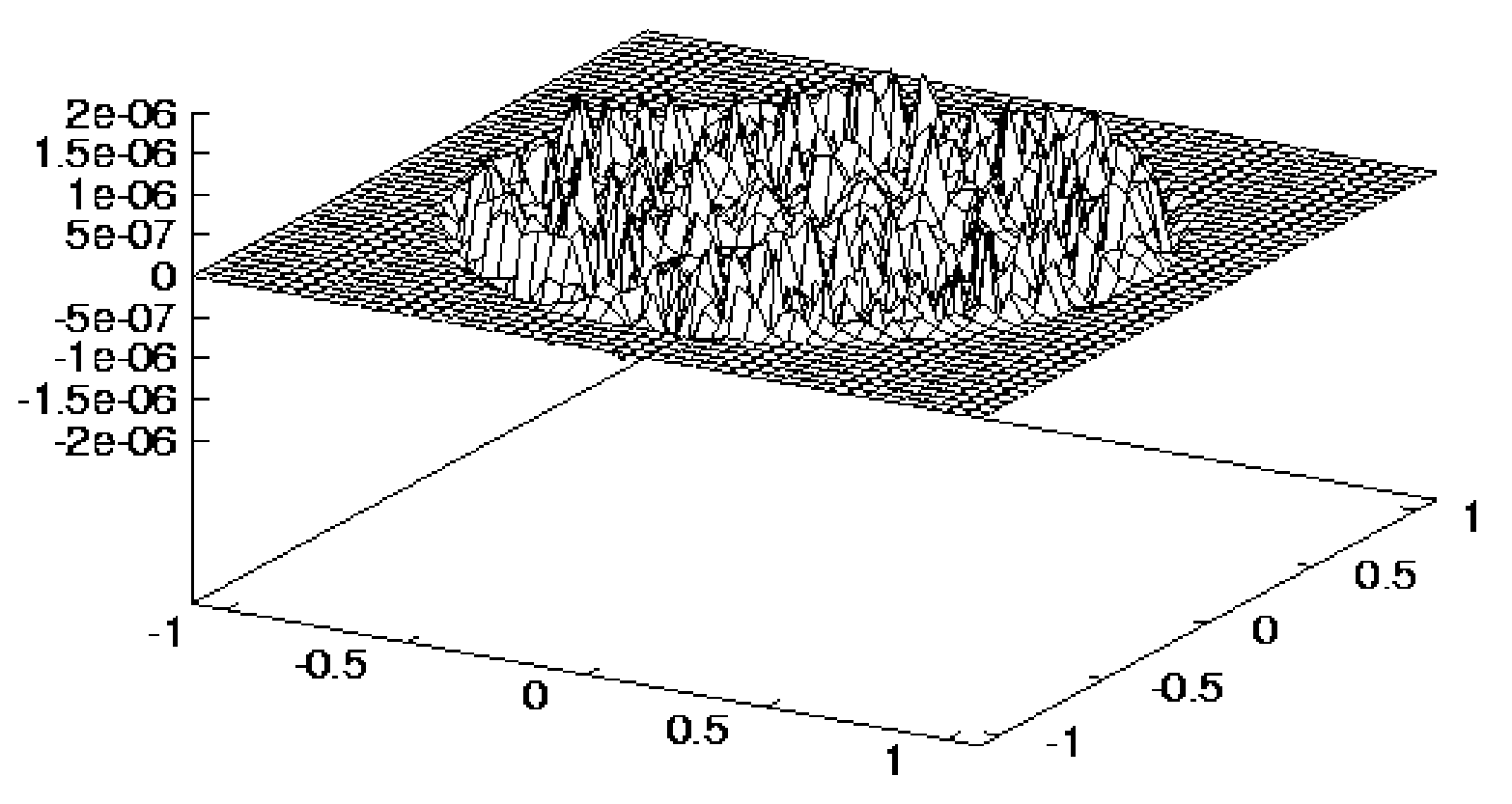,height=1.3in,width=2.2in}}}}
\caption{Cauchy run with analytic extraction data, numeric characteristic evolution and  injection.
All plots represent $z=0$ slices of the metric component ${}^{(3)}g_{xy}$. Top left: the Cauchy data is
shown after one time-step for a grid size of $32^3$. The same run, at time $t=20$ can be seen 
on the top right plot. The bottom left and right plots are the first and the last time-step of a
$48^3$ grid run up to $t=20$. }
\label{fig:stability.nadm.plots}
\end{figure}

In the third stage characteristic evolution and injection are numeric. Thus the discretization
noise of the characteristic evolution and the injection are fed into the Cauchy evolution. Extraction
data in these runs is analytic. 
In other words the extraction module is replaced by a set of 
functions that provide the extraction results analytically.
These analytic results are used as boundary 
data for the characteristic evolution and in the injection.
A scheme of the setup can be seen in
Figure~\ref{fig:stability.nadm.setup}, on the left. Runs were made 
up to $t=20$,
for grid sizes of 
$32^3, 48^3$, and $64^3$.
Figure~\ref{fig:stability.nadm.plots} 
contains surface plots of ${}^{(3)}g_{xy}$ at the beginning and the end
of the runs. The runs did not reveal  short time-scale instabilities.

\subsection{Extraction from analytic Cauchy data}

\begin{figure}
\centerline{
\psfig{figure=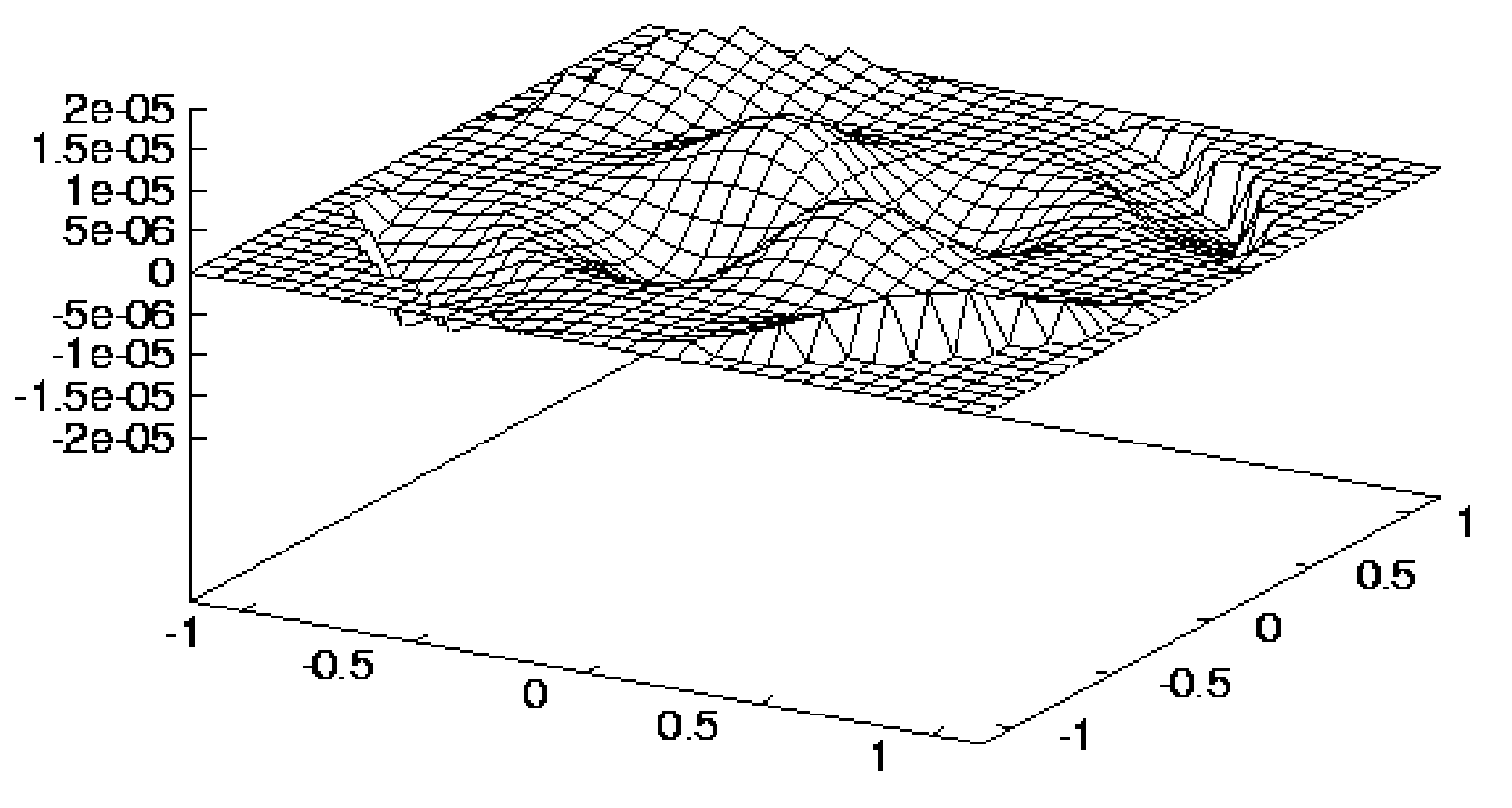,height=1.3in,width=2.2in}
\hspace{0.1in}
\psfig{figure=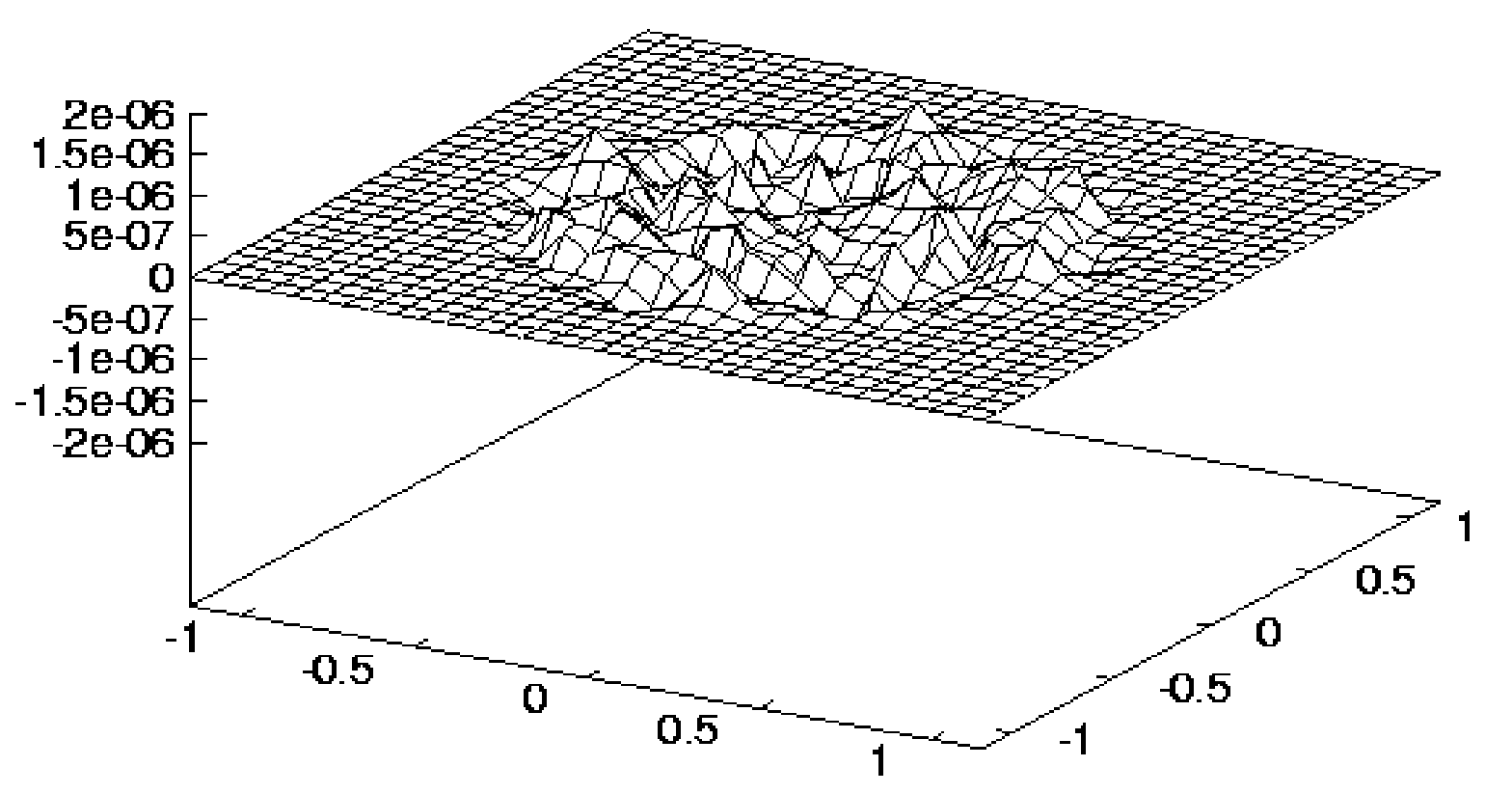,height=1.3in,width=2.2in}}
\caption{Cauchy run with numeric extraction data, with the extraction module fed by analytic
Cauchy data. Characteristic evolution and injection are numeric as well.
All plots represent $z=0$ slices of the metric component ${}^{(3)}g_{xy}$. Left: the Cauchy data is
shown after one time-step for a grid size of $32^3$. The same run at time $t=20$ can be seen 
on the right plot. }
\label{fig:stability.eadm.plots}
\end{figure}

In the next stage the extraction,
characteristic evolution and injection are numeric.
However, in order
to avoid a closed loop, input data for the extraction module is provided by a Cartesian grid
containing analytic Cauchy data. 
Thus, in addition to the previous run, the numerical noise
coming from the extraction module is fed into the characteristic evolution and the injection.
The right hand side of Figure~\ref{fig:stability.eadm.setup} shows the setup of the runs. Grid sizes were
$32^3$ and $40^3$. Surface plots of ${}^{(3)}g_{xy}$ at $z=0$ are provided in 
Figure~\ref{fig:stability.eadm.plots}. These runs revealed no exponentially 
growing modes.

\subsection{Extraction from decoupled numeric Cauchy data}

\begin{figure}
\centerline{
\epsfxsize=2.1in\epsfbox{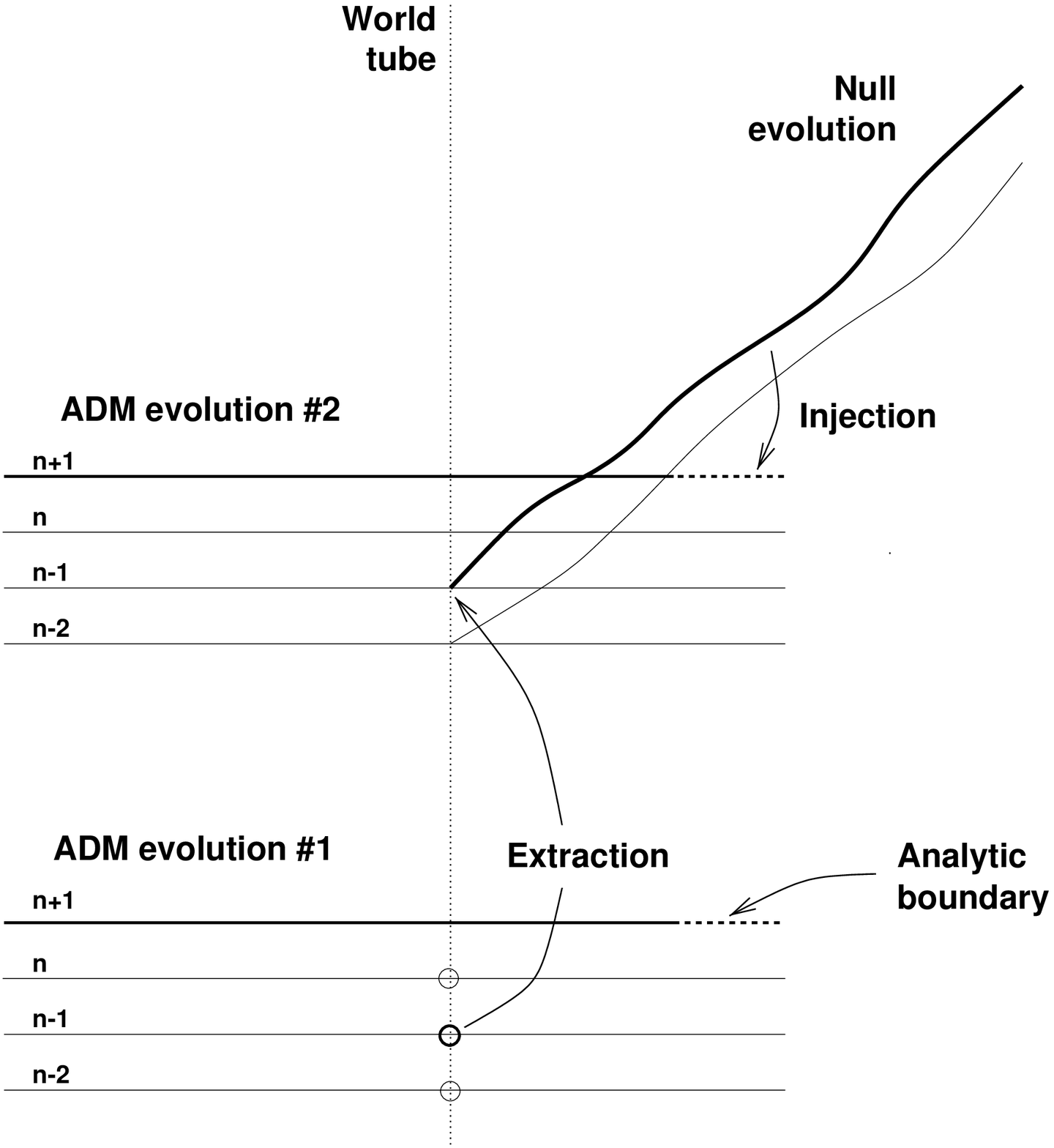}}
\caption{Double Cauchy evolution. The first code serves as a numeric input for the 
CCM routines attached to the second Cauchy code.}
\label{fig:stability.2adm.setup}
\end{figure}

\begin{figure}
\centerline{\vbox{\hbox{
\psfig{figure=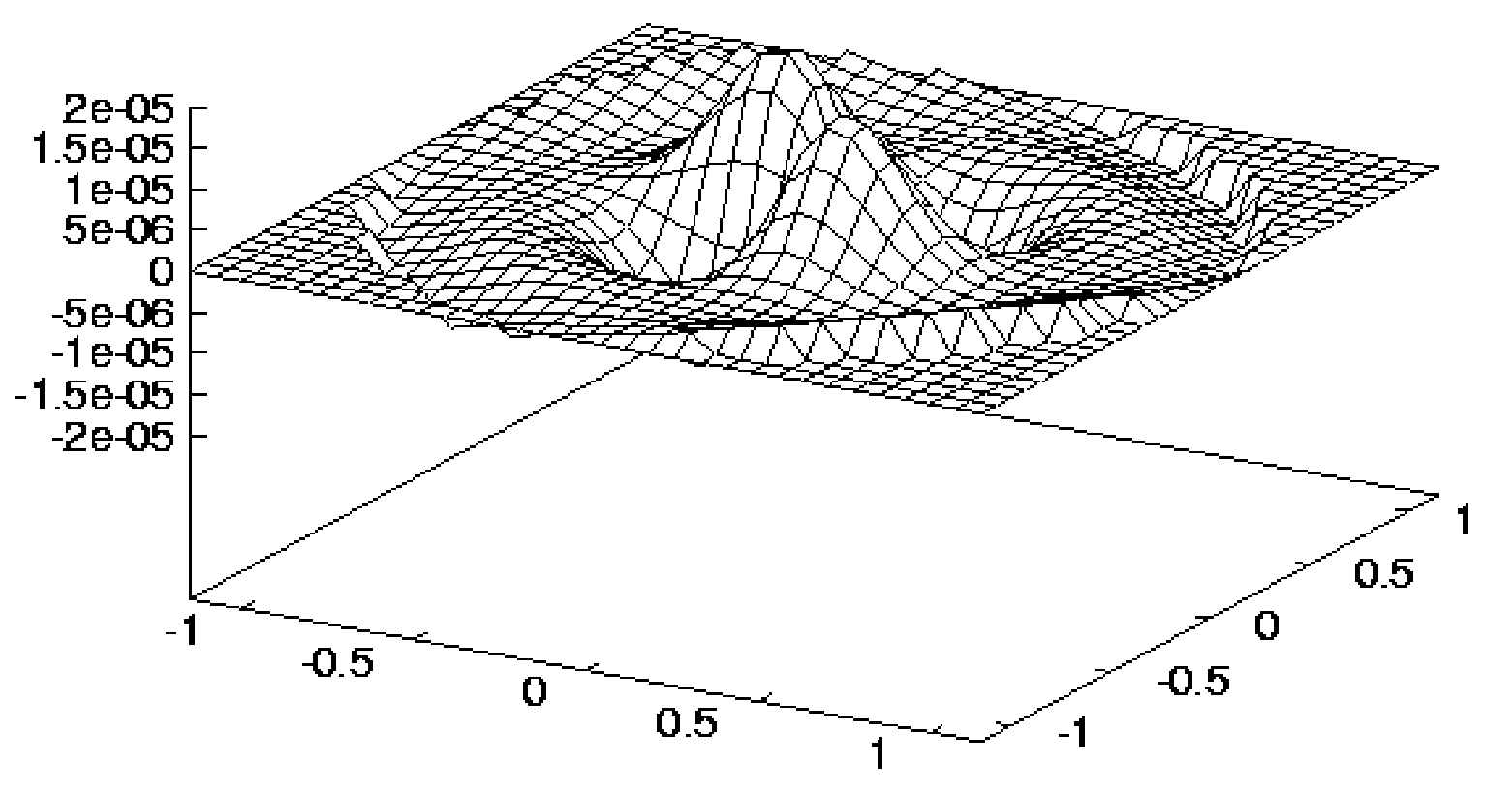,height=1.3in,width=2.2in}
\hspace{0.1in}
\psfig{figure=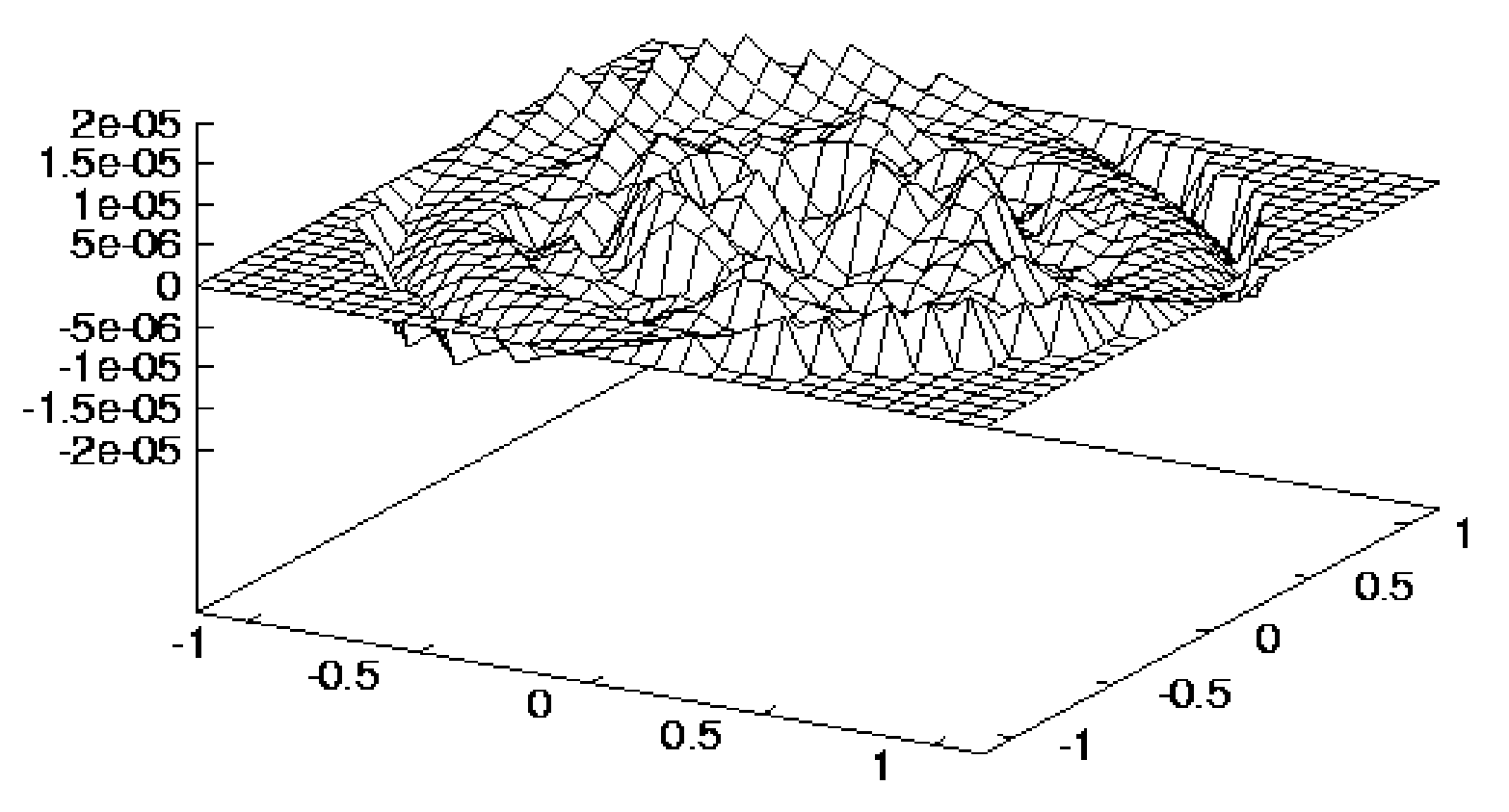,height=1.3in,width=2.2in}}
\vspace{0.5in}
\hbox{
\psfig{figure=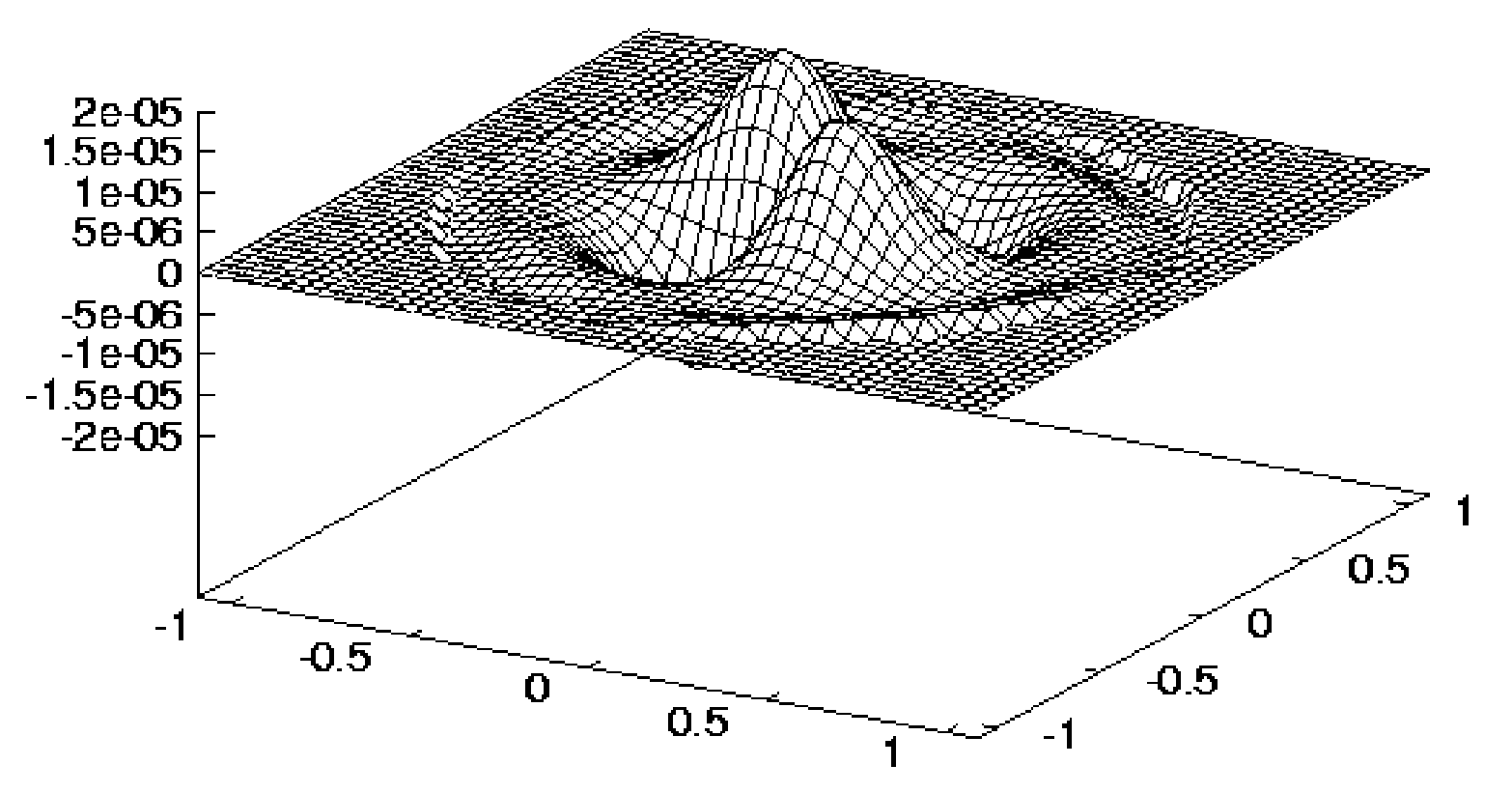,height=1.3in,width=2.2in}
\hspace{0.1in}
\psfig{figure=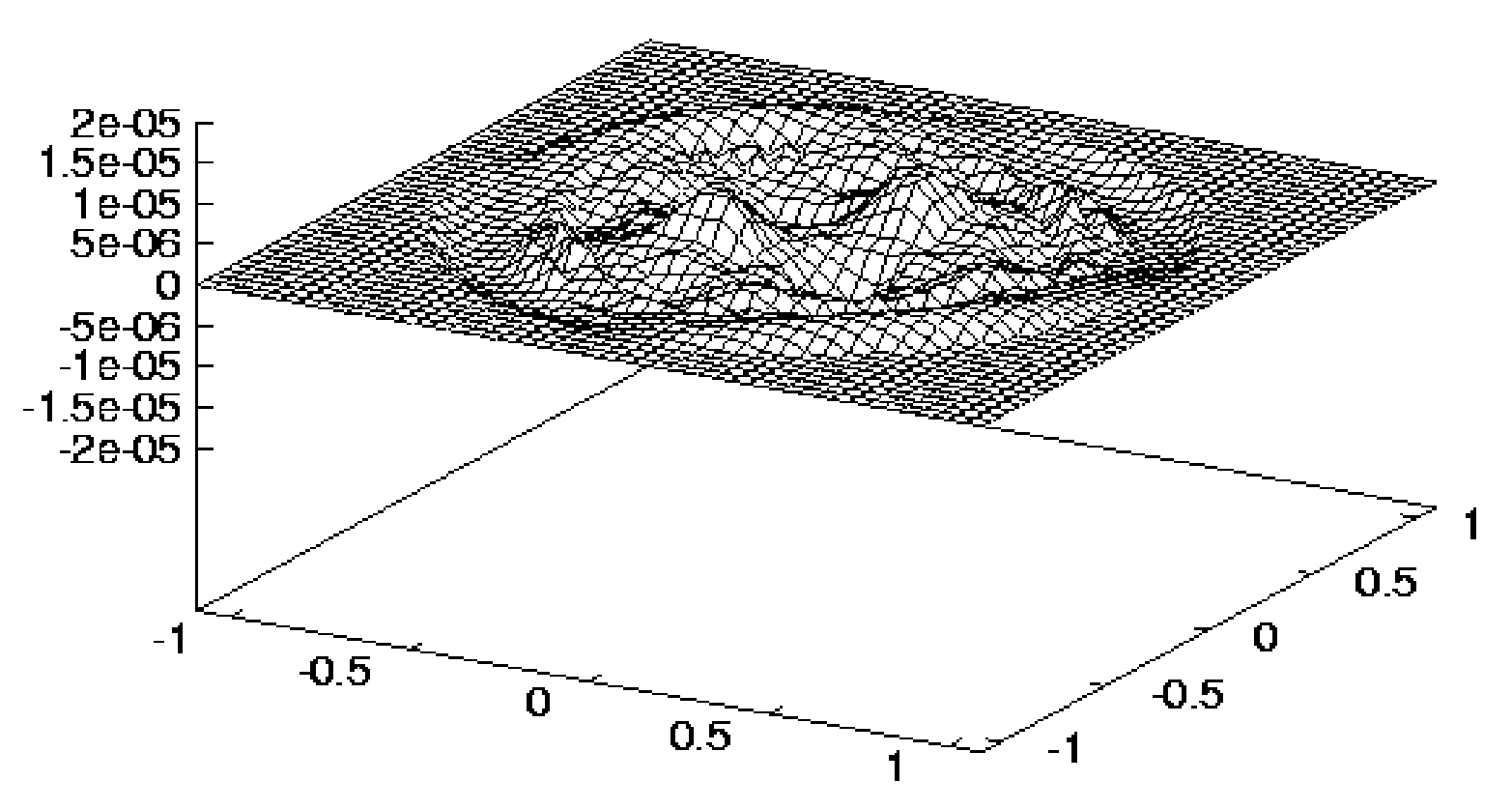,height=1.3in,width=2.2in}}}}
\caption{Results for the double set of Cauchy runs.
All plots represent $z=0$ slices of the metric component ${}^{(3)}g_{xy}$. Top left: the Cauchy data is
shown after one time-step for a grid size of $32^3$. The same run at time $t=10$ can be seen 
on the top right plot. The bottom left and right plots are the first and the last time-step of a
$48^3$ grid run up to $t=10$. }
\label{fig:stability.2adm.plots}
\end{figure}

In the last set of runs there are two Cauchy codes involved. The first is provided
analytic boundary data, and it serves as numeric input for the extraction module, 
characteristic evolution, and injection that provide boundary data for the second Cauchy code.
A schematic diagram illustrating the numerical setup can be seen 
on Figure~\ref{fig:stability.2adm.setup}. 
Grid sizes were $32^3$ and $48^3$. Due to the extensive hard-disk usage of the 
experiment, the runs lasted only up to $t=10$. Surface plots of the metric
component ${}^{(3)}g_{xy}$ at $z=0$ can be seen on Figure~\ref{fig:stability.2adm.plots}.
Although the numerical noise left in the grid is larger in this case then
for the previous runs, no numerical instabilities were seen at the time-scale
of the run.

\section{The fully coupled problem}

\begin{figure}
\centerline{
\epsfxsize=3.0in\epsfbox{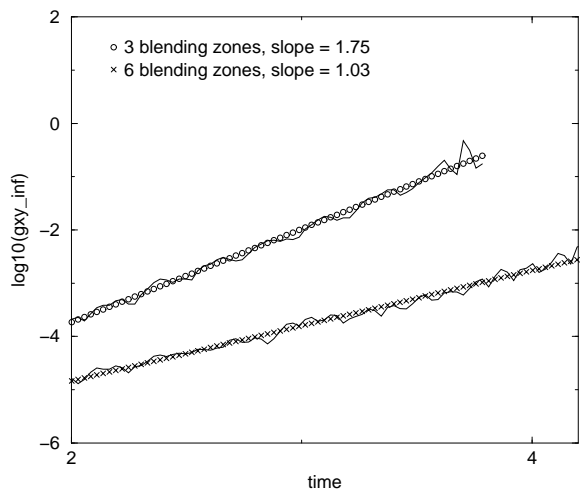}}
\caption{The runs with the fully coupled Cauchy+Characteristic codes
demonstrate exponential growth versus the number of blending zones.
Grid size for both runs is $32^3$.}
\label{fig:stability.blowup}
\end{figure}

The main conclusion of the runs from the previous section is that numerical
noise by itself will not destabilize the Cauchy-CCM system on a short term.
The question  then is: 
What is destabilizing the full Cauchy-CCM system?
In order to shed light on the issue 
a run was performed using the fully coupled system,
a Cartesian grid of $32^3$ points,
with a blending width  of $3 \Delta x$. Then 
the run was repeated using the same grid-spacing $\Delta x$ 
and same world-tube radius but with 
blending over a region of $6 \Delta x$.
The result from the two runs is plotted in 
Figure~\ref{fig:stability.blowup}.  As it can be seen, the smaller the cavity
between the extraction world-tube $\Gamma$ and the outermost injection world-tube
$\Sigma\; ( =\Sigma^{[M]})$, the steeper the exponential time dependence of 
the $\ell_\infty$ norm of ${}^{(3)}g_{ij}$.
But even if one uses a large $\Gamma-\Sigma$ cavity,
the exponential mode is there. Furthermore
the convergence radius of the
$\lambda$-expansion used in the injection module is limited, 
and thus one must
not separate the two world-tubes excessively.

An alternative to the Dirichlet data used at the Cauchy boundary
is a Sommerfeld outgoing radiation condition. It performed better then the Dirichlet
condition, so a ``modified Sommerfeld''
version of the injection module has been implemented.\footnote{The
modified Sommerfeld version in a CCM context is written in the form
\begin{equation}
(\partial_t + \partial_r)\,{}^{(3)}g_{ij}^{Cauchy} = 
(\partial_t + \partial_r)\,{}^{(3)}g_{ij}^{characteristic}.
\end{equation} 
See  \cite{Bishop97b} for implementation of the modified
Sommerfeld condition for the case of SWE.}
The details of implementation will not be given, for not even
the outgoing radiation condition has solved the problem. During
the period of time when these experiments were done, 
it was found that the iterative Crank-Nicholson evolution 
scheme\footnote{See Section \ref{sec:ladm-cart.ICN1d} for a description of the
iterative Crank-Nicholson evolution 
scheme.} 
is not unconditionally stable with respect to the number of
iterations performed. One must do two or three, or six or seven, etc. iterations
but not four or five iterations, etc. \cite{Teukolsky99}. Still,
with the lessons applied, the boundary module did not achieve stability.

In order to shed light on the problem, we undertook
a systematic study of
the linearized ADM equations and  properties
of a variety of evolution codes and boundary conditions.
Chapters \ref{chap:ladm-cart} and \ref{chap:ladm-sph} describe this study.
As it is shown, the coupled 
set of partial-differential equations that form the principal
part of the ADM equations needs appropriate boundary conditions.
In particular, one should not specify boundary-values to six metric components,
but provide the two radiation degrees of freedom and use
a set of boundary constraints that determine the remaining 
four components of the metric tensor at the boundary.
Thus
the fact that CCM (as well as all other boundary conditions used 
so far with the ADM code) 
sets boundary values to all six metric components gives rise
to an ill-posed problem. This accounts for at least one 
source of the instabilities that are seen.  There might be other numerical
problems, but those cannot be studied before implementation of
a consistent spherical boundary is worked out for a Cartesian Cauchy code.

\chapter{Linearized Cartesian Cauchy Evolution}
\label{chap:ladm-cart}

This chapter discusses the stability
properties of numerical implementations of
 the linearized ADM equations in Cartesian coordinates, as defined in 
Section~\ref{sec:ladm-cart.fieldeq}. 
Throughout this chapter the slicing is geodesic, that is
the lapse and shift are fixed by 
\begin{equation}
\alpha = 1, \;\;\; \beta^i = 0.
\label{eq:ladm-cart.slicing}
\end{equation}

Four different finite-difference algorithms are compared. The issue
of consistent boundary conditions is also  addressed. A 
robust stability test is defined and used to check the stability
properties of the various evolution algorithms with various boundary conditions.
Some of the results presented in this chapter can be also found in 
\cite{Szilagyi99}.

Throughout this Chapter
we use Greek letters for space-time indices and Latin letters for spatial
indices. Four dimensional geometric quantities are explicitly indicated, such
as ${}^{(4)}R_{\alpha\beta}$ and ${}^{(4)}G_{\alpha\beta}$ for the Ricci and
Einstein tensors of the space-time, whereas $R_{ij}$ and $R$ refer to the
Ricci tensor and Ricci scalar of the Cauchy hypersurfaces. 
These quantities are computed to linear order throughout this and the following
Chapter.

The main question we ask for a number of numerical codes is
whether they are ``stable'' or ``unstable''. In the case of a
scalar field we call an evolution unstable if the $\ell_\infty$ norm
of the field grows exponentially with time. Otherwise the evolution 
is called ``stable''.  In the case of the Linearized Gravitational (LG)
equations we use the Hamiltonian constraint to monitor if
exponential modes are present in the evolution. 
An exponentially growing Hamiltonian constraint is the
signature of an unstable run.

\section{Linearized field equations in Cartesian coordinates}
\label{sec:ladm-cart.fieldeq}

The space-time  is treated as a perturbation
around flat space. In Cartesian coordinates
$(t,x,y,z)$ the linearized metric can be written as
\begin{eqnarray}
{}^{(4)}g_{\alpha \beta} = {}^{(4)}\eta_{\alpha \beta} +  {}^{(4)}h_{\alpha \beta},
\end{eqnarray}
where ${}^{(4)}\eta_{\alpha \beta}$ is the Minkowski metric  
${}^{(4)}\eta_{\alpha \beta} = \mbox{diag} (-1, 1, 1, 1)$.
The gauge condition Eq.~(\ref{eq:ladm-cart.slicing}) implies
${}^{(4)}h_{t\alpha} = 0.$

We also use the notation
\begin{eqnarray}
\dot f &\equiv& \partial_t f\,, \\
h &\equiv& \eta^{ij} \,h_{ij} = h^i_i\,, \\
h^{ij} &\equiv& \eta^{in} \,\eta^{jm}\, h_{nm}\,.
\end{eqnarray}

The Hamiltonian and momentum constraints are
\begin{equation}
{\cal C} := {}^{(4)}G_{tt}\,, 
\;\; {\cal C}_i := -{}^{(4)} G_{ti}\,.
\end{equation}

The evolution equations ${\cal E}_{ij}$ are built from the 
spatial components of the linearized Ricci tensor, the background
metric  and the 
Hamiltonian constraint:
\begin{equation}
{\cal E}_{ij} := {}^{(4)} R_{ij} + 
\frac 12 \lambda \eta_{ij} {\cal C},
\label{eq:ladm-cart.eveq-Eij}
\end{equation}
where the case $\lambda = -1$  corresponds to the spatial
components of the
Einstein tensor while $\lambda = 0$  corresponds to the 
``Standard ADM'' system as described in Section~\ref{sec:intro.ADM}.
The parameter $\lambda$ must satisfy
$1+\lambda > 0$ for a well-posed hyperbolic initial value problem for 
the system of equations governing constraint evolution 
\cite{Frittelli97a,WinicourPersonal}. Indeed, the
linearized Bianchi identities 
$\partial_\beta\, {}^{(4)} G_\alpha^\beta \equiv 0$ imply that
\begin{eqnarray}
\dot {\cal C}^i + (1 + \lambda) 
\partial^i C + \partial_j {\cal E}^{ij} &\equiv& 0, \\
\dot {\cal C} + \partial_i {\cal C}^i &\equiv& 0.
\end{eqnarray}
When the evolution equations are satisfied the Hamiltonian constraint
satisfies the wave equation
\begin{equation}
\ddot {\cal C} - (1 + \lambda) \partial^k \partial_k {\cal C} = 0,
\label{eq:ladm-cart.HC-SWE}
\end{equation}
propagating with speed $v_{\cal C} = \sqrt{1+\lambda}$. Early codes
\cite{York79} have used the case $\lambda=0$ (in which the Hamiltonian
constraint propagates along the light cone). Here 
the behavior of the Hamiltonian constraint 
${\cal C}$ in codes running with different 
values of $\lambda$ is studied.

The Ricci tensor components ${}^{(4)}R_{ij}$ can be written as
\begin{equation}
{}^{(4)} R_{ij} =  R_{ij} + \frac 12 \ddot h_{ij}\,,
\label{eq:ladm-cart.3Rijdef}
\end{equation}
where the 3-D curvature tensor $R_{ij}$ is given by 
\begin{equation}
R_{ij} = \frac{1}{2} \left\{
\partial^k \partial_i h_{jk} +
\partial^k \partial_j h_{ik} -
\partial^k \partial_k h_{ij} -
\partial_i \partial_j h \right\}. \label{eq:ladm-cart.3Rijexp}
\end{equation}
The Hamiltonian and momentum constraints are
\begin{eqnarray}
{\cal{C}} & =& \frac{1}{2}
\left(\partial_i \partial_j h^{ij}
- \partial_i \partial_j \eta^{ij} h \right), 
\label{eq:ladm-cart.Hconst-exp}\\
{\cal{C}}^i &=& -\frac{1}{2}
\left(  \partial_j \dot h^{ij} - \eta^{ij}  \partial_j \dot h \right).
\label{eq:ladm-cart.momconst-exp}
\end{eqnarray}

In some of the  algorithms 
the evolution equations ${\cal E}_{ij}$
are written in a form that is  first-differential-order in time 
and  second-differential-order in space:
\begin{eqnarray}
\dot h_{ij} &=& -2 K_{ij}, \label{eq:ladm-cart.Kdef} \\
\dot K_{ij} &=& R_{ij} + \frac 12 \lambda \eta_{ij} {\cal C},
\label{eq:ladm-cart.Kdoteq}
\end{eqnarray}
where Eq.~(\ref{eq:ladm-cart.3Rijdef}) was substituted into
Eq.~(\ref{eq:ladm-cart.eveq-Eij}). 
Except for  numerical roundoff error, using the variables
$g_{ij}$  instead of $h_{ij}$ makes no difference, since the two tensors
differ only by a constant tensor $\eta_{ij} = \delta_{ij}$ (in Cartesian
coordinates). In Section~\ref{sec:ladm-cart.1d} the variable $g_{ij}$ 
is used
while the codes described in Section~\ref{sec:ladm-cart.3d} use $h_{ij}$.

Since evolving the linearized GR equations  implies
propagating scalar waves (along with gauge modes), the scalar wave equation
will provide a useful test bed in ruling out  bad choices of 
finite-difference approximations. Algorithms that fail
for the scalar wave equation (SWE) will necessarily fail for the (linearized) 
evolution equations describing gravity.

In addition,  finite difference
algorithms and boundary conditions were first tested in the 1-D case.  
This provides a numerically inexpensive environment for ruling out unstable
evolution schemes.

\section{Study of the 1-D case}
\label{sec:ladm-cart.1d}

This section describes a large number of 1-D numerical experiments.
Many of the results quoted in this section are 
superseded by the 
3-D results in Section 
\ref{sec:ladm-cart.3d}. The work done in 1-D is 
described in detail to indicate how lower dimensions
provide  a first criteria in ruling
out bad choices of finite-difference algorithms.

\subsection{Scalar wave equation}

In one dimension the SWE takes the form
\begin{equation}
    \partial_{tt}\phi- \partial_{xx}\phi =0. \label{eq:ladm-cart.1dswe2nd}
\end{equation}

Equation~(\ref{eq:ladm-cart.1dswe2nd}) can be split into
two equations that are first-order in time, and second-order in space: 
\begin{eqnarray}
    \partial_{t}\phi &=& \xi 
\label{eq:ladm-cart.1dswe1st1},  \\
    \partial_{t}\xi &=& \partial_{xx}\phi.
\label{eq:ladm-cart.1dswe1st2}
\end{eqnarray}

Note that $\phi(x,t) = t$ is a solution of the SWE.
To rule it out one needs to use proper 
 initial and/or boundary data. However, such secular growth should not
be interpreted as an instability.

\subsection{1-D linearized gravity}

Plane-wave solutions for a 1-dimensional version of gravitational equations
were  also studied. 
The Cartesian metric tensor ${}^{(4)}g_{\mu \nu}$ takes the form 
\begin{equation}
{}^{(4)}g_{\mu \nu}(t,x) = {}^{(4)}\eta_{\mu \nu} +  {}^{(4)}h_{\mu \nu}(t,x).
\end{equation}
In this 1-D study
the parameter $\lambda$ in
Eq.~(\ref{eq:ladm-cart.eveq-Eij}) is set to zero, so that
the evolution equations (\ref{eq:ladm-cart.eveq-Eij}) reduce to
\begin{equation}
{}^{(4)} R_{ij} = R_{ij} + \frac 12 \ddot g_{ij} = 0.
\label{eq:ladm-cart.eveq-1d}
\end{equation}
The first-differential-order in time form of Eq.~(\ref{eq:ladm-cart.eveq-1d}) is
\begin{eqnarray}
\dot g_{ij} &=& - 2 K_{ij}, 
\label{eq:ladm-cart.eveq-1d.gdot} \\
\dot K_{ij} &=& R_{ij}.
\label{eq:ladm-cart.eveq-1d.Kdot}
\end{eqnarray}
Recall from Section~\ref{sec:intro.ADM} that $K_{ij}$ stands
for the extrinsic curvature of the $t=$ constant Cauchy
slice.

The constraint equations
(\ref{eq:ladm-cart.Hconst-exp}) - (\ref{eq:ladm-cart.momconst-exp})  
are

\begin{eqnarray}
g_{yy,xx} + g_{zz,xx}  = 0, \label{eq:ladm-cart.1dlgrc1} \\
g_{xy,xt} = g_{xz,xt} = 0,
\label{eq:ladm-cart.1dlgrc2} \\
g_{yy,xt} + g_{zz,xt}  = 0. \label{eq:ladm-cart.1dlgrc3}
\end{eqnarray}
These are not imposed in the 
evolution code; therefore this is  a case of  unconstrained evolution.
The  tensor $R_{ij}$ is computed via 
Eq.~(\ref{eq:ladm-cart.3Rijexp}):

\begin{eqnarray}
  R_{xx} &=& -\frac{1}{2}\left[ g_{yy,xx} + g_{zz,xx} \right],
  \nonumber \\
  R_{xy} &=& 0,
  \nonumber \\
  R_{xz} &=& 0,
  \nonumber \\
  R_{yy} &=& -\frac{1}{2} g_{yy,xx}\,,
  \nonumber \\
  R_{yz} &=& -\frac{1}{2} g_{yz,xx}\,,
  \nonumber \\
  R_{zz} &=& -\frac{1}{2} g_{zz,xx}\,.
  \label{eq:ladm-cart.lgrR1d}
\end{eqnarray}

The mode $g_{ij} = 
\eta_{ij} + a_{ij} t$ is 
a solution of Eq.~(\ref{eq:ladm-cart.eveq-1d}) 
with $a_{ij}$ symmetric and constant. 
Since the corresponding Ricci tensor is zero this is
a pure gauge mode that, if present, 
is  introduced by initial data. Another way
of seeing that a pure gauge transformation can lead to linearly growing
modes is to apply the infinitesimal gauge (coordinate) transformation 
\begin{equation}
\delta g^{\alpha \beta} = 
g^{\alpha \mu} \zeta^\beta_{,\mu} + 
g^{\beta \mu} \zeta^\alpha_{,\mu}
\end{equation}
with the choice
\begin{equation}
\zeta^\alpha = \left( \varepsilon x y, 
\varepsilon y t, \varepsilon x t, 0 \right). \label{eq:ladm-cart.lgr1dzeta}
\end{equation}
This immediately gives
\begin{equation}
\delta g^{00} = \delta g^{0i} = 0,
\end{equation}
so that the gauge transformation (\ref{eq:ladm-cart.lgr1dzeta}) 
is consistent with the geodesic gauge condition 
Eq.~(\ref{eq:ladm-cart.slicing}).
For the spatial components the result is 
\begin{equation}
\delta g^{ij} = \
\left( 
\begin{array}{ccc}
0 &2\,\varepsilon\,t  &0 \\
2\,\varepsilon\,t &0  &0 \\
0     &0    &0
\end{array}
\right).
\end{equation}

\vspace{3em}

For the purpose of defining various (discretized) evolution schemes for the
SWE and the linearized gravitational
 equations in 1-D, physical fields are represented
on a set of equally spaced points.
Grid conventions are as follows: 
$$
x_I =I\Delta x,\,\,\, t_N =N\Delta t,\,\,\,
\phi(x_I,t_N)=\phi_{[I]}^{[N]},
$$
 where $I=1\ldots I_{max}$ and $N\geq 0$.

\subsection{Non-staggered leap-frog (LF1) for the 1-D SWE}

This section describes the non-staggered leap-frog evolution
algorithm for  
Eqs.~(\ref{eq:ladm-cart.1dswe1st1}) - (\ref{eq:ladm-cart.1dswe1st2}).
 
Given  $\phi^{[0]}$ and $\xi^{[0]}$ (i.e. initial data at $t=0$)
the quantities $\phi^{[1]}$ and $\xi^{[1]}$ are updated
as follows:
\begin{eqnarray}
        \phi^{[1]}_{[I]} &=& \phi^{[0]}_{[I]}
 +  \Delta t \cdot \xi^{[0]}_{[I]}\,, \\
        \xi^{[1]}_{[I]}  &=& \xi^{[0]}_{[I]} + 
\frac{\Delta t}{(\Delta x)^2} \cdot 
\left[\phi^{[0]}_{[I+1]} - 2 \phi^{[0]}_{[I]}
 + \phi^{[0]}_{[I-1]} \right],
\end{eqnarray}
which amounts to a time-step first-order in time from level 
$N=0$ to level $N=1$.
The explicit form of the evolution scheme is: 
\begin{eqnarray}
        \phi^{[N]}_{[I]} &=& \phi^{[N-2]}_{[I]}
 +  2 \Delta t \cdot \xi^{[N-1]}_{[I]}\,, 
\label{eq:ladm-cart.leap1}\\
        \xi^{[N]}_{[I]}  &=& \xi^{[N-2]}_{[I]} + 
\frac{2\Delta t}{(\Delta x)^2} \cdot 
\left[\phi^{[N-1]}_{[I+1]} - 2 \phi^{[N-1]}_{[I]}
 + \phi^{[N-1]}_{[I-1]} \right].
\label{eq:ladm-cart.leap2}
\end{eqnarray}

\subsection{Second-order evolution scheme (2ND)}

\subsubsection{Scalar wave equation}

The second-order in time 
form of the wave-equation (\ref{eq:ladm-cart.1dswe2nd})
is discretized in the following way:

\begin{equation}
\phi^{[N]}_{[I]}=2 \phi^{[N-1]}_{[I]} -\phi^{[N-2]}_{[I]} 
+ \left(\frac{\Delta t}{\Delta x} \right)^2 \cdot \left[
\phi^{[N-1]}_{[I+1]} - 2 \phi^{[N-1]}_{[I]} +  \phi^{[N-1]}_{[I-1]}\right]
\label{eq:ladm-cart.3level}
\end{equation} 
Initial data is $\phi^{[0]}=\phi(x,t=0)$ and $\phi^{[1]}=\phi(x,t=\Delta t)$.

\subsubsection{Linearized gravity}

The  LG  equations~(\ref{eq:ladm-cart.eveq-1d}) 
are discretized in the following way:

\begin{eqnarray}
g^{[N]}_{xx[I]} &=& 2 g^{[N-1]}_{xx[I]} - g^{[N-2]}_{xx[I]} +
\nonumber \\
&+& \left(\frac{\Delta t}{\Delta x}\right)^2  \cdot 
\left[g^{[N-1]}_{yy[I+1]} - 2  g^{[N-1]}_{yy[I]} + g^{[N-1]}_{yy[I-1]} 
\right.
\nonumber \\
&&
\left.
+ g^{[N-1]}_{zz[I+1]} - 2 g^{[N-1]}_{zz[I]} + g^{[N-1]}_{zz[I-1]}\right],
\nonumber \\
g^{[N]}_{xy[I]} &=& 2 g^{[N-1]}_{xy[I]} - g^{[N-2]}_{xy[I]},
\nonumber \\
g^{[N]}_{xz[I]} &=& 2 g^{[N-1]}_{xz[I]} - g^{[N-2]}_{xz[I]},
\nonumber \\
g^{[N]}_{yy[I]} &=& 2 g^{[N-1]}_{yy[I]} - g^{[N-2]}_{yy[I]} 
+ \left(\frac{\Delta t}{\Delta x}\right)^2    \cdot 
\left[g^{[N-1]}_{yy[I+1]} - 2 g^{[N-1]}_{yy[I]} + g^{[N-1]}_{yy[I-1]}\right],
\nonumber \\
g^{[N]}_{yz[I]} &=& 2 g^{[N-1]}_{yz[I]} - g^{[N-2]}_{yz[I]} 
+ \left(\frac{\Delta t}{\Delta x}\right)^2   \cdot 
\left[g^{[N-1]}_{yz[I+1]} - 2 g^{[N-1]}_{yz[I]} + g^{[N-1]}_{yz[I-1]}\right],
\nonumber \\
g^{[N]}_{zz[I]} &=& 2 g^{[N-1]}_{zz[I]} - g^{[N-2]}_{zz[I]} 
+ \left(\frac{\Delta t}{\Delta x}\right)^2  \cdot 
\left[g^{[N-1]}_{zz[I+1]} - 2 g^{[N-1]}_{zz[I]} + g^{[N-1]}_{zz[I-1]}\right].
\nonumber \\
\end{eqnarray}

Required initial data is $g_{ij}^{[0]}$ and $g_{ij}^{[1]}$.

\subsection{Iterative Crank-Nicholson evolution scheme (ICN)}
\label{sec:ladm-cart.ICN1d}

\subsubsection{Scalar wave equation}

The wave-equation is solved
in its first-order form given by
Eqs.~(\ref{eq:ladm-cart.1dswe1st1}) - (\ref{eq:ladm-cart.1dswe1st2}).

For each time-step one performs a given number of Crank-Nicholson
iterations. The first one is first-order in time (counted for as 
iteration $0$), and the rest of them are second-order in time.
The sequence of operations executed for each 
time-step is the following:

\begin{enumerate}
\item Compute $\stackrel{(0)}{\phi}\!{}^{[N+1]}$ 
and $\stackrel{(0)}{\xi}\!{}^{[N+1]}$ first-order in time:
\begin{eqnarray}
        \stackrel{(0)}{\phi}\!{}^{[N+1]}_{[I]} 
&=& \phi^{[N]}_{[I]}
 +  \Delta t \cdot \xi^{[N]}_{[I]}, \\
        \stackrel{(0)}{\xi}\!{}^{[N+1]}_{[I]}  
&=& \xi^{[N]}_{[I]} + 
\frac{\Delta t}{(\Delta x)^2} \cdot 
\left\{\phi^{[N]}_{[I+1]} - 2 \phi^{[N]}_{[I]}
 + \phi^{[N]}_{[I-1]} \right\}.
\end{eqnarray}

\item \label{item:ladm-cart.1dsweCNloopstart} Compute 
$\stackrel{(i)}{\phi}\!{}^{[N+1/2]}$ and
$\stackrel{(i)}{\xi}\!{}^{[N+1/2]}$ 
by averaging:
\begin{eqnarray}
\stackrel{(i)}{\phi}\!{}^{[N+1/2]} 
&=& \frac{1}{2} \left\{ \phi^{[N]} + 
\stackrel{(i)}{\phi}\!{}^{[N+1]} \right\},\\
\stackrel{(i)}{\xi}\!{}^{[N+1/2]} 
&=& \frac{1}{2} \left\{ \xi^{[N]} +  
\stackrel{(i)}{\xi}\!{}^{[N+1]} \right\}.
\end{eqnarray}

\item Compute  
$\stackrel{(i+1)}{\phi}\!{}^{[N+1]}, 
\stackrel{(i+1)}{\xi}\!{}^{[N+1]}$ using
levels $N$ and $N+1/2$:

\begin{eqnarray}
\stackrel{(i+1)}{\phi}\!{}^{[N+1]}_{[I]}
         &=& \phi^{[N]}_{[I]}
         +  \Delta t \cdot 
         \stackrel{(i)}{\xi}\!{}^{[N+1/2]}_{[I]}, \\
\stackrel{(i+1)}{\xi}\!{}^{[N+1]}_{[I]}  &=& \xi^{[N]}_{[I]} + 
\frac{\Delta t}{(\Delta x)^2} \cdot 
\left\{
\stackrel{(i)}{\phi}\!{}^{[N+1/2]}_{[I+1]} \right.
\nonumber \\ && \left. 
- 2 \stackrel{(i)}{\phi}\!{}^{[N+1/2]}_{[I]}
+ \stackrel{(i)}{\phi}\!{}^{[N+1/2]}_{[I-1]} 
\right\}.
\end{eqnarray}
\item Increment $i$ by one, then go back to 
step \ref{item:ladm-cart.1dsweCNloopstart}, until the desired number
of iterations is reached.
\end{enumerate}

Initial data is $\phi^{[0]}$ and $\xi^{[0]}$.

A discretized stability analysis of the evolution scheme 
shows \cite{Teukolsky99} that the algorithm is stable for
two and three iterations, unstable for four and five iterations, 
stable for six and seven iterations, etc.

\subsubsection{Linearized gravity}

Starting from 
Eqs.~(\ref{eq:ladm-cart.eveq-1d.gdot}) - (\ref{eq:ladm-cart.eveq-1d.Kdot}),
the ICN algorithm is:

\begin{enumerate}
\item Compute $\stackrel{(0)}{g_{ij}}\!{}^{[N+1]}$ 
and $\stackrel{(0)}{K_{ij}}\!{}^{[N+1]}$ first-order in time:
\begin{eqnarray}
        \stackrel{(0)}{g_{ij}}\!{}^{[N+1]}_{[I]} 
&=& g^{[N]}_{ij\,[I]}
 - 2 \, \Delta t \cdot K^{[N]}_{ij\,[I]},
\nonumber \\
        \stackrel{(0)}{K}\!{}^{[N+1]}_{xx\,[I]}  
&=& K^{[N]}_{xx\,[I]} - \frac{1}{2} 
\frac{\Delta t}{(\Delta x)^2} \times \nonumber \\ & \times & 
\left\{
   g^{[N]}_{yy\,[I+1]}
 - 2 g^{[N]}_{yy\,[I]}
 + g^{[N]}_{yy\,[I-1]}  \right. \nonumber \\ && \left.
 + g^{[N]}_{zz\,[I+1]}
 - 2 g^{[N]}_{zz\,[I]}
 + g^{[N]}_{zz\,[I-1]} \right\},
\nonumber \\
        \stackrel{(0)}{K}\!{}^{[N+1]}_{xy\,[I]} &=& K^{[N]}_{xy\,[I]},
\nonumber \\
        \stackrel{(0)}{K}\!{}^{[N+1]}_{xz\,[I]} &=& K^{[N]}_{xz\,[I]},
\nonumber \\
        \stackrel{(0)}{K}\!{}^{[N+1]}_{yy\,[I]}  
&=& K^{[N]}_{yy\,[I]} - \frac{1}{2} 
\frac{\Delta t}{(\Delta x)^2} \cdot 
\left\{g^{[N]}_{yy\,[I+1]} - 2 g^{[N]}_{yy\,[I]}
 + g^{[N]}_{yy\,[I-1]} \right\},
\nonumber \\
        \stackrel{(0)}{K}\!{}^{[N+1]}_{yz\,[I]}
&=& K^{[N]}_{yz\,[I]} - \frac{1}{2} 
\frac{\Delta t}{(\Delta x)^2} \cdot 
\left\{g^{[N]}_{yz\,[I+1]} - 2 g^{[N]}_{yz\,[I]}
 + g^{[N]}_{yz\,[I-1]} \right\},
\nonumber \\
        \stackrel{(0)}{K}\!{}^{[N+1]}_{zz\,[I]}  
&=& K^{[N]}_{zz\,[I]} - \frac{1}{2} 
\frac{\Delta t}{(\Delta x)^2} \cdot 
\left\{g^{[N]}_{zz\,[I+1]} - 2 g^{[N]}_{zz\,[I]}
 + g^{[N]}_{zz\,[I-1]} \right\}.
\nonumber \\
\end{eqnarray}

\item \label{item:ladm-cart.lgrCNloopstart} Compute 
$\stackrel{(i)}{\phi}\!{}^{[N+1/2]}$ and
$\stackrel{(i)}{\xi}\!{}^{[N+1/2]}$ 
by averaging:
\begin{eqnarray}
\stackrel{(i)}{g_{ij}}\!{}^{[N+1/2]} 
&=& \frac{1}{2} \left\{ g_{ij}^{[N]} + 
\stackrel{(i)}{g_{ij}}\!{}^{[N+1]} \right\},\\
\stackrel{(i)}{K_{ij}}\!{}^{[N+1/2]} 
&=& \frac{1}{2} \left\{ K_{ij}^{[N]} +  
\stackrel{(i)}{K_{ij}}\!{}^{[N+1]} \right\}.
\end{eqnarray}

\item Compute  
$\stackrel{(i+1)}{g_{ij}}\!{}^{[N+1]}, 
\stackrel{(i+1)}{K_{ij}}\!{}^{[N+1]}$ using
levels $N$ and $N+1/2$:

\begin{eqnarray}
\stackrel{(i+1)}{g_{ij}}\!{}^{[N+1]}_{[I]}
         &=& g^{[N]}_{ij\,[I]}
         - 2 \, \Delta t \cdot 
         \stackrel{(i)}{K_{ij}}\!{}^{[N+1/2]}_{[I]},
\nonumber \\
\stackrel{(i+1)}{K_{xx}}\!{}^{[N+1]}_{[I]}  &=& K^{[N]}_{xx\,[I]} - 
\frac{1}{2} \frac{\Delta t}{(\Delta x)^2} \nonumber \\
&\times& 
\left\{
\stackrel{(i)}{g_{yy}}\!{}^{[N+1/2]}_{[I+1]} 
- 2 \stackrel{(i)}{g_{yy}}\!{}^{[N+1/2]}_{[I]}
+ \stackrel{(i)}{g_{yy}}\!{}^{[N+1/2]}_{[I-1]} 
\nonumber \right. \\ && \left.
+ \stackrel{(i)}{g_{zz}}\!{}^{[N+1/2]}_{[I+1]} 
- 2 \stackrel{(i)}{g_{zz}}\!{}^{[N+1/2]}_{[I]}
+ \stackrel{(i)}{g_{zz}}\!{}^{[N+1/2]}_{[I-1]} \right\},
\nonumber \\
\stackrel{(i+1)}{K_{xy}}\!{}^{[N+1]}_{[I]}  &=& K^{[N]}_{xy\,[I]},
\nonumber \\
\stackrel{(i+1)}{K_{xz}}\!{}^{[N+1]}_{[I]}  &=& K^{[N]}_{xz\,[I]},
\nonumber  \\ 
\stackrel{(i+1)}{K_{yy}}\!{}^{[N+1]}_{[I]}  &=& K^{[N]}_{yy\,[I]} - 
\frac{1}{2} \frac{\Delta t}{(\Delta x)^2} \times \nonumber \\ &\times & 
\left\{
\stackrel{(i)}{g_{yy}}\!{}^{[N+1/2]}_{[I+1]} 
- 2 \stackrel{(i)}{g_{yy}}\!{}^{[N+1/2]}_{[I]}
+ \stackrel{(i)}{g_{yy}}\!{}^{[N+1/2]}_{[I-1]}
 \right\},
\nonumber \\
\stackrel{(i+1)}{K_{yz}}\!{}^{[N+1]}_{[I]}  &=& K^{[N]}_{yz\,[I]} - 
\frac{1}{2} \frac{\Delta t}{(\Delta x)^2} \times \nonumber \\ &\times & 
\left\{
\stackrel{(i)}{g_{yz}}\!{}^{[N+1/2]}_{[I+1]} 
- 2 \stackrel{(i)}{g_{yz}}\!{}^{[N+1/2]}_{[I]}
+ \stackrel{(i)}{g_{yz}}\!{}^{[N+1/2]}_{[I-1]} \right\},
\nonumber \\
\stackrel{(i+1)}{K_{zz}}\!{}^{[N+1]}_{[I]}  &=& K^{[N]}_{zz\,[I]} - 
\frac{1}{2} \frac{\Delta t}{(\Delta x)^2} \times \nonumber \\ &\times & 
\left\{
\stackrel{(i)}{g_{zz}}\!{}^{[N+1/2]}_{[I+1]} 
- 2 \stackrel{(i)}{g_{zz}}\!{}^{[N+1/2]}_{[I]}
+ \stackrel{(i)}{g_{zz}}\!{}^{[N+1/2]}_{[I-1]} \right\}.
\nonumber \\
\end{eqnarray}
\item Increment $i$ by one, then go back to 
step \ref{item:ladm-cart.lgrCNloopstart}, until the desired number
of iterations is reached.
\end{enumerate}

Initial data is 
$g_{ij}^{[0]}$ and $K_{ij}^{[0]}$.

\subsection{Boundary conditions}

\subsubsection{Analytic formulation}
\label{seq:ladm-cart.analbdry}

Let $L$ be the size of the evolution domain:
\begin{equation}
  x \in [0, L].
\end{equation}
Codes were  run with the following types of boundary conditions:
\begin{itemize}
\item Periodic boundaries:
\begin{equation}
\phi(x+L,t) = \phi(x,t)
\end{equation}

\item Reflecting boundaries:
\begin{equation}
\phi(0,t) = \phi(L,t) = 0
\end{equation}

\item Sommerfeld boundaries (outgoing radiation condition): 
based on the assumption that there is no incoming radiation
at the boundaries, one obtains
\begin{eqnarray}
\left[(\partial_t - \partial_x) \phi \right]_{|x=0} = 0, \\
\left[(\partial_t + \partial_x) \phi \right]_{|x=L} = 0.
\end{eqnarray}

\end{itemize}

\subsubsection{Discretized formulation}

Periodic boundary conditions consist of the identification
\begin{eqnarray}
\phi_{[0]} &\equiv& \phi_{[I_{max}]}, \\
\phi_{[I_{max}+1]} &\equiv& \phi_{[1]}.
\end{eqnarray}
Reflecting  boundary conditions are imposed by setting
\begin{equation}
\phi_{[1]} = \phi_{[I_{max}]} = 0.
\end{equation}
Sommerfeld boundary conditions are tested in first- and in second-order 
form. The first-order form of the 1-D outgoing
boundary condition is:

\begin{eqnarray}
\phi^{[N]}_{[1]} &=&  \frac{1}{\Delta t + \Delta x}
 \left[ \Delta t \cdot  \phi^{[N]}_{[2]} 
+ \Delta x \cdot  \phi^{[N-1]}_{[1]}\right],
\label{eq:ladm-cart.somm1a}\\
\phi^{[N]}_{[I_{max}]} &=&  \frac{1}{\Delta t + \Delta x}
 \left[ \Delta t \cdot  \phi^{[N]}_{[I_{max}-1]} 
+ \Delta x \cdot  \phi^{[N-1]}_{[I_{max}]}\right],
\label{eq:ladm-cart.somm1b}
\end{eqnarray}
while the second-order form is:
\begin{eqnarray}
\phi^{[N]}_{[1]}  &=& \frac{1}{3(\Delta t+\Delta x)}
\times 
\left[4 \Delta t \cdot 
\phi^{[N]}_{[2]}  
- \Delta t \cdot \phi^{[N]}_{[3]}    
\right. \nonumber \\ && \left.
+ 4 \Delta x \cdot \phi^{[N-1]}_{[1]}  
- \Delta x \cdot \phi^{[N-2]}_{[1]}  \right],
\label{eq:ladm-cart.somm2a}\\
\phi^{[N]}_{[I_{max}]}  &=& \frac{1}{3(\Delta t+\Delta x)}
\times 
\left[4 \Delta t \cdot 
\phi^{[N]}_{[I_{max}-1]}  
- \Delta t \cdot \phi^{[N]}_{[I_{max}-2]}    
\right. \nonumber \\ && \left.
+ 4 \Delta x \cdot \phi^{[N-1]}_{[I_{max}]}  
- \Delta x \cdot \phi^{[N-2]}_{[I_{max}]}  \right].
\label{eq:ladm-cart.somm2b}
\end{eqnarray}

Non-periodic boundary conditions were  imposed either on 
$\phi$ or on $\xi$. Since $\xi$ is computed as a spatial derivative
of $\phi$, if a boundary condition is imposed on
$\phi_{[1]}$ and $\phi_{[I_{max}]}$, the field $\xi$ is evolved
only in the points $x_{[2]}\ldots x_{[I_{max-1}]}$.  
Since $\phi$ is computed as a time derivative of $\xi$, imposing boundary
conditions on $\xi_{[1]}$ and $\xi_{[I_{max}]}$ allows evolution of
the field $\phi$ in the whole numerical grid.

\subsection{Code tests for the 1-D scalar wave equation}

This section investigates the stability properties of different 
evolution schemes with periodic, reflecting, and 
outgoing (Sommerfeld) boundary data. 
The different codes were tested qualitatively 
with a pulse of compact support:

\begin{equation}
\phi(x,t) = \left\{ 
\begin{array}{ll}
\phi(x,t) = A \cdot \left[\frac{(t - \epsilon \cdot x)^2}{w^2} - 1 \right]^4,
 &  |\epsilon \cdot x-t| < w \\
0,   &   |\epsilon \cdot x-t|\geq w.
\label{eq:ladm-cart.compinidata}
\end{array}
\right.
\end{equation}

These runs were done for $10$ crossing times 
on a grid that goes from $-1$ to $1$  with the choices
$$
A=1,\;w=0.75,\;\epsilon=\pm 1.
$$
The codes 
did not show any unexpected qualitative properties. The amount of 
dispersion was insignificant.

The stability properties of the codes were tested
for $\sim 10^3$ crossing times. Initial data was 
generated by multiplying the right-hand-side of 
Eq.~(\ref{eq:ladm-cart.compinidata}) with a set of random numbers distributed
between $-0.5$ and  $0.5$.
When running the SWE with random initial data, one  
introduces  the mode $\phi = t$.
However, when applying freezing or Sommerfeld
boundary conditions this mode is  ruled out.

\subsubsection{LF1 evolution scheme}

\begin{figure}
\centerline{\hbox{
\psfig{figure=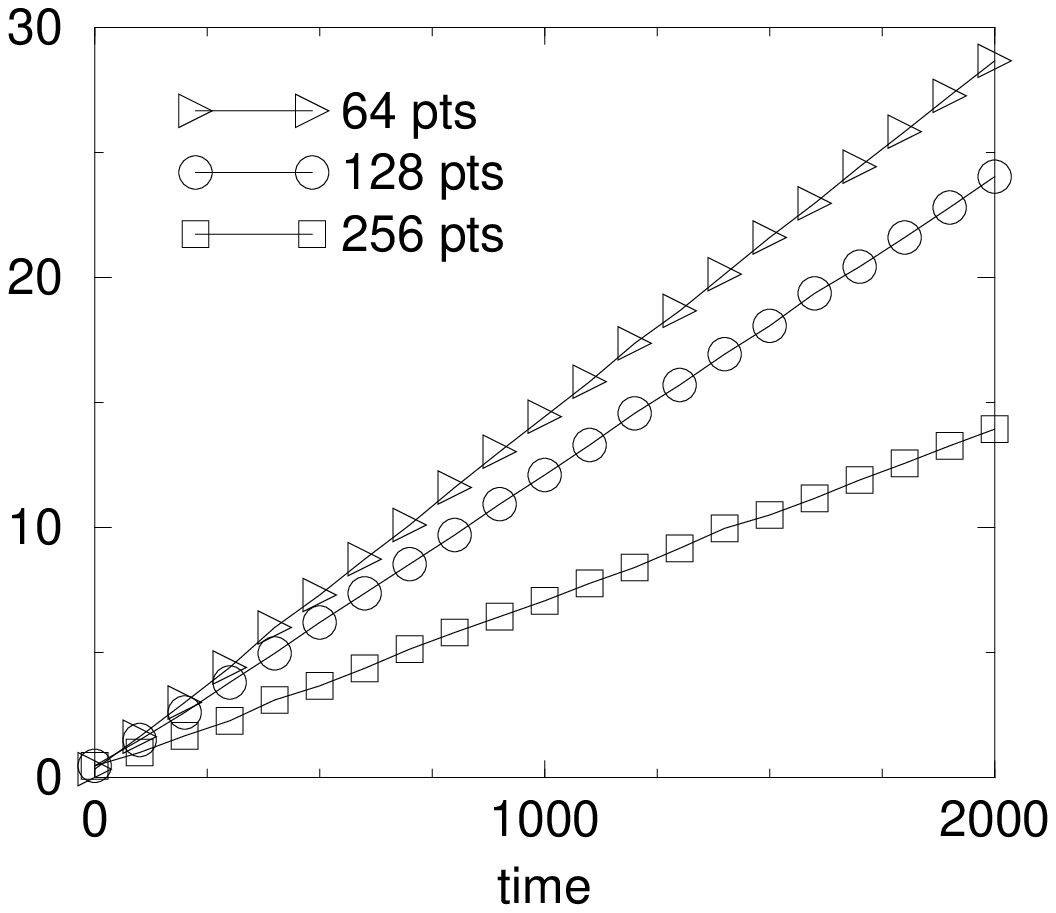,height=1.40in,width=1.40in}
\hspace{0.1in}
\psfig{figure=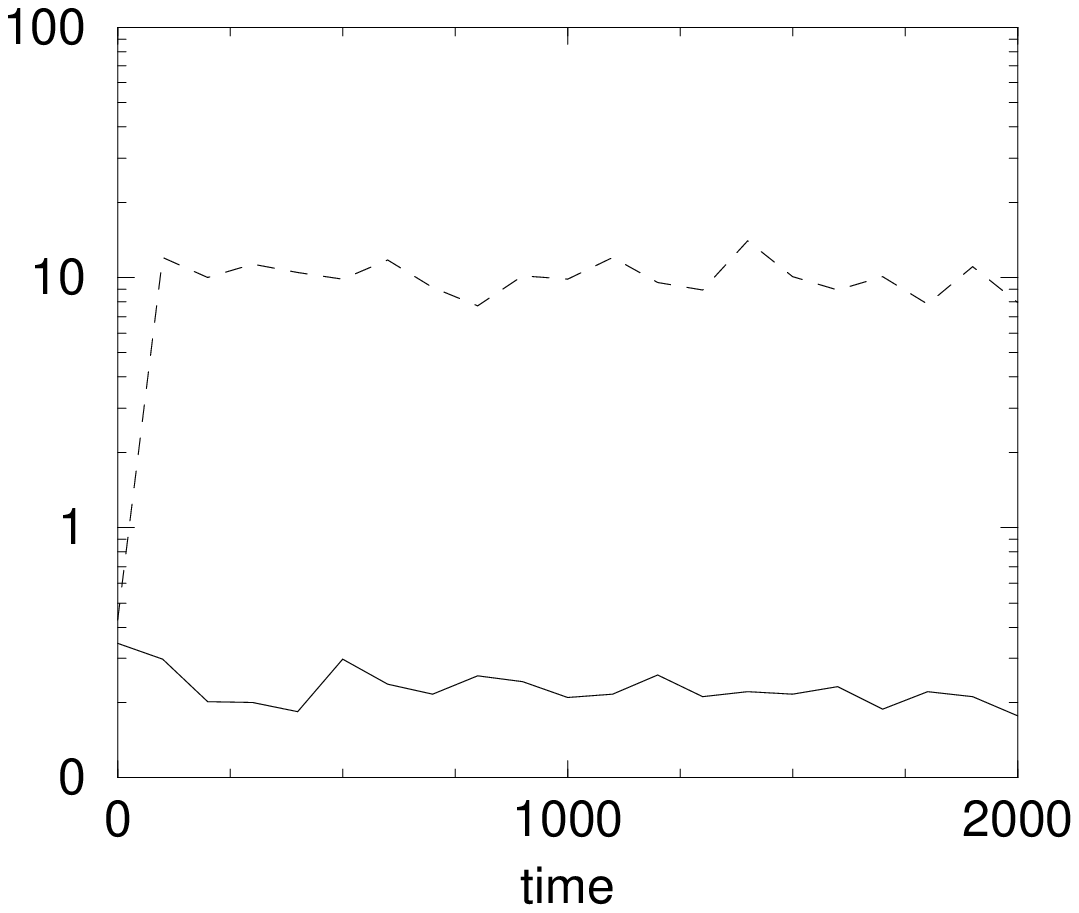,height=1.40in,width=1.40in}
\hspace{0.1in}
\psfig{figure=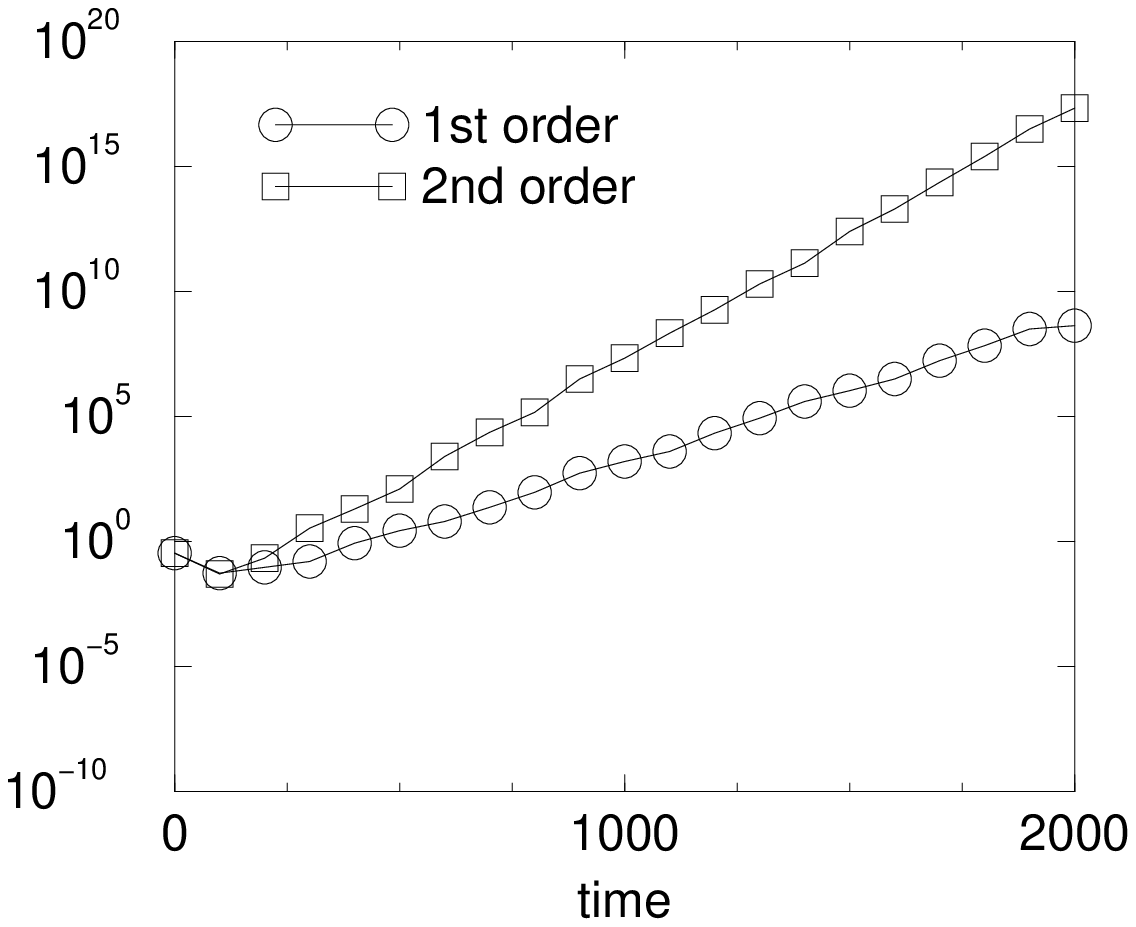,height=1.40in,width=1.40in}}}
\caption{Stability runs with the LF1 evolution scheme.
Left:  $||\phi||_{\infty}$
as a function of time for different grid sizes, 
using periodic boundary conditions.
Middle: $||\phi||_{\infty}$ (dashed line) 
and $||\xi||_{\infty}$ (continuous line) 
for a grid size of 64, using reflecting 
boundary conditions applied to $\xi$.
Right: $||\phi||_{\infty}$
as a function of time for 64 points, using first- and second-order
radiation boundary condition applied to $\phi$.
}\label{fig:ladm-cart.swe.1d.NSruns}
\end{figure}

Runs were made at a CFL factor of $\Delta t / \Delta x = 0.25$ (which is 
at half of the CFL limit), using grid sizes of $64, 128, 256$. 
Results are as follows:
\begin{itemize}

\item Periodic boundaries display linearly growing
 $||\phi||_{\infty}$ as a function of time, 
while $||\xi||_{\infty}$ stays bounded, i.e. $||\xi||_{\infty} \leq C$ with $C$ constant.

\item Reflecting boundary conditions applied  to either $\phi$ or $\xi$
result in stable runs. Moreover, both $||\phi||_\infty$ and $||\xi||_\infty$ stay bounded.

\item The outgoing
radiation condition in first-order or  second-order form 
applied to $\phi$ or to $\xi$
gives {\em unstable} runs, showing
an exponentially growing time dependence of both $||\phi||_{\infty}$ and
$||\xi||_{\infty}$.

\end{itemize}
Plots of these three sets of runs are shown in 
Figure~\ref{fig:ladm-cart.swe.1d.NSruns}.

\subsubsection{2ND evolution scheme}

\begin{figure}
\centerline{\hbox{
\psfig{figure=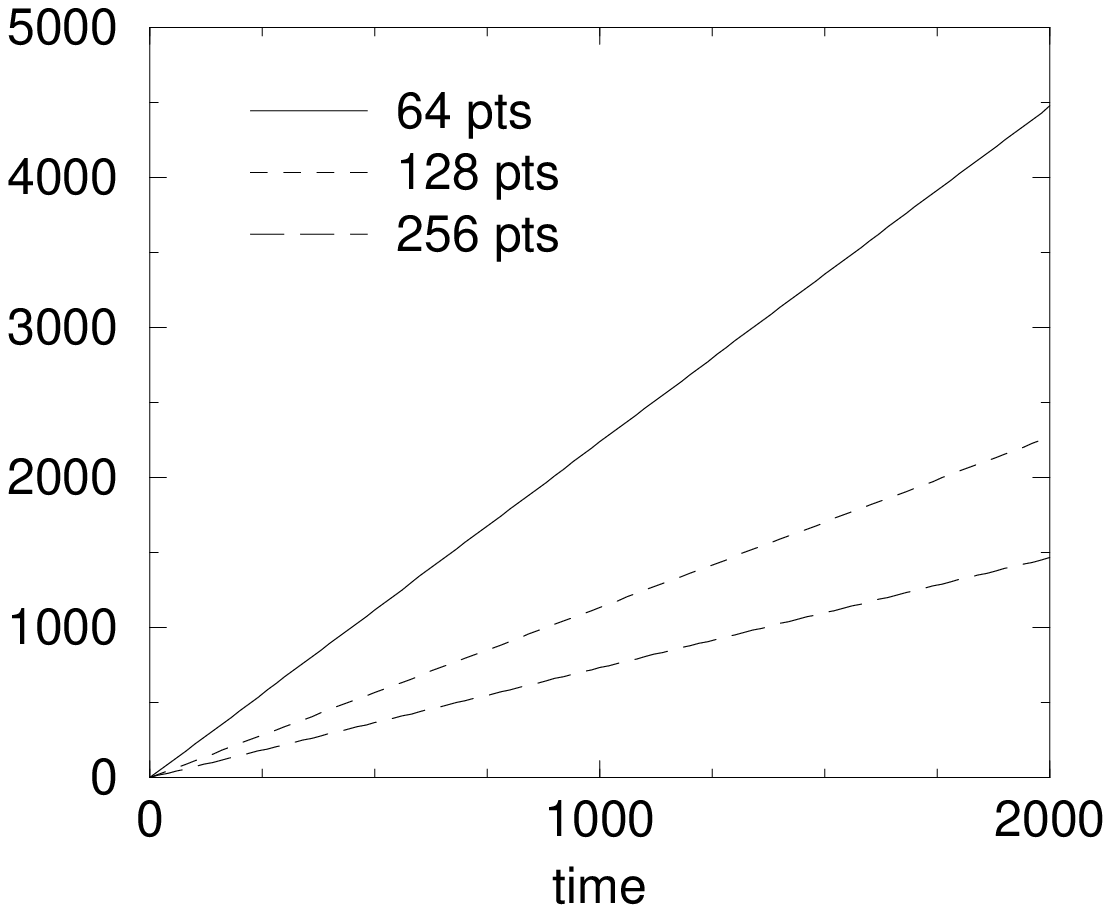,height=1.40in,width=1.40in}
\hspace{0.1in}
\psfig{figure=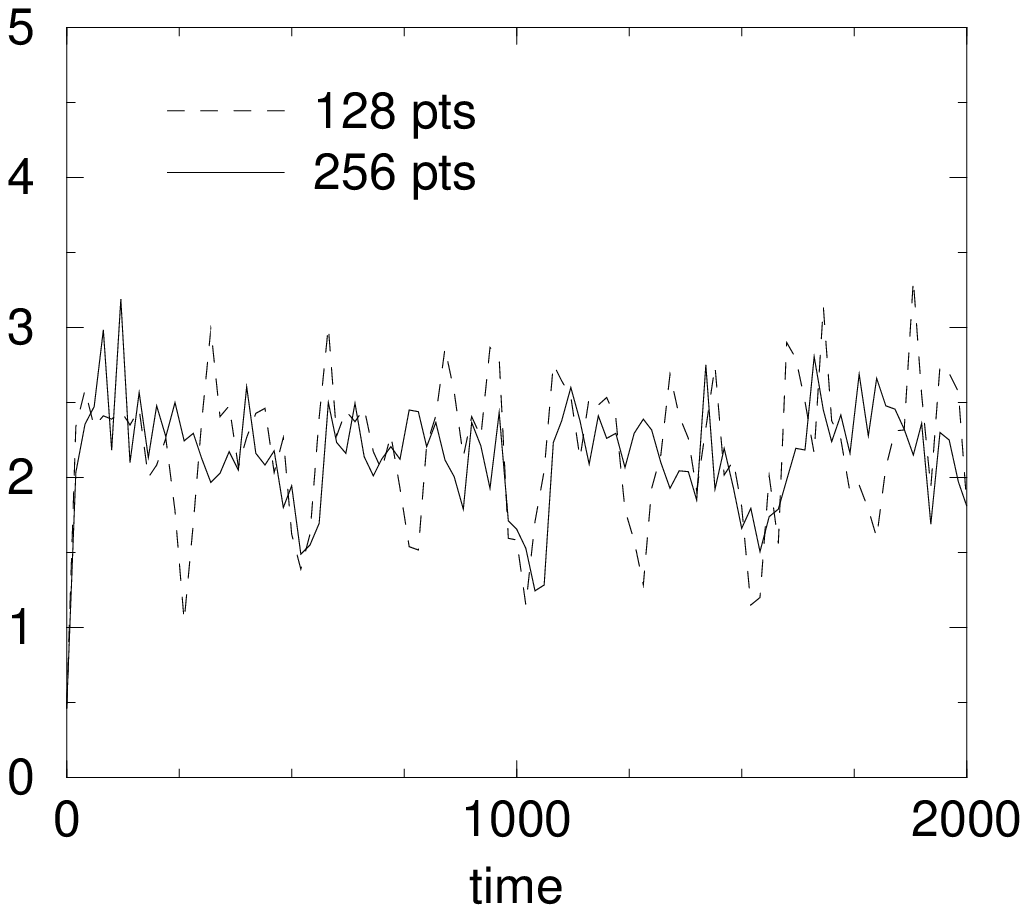,height=1.40in,width=1.40in}
\hspace{0.1in}
\psfig{figure=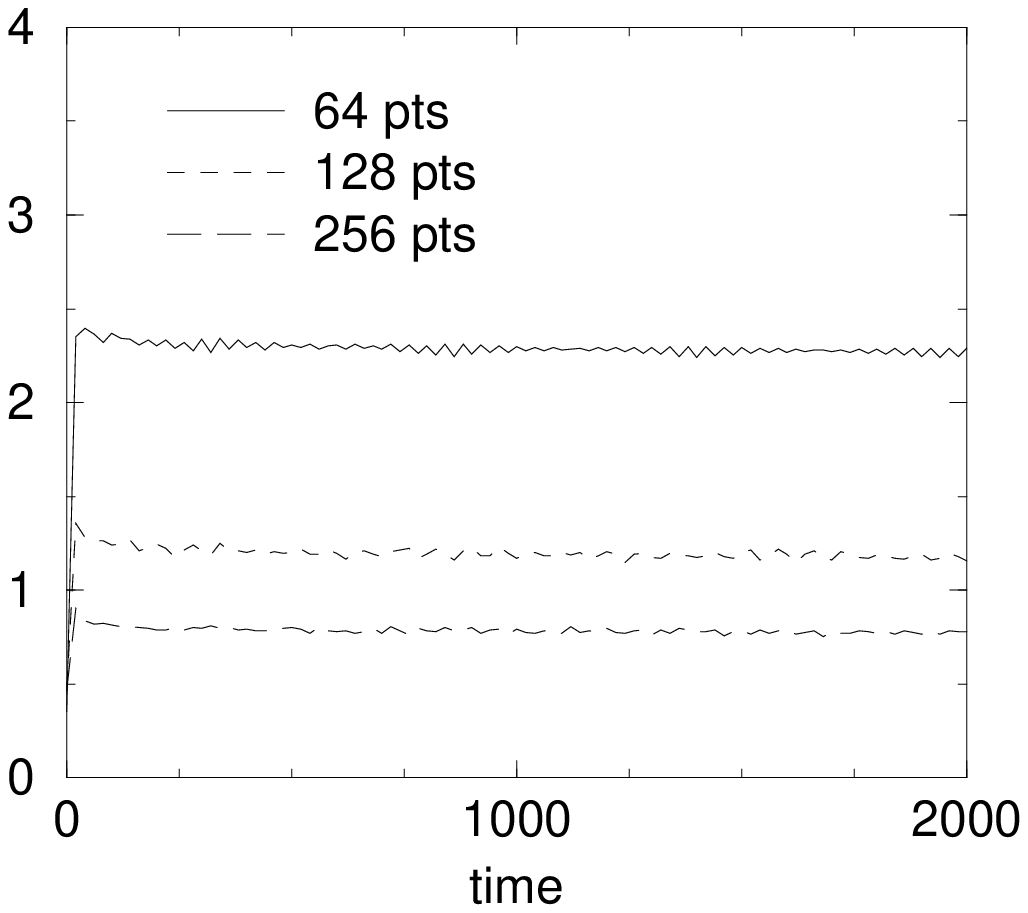,height=1.40in,width=1.40in}
}}
\caption{
Stability runs for the 2ND evolution scheme.
$||\phi||_{\infty}$ is shown 
as a function of time for different grid sizes, 
using the following boundary conditions: periodic (on the left), 
reflecting (in the middle), and first-order outgoing 
radiation (on the right).
\label{fig:ladm-cart.swe.1d.2NDruns}}
\end{figure}

Runs 
were made for 1000 crossing times (i.e. up to $t=2000$), for 
grid sizes of $64,\;128, \;256$. The  CFL factor was 
 $\Delta t / \Delta x = 0.50$ 
(which is at half of the CFL limit).
Boundary conditions were  
periodic, reflecting, first- and second-order outgoing radiation condition,
with the non-periodic boundary conditions being applied 
either to $\phi$ or to $\xi$.
The code proved to be stable in all cases (see 
Figure~\ref{fig:ladm-cart.swe.1d.2NDruns}).
Periodic boundary conditions allowed for a linear growth of 
$||\phi||_\infty$.
In terms of stability
there was no significant difference between 
applying first- or second-order Sommerfeld
boundary conditions. 

\subsubsection{ICN evolution scheme}

\begin{figure}
\centerline{\hbox{
\psfig{figure=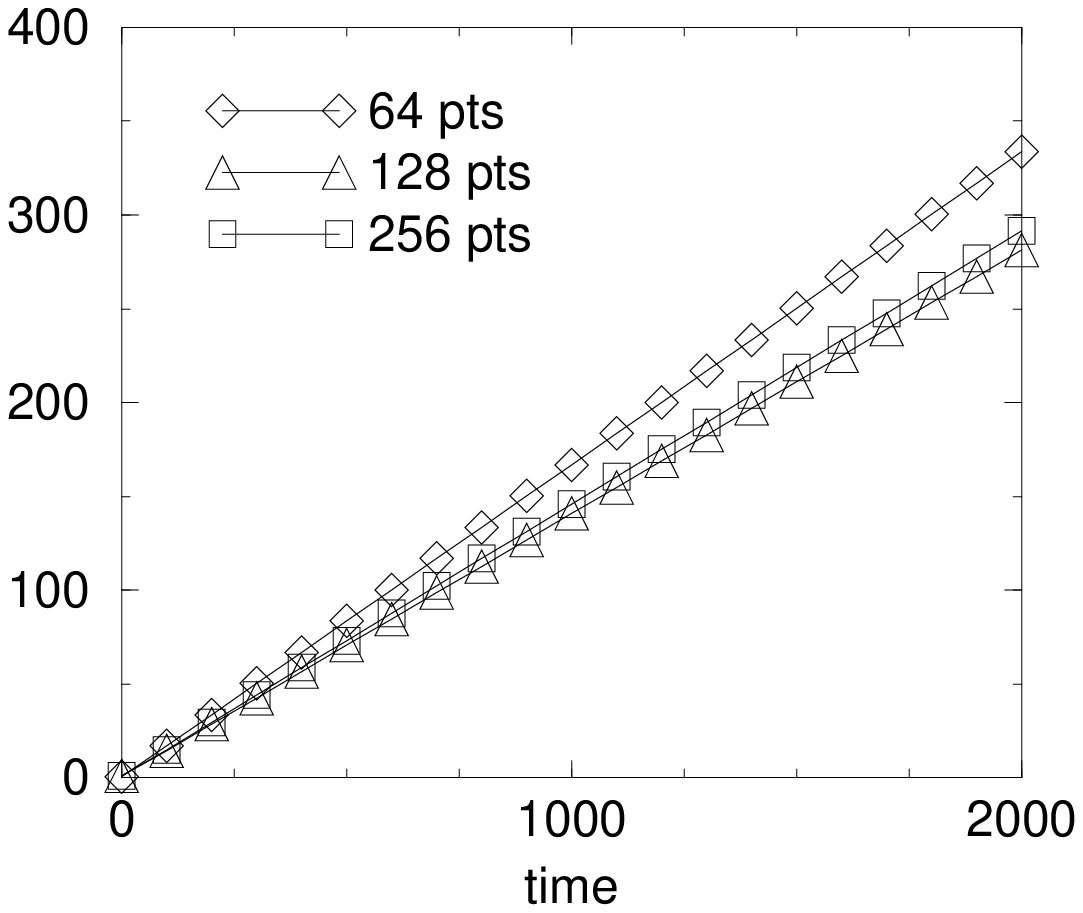,height=1.40in,width=1.40in}
\hspace{0.1in}
\psfig{figure=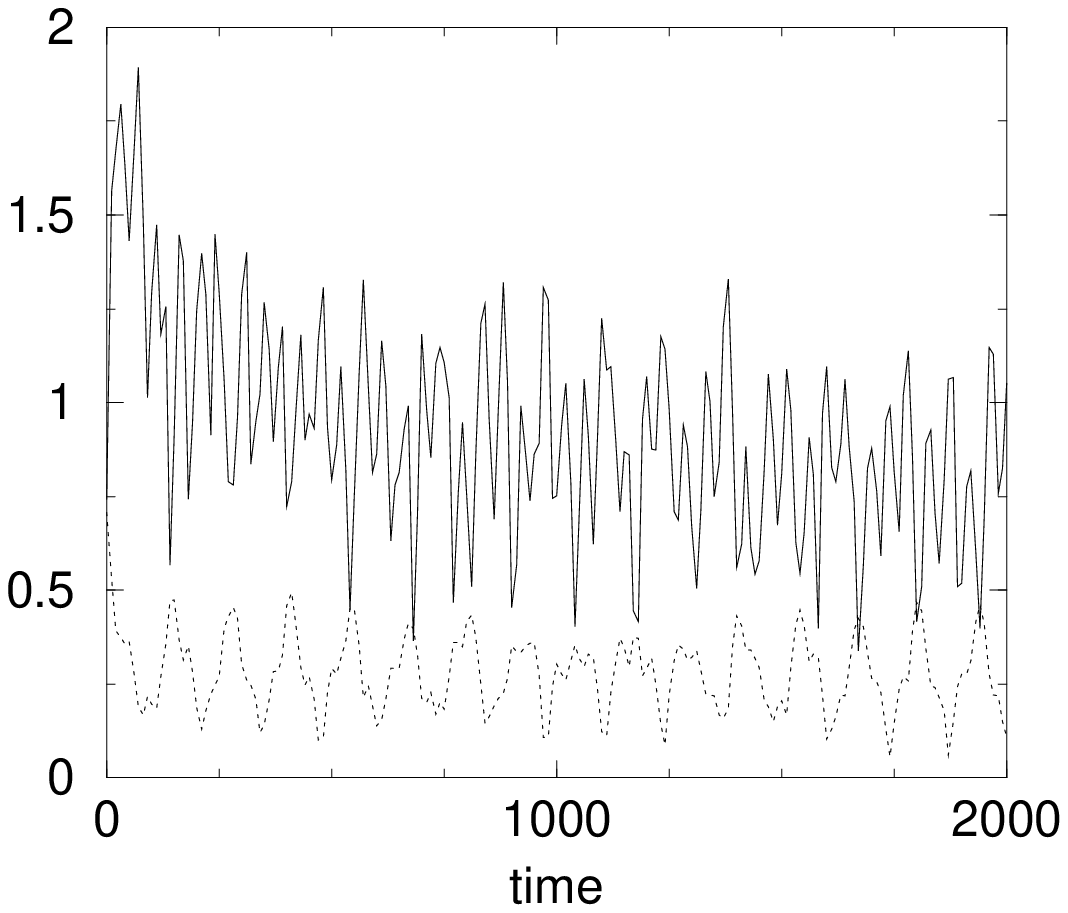,height=1.40in,width=1.40in}
\hspace{0.1in}
\psfig{figure=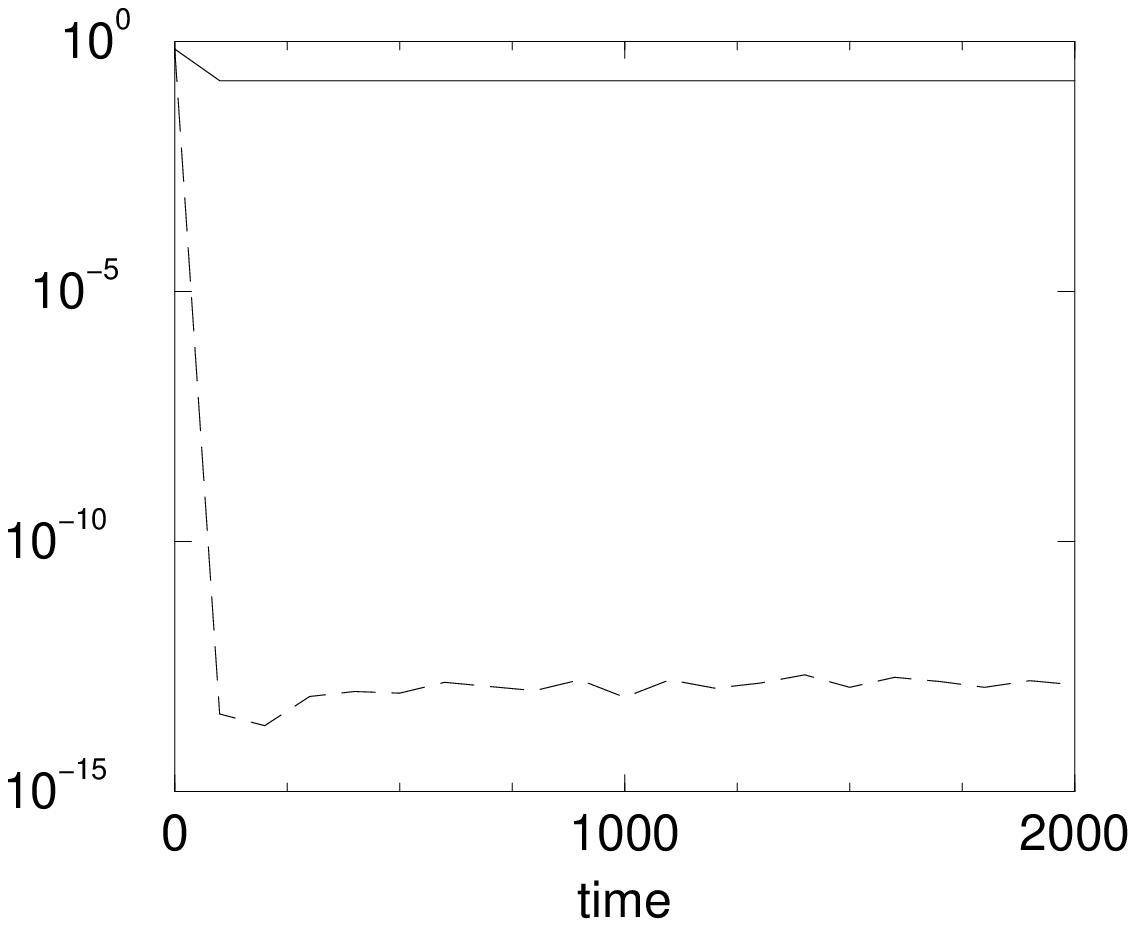,height=1.40in,width=1.40in}
}}
\caption{Stability runs with an ICN evolution scheme.
Left: $||\phi||_{\infty}$ as 
a function of time for different grid sizes, 
 periodic boundary conditions. 
Middle: $||\phi||_{\infty}$ (dotted line) and 
$||\xi||_{\infty}$ (continuous line) 
for a grid size of 64, using reflecting 
boundary conditions applied to $\phi$.
Right: $||\phi||_{\infty}$ (continuous line) and 
$||\xi||_{\infty}$ (dotted line) for a grid size of 64, using 
second-order outgoing radiation 
boundary conditions applied to $\xi$.
\label{fig:ladm-cart.swe.1d.CNruns}}
\end{figure}

The setup of the
 stability runs was similar to those for the 2ND
scheme except for the CFL factor, which in this case was
$\Delta t / \Delta x = 0.25$.
Boundary conditions applied were periodic, reflecting, 
and first- and second-order outgoing radiation condition. The code was
found to be stable for all of these cases, as shown
in Figure~\ref{fig:ladm-cart.swe.1d.CNruns}.
None of the runs produced exponentially growing 
time-dependences.

\subsubsection{The roundoff problem}
\label{seq:ladm-cart.roundoff}

Next the SWE ICN code was run with periodic boundary conditions,
grid sizes of 64, 128, 256, using the following initial data:
\begin{eqnarray}
\phi(x,0) &=& 1 + A \cdot \varrho_c, \\
\xi(x,0) &=& A \cdot  \varrho_c,
\end{eqnarray}
where $\varrho_c$ stands for random data of compact support.

\begin{figure}
\centerline{\hbox{
\psfig{figure=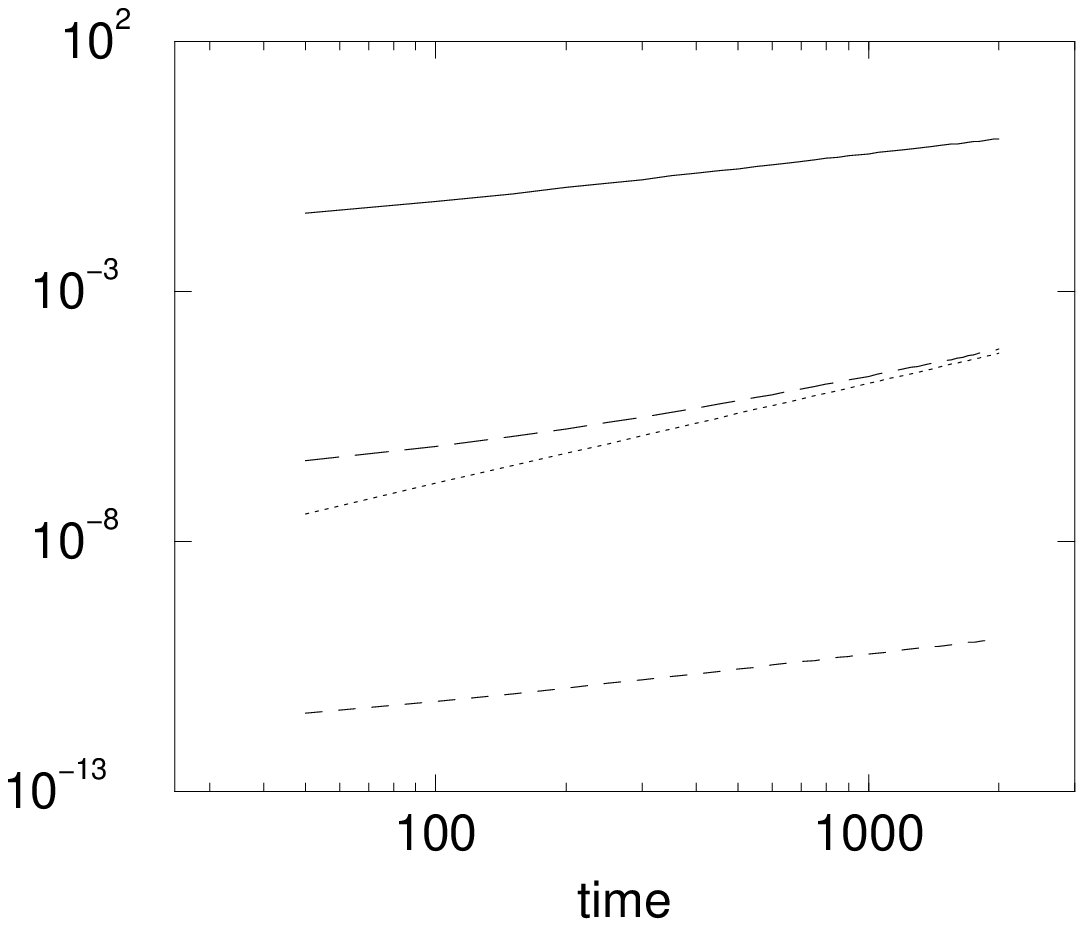,height=1.40in,width=1.40in}
\hspace{0.1in}
\psfig{figure=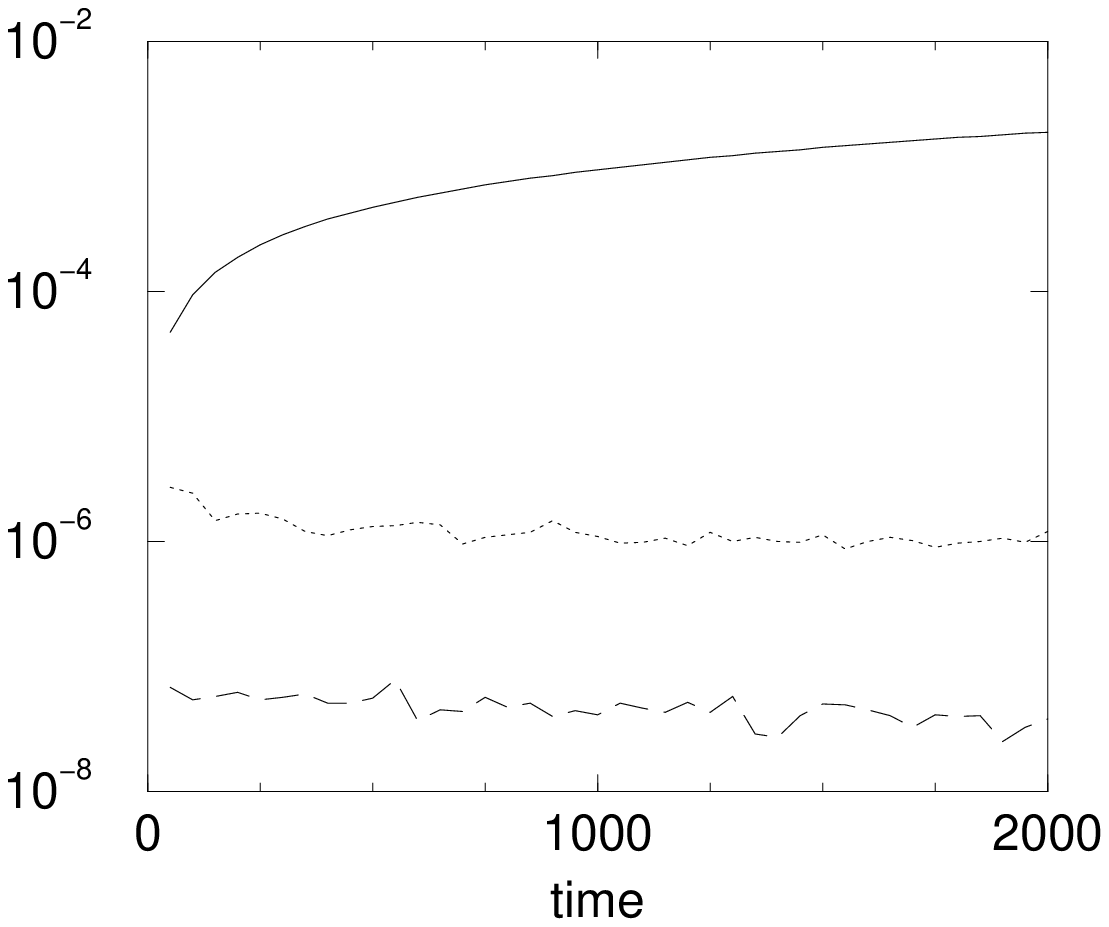,height=1.40in,width=1.40in}
\hspace{0.1in}
\psfig{figure=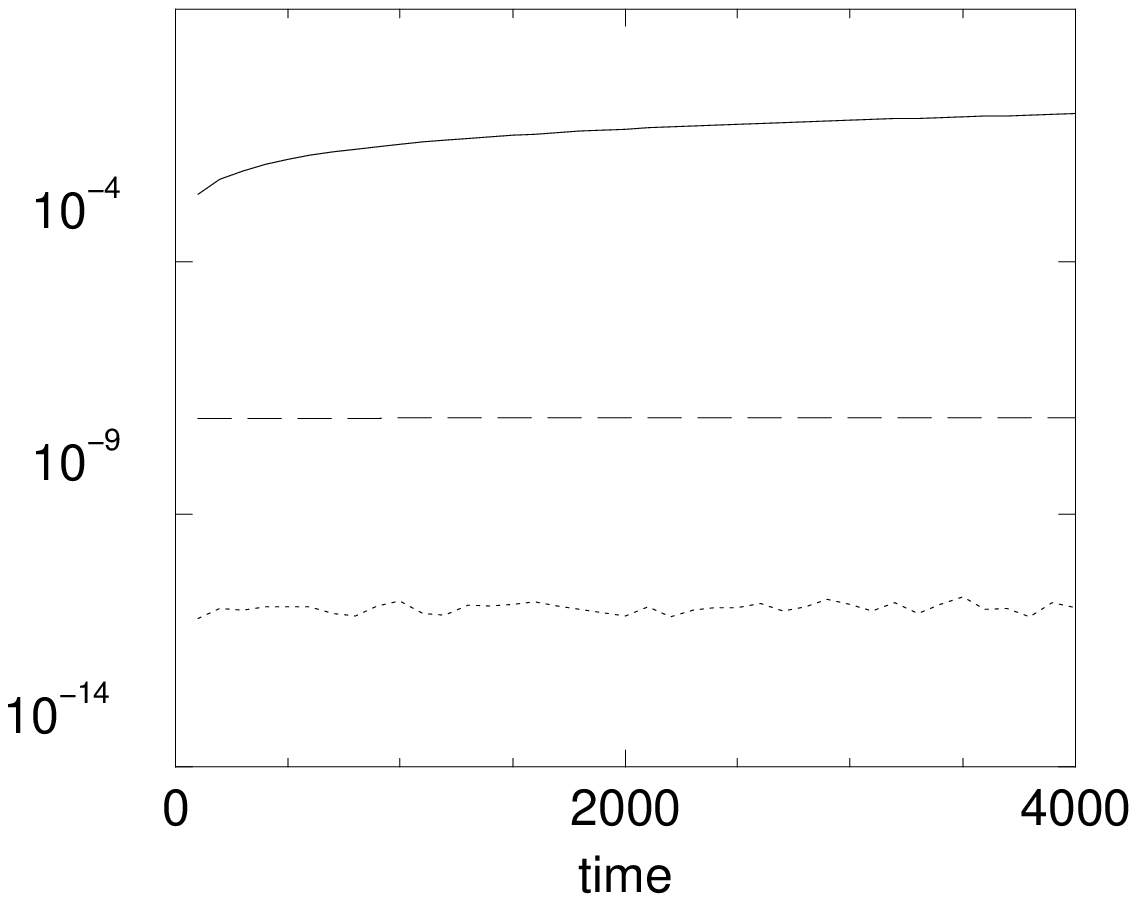,height=1.40in,width=1.40in}
}}
\caption{Stability runs with an ICN evolution scheme.
Left: $||g_{yy}-1||_{\infty}$ using periodic boundary condition, 
256 points. The  top three curves correspond
to solving the equations for $g_{ij}$, using as initial data 
$g_{yy}=1+O(10^{-1})$  (solid line),  
$g_{yy}=1+O(10^{-6})$ (long dashed line),
$g_{yy}=1+O(10^{-11})$ (dotted line);
the bottom curve (dashed line) corresponds to solving the equations for 
$h_{ij}$, using $h_{yy} = O(10^{-11})$ as initial data.
Middle: reflecting boundary conditions applied to $g_{ij}$, showing
$||g_{xy}||_{\infty}$ in the top (solid line), 
$||g_{yz}||_{\infty}$ in the bottom (dashed line),
$||K_{xx}||_{\infty}$ in the middle (dotted line).
Right:  first-order Sommerfeld boundary conditions applied to $K_{ij}$ 
using 128 gridpoints. The top (continuous) curve 
shows $||g_{xx}-1||_{\infty}$, 
the middle (dashed) line corresponds to  $||g_{zz}-1||_{\infty}$ while 
the bottom (dotted) curve represents $||k_{zz}||_{\infty}$.
\label{fig:ladm-cart.lgr.1d.CNruns}}
\end{figure}

When choosing $A = 10^{-1}$ the behavior of $\xi(x,t)$ is linear with respect to time.
For $A = 10^{-11}$ the function $||\xi(x,t)||_{\infty}$ becomes a higher order
 polynomial as a function of time:
\begin{eqnarray}
\mbox{64 pts:  } ||\xi||_{\infty}
&=& 1 + O(10^{-12}) + O(10^{-13})\cdot t 
\nonumber \\ &&
  + O(10^{-13})\cdot t^2 + O(10^{-19})\cdot t^3 \ldots \nonumber \\
\mbox{128 pts:  } ||\xi||_{\infty}
&=& 1 + O(10^{-10}) + O(10^{-12})\cdot t 
\nonumber \\ &&
  + O(10^{-12})\cdot t^2 + O(10^{-19})\cdot t^3 \ldots \nonumber\\
\mbox{256 pts:  } ||\xi||_{\infty}
&=& 1 + O(10^{-10}) + O(10^{-12})\cdot t 
\nonumber \\ &&
  + O(10^{-11})\cdot t^2 + O(10^{-18})\cdot t^3 \ldots \nonumber
\end{eqnarray}
This quadratic time dependence was displayed as expected by 
 the functions $g_{yy}$ and $g_{zz}$
when solving the LG  equations for $g_{ij}$ with 
an ICN evolution scheme. Plots of 
$g_{yy}$ are shown in Figure~\ref{fig:ladm-cart.lgr.1d.CNruns} (left).

This higher order polynomial behavior is not present when using 
the 2ND evolution scheme, nor is it present when evolving 
the SWE for the function $\phi(x,0)$ where the constant 1 does not enter.

All these runs were made on a {\sc Cray C90} architecture
in single precision
which corresponds to a roundoff error of $O(10^{-14})$ when representing
an  $O(1)$ quantity.

\subsection{Code tests for the 1-D LG equations}

The different codes were tested for proper qualitative
behavior 
using a pulse of compact support:

\begin{equation}
g_{ij} = \left( 
\begin{array}{ccc}
1 & 0 & 0 \\
0 & 1+\phi(x,t) & 0 \\
0 & 0 & 1-\phi(x,t)
\end{array}
\right)
\end{equation}
with the function $\phi(x,t)$  defined in  
Eq.~(\ref{eq:ladm-cart.compinidata}).
These runs were done for $\sim 10$ crossing times, 
on a grid that goes from $-1$ to $1$  with the choices
$$
A=10^{-6},\;w=0.75,\;\epsilon=\pm 1.
$$
No significant dispersion was observed.

Stability of the codes was tested
for $\sim 10^3$ crossing times. Initial data for each component of 
$g_{ij}-\eta_{ij}$ was 
generated by multiplying the function $\phi(x,t)$ defined in  
Eq.~(\ref{eq:ladm-cart.compinidata}) with a set of random numbers distributed
between $-0.5$ and  $0.5$. The same procedure was followed for initializing
$K_{ij}$ when using the ICN evolution scheme.

\subsubsection{2ND evolution scheme}

Runs 
were made for 1000 crossing times, with 
grid sizes of 
$64,\;128$, and $256$, and a CFL factor of $\Delta t / \Delta x = 0.50$. 
The results are as follows:
\begin{itemize}
\item Periodic boundaries: $||h_{ij}||_{\infty} = ||g_{ij}-\eta_{ij}||_{\infty}$ 
grew linearly in time. 

\item Reflecting (freezing) boundaries: 
\begin{itemize}
\item $h_{xx}, h_{xy}, h_{xz}$: the $\ell_{\infty}$ norm 
grew like $t \cdot O(10^{-4})$,
\item $h_{yy}, h_{yz}, h_{zz}$: these functions stay 
$O(10^{-6})$. 
\end{itemize}

\item Sommerfeld (first- and second-order): 
\begin{itemize}
\item $h_{xx}, h_{xy}, h_{xz}$: the $\ell_{\infty}$ norm 
grew linearly in time with a slope $O(10^{-4})$,
\item $h_{yy}, h_{yz}, h_{zz}$: these functions slowly 
decreased with time, being $O(10^{-6})$. 
\end{itemize}
There was no qualitative difference between the runs with  first 
and second order Sommerfeld boundary condition.

\end{itemize}

\begin{figure}
\centerline{\hbox{
\psfig{figure=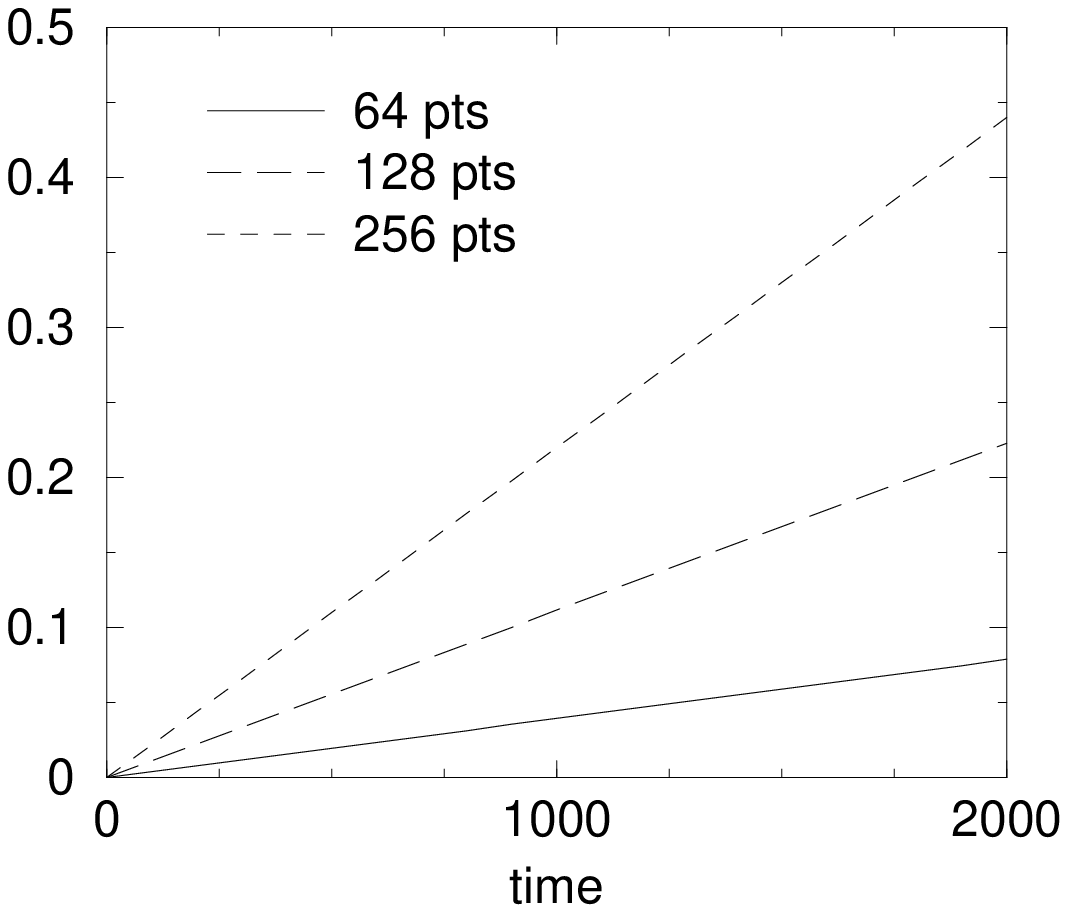,height=1.40in,width=1.40in}
\hspace{0.1in}
\psfig{figure=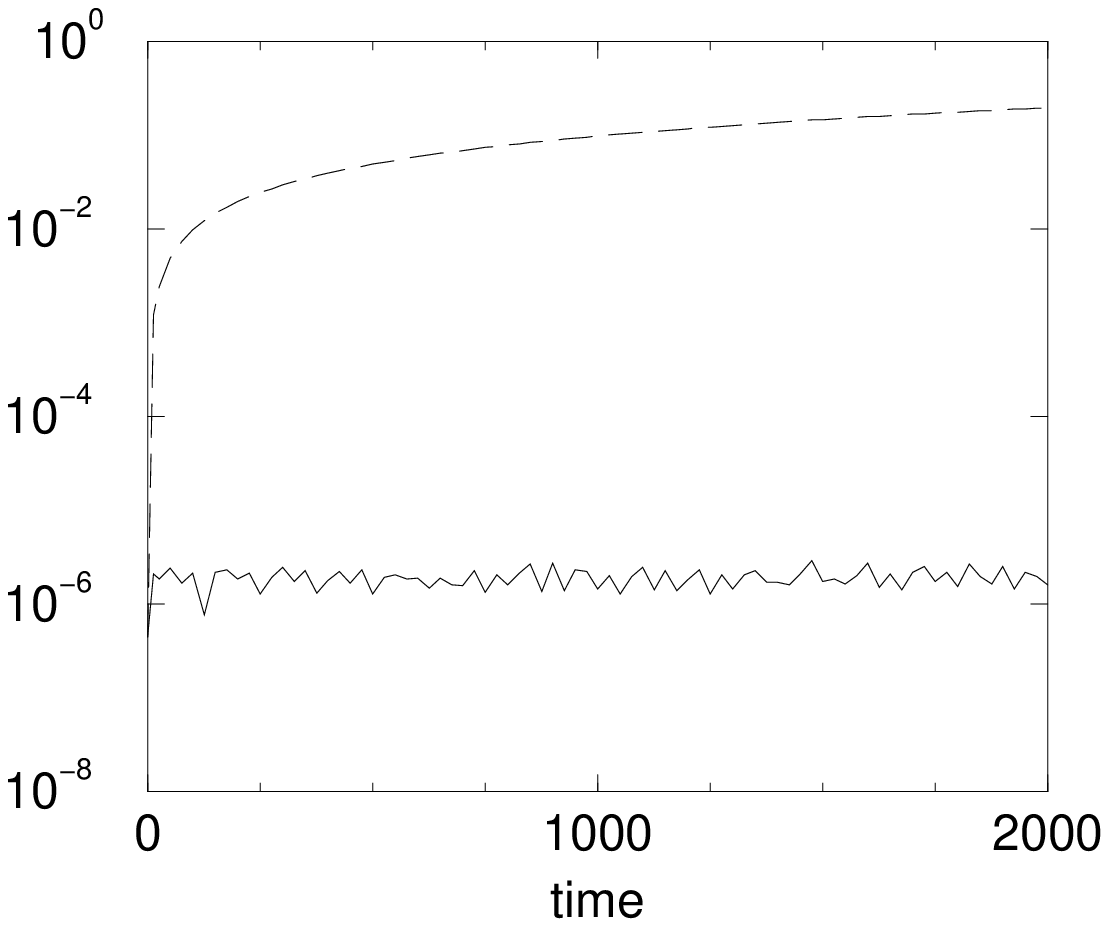,height=1.40in,width=1.40in}
\hspace{0.1in}
\psfig{figure=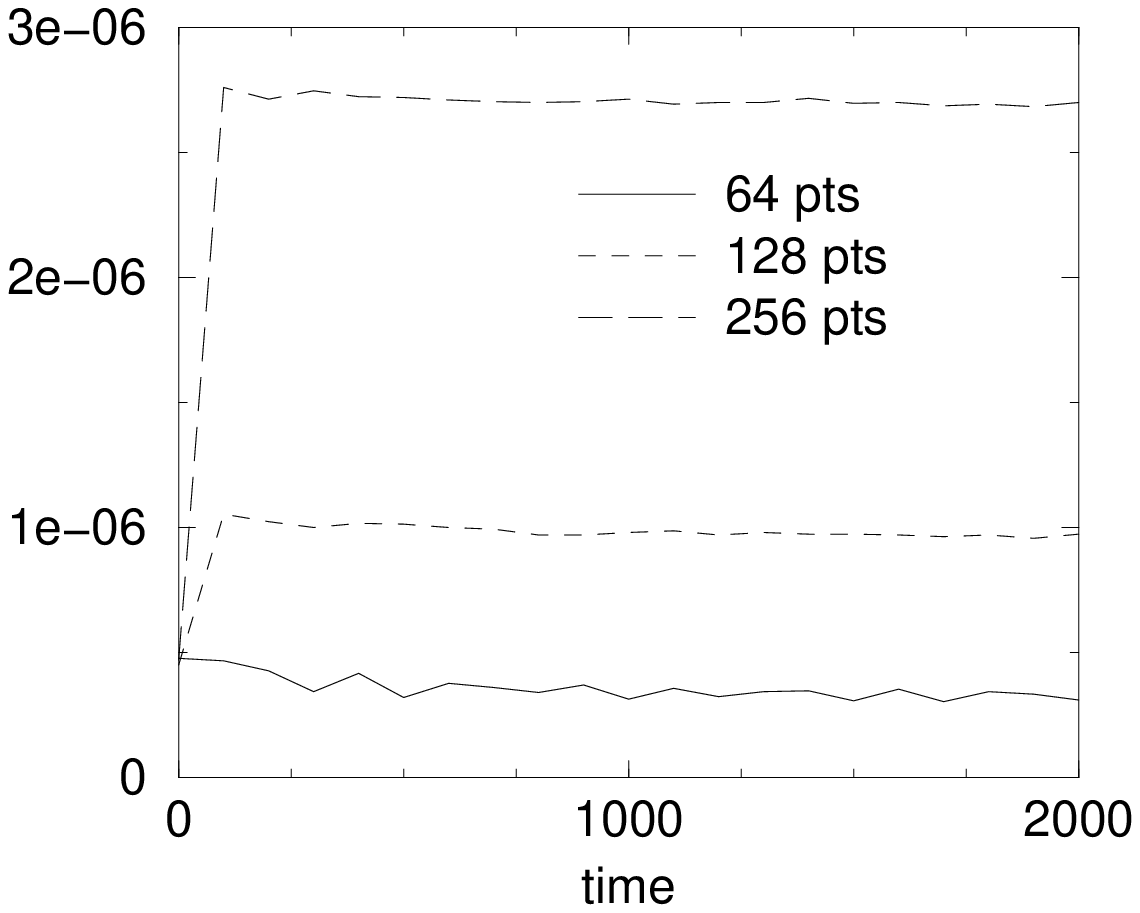,height=1.40in,width=1.40in}
}}
\caption{Stability runs with a second-order evolution scheme.
Left: $||h_{xy}||_{\infty}$ as 
a function of time for different grid sizes, 
using periodic boundary conditions. 
Middle: $||h_{xy}||_{\infty}$ (dashed line) and 
$||h_{yz}||_{\infty}$ (continuous line) 
for a grid size of 128, using reflecting boundary conditions.
Right: $||h_{yz}||_{\infty}$ as 
a function of time for different grid sizes, 
using first-order Sommerfeld boundary conditions.
\label{fig:ladm-cart.lgr.1d.2NDruns}}
\end{figure}

\subsubsection{ICN evolution scheme}

Runs 
were made for 1000 crossing times, with 
grid sizes of 
$64,\;128$, and $256$. The CFL factor was $\Delta t / \Delta x = 0.25$.
At each time-step the code performed two Crank-Nicholson iterations.
The results are as follows:
\begin{itemize}
\item Periodic boundaries:
\begin{itemize}
\item $h_{xx}, h_{xy}, h_{xz}, h_{yz}$: the $\ell_\infty$ norms grew
linearly in time. The slope is   $O(10^{-6})$ 
for $h_{xi}$, and it is  $O(10^{-9})$ for $h_{yz}$.
\item $h_{yy}, h_{zz}$: the $\ell_\infty$ norm of 
these components showed a polynomial behavior as a function of time. 
This is the same roundoff behavior as the one described in Section 
\ref{seq:ladm-cart.roundoff}.
\end{itemize}

\item Reflecting (freezing) boundaries: 
\begin{itemize}

\item the $\ell_\infty$ norms of $h_{xx}, h_{xy}, h_{xz}$ grew linearly 
in time with a slope of $O(10^{-6})$,

\item $h_{yy}, h_{yz}, h_{zz}$ were slowly decreasing in time,

\item $K_{xx}, K_{yy}, K_{yz}, K_{zz}$ were slowly decreasing in time,
\item the $\ell_\infty$ norms of $K_{xy}, K_{xz}$ were time-independent.

\end{itemize}

Applying reflecting boundary conditions 
to $g_{ij}$ instead of $K_{ij}$ did not influence 
the stability properties of the code.

\item Sommerfeld boundaries (first- and second-order): 

Runs were made up to
$t=4000$.
Applying first or second order Sommerfeld 
boundary condition to the components of 
either  $g_{ij}$ or $K_{ij}$ gave stable runs. 
As expected from the form of the evolution equations (\ref{eq:ladm-cart.eveq-1d}),
the $\ell_{\infty}$ norms of $h_{xx}, h_{xy}$, and $h_{xz}$ 
showed a linear growth with time.

\end{itemize}

These results indicate that there
are several choices of evolution algorithms 
that seem to perform equally well with periodic and
 Dirichlet boundary conditions.  The non-staggered leap-frog algorithm
failed for the outgoing radiation boundaries (Sommerfeld conditions).

\section{SWE and LG equations in 3-D}
\label{sec:ladm-cart.3d}

This section studies the numerical stability of 3-D evolution codes
for linearized gravity
with various boundary conditions.
Evolution algorithms
are  described (and tested) first for the scalar wave equation.
Next the LG equations are studied for periodic boundaries.
Furthermore runs are described with a variety  of
boundary conditions on the $z=$ constant faces of the numerical grid,
keeping the $x$- and $y$-directions periodic.  These tests are
 first performed with the second-order evolution scheme.
  Then the best boundary algorithms are tested for
the ICN and for the leap-frog algorithms.
At the end a robustly stable algorithm 
is presented for evolving a region of space-time
contained within a cube, using random initial and random boundary data.

\subsection{Evolution schemes}

Four different evolution schemes were used for the scalar
wave equation
\begin{equation}
\partial^\mu \partial_\mu \phi = 0.
\label{eq:ladm-cart.3dswe-2nd}
\end{equation}
With the exception of the algorithm 2ND,
the numerical algorithms  evolve the SWE 
in first-order in time form 
\begin{eqnarray}
\partial_t \phi &=& \xi \label{eq:ladm-cart.3dswe1st1} \\
\partial_t \xi &=& \nabla^2 \phi . \label{eq:ladm-cart.3dswe1st2}
\end{eqnarray}

The generalization of the finite-difference algorithms to the 
LG equations is straightforward.

\subsubsection{Non-staggered leap-frog (LF1)}
The first algorithm, LF1, is
a standard leap-frog implementation of the 
Eqs.~(\ref{eq:ladm-cart.3dswe1st1}) - (\ref{eq:ladm-cart.3dswe1st2}):
\begin{eqnarray}
      \phi^{[N+1]}_{[I,J,K]} &=&\phi^{[N-1]}_{[I,J,K]} -4 \xi^n_{[I,J,K]} \Delta t
                   \nonumber \\
      \xi^{[N+1]}_{[I,J,K]}&=& \xi^{[N-1]}_{[I,J,K]} -\nabla^2 \phi^n_{[I,J,K]} \Delta t,
\end{eqnarray}
where $\nabla^2$ is the second-order accurate centered difference
approximation to the Laplacian. It is known that this algorithm has a
time-splitting instability in the presence of dissipative and non-linear
effects~\cite{New98}.

\subsubsection{Staggered  leap-frog (LF2)}

The second algorithm, LF2, is a staggered in time
leap-frog scheme which is not subject to the time-splitting instability:
\begin{eqnarray}
   \phi^{[N+1]}_{[I,J,K]} &=&\phi^{[N]}_{[I,J,K]} -2 \xi^{[N+1/2]}_{[I,J,K]}\Delta t
\label{eq:ladm-cart.3d.lf21}\\
   \xi^{[N+1/2]}_{[I,J,K]}&=& \xi^{[N-1/2]}_{[I,J,K]} - \frac{1}{2}
         \nabla^2 \phi^n_{[I,J,K]} \Delta t.
\label{eq:ladm-cart.3d.lf22}
\end{eqnarray}
Here $\xi$ is evaluated on the half grid. By subtracting the equation
\begin{eqnarray}
      \phi^{[N]}_{[I,J,K]} =\phi^{[N-1]}_{[I,J,K]} -2 \xi^{[N-1/2]}_{[I,J,K]}\Delta t
\end{eqnarray}
from Eq.~(\ref{eq:ladm-cart.3d.lf21}) and using Eq.~(\ref{eq:ladm-cart.3d.lf22}) to eliminate $\xi$, 
it can be seen that LF2 is equivalent to the standard leap-frog scheme for the 
second-differential-order in time form of the  
wave equation (\ref{eq:ladm-cart.3dswe-2nd}),
in which $\phi$ lies on integral time levels and $\xi$ is not introduced.

\subsubsection{Second-order scheme (2ND)}

The scalar wave equation~(\ref{eq:ladm-cart.3dswe-2nd})
is approximated by a
the three-level, second-differential-order  in time algorithm:
\begin{eqnarray}
\phi^{[N+1]}_{[I,J,K]} = 2 \phi^n_{[I,J,K]} - \phi^{[N-1]}_{[I,J,K]}
+ \Delta t^2 \nabla^2 \phi^{[N]}_{[I,J,K]}.
\end{eqnarray}

\subsubsection{Iterative Crank-Nicholson (ICN)}

The fourth algorithm is an iterative
Crank-Nicholson algorithm with two iterations. 
The following sequence of operations is executed
for each  time-step:
\begin{enumerate}
\item Compute the first-order accurate quantities
\begin{eqnarray}
     \stackrel{(0)}{\phi}\!{}^{[N+1]}_{[I,J,K]}
         &=& \phi^{[N]}_{[I,J,K]}-2 \xi^{[N]}_{[I,J,K]}\Delta t, \nonumber \\
     \stackrel{(0)}{\xi}\!{}^{[N+1]}_{[I,J,K]}
         &=& \xi^{[N]}_{[I,J,K]}-\frac{1}{2}\nabla^2\phi^{[N]}_{[I,J,K]}\Delta t.
\end{eqnarray}

\item \label{item:ladm-cart.sweCNloopstart} Compute the mid-level values
\begin{eqnarray}
    \stackrel{(i)}{\phi}\!{}^{[N+1/2]}_{[I,J,K]}
        &=& \frac{1}{2} \left(\phi^{[N]}_{[I,J,K]}+
                 \stackrel{(i)}{\phi}\!{}^{[N+1]}_{[I,J,K]} \right),
                   \nonumber \\
    \stackrel{(i)}{\xi}\!{}^{[N+1/2]}_{[I,J,K]}
        &=& \frac{1}{2} \left( \xi^{[N]}_{[I,J,K]} +
              \stackrel{(i)}{\xi}\!{}^{[N+1]}_{[I,J,K]} \right).
\end{eqnarray}

\item Update  using levels $n$ and $n+1/2$:
\begin{eqnarray}
   \stackrel{(i+1)}{\phi}\!{}^{[N+1]}_{[I,J,K]}
       &=& \phi^{[N]}_{[I,J,K]}
         -2\stackrel{(i)}{\xi}\!{}^{[N+1/2]}_{[I,J,K]} \Delta t,
                     \nonumber \\
   \stackrel{(i+1)}{\xi}\!{}^{[N+1]}_{[I,J,K]}
       &=& \xi^{[N]}_{[I,J,K]}
        -\frac{1}{2}\nabla^2 \stackrel{(i)}{\phi}\!{}^{[N+1/2]}_{[I,J,K]} \Delta t.
\end{eqnarray}
\item Increment $i$ by one and return to step \ref{item:ladm-cart.sweCNloopstart}
until $i=2$ is reached.
\end{enumerate}

\subsection{Periodic boundary conditions}

The LG equations (\ref{eq:ladm-cart.eveq-Eij}) were studied with
respect to the parameter $\lambda$, using
periodic boundary conditions and the 2ND evolution algorithm with a 
CFL factor of $0.25$.
Initial data for these runs was unconstrained (i.e. a set of random numbers, 
scattered between $-0.5 \cdot 10^{-6} \ldots 0.5 \cdot 10^{-6}$).
The grid size was $16^3$.
Non-exponentially growing runs were made up to 1000 crossing times (i.e.
up to $t=2000$, with a grid going from $-1$ to $1$.), unless stated otherwise.
Numerical experiments showed the following:

\begin{itemize}

\item Runs for $\lambda 
\in \left\{ -4.0, \, -2.0, \, -1.2, \, -1.0, -0.8, \, 
\, -0.4 \right\}$  were unstable. 
The choice $\lambda=-1$
gives a growth rate that is significantly smaller than the other values,
 yet it gives an unstable run. Runs 
for $\lambda \in \left\{ 4.4,  \,6.0,  \,8.0 \right\}$ were 
unstable as well. 

\item Runs for 
$\lambda   \in \left\{0.0, \, 0.8, \, 1.6, \, 2.4, \, 3.2, \, 4.0 \right\}$
were stable, as follows: 

\begin{itemize}
\item The case $\lambda=0$ gave a linear growth of the Hamiltonian constraint, 
with a slope of $O(10^{-3})$.
\item The case $\lambda=4$ gave a linear growth of the Hamiltonian constraint, 
with a slope of $O(10^{-5})$.
\item The cases $0<\lambda<4$ gave runs where the Hamiltonian constraint
remained bounded, with a magnitude of $O(10^{-4})$.

\end{itemize}
\end{itemize}

\begin{figure}
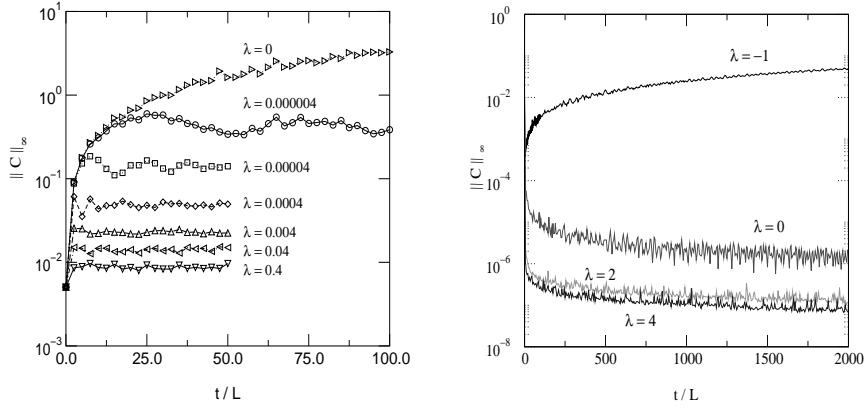

\centerline{\hbox{
\psfig{figure=lgr.3d.LF2.stage1.eps,height=2.1in,width=2.1in}
\hspace{0.2in}
\psfig{figure=lgr.3d.ICN.stage1.eps,height=2.1in,width=2.1in}
}}
\caption{
The $\ell_\infty$ norm of the Hamiltonian constraint
versus time (in crossing times)
for the algorithms 2ND (left) and 
ICN (right), periodic boundary conditions. 
Runs were made with $48$ gridpoints, random initial
data of $O(10^{-6})$, various choices of $\lambda$. 
The ratio $\Delta t / \Delta x$ was chosen
to be $0.25$.}
\label{fig:adm-cart.LF2.stage1}
\label{fig:adm-cart.ICN.stage1}
\end{figure}

Similar results were reproduced when using the schemes ICN, LF1, LF2.
In the case of ICN there was an additional
(isolated) value $\lambda = -1$ that resulted in a stable evolution.
Plots of the stable runs performed with ICN can be found in 
Figure~\ref{fig:adm-cart.ICN.stage1} (right).

To check that $\lambda=0$ is indeed the edge of the 
stability domain, runs were made
with values of $\lambda/4 = 10^{-1}, 10^{-2}, 
10^{-3}, 10^{-4}, 10^{-5}, 10^{-6}, -10^{-6}$ 
for a grid size of $48^3$, up to 100 crossing times. 
The run with negative $\lambda$ resulted
in a slow exponential growth of the Hamiltonian constraint. 
Runs with positive values of $\lambda$ are shown in 
 Figure~\ref{fig:adm-cart.LF2.stage1} (left), each of them being stable.

For the values $\lambda=0.0, 2.0,$ and  $4.0$, test runs were made with 
a grid size of $48^3$ and 2000 crossing times. The code revealed no unstable modes.

As suggested by Eq.~(\ref{eq:ladm-cart.HC-SWE}), 
the larger the value of $\lambda$ the faster 
certain quantities  propagate across the numerical grid. For any given
value of the $\Delta t / \Delta x$ ratio, there are values of $\lambda$ 
large enough such that the CFL condition is violated, e.g. the velocity
$v_{\cal C}$ of the Hamiltonian constraint  gets larger than the numerical
velocity intrinsic to the evolution scheme.
To support this argument a run was made with $\Delta t= \Delta
x/8$ (half the time-step of the standard runs) and $\lambda =20$, 
2000 crossing times.
 It showed no exponential growth.

\subsection{Dirichlet boundary conditions}

Next various boundary
conditions were tested for the SWE and LG equations.  
First a test bed was defined so as to 
efficiently reveal  unstable (boundary) algorithms.
Then this test bed was applied to the SWE and 
a number of boundary routines for the LG equations, using the 2ND 
evolution scheme.  
The boundary algorithms that performed best
were tested for the other three evolution schemes LF1, LF2, and ICN.
A stable algorithm is finally formulated for the LG equations
with a cubic boundary.

\subsubsection{Definition of test bed for Dirichlet boundary condition}

The stability test bed is defined as follows:
\begin{itemize}
\item random initial data for all dynamic quantities
\item random data for all free functions on the boundary
\item grid size of at least $48^3$, evolving the 
code for at least 2000 crossing times.
\end{itemize}
In this context random data means a set of numbers equally 
distributed in the interval $[-a, +a]$.
The value of the parameter $a$ was chosen to be $O(10^{-6})$.

If an unstable mode reveals itself for grid sizes smaller than $48^3$,
the larger grid size test becomes irrelevant.

\subsubsection{Scalar wave code with Dirichlet boundary condition}

The test bed was tried with a scalar wave code, 
2ND evolution algorithm, 
using periodic boundary conditions in the $x$- and 
$y$-directions, and 
random Dirichlet boundary condition on the $z=$ constant faces. 
For a grid size of $48^3$ and a CFL factor of $0.25$, the 
code ran stably for 2000 crossing times, showing a growth that is not
stronger than linear for the quantity 
$||\phi||_{\infty}$ as a function of time.

\subsubsection{The ADM equations with ``plain'' Dirichlet boundary condition}

The term ``plain'' Dirichlet boundary condition  denotes
setting all six metric functions to 
random numbers on the $z=$ constant faces of the 
evolution domain with the $x$- and $y$-directions kept periodic.
This boundary condition was unstable for the choices $\lambda = 0.0, 
1.0, 2.0$ or $3.0$. The grid size was
$16^3$ and the runs lasted less than 1000 crossing times.

The choice of $\lambda=4.0$ was more robust, but for a grid size
of $48^3$ the instability emerged after only 150 crossing times.

The following analytic argument indicates what 
goes wrong  \cite{WinicourPersonal}.  Let us define
\begin{equation}
\Psi := \partial_A \partial^A h^B_B - \partial^A \partial^B h_{AB}.
\end{equation}
The evolution equations $^{(4)}R_{ij} = 0$  imply
that the quantity $\Psi$  propagates as a scalar wave, i.e.
$\Box \Psi = 0$. The question arises: what boundary condition is applied
to $\Psi$ when  $h_{ij}$ is set to zero at the boundary? Clearly
$\left. h_{ij} \right|_{bdry} = 0$ implies $\left.\Psi\right|_{bdry}  = 0$.
On the other hand
\begin{equation}
   2 \partial_A {}^{(4)}G_z^A \;\;=\;\;\partial_A\ddot  h_z^A -
\partial_A \partial_B(\partial^B h_z^A-\partial^A h_z^B)
         -\partial_z \Psi  \;\;=\;\;0.
\end{equation}
One can see that setting $h_{ij}$ to zero at the boundary implies
an {\em additional} condition on the boundary value of $\Psi$, namely
$\left.\partial_z\Psi\right|_{bdry}  = 0$. Using Dirichlet and Neumann
conditions as simultaneous boundary conditions on a scalar wave is
over-determined. {\em Therefore setting
$\left.h_{ij} \right|_{bdry} = 0$
gives rise to an inconsistent boundary condition.}

\subsubsection{The characteristic code with ``plain'' Dirichlet boundaries}

For comparison a similar test was 
performed with the characteristic code, 
as described in Chapter~\ref{chap:nullcode}.
 Since non-linear 
terms are included in the
code, the amplitude of the initial and boundary data was set to  
$a = 10^{-6} \cdot (\Delta t)^2$, and the non-linear term $\beta_{,r}$ 
was set to zero on the
initial slice as well as on the inner boundary. The code revealed no unstable modes. 
Not even a linearly growing mode was present. The radial grid size
was set to 49, the 2-D angular grid using $55^2$ points 
per stereographic patch. The  
inner boundary was placed at $r = 1$ at which $\Delta t / \Delta r \simeq 0.5$.
The run was made up to $t=4000$.

\subsubsection{A systematic search for a stable, Dirichlet boundary condition}

Although the question of initial boundary value problem
for  symmetric hyperbolic systems  is discussed
in  \cite{Friedrich98,Stewart98}, the ADM evolution equations
are neither first-order nor hyperbolic. 
Thus we are left with the option of studying the 
boundary problem via systematic experiments.

\pagebreak[3]
\vspace{1em}
\noindent
{\em Analytic discussion}
\vspace{0.5em}

\noindent
We first studied reflecting boundary conditions on 
the $z=$ constant faces of a cube that has
periodic boundary conditions in the $x$- and $y$-directions.

The strategy was to apply Dirichlet boundary conditions to only
some of the components of the metric tensor $h_{ij}$, 
 the others were constrained by
demanding that  various combinations of the following quantities vanish 
on the boundary:
\begin{eqnarray}
\label{eq:ladm-cart.bdry-eqs.FIRST}
-2 \, {\cal C}^z &\equiv& \partial_j \dot h^{zj} - \partial^z \dot h ; \\
\label{eq:ladm-cart.bdry-eqs.SECOND}
-2 \, {\cal C}^A &\equiv& \partial_j \dot h^{Aj} -  \partial^A \dot h ; \\
2 \, {}^{(4)}R^t_t &\equiv& \ddot h; 
\label{eq:ladm-cart.bdry-eqs.Rtt} \\
  2\, {}^{(4)}G_{zz} &\equiv& -\ddot h^A_A +\partial^B\partial_B h^A_A
             - \partial_A\partial_Bh^{AB}; \label{eq:ladm-cart.bdry-eqs.Gzz}\\
   2\, {}^{(4)}G_z^A &\equiv& \ddot h_z^A 
-\partial_B(\partial^B h_z^A-\partial^A h_z^B)
\nonumber \\ &&
         -\partial_z\partial^A h^B_B +\partial_z\partial^B h^A_B; \;\;
 \label{eq:ladm-cart.bdry-eqs.GzA} \\
2 \left( - \dot {\cal C}^A + \partial^A \,  {}^{(4)}R^t_t\right) 
&:=&  \partial_j  \ddot h^{Aj} ; 
\label{eq:ladm-cart.bdry-eqs.mmconst.A}\\
2 \left( - \dot {\cal C}^z + \partial^z \,  {}^{(4)}R^t_t\right) 
&:=&  \partial_j  \ddot h^{zj} .
\label{eq:ladm-cart.bdry-eqs.mmconst.z}
\label{eq:ladm-cart.bdry-eqs.LAST}
\end{eqnarray}
Here uppercase Latin letters  indicate the directions in the plane
of the boundary (i.e. the $x$-  and $y$-directions).
In formulating sets of boundary constraints one should recall that
use of the Hamiltonian constraint and of the three momentum constraints
does not give four independent conditions \cite{Alcubierre99c}, these
quantities being related via the Bianchi identities.

The following  boundary conditions were tested:

\begin{enumerate}
\item \label{eq:ladm-cart.constob1}
\begin{eqnarray}
h^{xx} - h^{yy} = h^{xy} &=& 0, \\
 \dot h^{ij}_{,j} - \eta^{ij} \dot h_{,j} &=& 0, 
\label{eq:ladm-cart.cond22}\\
 \ddot h &=& 0.
\end{eqnarray}
\item\label{eq:ladm-cart.constob2}
\begin{eqnarray}
h^{xx} = h^{yy} = h^{xy} &=& 0,  \label{eq:ladm-cart.const21}\\
 \ddot h^{ij}_{,j}  &=& 0. \label{eq:ladm-cart.const22}
\end{eqnarray}
\item\label{eq:ladm-cart.constob3}
\begin{eqnarray}
h^{xx} - h^{yy} = h^{xy} &=& 0,  \label{eq:ladm-cart.const31} \\
\ddot h^{ij}_{,j}  &=& 0,        \label{eq:ladm-cart.const32} \\
\ddot h &=& 0.                  \label{eq:ladm-cart.const33}
\end{eqnarray}
\item \label{eq:ladm-cart.constob4}
\begin{eqnarray}
h^{xx} = h^{yy} = h^{xy} &=& 0, \\
\dot h^{Ai}_{,i} - \eta^{Ai} \dot h_{,i} &=& 0,\\
\ddot h^{zi}_{,i} &=& 0.\\
\end{eqnarray}
\item \label{eq:ladm-cart.constob5}
\begin{eqnarray}
h^{xx} - h^{yy} = h^{xy} &=& 0, \\
\dot h^{Ai}_{,i} - \eta^{Ai} \dot h_{,i} &=& 0,\\
\ddot h^{zi}_{,i} &=& 0,\\
\ddot h &=& 0.
\end{eqnarray}
\item\label{eq:ladm-cart.constob6}
\begin{eqnarray}
h^{xx} = h^{yy} = h^{xy} &=& 0, \\
\ddot h^{ij}_{,j}  - \frac{1}{2} \eta^{ij} \, \ddot h_{,j} &=& 0. \label{eq:const52}
\end{eqnarray}
\end{enumerate}

\pagebreak[3]
\vspace{1em}
\noindent
{\em Finite differencing}
\vspace{0.5em}

\noindent
Applying the various  proposed algorithms requires
solving finite-difference equations that involve 
 time  and  space derivatives.
A function $F$ is represented 
by a discrete set of values $F^{[N]}_{[I,J,K]}$ where $N$ labels the time level,
and $I,J,K$ correspond to grid indices along the coordinates $x,y,z$. 

The $x$- and $y$-derivatives are  either centered at the
gridpoints 
\begin{eqnarray}
\partial_x F_{[I,J,K]} = \frac{1}{2 h} \left[ F_{[I+1,J,K]} - F_{[I-1,J,K]} \right] 
+ O(\Delta^2),\\
\partial_y F_{[I,J,K]} = \frac{1}{2 h} \left[ F_{[I,J+1,K]} - F_{[I,J-1,K]} \right] 
+ O(\Delta^2),
\end{eqnarray}
or in the middle of the 2-D cells:
\begin{eqnarray}
\partial_x F_{[I+1/2,J,K]} = \frac{1}{h} \left[ F_{[I+1,J,K]} - F_{[I,J,K]} \right] 
+ O(\Delta^2), \\
\partial_y F_{[I,J+1/2,K]} = \frac{1}{h} \left[ F_{[I,J+1,K]} - F_{[I,J,K]} \right] 
+ O(\Delta^2).
\end{eqnarray}
The $z$-derivatives are computed either at the boundary:
\begin{eqnarray}
\label{eq:ladm-cart.side-z-deriv.1}
\partial_z F_{[I,J,1]} &=& \frac{1}{2 h} 
\left[ -3 F_{[I,J,1]} + 4  F_{[I,J,2]} - F_{[I,J,3]} \right] 
+ O(\Delta^2), \\
\label{eq:ladm-cart.side-z-deriv.2}
\partial_z F_{[I,J,K_{max}]} &=& 
\frac{1}{2 h} \left[  F_{[I,J,K_{max}-2]} 
-4 F_{[I,J,K_{max}-1]} +3 F_{[I,J,K_{max}]} \right] 
 \nonumber \\ &&
+ O(\Delta^2),
\end{eqnarray}
or at the point next to the boundary:
\begin{eqnarray}
\partial_z F_{[I,J,2]} &=& \frac{1}{2 h} \left[ F_{[I,J,3]} - F_{[I,J,1]} \right] 
+ O(\Delta^2), \\
\partial_z F_{[I,J,K_{max}-1]} &=& 
\frac{1}{2 h} \left[ F_{[I,J,K_{max}]} - F_{[I,J,K_{max}-2]} \right]  
+ O(\Delta^2),
\nonumber \\
\end{eqnarray}
or  
between the boundary point and its' nearest neighbor:
\begin{eqnarray}
\partial_z F_{[I,J,1+1/2]} &=& \frac{1}{h} \left[ F_{[I,J,2]} - F_{[I,J,1]} \right] 
+ O(\Delta^2), \\
\partial_z F_{[I,J,K_{max}-1/2]} &=& 
\frac{1}{h} \left[ F_{[I,J,K_{max}]} - F_{[I,J,K_{max}-1]} \right]  
+ O(\Delta^2).
\nonumber \\
\end{eqnarray}
Runs were made with various choices of $\lambda$, using the 2ND evolution
algorithm. Initial data was a set of random
numbers of $O(10^{-6})$ in a domain of compact support, unless 
stated otherwise.

All of the boundary constraints listed above have the form
$\dot {\cal F}(h^{ij}, h^{ij}_{,k})  = 0$ or
$\ddot {\cal F}(h^{ij}, h^{ij}_{,k})  = 0$. 
Since initial data in the neighborhood of the boundary is zero,
the functions ${\cal F}$ vanish for the first
few time levels. Analytically the function ${\cal F}$
and its first time derivative are initially zero (in the
neighborhood of the boundary). Thus imposing $\dot {\cal F} = 0$
or $\ddot {\cal F} = 0$ amounts to asking that ${\cal F} = 0$ at all times.
In other words, the use of initial data of compact support
implies that the time derivatives
in the various boundary conditions play no effective role. 
For this reason all runs were made with
the same stencils for first-order  time derivatives, 
centered between the last two time levels.

\pagebreak[3]
\vspace{1em}
\noindent
{\em Boundary algorithm}
\vspace{0.5em}

\noindent
In order to understand the implementation of the above boundary
constraints consider the case defined by 
Eqs.~(\ref{eq:ladm-cart.const21}) - (\ref{eq:ladm-cart.const22}). Written out in
an explicit form, the boundary constraints (\ref{eq:ladm-cart.const22}) become
\begin{eqnarray}
\label{eq:ladm-cart.hij_j-exp.FIRST}
\ddot h^{xx}_{,x} + \ddot h^{xy}_{,y}  + \ddot h^{xz}_{,z} &=& 0, \\
\label{eq:ladm-cart.hij_j-exp.SECOND}
\ddot h^{yx}_{,x} + \ddot h^{yy}_{,y} +  \ddot h^{yz}_{,z} &=& 0, \\
\ddot h^{zx}_{,x} + \ddot h^{zy}_{,y}  + \ddot h^{zz}_{,z} &=& 0.
\label{eq:ladm-cart.hij_j-exp.LAST}
\end{eqnarray}

With the boundary values of $h^{xx}, h^{xy}$ and $h^{yy}$ set to zero,
the boundary constraints~(\ref{eq:ladm-cart.hij_j-exp.FIRST}) - 
(\ref{eq:ladm-cart.hij_j-exp.SECOND})
serve as Neumann-type conditions on $h^{xz}$ and $h^{yz}$.
Having determined these two functions at the boundary point, 
one can use Eq.~(\ref{eq:ladm-cart.hij_j-exp.LAST}) to 
compute the boundary value for $h^{zz}$.

The other boundary systems do not provide such a clean method of updating
boundary points: the boundary equations are a set of coupled PDEs, 
involve quantities whose boundary
values are not yet known. For example, the 
system~(\ref{eq:ladm-cart.const31}) - (\ref{eq:ladm-cart.const33})
 provides only two functions at the boundary, $h^{xx}-h^{yy}$ and 
$h^{xy}$ with the additional constraint 
$\ddot h^{xx} + \ddot h^{yy} + \ddot h^{zz} = 0$.
This means that the boundary values of $h^{xx}$ and $h^{yy}$ 
are functions of $h^{zz}$, which
is determined by  Eq.~(\ref{eq:ladm-cart.const32}) which also involves
the boundary values of $h^{xx}$ and $h^{yy}$. 
An iterative approach was adopted to solve such coupled systems.

In fact, a non-iterative approach could be used only for the 
system~(\ref{eq:ladm-cart.const21}) - (\ref{eq:ladm-cart.const22}), 
solved with $x$- and $y$-derivatives centered on the gridpoints.

\pagebreak[3]
\vspace{1em}
\noindent
{\em Numeric results}
\vspace{0.5em}

\noindent
The first set of runs was made for a grid size of $16^3$, 
with a CFL factor of $0.25$, for  500 crossing times 
(i.e. on a grid of $-1...1$ up to $t=1000$), with the choice of 
$\lambda=4$ as defined in Eq.~(\ref{eq:ladm-cart.eveq-Eij}).

The results were:

\begin{itemize}
\item The condition 
\begin{eqnarray}
h^{xx}\;=\; h^{yy}\;=\;h^{xx}&=&0, \label{eq:ladm-cart.stable11}\\
\ddot h^{ij}_{,j} &=&0\label{eq:ladm-cart.stable12} 
\end{eqnarray}
showed no signs of instability for the following two stencils:
\begin{enumerate}
\item $z$-derivative computed at the boundary, 
$x$- and $y$-derivatives centered on the point;
\item $z$-derivative centered between the last two points,
$x$- and $y$-derivatives centered in the 2-D cells.
\end{enumerate}

\item The condition 
\begin{eqnarray}
h^{xx}- h^{yy}\;=\;h^{xx}&=&0, \label{eq:ladm-cart.stable21}\\
\ddot h^{ij}_{,j} &=&0,\label{eq:ladm-cart.stable22} \\
\ddot h &=&0\label{eq:ladm-cart.stable23}
\end{eqnarray}
showed no signs of instability for the stencil in which 
$z$-derivative is centered between the last two points,
and the $x$- and $y$-derivatives are centered in the 2-D cells.

\item All other (non-periodic) boundary conditions 
gave unstable runs.

\end{itemize}

Next the three 
well-behaved cases were tested 
with homogeneous boundary data and  random initial 
data (without compact support).

The runs with Eqs.~(\ref{eq:ladm-cart.stable11}) - (\ref{eq:ladm-cart.stable12}) 
showed no qualitative
difference. The Hamiltonian constraint was still a linear function of time.

The runs with Eqs.~(\ref{eq:ladm-cart.stable21}) - (\ref{eq:ladm-cart.stable23}) 
resulted in
\begin{eqnarray}
||{\cal C}||_{\infty} &=& 
O(10^{-2}) + t \cdot O(10^{-4}) + t^2\cdot  O(10^{-4})
\nonumber \\ &&
+ t^3 \cdot O(10^{-3}) + t^4 \cdot O(10^{-12}) + \cdots \nonumber 
\end{eqnarray}

Running with  $32^3$ gridpoints gave similar results.
When running with $48^3$ points for $1000$ crossing times, with 
 derivatives centered on the cell, both the condition 
Eqs.~(\ref{eq:ladm-cart.stable11}) - (\ref{eq:ladm-cart.stable12}) and 
Eqs.~(\ref{eq:ladm-cart.stable21}) - (\ref{eq:ladm-cart.stable23}) 
revealed an unstable mode.
The only boundary condition 
surviving the stability tests is  
Eqs.~(\ref{eq:ladm-cart.stable11}) - (\ref{eq:ladm-cart.stable12}), using 
$z$-derivatives computed at the boundary and
$x$- and $y$-derivatives centered on the point. 

Next,   boundary values for 
$h^{xx}, h^{yy}$ and $h^{xy}$ were set to random numbers.
The Hamiltonian constraint remained a linear function of time.

Even though the results were significantly better than for
the ``plain'' Dirichlet boundary condition, the picture is still unclear.
Ideally one should have to specify two functions at the
boundary (the two polarization modes). Then, starting from these, one 
should be able to reconstruct boundary values for all
 six metric functions by use of
four boundary constraints. 

The following boundary condition is proposed:
\begin{itemize}
\item provide $h^{xy}$ and $h^{xx} - h^{yy}$ freely (i.e. set the 
polarization modes to random numbers), and

\item compute $h^{xx} + h^{yy}, h^{zx}, h^{zy}, h^{zz}$ using
Eqs.~(\ref{eq:ladm-cart.bdry-eqs.Gzz}), 
(\ref{eq:ladm-cart.bdry-eqs.mmconst.A}) 
and (\ref{eq:ladm-cart.bdry-eqs.mmconst.z}).
\end{itemize}

Initial data is a  set of random numbers (without compact support), 
for each metric function. The code 
was run for 2000 crossing times.

This boundary condition caused exponential 
growth of the Hamiltonian constraint
after $1000$ crossing times  for a grid size of 
$48^3$  with the value of  $\lambda = 4.0$.
The value $\lambda = 3.0$ was unstable after  
$800$ crossing times.

\subsubsection{Convergence test}

In order to study its convergence properties, 
the above  boundary algorithm is extended to
the edges and corners of the cube-shaped boundary.
This is done done
using the following rules:
\begin{enumerate}
\item On edges and corners use the equation
\begin{equation}
\ddot h = 0
\end{equation}
together with the two ``+''-modes given on the neighboring sides to compute
the diagonal components of the metric.
\item On edges use the equation
\begin{equation}
\ddot h^{ij}_{j} = 0
\end{equation}
with sideways finite difference derivatives to compute the missing non-diagonal 
component.
\end{enumerate}

Although the boundary algorithm is unstable, the exponential 
growth is slow enough to allow a short time
convergence test. This was done
using  $\ell=4$ time-symmetric linear waves \cite{Teukolsky82} of amplitude $10^{-6}$, width $1$, at t=8,
bounding box $-4\ldots4$, grid sizes $50, 60, 70, 80$. 
The measured convergence rates of the six metric functions were better than 
$O(h^{2.04})$.

\subsection{Boundary constraints and the first-order schemes}

The previous sections have identified a few algorithms that work significantly
better  than the inconsistent 
choice $h_{ij}|_{bdry} = 0$.
As a next step, a number of boundary algorithms were tested  with the other
 evolution schemes: LF1, LF2, and ICN.

First the the boundary constraints 
Eqs.~(\ref{eq:ladm-cart.bdry-eqs.SECOND}) - (\ref{eq:ladm-cart.bdry-eqs.LAST})
are rewritten in terms of the variables 
$h_{ij}$ and $K_{ij} = - \frac 12 \dot h_{ij}$:
\begin{eqnarray}
\label{eq:ladm-cart.bdry-eqs.2.FIRST}
\label{eq:ladm-cart.bdry-eqs.2.Gzz}
  2\, {}^{(4)}G_{zz} &=& 2 \dot K^A_A +\partial^B\partial_B h^A_A
             - \partial_A\partial_Bh^{AB} ,
\\
\label{eq:ladm-cart.bdry-eqs.2.mmconst.A}
 \dot {\cal C}^A - \partial^A \,  {}^{(4)}R^t_t 
&=&  \partial_z  \dot K^{Az} + \partial_B  \dot K^{AB} , 
\\
\label{eq:ladm-cart.bdry-eqs.2.mconst.A}
\, {\cal C}^A &=& \partial_z K^{Az} + \partial_B K^{AB} -  \partial^A K , 
\\
\label{eq:ladm-cart.bdry-eqs.2.GzA} 
   2\, {}^{(4)}G_z^A &=& -2 \dot K_z^A 
-\partial_B(\partial^B h_z^A-\partial^A h_z^B)
\nonumber \\ &&
         -\partial_z\partial^A h^B_B +\partial_z\partial^B h^A_B ,
\\
\label{eq:ladm-cart.bdry-eqs.2.Rtt} 
\, {}^{(4)}R^t_t &=& - \dot K, 
\\
\label{eq:ladm-cart.bdry-eqs.2.mmconst.z}
\dot {\cal C}^z - \partial^z \,  {}^{(4)}R^t_t 
&=&  \partial_z  \dot K^{zz} + \partial_B  \dot K^{zB} .
\label{eq:ladm-cart.bdry-eqs.2.LAST}
\end{eqnarray}

Given the $h_{TT}$ components\footnote{These functions correspond to 
the gravitational wave that propagates (in this case) in the 
$z$-direction, two polarization modes. (See Section~\ref{sec:null.news}.)}
 $h^{xy}$ and $h^{xx}-h^{yy}$,
equation (\ref{eq:ladm-cart.bdry-eqs.2.Gzz})
 determines the missing  component of the 2-tensor $h_{AB}$.
The $h^{zA}$ components can then be determined via any of the 
equations~(\ref{eq:ladm-cart.bdry-eqs.2.mmconst.A}) - 
(\ref{eq:ladm-cart.bdry-eqs.2.GzA}).
Either of the remaining  expressions,  
(\ref{eq:ladm-cart.bdry-eqs.2.Rtt}) or 
(\ref{eq:ladm-cart.bdry-eqs.2.mmconst.z}), can be used to compute
 the component $h_{zz}$.
These equations provide  
five\footnote{There is one more combination of constraints, 
$(\ref{eq:ladm-cart.bdry-eqs.2.Gzz})
,(\ref{eq:ladm-cart.bdry-eqs.2.mconst.A})$, and
$(\ref{eq:ladm-cart.bdry-eqs.2.mmconst.z})$,
 whose numerical implementation was not clear and so we have not
insisted in working it out.} alternative sets of boundary constraints:
\pagebreak[3]
\begin{tabbing}
mmmmmm \= mmmmmm \= \kill \\
\> No. \> Equations involved: \\
\> 1)  \> 
$(\ref{eq:ladm-cart.bdry-eqs.2.Gzz}),
(\ref{eq:ladm-cart.bdry-eqs.2.mconst.A})
\;\; \mbox{and}\;\; (\ref{eq:ladm-cart.bdry-eqs.2.Rtt})$, \\
\> 2) \>  
$(\ref{eq:ladm-cart.bdry-eqs.2.Gzz}),
(\ref{eq:ladm-cart.bdry-eqs.2.mmconst.A})
\;\; \mbox{and}\;\; (\ref{eq:ladm-cart.bdry-eqs.2.Rtt})$,\\
\> 3)\>  
$(\ref{eq:ladm-cart.bdry-eqs.2.Gzz}),
(\ref{eq:ladm-cart.bdry-eqs.2.mmconst.A})
\;\; \mbox{and}\;\; (\ref{eq:ladm-cart.bdry-eqs.2.mmconst.z})$,\\
\> 4) \>  
$(\ref{eq:ladm-cart.bdry-eqs.2.Gzz}),
(\ref{eq:ladm-cart.bdry-eqs.2.GzA})
\;\; \mbox{and}\;\; (\ref{eq:ladm-cart.bdry-eqs.2.Rtt})$,\\
\> 5) \>  
$(\ref{eq:ladm-cart.bdry-eqs.2.Gzz}),
(\ref{eq:ladm-cart.bdry-eqs.2.GzA})
\;\; \mbox{and}\;\; (\ref{eq:ladm-cart.bdry-eqs.2.mmconst.z})$.
\end{tabbing}

The linearized ADM equations 
(\ref{eq:ladm-cart.Kdef})-(\ref{eq:ladm-cart.Kdoteq}) 
have the form
\begin{eqnarray}
\dot h_{ij} &=& -2 K_{ij}, \\
K_{ij}  &=& \sum_{k,l,m,n} c[k,l,m,n;i,j] \partial_k \partial_l h_{mn},
\end{eqnarray}
with $c[k,l,m,n;i,j]$ being a set of numeric coefficients.
The metric $h_{ij}$ can be updated 
at each gridpoint where $K_{ij}$ is known. In other words
there is no need for boundary data for $h_{ij}$. On the other
hand updating $K_{ij}$ involves finite-difference expressions
involving $h_{ij}$. This implies that the tensor $K_{ij}$ cannot
be updated at the boundary via the evolution algorithm. A separate
boundary algorithm is necessary. This algorithm uses a
set of boundary constraints and the free functions $K_{TT}$.

When approximating the boundary constraints by 
finite-difference equations,
those derivatives that are parallel to the boundary are centered on
the gridpoint. Derivatives  perpendicular 
to the boundary (the $z$-derivatives)
are computed using 3-point, second-order formulae 
(see Eqs.~(\ref{eq:ladm-cart.side-z-deriv.1}) 
- (\ref{eq:ladm-cart.side-z-deriv.2})).
Time derivatives are centered in the mid-level $N+1/2$. Since the 
momentum constraint  ${\cal C}^A$ does not contain explicit time derivatives,
it is imposed on the time level $t_N$. These rules 
furnish  finite-difference approximations for any of the boundary constraints.
For example, Eq.~(\ref{eq:ladm-cart.bdry-eqs.2.Gzz}) is approximated by
\begin{eqnarray}
         \frac{1}{\Delta t} \left[\left(K^A_A \right)^{[N+1]}_{[I,J,1]} \right.
&-&\left.\left(K^A_A\right)^{[N]}_{[I,J,1]}\right]
 =  \frac{1}{4 (\Delta x)^2} \times \nonumber \\ 
&\Big(&  
- 2 h_{xx [I,J-1,1]}^{[N+1/2]}   + 4 h_{xx [I,J,1]}^{[N+1/2]} - 2 h_{xx [I,J+1,1]}^{[N+1/2]} \nonumber \\ &&
- 2 h_{yy [I-1,J,1]}^{[N+1/2]}   + 4 h_{yy [I,J,1]}^{[N+1/2]} - 2 h_{yy [I+1,J,1]}^{[N+1/2]} \nonumber \\ &&
-     h_{xy [I+1,J-1,1]}^{[N+1/2]} -     h_{xy [I-1,J+1,1]}^{[N+1/2]}  \nonumber \\ &&
+     h_{xy [I+1,J+1,1]}^{[N+1/2]} +     h_{xy [I-1,J-1,1]}^{[N+1/2]} \;\; \Big) 
\end{eqnarray}
and used to update $\left(K^A_A\right)^{[N+1]}_{[I,J,1]}\; ; \;$ similarly, 
Eq.~(\ref{eq:ladm-cart.bdry-eqs.2.mmconst.z}) is approximated by
\begin{eqnarray}
\frac{1}{2 \Delta t  \Delta x} \Big\{
\left[
3 K^{zz [N+1]}_{[I,J,1]} - 4  K^{zz [N+1]}_{[I,J,2]}
+  K^{zz [N+1]}_{[I,J,3]} \right] -&& \nonumber \\
\left[
3 K^{zz [N]}_{[I,J,1]} - 4  K^{zz [N]}_{[I,J,2]}
+  K^{zz [N]}_{[I,J,3]} \right] \Big\} &+& \nonumber \\
\frac{1}{2 \Delta t \Delta x} \Big\{
\left[ K^{xz [N+1]}_{[I+1,J,1]} -  K^{xz [N+1]}_{[I-1,J,1]} \right]- 
\left[ K^{xz [N  ]}_{[I+1,J,1]} -  K^{xz [N  ]}_{[I-1,J,1]} \right]
\Big\} &+&  \nonumber  \\
\frac{1}{2 \Delta t \Delta x} \Big\{
\left[ K^{yz [N+1]}_{[I,J+1,1]} -  K^{yz [N+1]}_{[I,J-1,1]} \right]- 
\left[ K^{yz [N  ]}_{[I,J+1,1]} -  K^{yz [N  ]}_{[I,J-1,1]} \right]
\Big\}  &=& 0
\nonumber \\
\end{eqnarray}
and used to update  $K^{zz [N+1]}_{[I,J,1]}$.

Stability tests were made for 2000 crossing times, grid size of $48^3$,
$\Delta t = \Delta x / 4$,
 random initial data and  random boundary values
for the $K_{TT}$ functions. 
All five systems were tested for
$\lambda = 2$, with all evolution algorithms. Only the runs with  ICN were stable.
In addition system (1) was tested using $\lambda=0$ and $\lambda=4$, with all
evolution algorithms. Again, all runs became unstable except for  ICN.
System (5) was tested for $\lambda=0$ and $\lambda=4$ as well; it remained stable
for ICN and became unstable in a short time for the leap-frog algorithms.

\begin{figure}
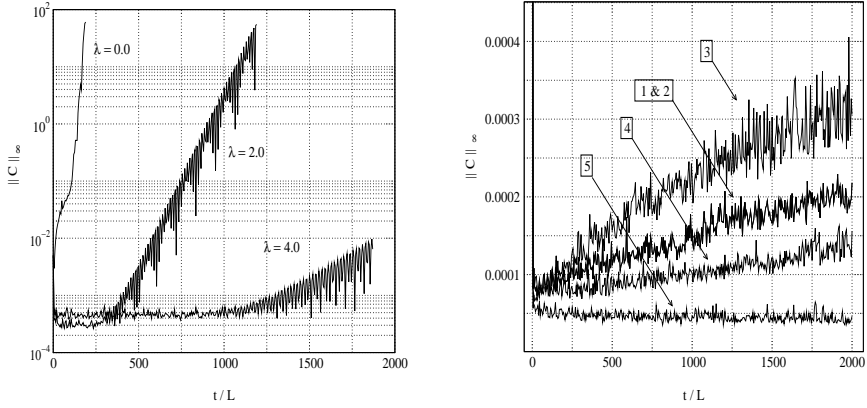

\centerline{\hbox{
\psfig{figure=lgr.3d.LF2.stage2.eps,height=2.1in,width=2.1in}
\hspace{0.2in}
\psfig{figure=lgr.3d.ICN.stage2.eps,height=2.1in,width=2.1in}
}}
\caption{
The $\ell_\infty$ norm of the Hamiltonian constraint
versus time (in crossing times)
for stability runs with random initial and  $h_{TT}$ boundary
 data of $O(10^{-6})$, 
with $\Delta t = \Delta x / 4$. The $x$- and $y$-directions
have periodic boundary condition.
Left: algorithm LF2. The boundary constraints
are given by system (1), with runs being made for $\lambda = 0, 2$, and $4$.
Right: algorithm ICN. $\lambda = 2$, for all
five boundary-constraint systems.
}
\label{fig:adm-cart.LF2.stage2}
\label{fig:adm-cart.ICN.stage2}
\end{figure}

Five stable boundary-algorithms have been found for 
the LG equations evolved via an ICN scheme. 
None of these
boundary algorithms were stable
when evolved with the leap-frog schemes. 
 Figure~\ref{fig:adm-cart.LF2.stage2}
plots
of the $\ell_\infty$  norm of the  Hamiltonian constraint as a function
of time for ICN and LF2.

\subsection{Outgoing radiation boundary conditions}

A few runs were performed with the ``modified'' Sommerfeld boundary conditions. 
 ``Modified'' in the present context means
\begin{equation}
 \left[ (\partial_t - \partial_x) F\right]_{|x_{min}} = S_{|x_{min}}, \;\;\;
 \left[ (\partial_t + \partial_x) F\right]_{|x_{max}} = S_{|x_{max}}
\end{equation}
where $S$ is a source, describing incoming radiation. In the numerical
experiments we set $S$ to random numbers.

Our efforts in this direction were limited.
We have not been able to find a stable boundary algorithm 
for the LG equations, when using Sommerfeld radiation conditions.
In particular, we applied Sommerfeld condition on the radiation degrees of 
freedom combined with different sets of boundary constraints. These
experiments produced unstable runs.

\subsection{Evolution of a bounded space-time region}

Next we want to extend the boundary algorithm to all faces of a cube.
The edges and corners must be handled separately.

Algorithm (5) was used on all faces.
The two components $K_{TT} = -\frac{1}{2} \dot h_{TT}$
are treated as free data
(they are specified randomly)
on all faces, edges, and  corners. While
this means two free quantities and four constraints on the faces,
there are four free quantities on the edges, so only two constraints are needed.
Similarly, on the corners,
there are five free quantities, for the identity
$ [K_{xx}-K_{yy}] + [K_{yy}-K_{zz}] + [K_{zz}-K_{xx}] = 0$ reduces
the total number of six $TT$ components
to five that are independent. Thus only one constraint is needed
at the corners.

All non-diagonal components are provided at the corners.
Given $[K_{xx}-K_{yy}]$ and $[K_{zz}-K_{xx}]$ the missing 
 diagonal component is $K_{xx}$,   which can be computed from
$$3K_{xx}=K+[K_{xx}-K_{yy}]+[K_{xx}-K_{zz}].$$ The trace $K$ is updated using the
condition $${}^{(4)}R^t_t =-\dot K=0.$$

On the edges parallel to the $x$-axis one already
has $K_{xy}$ and $K_{xz}$ as boundary
data. The missing boundary data $K_{yz}$ 
is computed using $^{(4)}G_{yz}$, an equation
that is also used on both neighboring faces. Derivatives
in the $y$- and $z$-directions are
computed by sideways, 3-point, finite difference formulae.
The diagonal components of the 3-metric are computed the same way as
on the corners.

Note that the routine that solves the constraint
$$ (- \dot {\cal C}^n + \partial^n{}^{(4)}R^t_t) = 0$$
on a face of the cube (with normal in the $n$-direction)
must be called {\em after} the missing non-diagonal components
have been updated on the edges surrounding that face.
Otherwise, in the case of the $z=\mbox{constant}$ face,
when computing the quantity
$K_{yz,y}$ on the top time level,
 with centered finite differencing, one would be using
values of $K_{yz}$ on the edge parallel to the $x$-axis
that had not yet been updated.

\begin{figure}
\centerline{
\psfig{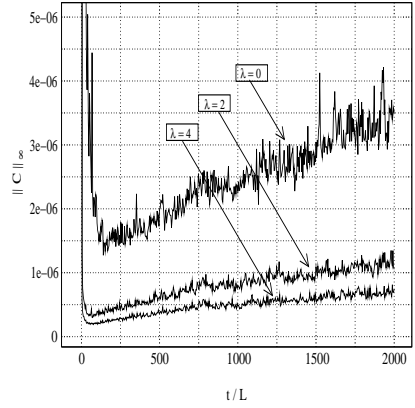}}
\caption{
Results of the stability tests for a code
evolving the interior region of a cube with constrained
boundaries at the faces, edges, and corners of the cube.
The Hamiltonian constraint is measured  via the  $\ell_\infty$ norm.
Initial data and the $h_{TT}$ boundary data  are random of
 $O(10^{-6})$. The CFL ratio is $\Delta t / \Delta x = 0.25$. 
Runs were performed for 2000 crossing times, using, $\lambda = 0, 
2$, and $4$.}
\label{fig:adm-cart.ICN.stage3}
\end{figure}

Runs performed with $\lambda = 0, 2, 4$.
showed that the above algorithm is robustly stable.
All three runs were given random initial and boundary data. A graph
showing the Hamiltonian constraint as a function of time is
shown in Figure~\ref{fig:adm-cart.ICN.stage3}.

\chapter{Spherical Boundaries}
\label{chap:ladm-sph}

The previous chapter described how to
stably evolve 
the linearized ADM system in Cartesian coordinates,
with boundaries on the faces of a cube. However the Cauchy 
boundary required by CCM is a sphere; 
it is not aligned to the Cartesian-grid-structure. 
We need to adapt the boundary 
algorithm found for the faces of the cube
to the case of a spherical boundary.

The approach we present here uses a spherical grid
that  forms a boundary for the   Cartesian evolution domain.
This spherical boundary is connected to the 
evolution points via an interpolation algorithm. 
First we describe how such an approach works for the  SWE case. Next the 
linearized ADM system is analyzed.
Throughout the chapter the evolution scheme  is ICN.

\section{The scalar wave problem}
\label{sec:ladm-sph.SWE}

\begin{figure}
\centerline{\epsfxsize=3.5in\epsfbox{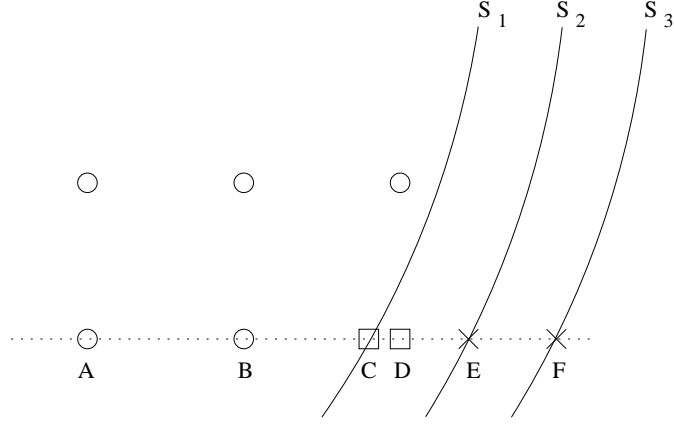}}
\caption{Interpolation scheme providing spherical boundary
to a Cartesian code.}
\label{fig:ladm-sph.interp}
\end{figure}

On a Cartesian set of grid-points 
$\left \{ x_I, y_J, z_K \right \}$ that lie inside a sphere of radius $R$,
the scalar wave equation takes the first-order in time discretized form
\begin{equation}
\begin{array}{rrccl}
\left( \partial_t \phi \right)_{[I,J,K]}&=&\xi_{[I,J,K]},&\\
\left( \partial_t \xi \right)_{[I,J,K]}&=&
\frac {1}{(\Delta x)^2}\; (& 
\phi_{[I+1,J,K]} - 2 \phi_{[I,J,K]} +  \phi_{[I-1,J,K]}\\
&&&
\phi_{[I,J+1,K]} - 2 \phi_{[I,J,K]} +  \phi_{[I,J-1,K]}\\
&&&
\phi_{[I,J,K+1]} - 2 \phi_{[I,J,K]} +  \phi_{[I,J,K-1]}  
&)+ O(\Delta^2).
\end{array}
\end{equation}
In order to properly update the point $[I,J,K]$, one needs
values of $\phi$ at the neighboring points 
$[I\pm 1,J,K], [I,J \pm 1,K], [I,J,K \pm 1]$. These points are at most 
a distance $\Delta x$ outside of the sphere $R$. 
The boundary algorithm adopted from \cite{Bishop97b} is the following:
\begin{itemize}
\item Let $S_1$ denote the spherical boundary of the evolution domain, 
with radius $R_1$. Let $S_2$ and $S_3$ be two additional spheres, concentric
with $S_1$, radii $R_2 = R_1 + \Delta x,\;\; R_3 = R_1 + 2 \Delta x$.
The spheres are described by a stereographic grid-structure 
corresponding to the coordinate patches $\zeta_N, \zeta_S$. 
\item Let $D$ be a point outside $S_1$ with at least one of its 
nearest neighbors in the Cartesian evolution domain. 
\item We define the most normal direction through $D$ as  
the direction given by a vector $\vec n$ parallel to either the $x$-,
or the $y$-, or the $z$-axis, chosen such that it is closest to the radial
direction through $D$.
\item Let $C$, $E$, and $F$ be  the intersection points of the most
normal direction through $D$  and the spheres $S_1$, $S_2$ and $S_3$. Let $A$ and $B$ be
the nearest two Cartesian evolution grid-points of $D$ in the most normal direction.
\item Given any smooth function $\phi$ on the Cartesian grid-points
inside $S_1$ and on the spherical gridpoints of $S_2$ and $S_3$,
one can obtain $\phi_C$ and $\phi_D$ by interpolation:
\begin{itemize}
\item The  points $E$ and $F$ do not necessarily coincide with stereographic
gridpoints, and so
 the values $\phi_E$ and $\phi_F$ are constructed
using 2-D quadratic interpolators in the plane $q=\Re(\zeta), p=\Im(\zeta)$.
\item Next, using a 1-D quadratic interpolator in the most normal direction,
the values of 
$\phi_C$ and $\phi_D$ are obtained
from the values of $\phi$ at $A,B,E$, and $F$.
\end{itemize}
An illustration of the interpolation
algorithm can be seen in Figure~\ref{fig:ladm-sph.interp}.
\end{itemize}

The algorithm presented above requires knowledge of 
$\phi$ on the Cartesian evolution gridpoints as well
as on the outer two spheres.  The sphere $S_3$ is the boundary
of the system, the field $\phi$ on  can be specified
arbitrarily on $S_3$. The value of  $\phi$ on
the sphere $S_2$ is updated using the scalar wave equation,
expressed in spherical coordinates.  
The field $\phi$ 
on the sphere $S_1$ is updated by the interpolation algorithm and 
is used in the evolution of $\phi$ on  $S_2$.  

The algorithm was found to be numerically stable with the usual settings
of $48^3$ Cartesian grid-points, 2000 crossing times, 
a CFL ratio of $\Delta t / \Delta x = 0.25$, random initial data and random
boundary data on $S_3$.

\section{The linearized ADM system in spherical coordinates}

\subsection{Conventions}

We introduce spherical 
coordinates  $[t, r, q, p]$, where
stereographic patching is used for the  angular coordinates $(q,p)$. 
The space-time metric  takes the form
\begin{eqnarray}
&& {}^{(4)}g_{\mu \nu} = {}^{(4)}\eta_{\mu \nu} +  {}^{(4)}h_{\mu \nu} = 
\\ &&=
\left [\begin {array}{cccc}
-1&0&0&0 \\\noalign{\medskip}
0&1-{\it W}r &
-{\frac {2 r^2 \Re(U)}{P}} &
-{\frac {2 r^2 \Im(U)}P} 
\\\noalign{\medskip}
0&-{\frac {2 r^2 \Re(U)}{P}} &
4\,{\frac {r^2}{P^2}} \left[ 1 +  K +  \Re(J) \right] &
4\,{\frac {r^2}{P^2}}   \Im(J) 
\\\noalign{\medskip}
0&-{\frac {2 r^2 \Im(U)}P} &
4\,{\frac {r^2}{P^2}}   \Im(J) &
4\,{\frac {r^2}{P^2}} \left[ 1 +  K -  \Re(J) \right] 
\end {array}\right ] \,,\nonumber
 \end{eqnarray}
where $P = 1+q^2+p^2$ and $\eta_{\mu \nu} = 
\mbox{diag}(-1, 1, \frac{4 r^2}{P^2}, \frac{4 r^2}{P^2})$.

We define the  vectors:

\begin{eqnarray}
q^{\mu} &=& \left[ 0, 0, \frac{P}{2}, I \frac{P}{2} \right ], \\
r^{\mu} &=& \left[ 0, 1, 0,0 \right]. 
\end{eqnarray}

\subsection{Evolution equations}

The ADM equations ${}^{(4)}R_{ij} = 0$ decompose into spin-weighted equations
that provide the evolution equations for $J,K,U,W$:

\begin{eqnarray}
{\cal E}_W &:=& r^\mu r^\nu \,{}^{(4)}R_{\mu \nu} = 
\frac{1}{2 r^2}\Big \{
r^3 \ddot W + 2 r (r W)_{,r} 
\nonumber \\
&&  
- r \eth \bar \eth W 
+ \left[ r^2 \left(\bar \eth U + \eth \bar U \right) \right]_{,r}
+ 2 [ r^2 K_{,r} ]_{,r}
\Big \} = 0,\label{eq:ladm-sph.Weq}\\
{\cal E}_K &:=& 
q^\mu \bar q^\nu  \,{}^{(4)}R_{\mu \nu} = \nonumber \\
&& \frac{1}{2 r^2} \Big\{
-2 r^4 \ddot K
+2 r^2 (r^2 K)_{,rr} 
+ 2 r^2 \eth \bar \eth K 
- r^2 ( \eth \eth \bar J+ \bar \eth \bar \eth J ) \nonumber \\
&& + \left[ r^4 ( \eth \bar U + \bar \eth U) \right]_{,r}
+ 2 r (r^3 W)_{,r} 
- r^3 \eth \bar \eth W \Big \} = 0,
\label{eq:ladm-sph.Keq}\\
{\cal E}_U &:=& q^\mu r^\nu  \,{}^{(4)}R_{\mu \nu}  = 
\frac{1}{4} \Big\{ 
2 r^2 \ddot U
- 4 U
\nonumber \\
&& 
+ \eth \eth \bar U
- \eth \bar \eth U
- 2 \bar \eth J_{,r}
+ 2 \eth K_{,r}
+ 2 \eth W
\Big\} = 0, \label{eq:ladm-sph.Ueq}\\ 
{\cal E}_J &:=&q^\mu q^\nu  \,{}^{(4)}R_{\mu \nu} =  - r^2 \ddot J 
+ (r^2 J_{,r})_{,r} - \frac 1 2 \; r \; \eth \eth W
+ (r^2\; \eth U)_{,r} = 0.
\nonumber \\
\label{eq:ladm-sph.Jeq}
\end{eqnarray}

\subsection{Multipole expansions}
Next the metric functions $K,W,J,U$ are written as an expansion in terms of the spherical
harmonics:
\begin{eqnarray}
K &=& \sum_{\ell=0}^\infty \sum_{m=-\ell}^\ell 
 k^{(\ell,m)}(t,r)\, Y^\ell_m(\theta, \phi), \label{eq:ladm-sph.spwexp.YlmFIRST}  \\
W &=& \sum_{\ell=0}^\infty \sum_{m=-\ell}^\ell
 w^{(\ell,m)}(t,r)\, Y^\ell_m(\theta, \phi), \\
J &=& \sum_{\ell=0}^\infty \sum_{m=-\ell}^\ell
 j^{(\ell,m)}(t,r)\, \eth \eth \,Y^\ell_m(\theta, \phi), \\
U &=& \sum_{\ell=0}^\infty \sum_{m=-\ell}^\ell
 u^{(\ell,m)}(t,r)\, \eth \, Y^\ell_m(\theta, \phi),\label{eq:ladm-sph.spwexp.YlmLAST}
\end{eqnarray}
with $(\theta,\phi) = f(q,p)$ defined by
\begin{eqnarray}
\zeta_{North} &=& q_N + I p_N = \sqrt{\frac{1-\cos \theta}{1+\cos \theta}}e^{I \phi},\\
\zeta_{South} &=& q_S + I p_S = \sqrt{\frac{1+\cos \theta}{1-\cos \theta}}e^{-I \phi}.
\end{eqnarray}

The identities
\begin{eqnarray}
\eth Y^0_0 &=& 0, \\
\eth \eth Y^1_0 \;= \; \eth \eth Y^1_{\pm 1} &=& 0
\end{eqnarray} 
imply that no $\ell=0$ mode is present in $U$ and that the lowest non-zero term
in the expansion of $J$ is a quadrupole.

The equations (\ref{eq:ladm-sph.Weq}) - (\ref{eq:ladm-sph.Jeq}),
written for the expanded metric variables
(\ref{eq:ladm-sph.spwexp.YlmFIRST}) - (\ref{eq:ladm-sph.spwexp.YlmLAST}),
 become a series of equations for the coefficients
$$\left \{ k^{(\ell,m)}, w^{(\ell,m)}, j^{(\ell,m)}, u^{(\ell,m)} \right \},$$ 
with no coupling  between 
 different spherical harmonics.

\subsection{Stability}

A numerically stable algorithm 
must satisfy the von Neumann stability criterion.  The von 
Neumann analysis assumes that the constant coefficient version of the evolution
equations has no exponentially growing spatially periodic modes.
In order to analyze such modes, we write the evolution equations corresponding to the
multipole modes in a constant-coefficient form and
study their solutions.
The analysis proceeds in the neighborhood of some shell $r=r_0$,
in order to avoid the coordinate singularity at $r=0$
(which is irrelevant for the purpose of CCM).

The methodology we adopt is the following:
Consider first the monopole mode $\ell = 0$.
This provides a set of two PDEs in the variables $(t,r)$.
Assume that the fundamental metric variables of the system
are of the form $\tilde W = r^m W, \tilde K = r^n K$.
Next write down the evolution equations for the coefficients 
$\tilde k^{(0,0)}$ and $\tilde w^{(0,0)}$ around $r=r_0$.
Assuming periodic boundary conditions in the radial direction,
choose the values of $(n,m)$ such that exponentially growing modes
are ruled out.

Given the values $(n,m)$ from the $\ell =0$ analysis
repeat the procedure for $\ell = 1, m = 0$. This results
in a system of three PDEs for the coefficients $ k^{(1,0)}, 
w^{(1,0)}$ and $u^{(1,0)}$.  Assume that the fundamental
variables of the system are the $\tilde K$ and $\tilde W$ defined by
the $\ell=0$ analysis, and $\tilde U = r^s U$. As it is shown,
assuming periodic $r$-dependence around $r=r_0$ will
necessarily imply the presence of exponentially growing modes.

\subsubsection{Monopole terms}
\label{sec:ladm-sph.monopole}

The monopole amplitudes $\left\{ k^{(0,0)}, w^{(0,0)} \right\}$ 
evolve according to  the equations 

\begin{eqnarray}
{\cal E}^{(0,0)}_K &=&  \frac 1 r \left[
-r^3 \, \ddot k^{(0,0)} + r \, \left( r^2 \, k^{(0,0)} \right)_{,rr}
+  \left(r^3 \, w^{(0,0)} \right)_{,r} \right ] = 0,
\nonumber \\
\label{eq:ladm-sph.k0eq}\\
{\cal E}^{(0,0)}_{W} &=& \frac 1 {2 r^2} \left [
r^3 \, \ddot w^{(0,0)} + 2\, r\,  \left(r\,w^{(0,0)} \right)_{,r}
+ 2 \,  \left(r^2 \,k^{(0,0)}_{,r} \right)_{,r}
\right ] = 0. \nonumber \\
\label{eq:ladm-sph.w0eq}
\end{eqnarray}
up to some overall numerical factors that from now on are neglected.

In order to select an appropriate choice of fundamental variables,
the evolution equations 
(\ref{eq:ladm-sph.k0eq}) - (\ref{eq:ladm-sph.w0eq}) are
rewritten in terms of 
$ \tilde W = r^m \cdot W, \tilde K = r^n \cdot K $:

\begin{eqnarray}
\left.{\cal E}_{\tilde K}\right|_{r=r_0} &=&
-{r_0}^{2-n}{\tilde k^{(0,0)}_{,tt}}
+{r_0}^{2-n}{\tilde k^{(0,0)}_{,rr}}
-2\,{r_0}^{1-n}\left (n-2\right ){\tilde k^{(0,0)}_{,r}}\nonumber \\
&\,&
+{r_0}^{-n}\left (n-1\right )\left (n-2\right ){\tilde k^{(0,0)}}
\nonumber \\&\,&
+{r_0}^{-m+2}{\tilde w^{(0,0)}_{,r}}
-{r_0}^{1-m}\left (m-3\right ){\tilde w^{(0,0)}} = 0, \label{eq:ladm-sph.kteq}
\\
\left.{\cal E}_{\tilde W}\right|_{r=r_0} &=& \frac{1}{2 r}  \left[
{\tilde w^{(0,0)}_{,tt}}\,{r_0}^{-m+2}
+2\,{r_0}^{1-m}{\tilde w^{(0,0)}_{,r}}
-2\,{r_0}^{-m}\left (m-1\right ){\tilde w^{(0,0)}}\right.\nonumber \\
&\,&\left.
+2\,{r_0}^{1-n}{\tilde k^{(0,0)}_{,rr}}
-4\,{r_0}^{-n}\left (n-1\right ){\tilde k^{(0,0)}_{,r}}\right.\nonumber \\
&\,&\left.
+2\,{r_0}^{-n-1}n\left (n-1\right ){\tilde k^{(0,0)}}\right] = 0.
\label{eq:ladm-sph.wteq}
\end{eqnarray}

Assume the following behavior for $\tilde k^{(0,0)}$ and $\tilde w^{(0,0)}$:
\begin{eqnarray}
\tilde k^{(0,0)}(t,r) &=& e^{ I (\alpha t +  \beta r) } k_0, \label{eq:ladm-sph.Kexp0}\\
\tilde w^{(0,0)}(t,r) &=& e^{ I (\alpha t +  \beta r) } w_0. \label{eq:ladm-sph.Wexp0}
\end{eqnarray}
The evolution equations 
(\ref{eq:ladm-sph.kteq}) - (\ref{eq:ladm-sph.wteq})  then impose 
 conditions on $\alpha, \beta, k_0$ and $w_0$:
\begin{eqnarray}
\left [I{r_0}^{-m+2}\beta-{r_0}^{1-m}\left (m-3\right )
\right ]&{w_0}&\nonumber \\
+\left [{r_0}^{2-n}{\alpha}^{2}-{r_0}^{2-n}{\beta}^{
2}-2\,I{r_0}^{1-n}\left (-2+n\right )\beta \right. && \nonumber \\
\left . +{r_0}^{-n}\left
(n-1\right )\left (-2+n\right )\right ]&{k_0}&=0,
\nonumber \\
\label{eq:ladm-sph.k1cond}\\
\left [-{\alpha}^{2}{r_0}^{-m+2}+2\,I{r_0}^{1-m}\beta-2\,{
r_0}^{-m}\left (m-1\right )\right ]&{w_0}&\nonumber \\
+\left [-2\,{r_0}^{1-n}{
\beta}^{2}-4\,I{r_0}^{-n}\left (n-1\right )\beta+2\,{r_0}^
{-n-1}n\left (n-1\right )\right ]&{k_0}&=0.
\nonumber \\
\label{eq:ladm-sph.w1cond} 
\end{eqnarray}
The system of equations (\ref{eq:ladm-sph.k1cond}) - (\ref{eq:ladm-sph.w1cond}) 
admits nontrivial solutions ($k_0, w_0$)
if and only if the determinant of the Jacobian of the system vanishes, that is
\begin{eqnarray}
\frac{1}{r_0^{m+n}}\left\{
{\alpha}^{4}{r_0}^{4} 
+\left [-{r_0}^{4}{\beta}^{2}-2\,I{
r_0}^{3}\left (n-1\right )\beta+{r_0}^{2}\left (-3\,n+{n}^{2}+2\,m
\right )\right ]{\alpha}^{2} \right.\nonumber \\
\left. +4\,I r_0\,\left (m-n\right )
\beta-4\,\left (n-1\right )\left (m-n-1\right ) \right \} = 0.
\nonumber \\
\label{eq:ladm-sph.detcond0}
\end{eqnarray}
Assuming periodic $r$-dependence implies a real value for $\beta$.
Then, unless the imaginary part 
of $\alpha = f(\beta, r_0)$ is non-negative, 
the evolution equations
(\ref{eq:ladm-sph.kteq}) - (\ref{eq:ladm-sph.wteq})
 allow exponentially growing modes.
The determinant condition (\ref{eq:ladm-sph.detcond0}) 
is a second-order polynomial
in terms of $\alpha^2$, with solutions of the form
\begin{equation}
\alpha = \pm \sqrt{A \pm \sqrt{B}}.
\end{equation}
The condition $\Im (\alpha) \geq 0$ implies 
that all coefficients of Eq.~(\ref{eq:ladm-sph.detcond0}) must be real.
This implies $n=m=1$. Thus the determinant condition 
(\ref{eq:ladm-sph.detcond0})  leads to   
$$
\alpha_{1,2} = 0,\;\;\; \alpha_{3,4} =  \pm \beta.
$$

The fundamental variables identified so far are $\tilde W = r \cdot W$
and $\tilde K = r \cdot K$.

\subsubsection{Dipole terms}
\label{sec:ladm-sph.dipole}

The equations evolving the dipole coefficients 
$\left\{ u^{(1,0)}, k^{(1,0)}, w^{(1,0)} \right\}$ are 

\begin{eqnarray}
{\cal E}^{(1,0)}_U &=& 
\frac {\zeta}{1+\zeta \bar \zeta } 
\times \nonumber \\ &\times &
\left\{
- r^2 \, u^{(1,0)}_{,tt}  - k^{(1,0)}_{,r}
- w^{(1,0)}+ u^{(1,0)}  + \bar u^{(1,0)} 
\right\}=0,
\label{eq:ladm-sph.u10eq} \\
{\cal E}^{(1,0)}_K &=& 
\frac {1}{r^2} \;
\frac {1-\zeta \bar \zeta }{1+\zeta \bar \zeta }  
\times \nonumber \\ &\times &
\left\{
-r^4 k^{(1,0)}_{,tt} 
+ \left[ r^4 \left( k^{(1,0)}_{,r}  
+ w^{(1,0)} 
-  u^{(1,0)}
-  \bar u^{(1,0)}\right) \right]_{,r} 
\right\}=0,\nonumber \\
\label{eq:ladm-sph.k10eq} \\
{\cal E}^{(1,0)}_W &=&
\frac{1}{r^2}
\frac {1-\zeta \bar \zeta }{1+\zeta \bar \zeta } 
\times \nonumber \\ &\times &
\left\{
\frac 1 2 r^3 w^{(1,0)}_{,tt} 
+  \left[r^2 \left( 
k^{(1,0)}_{,r} + w^{(1,0)}
-   u^{(1,0)} 
-   \bar u^{(1,0)}\right)\right]_{,r} 
\right\}=0.\nonumber \\\label{eq:ladm-sph.w10eq}
\end{eqnarray}
As already mentioned, the dipole term in $J$ vanishes.

Repeating the procedure from the $\ell = 0 $ case;
the evolution 
equations (\ref{eq:ladm-sph.u10eq}) - (\ref{eq:ladm-sph.w10eq}) can 
be rewritten using  the variables 
$  \tilde U = r^s \cdot U, \tilde W = r \cdot W, \tilde K = r \cdot K$. 
A behavior 
of the form
\begin{eqnarray}
\tilde u^{(1,0)}(t,r) &=& e^{ I (\alpha t +  \beta r) } u_0 \label{eq:ladm-sph.Uexp10},\\
\tilde k^{(1,0)}(t,r) &=& e^{ I (\alpha t +  \beta r) } k_0 \label{eq:ladm-sph.Kexp10},\\
\tilde w^{(1,0)}(t,r) &=& e^{ I (\alpha t +  \beta r) } w_0 \label{eq:ladm-sph.Wexp10}
\end{eqnarray}
is assumed. 

The equations
$\left\{ {\cal E}^{(1,0)}_{\tilde U}, \bar {\cal E}^{(1,0)}_{\tilde U}, 
{\cal E}^{(1,0)}_{\tilde K}, {\cal E}^{(1,0)}_{\tilde W} \right\}_{r=r_0}$ 
are a linear system  for
the variables $\left\{ \Re(u_0), \Im(u_0), k_0,  w_0\right\}$.
The existence of
 nontrivial solutions leads to the determinant condition 
satisfied:
\begin{equation}
r_0^8 \alpha^8 - (2+r_0^2 \beta^2)r_0^6 \alpha^6 + 
\left( 12 - 6 s + 2 I (s-4) r_0 \beta - 2 r_0^2 \beta^2  \right)r_0^4 \alpha^4 = 0.
\label{eq:ladm-sph.detcond1}
\end{equation}
Similar to the case of Eq.~(\ref{eq:ladm-sph.detcond0}),  
we assume periodic $r$-dependence. Thus $\beta$ must be real.
A well-behaved solution requires that all 
coefficients of the 
polynomial Eq.~(\ref{eq:ladm-sph.detcond1})
be real, which implies $s=4$.
However, this still allows exponential growth.
For instance, taking $r_0=1, \beta=0$, a solution of 
Eq.~(\ref{eq:ladm-sph.detcond1})  is $-\sqrt{1-\sqrt{13}}$,
which has a negative imaginary part.

The failure of the previous  approach to  a spherical ADM code
indicates that one must adopt a more sophisticated strategy
to treat the powers of $r$ that arise in the transformation from Cartesian
to spherical coordinates.
 The fact that the same set of coupled PDEs
can be implemented stably using Cartesian coordinates indicates that there
should be a stable way to implement ADM in spherical coordinates.
That is the subject of future work.


\chapter{SUMMARY}

This thesis presents a concentrated effort to develop and
calibrate the implementation of the Cauchy-Characteristic Matching
problem for 3-dimensional strongly gravitating systems.  
The related problems of Characteristic and ADM evolution
and their boundaries, have been discussed.

First we give a brief description of the
Cauchy and Characteristic formulations of the equations of General Relativity,
as well as of the concept of Cauchy Characteristic Matching (CCM).
Next the Pitt Null Code is described. 
The underlying physics is presented, i.e. the 
characteristic slicing, the spin-weighted metric functions and the
equations describing the evolution of space-time. 
We also show  how one can use characteristic evolution to numerically evolve
black-hole space-times.

In the following we define the concept of Cauchy Characteristic Matching
first for a spherically symmetric scalar wave, 
next for a 3-D scalar field. Then 
the same concept is outlined for the case of general relativity. 
To make the understanding
of the details easier, first a geometrical description is given,
and then a detailed description follows.
As described,  the extraction 
first interpolates Cauchy data from the Cartesian 
grid onto the extraction world-tube, then computes the Jacobian
of the coordinate transformation from Cartesian 
to Bondi coordinates and lastly it
computes the Bondi metric functions and provides 
these as boundary data to the Characteristic
evolution.  The injection, in turn, uses the Jacobian obtained
in the extraction
to compute the reverse coordinate transformation 
from Bondi to Cartesian coordinates
and then performs a  4-D interpolation
 to transfer data from the characteristic grid onto the Cauchy boundary gridpoints.
Along with the presentation of the matching modules, calibration tests are provided 
to show proper second-order accuracy for a number of test-beds.

Next we study the stability  properties of CCM. 
As it is shown, the numerical noise of the individual modules do not excite any 
short-time instabilities. However, in order to analyze the
 long-term stability properties of matching, one needs to assure that both
the Cauchy and the characteristic evolution codes are able to deal with the 
discretization error that is inherent to numerical boundary algorithms. 
This issue is also addressed. 
As it is  shown, the characteristic code
is able to deal with constraint violating boundary modes of high frequency
without signs of numerical instabilities. 
However, the Cauchy code using the ADM equations is numerically unstable unless
the boundary conditions are treated properly. 
A major contribution of this thesis is that,
in the context of linearized
gravitational theory, we have elucidated the appropriate boundary conditions for the 
coupled set of partial-differential equations that form the principal part 
of the ADM equations. In particular, one should not specify boundary values
for six metric components but provide boundary data for 
the two radiation degrees of freedom and use a set of boundary
constraints to determine the remaining four components of
the spatial metric tensor.

The   stability of the  injection
module requires a spherical  boundary condition
for the Cartesian Cauchy code. For the ADM system
this implies use of boundary constraints
in spherical coordinates. The question of spherical
boundary constraints applied to a Cartesian grid
is a complicated  problem 
that includes implementation of a spherical ADM evolution code,
which provides the subject of future work.



\newpage
\appendix

\chapter{Algebraic Expressions For The Linearized Quadrupole Waves}
\label{app:teuk-ext}

Using the notations
\begin{eqnarray}
\label{eq:teuk-ext.e1}
e_1 &=& {\exp{\left[-\frac{(u - R_\Gamma)^2}{\varpi^4}\right]}},
\\
\label{eq:teuk-ext.e2}
e_2 &=& {\exp{\left[-\frac{(u + R_\Gamma)^2}{\varpi^4}\right]}}, 
\\
\label{eq:teuk-ext.e3}
e_3 &=& {\exp\left[{\frac{4 u R_\Gamma}{\varpi^2}}\right]},
\end{eqnarray}
the functions $\gamma_{1\ldots 11}$ referred to in 
Section~\ref{sec:extract.teuk.extract} have the following explicit form:
\begin{eqnarray}
    \gamma_1 &=& -12{\,}\Big(
12{\,}{R_\Gamma}^2{\,}{\varpi}^4{\,}u
+18{\,}{e_3}{\,}{\varpi}^4{\,}{R_\Gamma}{\,}u^2
-32{\,}{e_3}{\,}{R_\Gamma}^6{\,}u
\nonumber \\ &&
-12{\,}{e_3}{\,}{\varpi}^4{\,}{R_\Gamma}^2{\,}u
+8{\,}{e_3}{\,}{R_\Gamma}^3{\,}u^4
+32{\,}{R_\Gamma}^6{\,}u
\nonumber \\ &&
+16{\,}{R_\Gamma}^2{\,}{\varpi}^2{\,}u^3
-8{\,}{\varpi}^2{\,}{R_\Gamma}^5
+48{\,}{e_3}{\,}{R_\Gamma}^5{\,}u^2
\nonumber \\ &&
-16{\,}{e_3}{\,}{\varpi}^2{\,}{R_\Gamma}^2{\,}u^3
+24{\,}{e_3}{\,}{R_\Gamma}^3{\,}{\varpi}^2{\,}u^2
+8{\,}{R_\Gamma}^3{\,}u^4
\nonumber \\ &&
+8{\,}{e_3}{\,}{R_\Gamma}^7
-9{\,}{e_3}{\,}{\varpi}^6{\,}u
-8{\,}{e_3}{\,}{R_\Gamma}^5{\,}{\varpi}^2
\nonumber \\ &&
+48{\,}{R_\Gamma}^5{\,}u^2
+18{\,}{R_\Gamma}{\,}{\varpi}^4{\,}u^2
-32{\,}{e_3}{\,}{R_\Gamma}^4{\,}u^3
\nonumber \\ &&
+32{\,}{R_\Gamma}^4{\,}u^3
+9{\,}{\varpi}^6{\,}u
+24{\,}{R_\Gamma}^3{\,}{\varpi}^2{\,}u^2
\nonumber \\ &&
+8{\,}{R_\Gamma}^7\Big)
\cdot{e_2}/{\varpi}^8/{R_\Gamma}^5,
\\
    \gamma_2 &=& 24{\,}\Big(
-12{\,}{e_3}{\,}{R_\Gamma}^4{\,}{\varpi}^2{\,}u
+12{\,}{R_\Gamma}^4{\,}{\varpi}^2{\,}u
+16{\,}{R_\Gamma}^3{\,}u^4
\nonumber \\ &&
+96{\,}{R_\Gamma}^5{\,}u^2
+64{\,}{R_\Gamma}^4{\,}u^3
+64{\,}{R_\Gamma}^6{\,}u
\nonumber \\ &&
+16{\,}{e_3}{\,}{R_\Gamma}^7
+21{\,}{\varpi}^6{\,}u
-12{\,}{\varpi}^2{\,}{R_\Gamma}^5
\nonumber \\ &&
+16{\,}{R_\Gamma}^7
+60{\,}{R_\Gamma}^3{\,}{\varpi}^2{\,}u^2
+36{\,}{R_\Gamma}^2{\,}{\varpi}^2{\,}u^3
\nonumber \\ &&
+42{\,}{R_\Gamma}{\,}{\varpi}^4{\,}u^2
+30{\,}{R_\Gamma}^2{\,}{\varpi}^4{\,}u
+16{\,}{e_3}{\,}{R_\Gamma}^3{\,}u^4
\nonumber \\ &&
-21{\,}{e_3}{\,}{\varpi}^6{\,}u
-12{\,}{e_3}{\,}{R_\Gamma}^5{\,}{\varpi}^2
+96{\,}{e_3}{\,}{R_\Gamma}^5{\,}u^2
\nonumber \\ &&
-36{\,}{e_3}{\,}{\varpi}^2{\,}{R_\Gamma}^2{\,}u^3
+42{\,}{e_3}{\,}{\varpi}^4{\,}{R_\Gamma}{\,}u^2
-30{\,}{e_3}{\,}{\varpi}^4{\,}{R_\Gamma}^2{\,}u
\nonumber \\ &&
+60{\,}{e_3}{\,}{R_\Gamma}^3{\,}{\varpi}^2{\,}u^2
-64{\,}{e_3}{\,}{R_\Gamma}^6{\,}u
-64{\,}{e_3}{\,}{R_\Gamma}^4{\,}u^3\Big)
\nonumber \\ &&
\cdot{e_2}/{\varpi}^8/{R_\Gamma}^5,
\\
    \gamma_3 &=& \pm{\,}6{\,}\Big(
-4{\,}{e_3}{\,}{R_\Gamma}^2{\,}u^3
+4{\,}{R_\Gamma}^2{\,}u^3
+6{\,}{\varpi}^2{\,}{R_\Gamma}^2{\,}u
\nonumber \\ &&
+12{\,}{R_\Gamma}^3{\,}u^2
+12{\,}{e_3}{\,}{R_\Gamma}^3{\,}u^2
+6{\,}{\varpi}^2{\,}{R_\Gamma}{\,}u^2
\nonumber \\ &&
+6{\,}{\varpi}^2{\,}{e_3}{\,}{R_\Gamma}{\,}u^2
-12{\,}{e_3}{\,}{R_\Gamma}^4{\,}u
-3{\,}{\varpi}^4{\,}{e_3}{\,}u
\nonumber \\ &&
+4{\,}{R_\Gamma}^5
+3{\,}{\varpi}^4{\,}u
-6{\,}{\varpi}^2{\,}{e_3}{\,}{R_\Gamma}^2{\,}u
\nonumber \\ &&
+4{\,}{e_3}{\,}{R_\Gamma}^5
+12{\,}{R_\Gamma}^4{\,}u\Big)
\cdot{e_2}/{\varpi}^6{\,}/{R_\Gamma}^5,
\\
    \gamma_4 &=& \mp{\,}48{\,}\Big(
-2{\,}{\varpi}^2{\,}{R_\Gamma}^5
-2{\,}{e_3}{\,}{R_\Gamma}^5{\,}{\varpi}^2
-6{\,}{e_3}{\,}{R_\Gamma}^4{\,}{\varpi}^2{\,}u
\nonumber \\ &&
-9{\,}{e_3}{\,}{\varpi}^4{\,}{R_\Gamma}^2{\,}u
+18{\,}{e_3}{\,}{R_\Gamma}^3{\,}{\varpi}^2{\,}u^2
-10{\,}{e_3}{\,}{\varpi}^2{\,}{R_\Gamma}^2{\,}u^3
\nonumber \\ &&
+12{\,}{e_3}{\,}{\varpi}^4{\,}{R_\Gamma}{\,}u^2
+18{\,}{R_\Gamma}^3{\,}{\varpi}^2{\,}u^2
+10{\,}{R_\Gamma}^2{\,}{\varpi}^2{\,}u^3
\nonumber \\ &&
+12{\,}{R_\Gamma}{\,}{\varpi}^4{\,}u^2
-6{\,}{e_3}{\,}{\varpi}^6{\,}u
+6{\,}{R_\Gamma}^4{\,}{\varpi}^2{\,}u
\nonumber \\ &&
+6{\,}{\varpi}^6{\,}u
+4{\,}{e_3}{\,}{R_\Gamma}^7
+24{\,}{R_\Gamma}^5{\,}u^2
\nonumber \\ &&
+4{\,}{R_\Gamma}^3{\,}u^4
+16{\,}{R_\Gamma}^6{\,}u
+16{\,}{R_\Gamma}^4{\,}u^3
\nonumber \\ &&
+24{\,}{e_3}{\,}{R_\Gamma}^5{\,}u^2
-16{\,}{e_3}{\,}{R_\Gamma}^6{\,}u
+4{\,}{e_3}{\,}{R_\Gamma}^3{\,}u^4
\nonumber \\ &&
+4{\,}{R_\Gamma}^7
-16{\,}{e_3}{\,}{R_\Gamma}^4{\,}u^3
+9{\,}{R_\Gamma}^2{\,}{\varpi}^4{\,}u\Big)
\nonumber \\ &&
\cdot{e_2}/{\varpi}^8/{R_\Gamma}^5,
\\
    \gamma_5 &=& -24{\,}\Big(
15{\,}u{\,}{\varpi}^8{\,}{e_3}
-9{\,}{\varpi}^6{\,}{R_\Gamma}^3{\,}{e_3}
+40{\,}{\varpi}^2{\,}{R_\Gamma}^7
\nonumber \\ &&
-54{\,}{R_\Gamma}^2{\,}{\varpi}^4{\,}u^3
-42{\,}u^2{\,}{R_\Gamma}{\,}{\varpi}^6
-40{\,}{R_\Gamma}^3{\,}{\varpi}^2{\,}u^4
\nonumber \\ &&
+6{\,}{\varpi}^8{\,}{R_\Gamma}
-80{\,}{R_\Gamma}^8{\,}u
-160{\,}{R_\Gamma}^6{\,}u^3
\nonumber \\ &&
+80{\,}{R_\Gamma}^6{\,}{\varpi}^2{\,}u
+6{\,}{R_\Gamma}^2{\,}{\varpi}^4{\,}{e_3}{\,}u^3
-6{\,}{\varpi}^4{\,}{R_\Gamma}^5{\,}{e_3}
\nonumber \\ &&
-18{\,}{R_\Gamma}^3{\,}{\varpi}^4{\,}{e_3}{\,}u^2
+27{\,}u{\,}{R_\Gamma}^2{\,}{\varpi}^6{\,}{e_3}
-160{\,}{R_\Gamma}^7{\,}u^2
\nonumber \\ &&
+18{\,}{R_\Gamma}^4{\,}{\varpi}^4{\,}u
-15{\,}u{\,}{\varpi}^8
-3{\,}u{\,}{R_\Gamma}^2{\,}{\varpi}^6
\nonumber \\ &&
-16{\,}{R_\Gamma}^9
-16{\,}{R_\Gamma}^4{\,}u^5
-80{\,}{R_\Gamma}^5{\,}u^4
\nonumber \\ &&
+9{\,}{\varpi}^6{\,}{R_\Gamma}^3
+6{\,}{\varpi}^4{\,}{R_\Gamma}^5
-6{\,}{\varpi}^8{\,}{R_\Gamma}{\,}{e_3}
\nonumber \\ &&
-80{\,}{R_\Gamma}^4{\,}{\varpi}^2{\,}u^3
-42{\,}{R_\Gamma}^3{\,}{\varpi}^4{\,}u^2
+18{\,}{R_\Gamma}^4{\,}{\varpi}^4{\,}u{\,}{e_3}
\nonumber \\ &&
-18{\,}u^2{\,}{R_\Gamma}{\,}{\varpi}^6{\,}{e_3}\Big)
\cdot{e_2}/{\varpi}^{10}{\,}/{R_\Gamma}^6,
\\
    \gamma_6 &=& 24{\,}\Big(
-48{\,}{\varpi}^6{\,}{R_\Gamma}^3{\,}{e_3}
+224{\,}{R_\Gamma}^6{\,}{\varpi}^2{\,}u
+8{\,}{\varpi}^2{\,}{R_\Gamma}^7{\,}{e_3}
\nonumber \\ &&
-264{\,}{R_\Gamma}^2{\,}{\varpi}^4{\,}u^3
-240{\,}{R_\Gamma}^3{\,}{\varpi}^4{\,}u^2
-416{\,}{R_\Gamma}^4{\,}{\varpi}^2{\,}u^3
\nonumber \\ &&
-24{\,}u{\,}{R_\Gamma}^2{\,}{\varpi}^6
-210{\,}u^2{\,}{R_\Gamma}{\,}{\varpi}^6
+75{\,}u{\,}{\varpi}^8{\,}{e_3}
\nonumber \\ &&
-320{\,}{R_\Gamma}^5{\,}u^4
-48{\,}{\varpi}^4{\,}{R_\Gamma}^5{\,}{e_3}
+72{\,}{R_\Gamma}^4{\,}{\varpi}^4{\,}u
\nonumber \\ &&
-184{\,}{R_\Gamma}^3{\,}{\varpi}^2{\,}u^4
-144{\,}{R_\Gamma}^5{\,}{\varpi}^2{\,}u^2
-96{\,}{R_\Gamma}^3{\,}{\varpi}^4{\,}{e_3}{\,}u^2
\nonumber \\ &&
-32{\,}{R_\Gamma}^6{\,}{\varpi}^2{\,}u{\,}{e_3}
+8{\,}{R_\Gamma}^3{\,}{\varpi}^2{\,}{e_3}{\,}u^4
-90{\,}u^2{\,}{R_\Gamma}{\,}{\varpi}^6{\,}{e_3}
\nonumber \\ &&
+120{\,}{R_\Gamma}^4{\,}{\varpi}^4{\,}u{\,}{e_3}
+144{\,}u{\,}{R_\Gamma}^2{\,}{\varpi}^6{\,}{e_3}
+48{\,}{R_\Gamma}^5{\,}{\varpi}^2{\,}u^2{\,}{e_3}
\nonumber \\ &&
-32{\,}{R_\Gamma}^4{\,}{\varpi}^2{\,}{e_3}{\,}u^3
+24{\,}{R_\Gamma}^2{\,}{\varpi}^4{\,}{e_3}{\,}u^3
+48{\,}{\varpi}^6{\,}{R_\Gamma}^3
\nonumber \\ &&
+30{\,}{\varpi}^8{\,}{R_\Gamma}
-64{\,}{R_\Gamma}^4{\,}u^5
-640{\,}{R_\Gamma}^6{\,}u^3
\nonumber \\ &&
-640{\,}{R_\Gamma}^7{\,}u^2
-30{\,}{\varpi}^8{\,}{R_\Gamma}{\,}{e_3}
+136{\,}{\varpi}^2{\,}{R_\Gamma}^7
\nonumber \\ &&
+48{\,}{\varpi}^4{\,}{R_\Gamma}^5
-64{\,}{R_\Gamma}^9
-320{\,}{R_\Gamma}^8{\,}u
\nonumber \\ &&
-75{\,}u{\,}{\varpi}^8\Big)
\cdot{e_2}/{\varpi}^{10}{\,}/{R_\Gamma}^6,
\\
    \gamma_7 &=& \pm{\,}6/{\varpi}^8{\,}\Big(
-48{\,}{R_\Gamma}^2{\,}{\varpi}^2{\,}u^3
-6{\,}{\varpi}^6{\,}{R_\Gamma}{\,}{e_3}
-12{\,}{\varpi}^4{\,}{R_\Gamma}^3{\,}{e_3}
\nonumber \\ &&
-72{\,}{R_\Gamma}^3{\,}{\varpi}^2{\,}u^2
-42{\,}{R_\Gamma}{\,}{\varpi}^4{\,}u^2
-32{\,}{e_3}{\,}{R_\Gamma}^4{\,}u^3
\nonumber \\ &&
-32{\,}{e_3}{\,}{R_\Gamma}^6{\,}u
+15{\,}{e_3}{\,}{\varpi}^6{\,}u
+8{\,}{e_3}{\,}{R_\Gamma}^3{\,}u^4
\nonumber \\ &&
+48{\,}{e_3}{\,}{R_\Gamma}^5{\,}u^2
-96{\,}{R_\Gamma}^4{\,}u^3
-15{\,}{\varpi}^6{\,}u
\nonumber \\ &&
+6{\,}{\varpi}^6{\,}{R_\Gamma}
+48{\,}{e_3}{\,}{R_\Gamma}^4{\,}{\varpi}^2{\,}u
-24{\,}{R_\Gamma}^7
\nonumber \\ &&
+12{\,}{\varpi}^4{\,}{R_\Gamma}^3
-24{\,}{e_3}{\,}{R_\Gamma}^3{\,}{\varpi}^2{\,}u^2
-96{\,}{R_\Gamma}^6{\,}u
\nonumber \\ &&
-144{\,}{R_\Gamma}^5{\,}u^2
+8{\,}{e_3}{\,}{R_\Gamma}^7
-24{\,}{R_\Gamma}^3{\,}u^4
\nonumber \\ &&
+36{\,}{e_3}{\,}{\varpi}^4{\,}{R_\Gamma}^2{\,}u
-18{\,}{e_3}{\,}{\varpi}^4{\,}{R_\Gamma}{\,}u^2
+24{\,}{\varpi}^2{\,}{R_\Gamma}^5
\nonumber \\ &&
-24{\,}{e_3}{\,}{R_\Gamma}^5{\,}{\varpi}^2
-12{\,}{R_\Gamma}^2{\,}{\varpi}^4{\,}u\Big)
\cdot{e_2}/{R_\Gamma}^6,
\\
    \gamma_8 &=& \mp{\,}24{\,}\Big(
56{\,}{\varpi}^2{\,}{R_\Gamma}^7
+12{\,}{R_\Gamma}^2{\,}{\varpi}^4{\,}{e_3}{\,}u^3
+90{\,}u{\,}{R_\Gamma}^2{\,}{\varpi}^6{\,}{e_3}
\nonumber \\ &&
+84{\,}{R_\Gamma}^4{\,}{\varpi}^4{\,}u{\,}{e_3}
-54{\,}u^2{\,}{R_\Gamma}{\,}{\varpi}^6{\,}{e_3}
+8{\,}{R_\Gamma}^3{\,}{\varpi}^2{\,}{e_3}{\,}u^4
\nonumber \\ &&
-32{\,}{R_\Gamma}^4{\,}{\varpi}^2{\,}{e_3}{\,}u^3
+48{\,}{R_\Gamma}^5{\,}{\varpi}^2{\,}u^2{\,}{e_3}
-60{\,}{R_\Gamma}^3{\,}{\varpi}^4{\,}{e_3}{\,}u^2
\nonumber \\ &&
-32{\,}{R_\Gamma}^6{\,}{\varpi}^2{\,}u{\,}{e_3}
+36{\,}{\varpi}^4{\,}{R_\Gamma}^5
-32{\,}{R_\Gamma}^9
\nonumber \\ &&
-36{\,}{\varpi}^4{\,}{R_\Gamma}^5{\,}{e_3}
-30{\,}{\varpi}^6{\,}{R_\Gamma}^3{\,}{e_3}
+36{\,}{R_\Gamma}^4{\,}{\varpi}^4{\,}u
\nonumber \\ &&
+64{\,}{R_\Gamma}^6{\,}{\varpi}^2{\,}u
-156{\,}{R_\Gamma}^3{\,}{\varpi}^4{\,}u^2
-256{\,}{R_\Gamma}^4{\,}{\varpi}^2{\,}u^3
\nonumber \\ &&
-18{\,}u{\,}{R_\Gamma}^2{\,}{\varpi}^6
-126{\,}u^2{\,}{R_\Gamma}{\,}{\varpi}^6
+45{\,}u{\,}{\varpi}^8{\,}{e_3}
\nonumber \\ &&
+8{\,}{\varpi}^2{\,}{R_\Gamma}^7{\,}{e_3}
-104{\,}{R_\Gamma}^3{\,}{\varpi}^2{\,}u^4
-156{\,}{R_\Gamma}^2{\,}{\varpi}^4{\,}u^3
\nonumber \\ &&
-320{\,}{R_\Gamma}^7{\,}u^2
-18{\,}{\varpi}^8{\,}{R_\Gamma}{\,}{e_3}
-32{\,}{R_\Gamma}^4{\,}u^5
\nonumber \\ &&
+30{\,}{\varpi}^6{\,}{R_\Gamma}^3
-144{\,}{R_\Gamma}^5{\,}{\varpi}^2{\,}u^2
-160{\,}{R_\Gamma}^5{\,}u^4
\nonumber \\ &&
-45{\,}u{\,}{\varpi}^8
-320{\,}{R_\Gamma}^6{\,}u^3
-160{\,}{R_\Gamma}^8{\,}u
\nonumber \\ &&
+18{\,}{\varpi}^8{\,}{R_\Gamma}\Big)
\cdot{e_2}/{\varpi}^{10}{\,}/{R_\Gamma}^6,
\\
    \gamma_9 &=& 6{\,}\Big(
-6{\,}{e_3}{\,}{\varpi}^4{\,}u^2
+12{\,}{\varpi}^2{\,}u^3{\,}{R_\Gamma}
+3{\,}{\varpi}^6{\,}{e_3}
\nonumber \\ &&
-48{\,}{R_\Gamma}^4{\,}{e_3}{\,}u^2
+6{\,}{\varpi}^4{\,}u^2
+12{\,}{\varpi}^2{\,}{R_\Gamma}^4{\,}{e_3}
\nonumber \\ &&
-8{\,}{R_\Gamma}^2{\,}{e_3}{\,}u^4
+48{\,}{R_\Gamma}^4{\,}u^2
+32{\,}{R_\Gamma}^3{\,}{e_3}{\,}u^3
\nonumber \\ &&
+32{\,}u{\,}{R_\Gamma}^5{\,}{e_3}
+32{\,}u{\,}{R_\Gamma}^5
+6{\,}{\varpi}^4{\,}{R_\Gamma}^2{\,}{e_3}
\nonumber \\ &&
-12{\,}{\varpi}^2{\,}u{\,}{R_\Gamma}^3
-6{\,}u{\,}{\varpi}^4{\,}{R_\Gamma}
-12{\,}{\varpi}^2{\,}{R_\Gamma}^4
\nonumber \\ &&
-3{\,}{\varpi}^6
+8{\,}{R_\Gamma}^6
-12{\,}{\varpi}^2{\,}u{\,}{R_\Gamma}^3{\,}{e_3}
\nonumber \\ &&
-6{\,}{R_\Gamma}^2{\,}{\varpi}^4
+8{\,}{R_\Gamma}^2{\,}u^4
-6{\,}{\varpi}^4{\,}u{\,}{e_3}{\,}{R_\Gamma}
\nonumber \\ &&
+12{\,}{\varpi}^2{\,}u^3{\,}{R_\Gamma}{\,}{e_3}
-12{\,}{\varpi}^2{\,}u^2{\,}{R_\Gamma}^2{\,}{e_3}
-8{\,}{R_\Gamma}^6{\,}{e_3}
\nonumber \\ &&
+32{\,}{R_\Gamma}^3{\,}u^3
+12{\,}{\varpi}^2{\,}u^2{\,}{R_\Gamma}^2)
\cdot{e_2}/{\varpi}^8/{R_\Gamma}^5{\,},
\\
    \gamma_{10} &=& 3{\,}\Big(
-6{\,}u^2{\,}{\varpi}^2{\,}{R_\Gamma}
-6{\,}{\varpi}^2{\,}u{\,}{R_\Gamma}^2
-4{\,}{e_3}{\,}{R_\Gamma}^5
\nonumber \\ &&
-12{\,}u{\,}{R_\Gamma}^4
-12{\,}u^2{\,}{R_\Gamma}^3
-3{\,}u{\,}{\varpi}^4
\nonumber \\ &&
-4{\,}u^3{\,}{R_\Gamma}^2
-6{\,}{e_3}{\,}u^2{\,}{\varpi}^2{\,}{R_\Gamma}
-12{\,}{e_3}{\,}u^2{\,}{R_\Gamma}^3
\nonumber \\ &&
+3{\,}{e_3}{\,}u{\,}{\varpi}^4
+4{\,}{e_3}{\,}u^3{\,}{R_\Gamma}^2
-4{\,}{R_\Gamma}^5
\nonumber \\ &&
+12{\,}{e_3}{\,}u{\,}{R_\Gamma}^4
+6{\,}{e_3}{\,}{\varpi}^2{\,}u{\,}{R_\Gamma}^2\Big)
\cdot{e_2}/{R_\Gamma}^4/{\varpi}^6{\,},
\\
    \gamma_{11} &=& 3{\,}\Big(
6{\,}{\varpi}^4{\,}{R_\Gamma}^3
-24{\,}{e_3}{\,}{\varpi}^2{\,}u^3{\,}{R_\Gamma}^2 
+18 {\,}{e_3}{\,}u^2{\,}{\varpi}^4{\,}{R_\Gamma}
\nonumber \\ &&
+16{\,}{e_3}{\,}{R_\Gamma}^7
-64{\,}{e_3} {\,}u^3{\,}{R_\Gamma} ^4
+24{\,}{e_3}{\,}u{\,}{R_\Gamma}^4{\,}{\varpi}^2
\nonumber \\ &&
+24{\,}{e_3}{\,}u^2 {\,}{\varpi}^2{\,}{R_\Gamma}^3 
-6{\,}{e_3}{\,}u{\,}{\varpi}^6
-64{\,}{e_3}{\,}u{\,}{R_\Gamma}^6
\nonumber \\ &&
-3 {\,}{e_3}{\,}{R_\Gamma}{\,}{\varpi}^6 
+16{\,}{e_3}{\,}u^4{\,}{R_\Gamma}^3
-6{\,}{e_3} {\,}{\varpi}^4{\,}{R_\Gamma}^3
\nonumber \\ &&
+96{\,}{e_3}{\,}u ^2{\,}{R_\Gamma}^5
-24{\,}{e_3}{\,}{\varpi}^2 {\,}{R_\Gamma}^5
+6{\,}u{\,}{\varpi}^6
\nonumber \\ &&
+3{\,}{R_\Gamma}{\,}{\varpi}^6 
+6{\,}u^2{\,}{\varpi}^4{\,}{R_\Gamma}
+12 {\,}{\varpi}^4{\,}u{\,}{R_\Gamma}^2\Big)
\nonumber \\ &&
\cdot{e_2}/{\varpi}^8/{R_\Gamma}^5{\,}.
\end{eqnarray}

In addition, the functions $\gamma_{12\ldots 17}$ used in expressing $J,U$ and $W$
are given by:
\begin{eqnarray}
    \gamma_{12} &=& 2{\,}\Big(
270{\,}{R_\Gamma}^{8}{\,}{\varpi}^{18}{\,}{e_1}{\,}u
-720{\,}{R_\Gamma}^{10}{\,}{\varpi}^{12}{\,}{e_1}{\,}u^{5}
\nonumber \\ &&
-135{\,}{R_\Gamma}^{6}{\,}{\varpi}^{20}{\,}{e_1}{\,}u
+3600{\,}{R_\Gamma}^{11}{\,}{\varpi}^{12}{\,}{e_1}{\,}u^{4}
\nonumber \\ &&
+7920{\,}{R_\Gamma}^{12}{\,}{\varpi}^{14}{\,}{e_1}{\,}u
+720{\,}{R_\Gamma}^{15}{\,}{\varpi}^{12}{\,}{e_2}
\nonumber \\ &&
+1080{\,}{R_\Gamma}^{11}{\,}{\varpi}^{16}{\,}{e_2}
+720{\,}{R_\Gamma}^{10}{\,}{\varpi}^{12}{\,}{e_2}{\,}u^{5}
\nonumber \\ &&
+3600{\,}{R_\Gamma}^{14}{\,}{\varpi}^{12}{\,}{e_2}{\,}u
+3600{\,}{R_\Gamma}^{11}{\,}{\varpi}^{12}{\,}{e_2}{\,}u^{4}
\nonumber \\ &&
+7200{\,}{R_\Gamma}^{12}{\,}{\varpi}^{12}{\,}{e_2}{\,}u^{3}
+7200{\,}{R_\Gamma}^{13}{\,}{\varpi}^{12}{\,}{e_2}{\,}u^{2}
\nonumber \\ &&
-7920{\,}{R_\Gamma}^{12}{\,}{\varpi}^{14}{\,}{e_2}{\,}u
+1080{\,}{R_\Gamma}^{11}{\,}{\varpi}^{16}{\,}{e_1}
\nonumber \\ &&
+540{\,}{R_\Gamma}^{8}{\,}{\varpi}^{16}{\,}{e_2}{\,}u^{3}
-540{\,}{R_\Gamma}^{9}{\,}{\varpi}^{16}{\,}{e_2}{\,}u^{2}
\nonumber \\ &&
-6480{\,}{R_\Gamma}^{11}{\,}{\varpi}^{14}{\,}{e_2}{\,}u^{2}
-720{\,}{R_\Gamma}^{10}{\,}{\varpi}^{14}{\,}{e_2}{\,}u^{3}
\nonumber \\ &&
+270{\,}{R_\Gamma}^{7}{\,}{\varpi}^{18}{\,}{e_2}{\,}u^{2}
+720{\,}{R_\Gamma}^{9}{\,}{\varpi}^{14}{\,}{e_2}{\,}u^{4}
\nonumber \\ &&
+135{\,}{R_\Gamma}^{6}{\,}{\varpi}^{20}{\,}{e_2}{\,}u
+270{\,}{R_\Gamma}^{7}{\,}{\varpi}^{18}{\,}{e_1}{\,}u^{2}
\nonumber \\ &&
-270{\,}{R_\Gamma}^{8}{\,}{\varpi}^{18}{\,}{e_2}{\,}u
-6480{\,}{R_\Gamma}^{11}{\,}{\varpi}^{14}{\,}{e_1}{\,}u^{2}
\nonumber \\ &&
+720{\,}{R_\Gamma}^{10}{\,}{\varpi}^{14}{\,}{e_1}{\,}u^{3}
+720{\,}{R_\Gamma}^{9}{\,}{\varpi}^{14}{\,}{e_1}{\,}u^{4}
\nonumber \\ &&
-540{\,}{R_\Gamma}^{8}{\,}{\varpi}^{16}{\,}{e_1}{\,}u^{3}
-540{\,}{R_\Gamma}^{9}{\,}{\varpi}^{16}{\,}{e_1}{\,}u^{2}
\nonumber \\ &&
-3600{\,}{R_\Gamma}^{14}{\,}{\varpi}^{12}{\,}{e_1}{\,}u
+7200{\,}{R_\Gamma}^{13}{\,}{\varpi}^{12}{\,}{e_1}{\,}u^{2}
\nonumber \\ &&
-7200{\,}{R_\Gamma}^{12}{\,}{\varpi}^{12}{\,}{e_1}{\,}u^{3}
+720{\,}{R_\Gamma}^{15}{\,}{\varpi}^{12}{\,}{e_1}
\nonumber \\ &&
-2880{\,}{R_\Gamma}^{13}{\,}{\varpi}^{14}{\,}{e_1}
-2880{\,}{R_\Gamma}^{13}{\,}{\varpi}^{14}{\,}{e_2}\Big)
\nonumber \\ &&
/{R_\Gamma}^{11}/{\varpi}^{22}/15,
\\
    \gamma_{13} &=& 2{\,}\Big(
18000{\,}{\varpi}^{12}{\,}{R_\Gamma}^{14}{\,}{e_2}
-16560{\,}{\varpi}^{14}{\,}{R_\Gamma}^{12}{\,}{e_2}
\nonumber \\ &&
+135{\,}{\varpi}^{20}{\,}{R_\Gamma}^{6}{\,}{e_2}
-2160{\,}{\varpi}^{16}{\,}{R_\Gamma}^{10}{\,}{e_2}
\nonumber \\ &&
-2160{\,}{\varpi}^{16}{\,}{R_\Gamma}^{10}{\,}{e_1}
-2880{\,}{\varpi}^{10}{\,}{R_\Gamma}^{16}{\,}{e_2}
\nonumber \\ &&
-3600{\,}{\varpi}^{12}{\,}{R_\Gamma}^{10}{\,}{e_1}{\,}u^{4}
+3600{\,}{\varpi}^{12}{\,}{R_\Gamma}^{13}{\,}{e_1}{\,}u
\nonumber \\ &&
-10800{\,}{\varpi}^{14}{\,}{R_\Gamma}^{11}{\,}{e_1}{\,}u
+7200{\,}{\varpi}^{12}{\,}{R_\Gamma}^{11}{\,}{e_1}{\,}u^{3}
\nonumber \\ &&
+720{\,}{\varpi}^{12}{\,}{R_\Gamma}^{9}{\,}{e_1}{\,}u^{5}
-1350{\,}{\varpi}^{18}{\,}{R_\Gamma}^{7}{\,}{e_1}{\,}u
\nonumber \\ &&
+1080{\,}{\varpi}^{18}{\,}{R_\Gamma}^{6}{\,}{e_1}{\,}u^{2}
-405{\,}{\varpi}^{20}{\,}{R_\Gamma}^{5}{\,}{e_1}{\,}u
\nonumber \\ &&
-7200{\,}{\varpi}^{12}{\,}{R_\Gamma}^{12}{\,}{e_1}{\,}u^{2}
+3600{\,}{\varpi}^{14}{\,}{R_\Gamma}^{12}{\,}{e_1}
\nonumber \\ &&
-57600{\,}{\varpi}^{10}{\,}{R_\Gamma}^{13}{\,}{e_2}{\,}u^{3}
-19440{\,}{\varpi}^{14}{\,}{R_\Gamma}^{11}{\,}{e_2}{\,}u
\nonumber \\ &&
-43200{\,}{\varpi}^{10}{\,}{R_\Gamma}^{14}{\,}{e_2}{\,}u^{2}
+3600{\,}{\varpi}^{12}{\,}{R_\Gamma}^{10}{\,}{e_2}{\,}u^{4}
\nonumber \\ &&
+540{\,}{\varpi}^{18}{\,}{R_\Gamma}^{6}{\,}{e_2}{\,}u^{2}
-2160{\,}{\varpi}^{14}{\,}{R_\Gamma}^{8}{\,}{e_2}{\,}u^{4}
\nonumber \\ &&
+4860{\,}{\varpi}^{16}{\,}{R_\Gamma}^{8}{\,}{e_2}{\,}u^{2}
-2880{\,}{\varpi}^{10}{\,}{R_\Gamma}^{10}{\,}{e_2}{\,}u^{6}
\nonumber \\ &&
-540{\,}{\varpi}^{16}{\,}{R_\Gamma}^{7}{\,}{e_2}{\,}u^{3}
-17280{\,}{\varpi}^{10}{\,}{R_\Gamma}^{15}{\,}{e_2}{\,}u
\nonumber \\ &&
+9360{\,}{\varpi}^{14}{\,}{R_\Gamma}^{9}{\,}{e_2}{\,}u^{3}
+68400{\,}{\varpi}^{12}{\,}{R_\Gamma}^{13}{\,}{e_2}{\,}u
\nonumber \\ &&
-2160{\,}{\varpi}^{16}{\,}{R_\Gamma}^{9}{\,}{e_2}{\,}u
-720{\,}{\varpi}^{12}{\,}{R_\Gamma}^{14}{\,}{e_1}
\nonumber \\ &&
-43200{\,}{\varpi}^{10}{\,}{R_\Gamma}^{12}{\,}{e_2}{\,}u^{4}
+93600{\,}{\varpi}^{12}{\,}{R_\Gamma}^{12}{\,}{e_2}{\,}u^{2}
\nonumber \\ &&
+8640{\,}{\varpi}^{14}{\,}{R_\Gamma}^{10}{\,}{e_2}{\,}u^{2}
+1890{\,}{\varpi}^{18}{\,}{R_\Gamma}^{7}{\,}{e_2}{\,}u
\nonumber \\ &&
+405{\,}{\varpi}^{20}{\,}{R_\Gamma}^{5}{\,}{e_2}{\,}u
-17280{\,}{\varpi}^{10}{\,}{R_\Gamma}^{11}{\,}{e_2}{\,}u^{5}
\nonumber \\ &&
+1080{\,}{\varpi}^{16}{\,}{R_\Gamma}^{9}{\,}{e_1}{\,}u
-540{\,}{\varpi}^{16}{\,}{R_\Gamma}^{7}{\,}{e_1}{\,}u^{3}
\nonumber \\ &&
+1620{\,}{\varpi}^{16}{\,}{R_\Gamma}^{8}{\,}{e_1}{\,}u^{2}
-3600{\,}{\varpi}^{14}{\,}{R_\Gamma}^{9}{\,}{e_1}{\,}u^{3}
\nonumber \\ &&
+10800{\,}{\varpi}^{14}{\,}{R_\Gamma}^{10}{\,}{e_1}{\,}u^{2}
-3600{\,}{\varpi}^{12}{\,}{R_\Gamma}^{9}{\,}{e_2}{\,}u^{5}
\nonumber \\ &&
+50400{\,}{\varpi}^{12}{\,}{R_\Gamma}^{11}{\,}{e_2}{\,}u^{3}
-135{\,}{\varpi}^{20}{\,}{R_\Gamma}^{6}{\,}{e_1}
\nonumber \\ &&
+270{\,}{\varpi}^{18}{\,}{R_\Gamma}^{8}{\,}{e_1}
-270{\,}{\varpi}^{18}{\,}{R_\Gamma}^{8}{\,}{e_2}\Big)
\nonumber \\ &&
/15/{R_\Gamma}^{11}/{\varpi}^{22},\\
    \gamma_{14} &=& 
-2{\,}\Big(
-540{\,}{R_\Gamma}^{11}{\,}{\varpi}^{14}{\,}{e_2}
-1620{\,}{R_\Gamma}^{9}{\,}{\varpi}^{14}{\,}{e_1}{\,}u^{2}
\nonumber \\ &&
+540{\,}{R_\Gamma}^{8}{\,}{\varpi}^{14}{\,}{e_1}{\,}u^{3}
+1620{\,}{R_\Gamma}^{10}{\,}{\varpi}^{14}{\,}{e_1}{\,}u
\nonumber \\ &&
-540{\,}{R_\Gamma}^{8}{\,}{\varpi}^{14}{\,}{e_2}{\,}u^{3}
-1620{\,}{R_\Gamma}^{9}{\,}{\varpi}^{14}{\,}{e_2}{\,}u^{2}
\nonumber \\ &&
-1620{\,}{R_\Gamma}^{10}{\,}{\varpi}^{14}{\,}{e_2}{\,}u
-810{\,}{R_\Gamma}^{7}{\,}{\varpi}^{16}{\,}{e_2}{\,}u^{2}
\nonumber \\ &&
-810{\,}{R_\Gamma}^{7}{\,}{\varpi}^{16}{\,}{e_1}{\,}u^{2}
-810{\,}{R_\Gamma}^{8}{\,}{\varpi}^{16}{\,}{e_2}{\,}u
\nonumber \\ &&
-405{\,}{R_\Gamma}^{6}{\,}{\varpi}^{18}{\,}{e_2}{\,}u
+810{\,}{R_\Gamma}^{8}{\,}{\varpi}^{16}{\,}{e_1}{\,}u
\nonumber \\ &&
-540{\,}{R_\Gamma}^{11}{\,}{\varpi}^{14}{\,}{e_1}
+405{\,}{R_\Gamma}^{6}{\,}{\varpi}^{18}{\,}{e_1}{\,}u\Big)
\nonumber \\ &&
/{R_\Gamma}^{12}/{\varpi}^{20}/15,
\\
    \gamma_{15} &=& 
-2{\,}\Big(
-14400{\,}{R_\Gamma}^{12}{\,}{\varpi}^{10}{\,}{e_2}{\,}u^{2}
-270{\,}{R_\Gamma}^{6}{\,}{\varpi}^{16}{\,}{e_2}{\,}u^{2}
\nonumber \\ &&
-7200{\,}{R_\Gamma}^{10}{\,}{\varpi}^{10}{\,}{e_2}{\,}u^{4}
+5040{\,}{R_\Gamma}^{12}{\,}{\varpi}^{12}{\,}{e_2}
\nonumber \\ &&
-1440{\,}{R_\Gamma}^{9}{\,}{\varpi}^{10}{\,}{e_2}{\,}u^{5}
-14400{\,}{R_\Gamma}^{11}{\,}{\varpi}^{10}{\,}{e_2}{\,}u^{3}
\nonumber \\ &&
-1440{\,}{R_\Gamma}^{9}{\,}{\varpi}^{12}{\,}{e_2}{\,}u^{3}
+8640{\,}{R_\Gamma}^{10}{\,}{\varpi}^{12}{\,}{e_2}{\,}u^{2}
\nonumber \\ &&
-2160{\,}{R_\Gamma}^{8}{\,}{\varpi}^{12}{\,}{e_2}{\,}u^{4}
+12960{\,}{R_\Gamma}^{11}{\,}{\varpi}^{12}{\,}{e_2}{\,}u
\nonumber \\ &&
-1440{\,}{R_\Gamma}^{14}{\,}{\varpi}^{10}{\,}{e_2}
+7200{\,}{R_\Gamma}^{10}{\,}{\varpi}^{10}{\,}{e_1}{\,}u^{4}
\nonumber \\ &&
+1440{\,}{R_\Gamma}^{14}{\,}{\varpi}^{10}{\,}{e_1}
+21600{\,}{R_\Gamma}^{11}{\,}{\varpi}^{12}{\,}{e_1}{\,}u
\nonumber \\ &&
+7200{\,}{R_\Gamma}^{9}{\,}{\varpi}^{12}{\,}{e_1}{\,}u^{3}
-1440{\,}{R_\Gamma}^{9}{\,}{\varpi}^{10}{\,}{e_1}{\,}u^{5}
\nonumber \\ &&
+1080{\,}{R_\Gamma}^{7}{\,}{\varpi}^{16}{\,}{e_2}{\,}u
-14400{\,}{R_\Gamma}^{11}{\,}{\varpi}^{10}{\,}{e_1}{\,}u^{3}
\nonumber \\ &&
-21600{\,}{R_\Gamma}^{10}{\,}{\varpi}^{12}{\,}{e_1}{\,}u^{2}
-5400{\,}{R_\Gamma}^{9}{\,}{\varpi}^{14}{\,}{e_1}{\,}u
\nonumber \\ &&
-7200{\,}{R_\Gamma}^{13}{\,}{\varpi}^{10}{\,}{e_1}{\,}u
+14400{\,}{R_\Gamma}^{12}{\,}{\varpi}^{10}{\,}{e_1}{\,}u^{2}
\nonumber \\ &&
+5400{\,}{R_\Gamma}^{10}{\,}{\varpi}^{14}{\,}{e_1}
-7200{\,}{R_\Gamma}^{12}{\,}{\varpi}^{12}{\,}{e_1}
\nonumber \\ &&
+270{\,}{R_\Gamma}^{6}{\,}{\varpi}^{16}{\,}{e_1}{\,}u^{2}
-540{\,}{R_\Gamma}^{7}{\,}{\varpi}^{16}{\,}{e_1}{\,}u
\nonumber \\ &&
-7200{\,}{R_\Gamma}^{13}{\,}{\varpi}^{10}{\,}{e_2}{\,}u
-135{\,}{R_\Gamma}^{6}{\,}{\varpi}^{18}{\,}{e_1}
\nonumber \\ &&
-270{\,}{R_\Gamma}^{8}{\,}{\varpi}^{16}{\,}{e_2}
+4320{\,}{R_\Gamma}^{9}{\,}{\varpi}^{14}{\,}{e_2}{\,}u
\nonumber \\ &&
-1080{\,}{R_\Gamma}^{7}{\,}{\varpi}^{14}{\,}{e_2}{\,}u^{3}
+3240{\,}{R_\Gamma}^{8}{\,}{\varpi}^{14}{\,}{e_2}{\,}u^{2}
\nonumber \\ &&
+270{\,}{R_\Gamma}^{8}{\,}{\varpi}^{16}{\,}{e_1}
+135{\,}{R_\Gamma}^{6}{\,}{\varpi}^{18}{\,}{e_2}\Big)
\nonumber \\ &&
/15/{R_\Gamma}^{12}/{\varpi}^{20},
\\
    \gamma_{16} &=& 
-\Big(
1080{\,}{R_\Gamma}^{8}{\,}{\varpi}^{12}{\,}{e_2}{\,}u^{3}
+810{\,}{R_\Gamma}^{7}{\,}{\varpi}^{14}{\,}{e_2}{\,}u^{2}
\nonumber \\ &&
+4320{\,}{R_\Gamma}^{11}{\,}{\varpi}^{10}{\,}{e_2}{\,}u^{2}
-1080{\,}{R_\Gamma}^{10}{\,}{\varpi}^{12}{\,}{e_2}{\,}u
\nonumber \\ &&
+270{\,}{R_\Gamma}^{7}{\,}{\varpi}^{14}{\,}{e_1}{\,}u^{2}
+720{\,}{R_\Gamma}^{9}{\,}{\varpi}^{10}{\,}{e_2}{\,}u^{4}
\nonumber \\ &&
+2880{\,}{R_\Gamma}^{12}{\,}{\varpi}^{10}{\,}{e_2}{\,}u
+135{\,}{R_\Gamma}^{7}{\,}{\varpi}^{16}{\,}{e_1}
\nonumber \\ &&
+270{\,}{R_\Gamma}^{6}{\,}{\varpi}^{16}{\,}{e_2}{\,}u
-270{\,}{R_\Gamma}^{6}{\,}{\varpi}^{16}{\,}{e_1}{\,}u
\nonumber \\ &&
+1080{\,}{R_\Gamma}^{9}{\,}{\varpi}^{12}{\,}{e_2}{\,}u^{2}
-540{\,}{R_\Gamma}^{8}{\,}{\varpi}^{14}{\,}{e_1}{\,}u
\nonumber \\ &&
+2880{\,}{R_\Gamma}^{10}{\,}{\varpi}^{10}{\,}{e_2}{\,}u^{3}
+720{\,}{R_\Gamma}^{13}{\,}{\varpi}^{10}{\,}{e_2}
\nonumber \\ &&
-1080{\,}{R_\Gamma}^{11}{\,}{\varpi}^{12}{\,}{e_2}
-135{\,}{R_\Gamma}^{7}{\,}{\varpi}^{16}{\,}{e_2}
\nonumber \\ &&
-270{\,}{R_\Gamma}^{9}{\,}{\varpi}^{14}{\,}{e_2}
+270{\,}{R_\Gamma}^{9}{\,}{\varpi}^{14}{\,}{e_1}\Big)
\nonumber \\ &&
/{\varpi}^{18}/{R_\Gamma}^{12}/15
\\
    \gamma_{17} &=& \Big(
-27{\,}{R_\Gamma}^{6}{\,}{\varpi}^{16}{\,}{e_1}
+48{\,}{R_\Gamma}^{8}{\,}{\varpi}^{10}{\,}{e_2}{\,}u^{4}
\nonumber \\ &&
-288{\,}{R_\Gamma}^{9}{\,}{\varpi}^{12}{\,}{e_2}{\,}u
+288{\,}{R_\Gamma}^{10}{\,}{\varpi}^{10}{\,}{e_2}{\,}u^{2}
\nonumber \\ &&
-144{\,}{R_\Gamma}^{8}{\,}{\varpi}^{12}{\,}{e_2}{\,}u^{2}
-18{\,}{R_\Gamma}^{6}{\,}{\varpi}^{14}{\,}{e_2}{\,}u^{2}
\nonumber \\ &&
-36{\,}{R_\Gamma}^{7}{\,}{\varpi}^{14}{\,}{e_2}{\,}u
+192{\,}{R_\Gamma}^{11}{\,}{\varpi}^{10}{\,}{e_2}{\,}u
\nonumber \\ &&
+192{\,}{R_\Gamma}^{9}{\,}{\varpi}^{10}{\,}{e_2}{\,}u^{3}
+90{\,}{R_\Gamma}^{6}{\,}{\varpi}^{14}{\,}{e_1}{\,}u^{2}
\nonumber \\ &&
+144{\,}{R_\Gamma}^{7}{\,}{\varpi}^{14}{\,}{e_1}{\,}u
-18{\,}{R_\Gamma}^{5}{\,}{\varpi}^{16}{\,}{e_1}{\,}u
\nonumber \\ &&
-216{\,}{R_\Gamma}^{7}{\,}{\varpi}^{12}{\,}{e_1}{\,}u^{3}
-216{\,}{R_\Gamma}^{8}{\,}{\varpi}^{12}{\,}{e_1}{\,}u^{2}
\nonumber \\ &&
+1920{\,}{R_\Gamma}^{12}{\,}{\varpi}^{8}{\,}{e_1}{\,}u^{2}
-192{\,}{R_\Gamma}^{9}{\,}{\varpi}^{8}{\,}{e_1}{\,}u^{5}
\nonumber \\ &&
+1728{\,}{R_\Gamma}^{11}{\,}{\varpi}^{10}{\,}{e_1}{\,}u
+18{\,}{R_\Gamma}^{5}{\,}{\varpi}^{16}{\,}{e_2}{\,}u
\nonumber \\ &&
+288{\,}{R_\Gamma}^{8}{\,}{\varpi}^{10}{\,}{e_1}{\,}u^{4}
+960{\,}{R_\Gamma}^{10}{\,}{\varpi}^{8}{\,}{e_1}{\,}u^{4}
\nonumber \\ &&
-960{\,}{R_\Gamma}^{13}{\,}{\varpi}^{8}{\,}{e_1}{\,}u
+360{\,}{R_\Gamma}^{9}{\,}{\varpi}^{12}{\,}{e_1}{\,}u
\nonumber \\ &&
-1920{\,}{R_\Gamma}^{11}{\,}{\varpi}^{8}{\,}{e_1}{\,}u^{3}
-192{\,}{R_\Gamma}^{9}{\,}{\varpi}^{10}{\,}{e_1}{\,}u^{3}
\nonumber \\ &&
-1152{\,}{R_\Gamma}^{10}{\,}{\varpi}^{10}{\,}{e_1}{\,}u^{2}
+192{\,}{R_\Gamma}^{14}{\,}{\varpi}^{8}{\,}{e_1}
\nonumber \\ &&
+72{\,}{R_\Gamma}^{10}{\,}{\varpi}^{12}{\,}{e_1}
+48{\,}{R_\Gamma}^{12}{\,}{\varpi}^{10}{\,}{e_2}
\nonumber \\ &&
-18{\,}{R_\Gamma}^{8}{\,}{\varpi}^{14}{\,}{e_1}
+27{\,}{R_\Gamma}^{6}{\,}{\varpi}^{16}{\,}{e_2}
\nonumber \\ &&
-672{\,}{R_\Gamma}^{12}{\,}{\varpi}^{10}{\,}{e_1}
+18{\,}{R_\Gamma}^{8}{\,}{\varpi}^{14}{\,}{e_2}
\nonumber \\ &&
-144{\,}{R_\Gamma}^{10}{\,}{\varpi}^{12}{\,}{e_2}\Big)
/{\varpi}^{18}/{R_\Gamma}^{12}.
\end{eqnarray}

\chapter{Bondi Functions For The Solution SIMPLE}
\label{app:simple}

Here  the formulae obtained for the static solution
SIMPLE, in the gauge described in Section \ref{page:extract:SIMPLE}
are listed.
Since the solution is time-independent, the functions $r, J, \beta, U, W$
and their $\lambda$-derivatives on the world-tube depend only on 
the angular variables $(q,p)$.
The function $r^{(0)}(q,p)$  is given by
\begin{eqnarray}
r^{(0)} &=&
{R_\Gamma}
-{\frac {1}{80}}\,R_\Gamma^5\,\gimel_1^2{a}^{4}
+{\frac {1}{112}}\,R_\Gamma^6\,\gimel_1^3{a}^{6} 
\nonumber \\ &&
-{\frac {11}{1920}}\,R_\Gamma^8\,\gimel_1^4{a}^{8}
+{\frac {179}{49280}}\,R_\Gamma^{10}\,\gimel_1^5{a}^{10} 
+ \cdots,
\end{eqnarray}
where
\begin{equation}
\gimel_1 = {\frac{\left (-4\,{p}^{2}+P^{2}\right )}{P^{2}}}, \;\;
P = 1+q^2+p^2.
\end{equation}
Using the same notations and further introducing
\begin{eqnarray}
\gimel_2 &=& 
\frac{\zeta^2+ 1 }{1 + \zeta \bar \zeta} =
\frac{(q + I p)^2 + 1}{P},
\\
\gimel_3 &=& 
\frac{\zeta - \bar \zeta}{1 + \zeta \bar \zeta} =
\frac{2 I p}{P},
\end{eqnarray} 
the Bondi functions and their 
$\lambda$-derivatives are given by
\begin{eqnarray}
J^{(0)} &=& 
-\frac {1}{2}\,\gimel_2^{2}R_\Gamma^{2}{a}^{2}
+\frac{3}{16}\,\bar \gimel_2\,\gimel_2^{3}R_\Gamma^{4}{a}^{4}
-{\frac {9}{80}}\,\bar \gimel_2^{2}\gimel_2^{4}R_\Gamma^{6}{a}^{6}
\nonumber \\ &&
+{\frac {1317}{17920}}\,\bar \gimel_2^{3}\gimel_2^{5}R_\Gamma^{8}{a}^{8}
-{\frac {26647}{537600}}\,\bar \gimel_2^{4}\gimel_2^{6}R_\Gamma^{10}{a}^{10}
+\cdots,
\\
J^{(0)}_{,\lambda} &=& 
-\gimel_2^{2}R_\Gamma{a}^{2}
+\frac{3}{4}\,\bar \gimel_2\,\gimel_2^{3}R_\Gamma^{3}{a}^{4}
-{\frac {27}{40}}\,\bar \gimel_2^{2}\gimel_2^{4}R_\Gamma^{5}{a}^{6}
\nonumber \\ &&
+{\frac {1317}{2240}}\,\bar \gimel_2^{3}\gimel_2^{5}R_\Gamma^{7}{a}^{8}
-{\frac {26647}{5376z0}}\,\bar \gimel_2^{4}\gimel_2^{6}R_\Gamma^{9}{a}^{10}
+\cdots,
\\
\beta^{(0)} &=& 
\frac{1}{32} R_\Gamma^4 \gimel_1^2 a^4
- \frac{1}{32} R_\Gamma^6 \gimel_1^3 a^6
+ \frac{137}{5120} R_\Gamma^8 \gimel_1^4 a^8
\nonumber \\ &&
- \frac{393}{17920} R_\Gamma^{10} \gimel_1^5 a^{10}
\cdots,
\\
\beta_{,\lambda}^{(0)} &=& 
\frac{1}{8} R_\Gamma^3 \gimel_1^2 a^4
-\frac{3}{16} R_\Gamma^5 \gimel_1^3 a^6
+\frac{137}{640} R_\Gamma^7 \gimel_1^4 a^8
\nonumber \\ &&
-\frac{393}{1792} R_\Gamma^9 \gimel_1^5 a^{10}
+ \cdots,
\\
\pagebreak[3]
U^{(0)} &=& 
\gimel_3\,\gimel_2\,R_\Gamma\,{a}^{2}
-\frac{1}{2}\,\gimel_3\,\bar \gimel_2\,\gimel_2^{2}R_\Gamma^{3}{a}^{4}
+{\frac {29}{80}}\,\gimel_3\,\bar \gimel_2^{2}\gimel_2^{3}R_\Gamma^{5}{a}^{6}
\nonumber \\ &&
-{\frac {319}{1120}}\,\gimel_3\,\bar \gimel_2^{3}\gimel_2^{4}R_\Gamma^{7}{a}^{8}
+{\frac {3103}{13440}}\,\gimel_3\,\bar \gimel_2^{4}\gimel_2^{5}R_\Gamma^{9}{a}^{10}
+ \cdots,
\\
U^{(0)}_{,\lambda} &=& 
\gimel_3\,\gimel_2\,{a}^{2}
-\frac{3}{2}\,\gimel_3\,\bar \gimel_2\,\gimel_2^{2}R_\Gamma^{2}{a}^{4}
+{\frac {29}{16}}\,\gimel_3\,\bar \gimel_2^{2}\gimel_2^{3}R_\Gamma^{4}{a}^{6}
\nonumber \\ &&
-{\frac {319}{160}}\,\gimel_3\,\bar \gimel_2^{3}\gimel_2^{4}R_\Gamma^{6}{a}^{8}
+{\frac {9309}{4480}}\,\gimel_3\,\bar \gimel_2^{4}\gimel_2^{5}R_\Gamma^{8}{a}^{10}
+ \cdots,
\\
W^{(0)} &=& 
\frac {1}{2}\left (-2+3\,\gimel_1\right )R_\Gamma\,{a}^{2}
-\frac{1}{8}\left (-4+5\,\gimel_1\right )\gimel_1\,R_\Gamma^{3}\,{a}^{4}
\nonumber \\ &&
+{\frac {1}{160}}\left (-58+67\,\gimel_1\right )\gimel_1^{2}\,R_\Gamma^{5}\,{a}^{6}
\nonumber \\ &&
-{\frac {1}{2240}}\left (-638+705\,\gimel_1\right )\gimel_1^{3}\,R_\Gamma^{7}\,{a}^{8}
\nonumber \\ &&
+{\frac {1}{26880}}\left (-6206+6669\,\gimel_1\right )\gimel_1^{4}\,R_\Gamma^{9}\,{a}^{10}
+\cdots, \\
W^{(0)}_{,\lambda} &=& 
\frac {1}{2}\left (-2+3\,\gimel_1\right ){a}^{2}
-\frac{3}{8}\left (-4+5\,\gimel_1\right )\gimel_1\,R_\Gamma^{2}\,{a}^{4}
\nonumber \\ &&
+{\frac {1}{32}}\left (-58+67\,\gimel_1\right )\gimel_1^{2}\,R_\Gamma^{4}\,{a}^{6}
\nonumber \\ &&
-{\frac {1}{320}}\left (-638+705\,\gimel_1\right )\gimel_1^{3}\,R_\Gamma^{6}\,{a}^{8}
\nonumber \\ &&
+{\frac {3}{8960}}\left (-6206+6669\,\gimel_1\right )\gimel_1^{4}\,R_\Gamma^{8}\,{a}^{10}
+\cdots.
\end{eqnarray}

\chapter{Linearized Quadrupole Wave Solution In Bondi Coordinates}
\label{app:teuk-null}

This appendix is a description of the algebraic steps performed
to obtain the linearized quadrupole wave solution in Bondi coordinates.
At the end explicit expressions are provided for the 
Bondi metric functions.
The work described below (and most of the work from the entire thesis)
could not have been accomplished without 
a significant input from
\cite{WinicourPersonal}.

The starting point is the linearized quadrupole solution
in coordinates $ x^\alpha = (t,r,x^A)$,
as given in Section~\ref{sec:extract.teuk.cauchy}.
The contravariant 
metric is written in the form
$g^{\alpha \beta} = \eta^{\alpha \beta} - h^{\alpha \beta}$, where 
computations are done up to order $O(h^2)$.

\section{Approximate null coordinates}
\label{sec:teuk-null.app-null}

The approximate null coordinates  $y'^{\mu'} = (u,\lambda, y^A)$ are defined
 by $u=t-r, \lambda = r - R_\Gamma$. This leads to a metric of the form

\begin{equation}
g'^{\mu'\nu'} = 
\frac{\partial y'^{\mu'}}{\partial x^\alpha}
\,\frac{\partial y'^{\nu'}}{\partial x^\beta} \, g^{\alpha \beta} = 
\eta'^{\mu'\nu'} - h'^{\mu'\nu'},
\end {equation}
where $h'^{\mu'\nu'}$ is
\begin{equation}
h'^{\mu'\nu'} = 
\frac{\partial y'^{\mu'}}{\partial x^\alpha}
\, \frac{\partial y'^{\nu'}}{\partial x^\beta} \, h^{\alpha \beta}
\end{equation}
and $\eta'^{\mu'\nu'}$ is the background null metric:
\begin{equation}
\eta'^{\mu'\nu'} =
\left( 
\begin{array}{cc}
\begin{array}{rr}
 0 & -1 \\
-1 &  1 \\
\end{array} & 0 \\
0 & \frac{q^{AB}}{(\lambda + R_\Gamma)^2}
\end{array}
\right).\label{eq:teuk-null.etadef}
\end{equation}

\section{Affine null coordinates}
\label{sec:teuk-null.aff-null}

As a further step the affine null coordinates are introduced in the following way:

\begin{equation}
\tilde{y}^\mu
= y'^\mu + \xi^\mu, \quad \mathrm{where} \quad \left. \xi^\mu \right|_{R_\Gamma} = 0.
\end{equation}
This leads to a metric that can be written as

\begin{equation}
\tilde{g}^{\tilde{\mu} \tilde{\nu}}(\tilde{y}) = 
\, \frac{\partial \tilde{y}^{\tilde{\mu}}}{\partial y'^{\alpha'}} 
\, \frac{\partial \tilde{y}^{\tilde{\nu}}}{\partial y'^{\beta'}}
\, g'^{\alpha' \beta'}(y') = 
\tilde{\eta}^{\tilde{\mu} \tilde{\nu}}(\tilde{y}) - 
\tilde{h}^{\tilde{\mu} \tilde{\nu}}.
\end{equation}
Here again $\tilde{\eta}^{\mu \nu}$ is the background metric in null coordinates, 
while 
\begin{equation}
\tilde{h}^{\tilde{\mu} \tilde{\nu}} = h'^{\mu' \nu'} - 
\xi^{\mu}_{, \alpha'} \eta^{\nu' \alpha'} - 
\xi^{\nu}_{, \alpha'} \eta^{\mu' \alpha'} +
\eta^{\mu' \nu'}_{, \alpha'} \xi^{\alpha}.
\end{equation}
The vector $\xi^{\mu}$ is fixed by the requirement that
 $\tilde{g}^{\tilde{\mu} \tilde{\nu}}$
be a null metric, that is

\begin{eqnarray}
\tilde{h}^{\tilde{u}\tilde{u}} = 0 & \Rightarrow & 
0 = h'^{u' u'} + 2 \; \xi^{u}_{,\lambda} \nonumber \\
& \Rightarrow & \xi^u 
\label{eq:teuk-null.int1}
= \int_0^\lambda \left[ - \frac{1}{2} h'^{u' u'} \right] d\lambda, \\
\tilde{h}^{\tilde{u}\tilde{A}} = 0 & \Rightarrow & 
0 = h'^{u' A'} - \xi^{u}_{,B} \; \eta^{A' B'} + \xi^{A}_{,\lambda}\nonumber \\
& \Rightarrow & \xi^A 
\label{eq:teuk-null.int2}
= \int_0^\lambda \left[ -  h'^{u' A'} + \xi^{u}_{,B'} \; \eta^{A' B'} \right] d\lambda, \\
\tilde{h}^{\tilde{u}\tilde{\lambda}} = 0 & \Rightarrow & 
0 = h'^{u' \lambda'} +  \xi^{u}_{,u} - \xi^{u}_{,\lambda} + \xi^{\lambda}_{,\lambda}\nonumber \\
& \Rightarrow & \xi^\lambda 
\label{eq:teuk-null.int3}
= \int_0^\lambda \left[ - h'^{u' \lambda'} - \xi^u_{,u} - \frac{1}{2}h^{u' u'}
\right] d\lambda. 
\end{eqnarray}

\section{Bondi coordinates}
\label{sec:teuk-null.bondi}

In order to get the null metric in Bondi coordinates $y^\mu = (u, r_b, y^A)$
the following coordinate transformation is defined
\begin{equation}
y^\mu = \tilde{y}^{\tilde{\mu}} + \eta^{\mu}, \quad \mathrm{where} \quad \eta^u = \eta^A = 0.
\end{equation}
The above coordinate transformation translates into 
\begin{equation}
r_b = \tilde{\lambda} + {R_\Gamma} + \eta^r.
\end{equation}
Carrying it out provides the following metric:
\begin{equation}
g_b^{\mu \nu}(y) = 
\frac{\partial y^\mu}{\partial \tilde{y}^{\tilde{\alpha}}} 
\frac{\partial y^\nu}{\partial \tilde{y}^{\tilde{\beta}}} 
\tilde{g}^{\tilde{\alpha}\tilde{\beta}} = 
\eta_b^{\mu \nu}(y) - h_b^{\mu \nu}(y),
\end{equation}
where $\eta_b^{\mu \nu}$ is the background metric in Bondi coordinates 
which is identical with the same metric in null coordinates. 
The expression for $h_b^{\mu \nu}$ is

\begin{equation}
h^{\mu \nu}_b = h^{\tilde{\mu}\tilde{\nu}} 
- \eta^{\mu}_{,\tilde{\alpha}} \tilde{\eta}^{\tilde{\alpha}\tilde{\nu}}
- \eta^{\nu}_{,\tilde{\alpha}} \tilde{\eta}^{\tilde{\alpha}\tilde{\mu}}
+ \eta^{\tilde{\mu}\tilde{\nu}}_{,\tilde{\alpha}} \eta^{\alpha}.
\label{eq:teuk-null.Bondi coordtrans}
\end{equation}
As a condition upon $\eta^r$ the following gauge condition is imposed:
\begin{equation}
\eta_{AB} h_b^{AB} = \eta_{AB} \tilde{h}^{AB}
 + \eta_{AB}\eta^{AB}_{,\lambda} \eta^r = 0.
\label{eq:teuk-null.gauge condition}
\end{equation}
Using 
\begin{displaymath}
\eta_{AB}\eta^{AB}_{,\lambda} = -\frac{4}{\lambda},
\end{displaymath}
 this translates into
\begin{equation}
\eta^r = \frac{\lambda}{4} \eta_{AB} \tilde{h}^{AB}.
\end{equation}
If the determinant condition is imposed to the Bondi metric $g_b^{AB}$, one obtains
\begin{eqnarray}
\frac{1}{r_b^4 q} 
&=& \left|  g_b^{AB}(y) \right| = \left|  \tilde{g}^{\tilde{A}\tilde{B}}(\tilde{y}) \right| 
= \left| \eta^{\tilde{A}\tilde{B}}(\tilde{y}) - \tilde{h}^{\tilde{A}\tilde{B}}(\tilde{y}) \right| \\ 
&=& \left| \eta^{\tilde{A}\tilde{B}}(\tilde{y}) \right| 
\left( 1 - \eta_{\tilde{A}\tilde{B}}(\tilde{y}) 
\; \tilde{h}^{\tilde{A}\tilde{B}}(\tilde{y})\right),
\end{eqnarray}
which is equivalent to satisfying the gauge condition
specified in Eq.~(\ref{eq:teuk-null.gauge condition}).

\section{Inverting the relation $r_b = r_b(\tilde{\lambda})$}
\label{sec:teuk-null.invert}

Once the metric is obtained in Bondi coordinates, 
computing the dependence $\tilde{\lambda}(r_b)$ is straightforward.
The expression for  $\partial r_b/ \partial \tilde{\lambda}$
 can be obtained directly from 
\begin{equation}
\partial r_b/ \partial \tilde{\lambda} = e^{- 2 \beta} = -g_b^{ru}.
\end{equation}
The value of $\partial^2 r_b/ \partial \tilde{\lambda}^2$ is determined 
by the value of $\beta_{,\tilde{\lambda}}$, which vanishes in linearized
theory (see Eq.~(\ref{eq:extract.nbetal})).
Thus 
$\partial^2 r_b/ \partial \tilde{\lambda}^2 = 0$, or 
in other words, the dependence $r_b = r_b(\tilde{\lambda})$ is linear.

\section{Explicit results for the Bondi functions}

Most of the coordinate transformations described in 
Sections~\ref{sec:teuk-null.app-null}-\ref{sec:teuk-null.invert}
are relatively easy to carry out with the exception of the 
integrals 
(\ref{eq:teuk-null.int1}) -  
(\ref{eq:teuk-null.int3}). These cannot be evaluated in a closed form,
and they were
approximated by a high-order expansion in the affine 
parameter $\lambda$. The resulting expressions for the Bondi 
variables $J,U,W$ as functions of $(u,\lambda,q,p)$
are listed below. As described in Section~\ref{sec:teuk-null.invert},
replacing the variable $\lambda$ with the Bondi $r$ is straightforward.

The function $J(u,\lambda,q,p)$ takes the form
\begin{eqnarray}
J &=& \varepsilon \left(  \frac{q^2-p^2+2 I q p }{P^2} \right) 
\times
\Big[ J^{(0)}(u)
+ \frac{\lambda^1}{1 !} J^{(1)}(u)
+  \nonumber \\ && 
+ \cdots + \frac{\lambda^6}{6 !} J^{(6)}(u)
+ O(\lambda^7) \Big]  + O(\varepsilon^2),
\end{eqnarray}
where  the coefficients $J^{(0)},\ldots J^{(6)}$ are given by


The function $U(u,\lambda,q,p)$ expanded in terms of $\lambda$ takes the form
\begin{eqnarray}
U &=& 
\varepsilon  \left( \frac{(P-2)(q+I\,p)}{P^2} \right)
\times \Big[
U^{(0)}(u)
+ \frac{\lambda^1}{1 !} U^{(1)}(u)
+ \nonumber \\ &&
+ \cdots
+ \frac{\lambda^6}{6 !} U^{(6)}(u)
+ O(\lambda^7) 
\Big] + O(\varepsilon^2),
\end{eqnarray}
where
%

Finally, the expansion of $W(u,\lambda,q,p)$ is given by
\begin{eqnarray}
W &=& \varepsilon  \left( \frac{6 - 6 P + P^2}{P^2} \right) \times  \Big[
W^{(0)}(u)
+ \frac{\lambda^1}{1 !} W^{(1)}(u) +
\nonumber \\ &&
+ \cdots
+ \frac{\lambda^6}{6 !} W^{(6)}(u)
+ O(\lambda^7)
\Big] + O(\varepsilon^2),
\end{eqnarray}
with
%

The symbols $e_1$ and $e_2$ are defined in 
Eqs.~(\ref{eq:teuk-ext.e1}) - (\ref{eq:teuk-ext.e2}).
The function $\beta$ is independent of $\lambda$ and is already given 
in Section~\ref{sec:extract.teuk}.

Evaluated at the extraction world-tube, the expressions for the Bondi
metric and its $\lambda$-derivative were 
identical to what was obtained
in Section~\ref{sec:extract.teuk}, using the 
algebraic procedure of the extraction module. 
This is quite reassuring since the two
sets of results were obtained in entirely different ways.



\newpage

\bibliography{references}
\bibliographystyle{prsty}

\end{document}